\documentclass{article}

\usepackage[numbers,sort&compress]{natbib}
\usepackage[utf8]{inputenc} % allow utf-8 input
\usepackage[T1]{fontenc}    % use 8-bit T1 fonts

\usepackage{amsmath}

\usepackage{arxiv}
\usepackage{natbib}
\usepackage{hyperref}       % hyperlinks
\usepackage{url}            % simple URL typesetting
\usepackage{booktabs}       % professional-quality tables
\usepackage{amsfonts}       % blackboard math symbols
\usepackage{nicefrac}       % compact symbols for 1/2, etc.
\usepackage{microtype}      % microtypography
\usepackage{lipsum}
\usepackage{graphicx}
\usepackage{pdflscape} % for landscape tables
\usepackage{array}     % for p{width} columns
\usepackage{booktabs}  % optional for prettier tables
\graphicspath{ {./images/} }
\usepackage{xcolor}
\usepackage{rotating}
\usepackage{comment}
\usepackage{pifont}
\usepackage{tcolorbox}
\definecolor{blue}{HTML}{486aa1} 
\definecolor{red}{HTML}{de5454} 
\definecolor{lightblue}{HTML}{e4eaf5} 
\definecolor{lightred}{HTML}{f5e4e4} 

\DeclareUnicodeCharacter{202F}{\,}

\usepackage{tabularx}
\usepackage{makecell}

\renewcommand{\today}{} % Remove today's date
\usepackage{fancyhdr}
\title{\textbf{Bridging Language Gaps: Advances in Cross-Lingual Information Retrieval with Multilingual LLMs}}

\author{
 Roksana Goworek\thanks{Equal contribution.}\\
  The Alan Turing Institute\\
  Queen Mary University of London\\
   \And
  Olivia Macmillan-Scott\footnotemark[1]  \\
  The Alan Turing Institute \\
  University College London \\
  \And
  Eda B. Özyiğit  \\
  The Alan Turing Institute\\
}

\begin{document}
\maketitle
\vspace{-3em}
\begin{center}

\texttt{\{rgowerek, omacmillan-scott, eozyigit\}@turing.ac.uk}
\end{center}
\vspace*{2em}

\begin{abstract}
Cross-lingual information retrieval (CLIR) addresses the challenge of retrieving relevant documents written in languages different from that of the original query. Research in this area has typically framed the task as monolingual retrieval augmented by translation, treating retrieval methods and cross-lingual capabilities in isolation. Both monolingual and cross-lingual retrieval usually follow a pipeline of query expansion, ranking, re-ranking and, increasingly, question answering. Recent advances, however, have shifted from translation-based methods toward embedding-based approaches and leverage multilingual large language models (LLMs), for which aligning representations across languages remains a central challenge. The emergence of cross-lingual embeddings and multilingual LLMs has introduced a new paradigm, offering improved retrieval performance and enabling answer generation. This survey provides a comprehensive overview of developments from early translation-based methods to state-of-the-art embedding-driven and generative techniques. It presents a structured account of core CLIR components, evaluation practices, and available resources. Persistent challenges such as data imbalance and linguistic variation are identified, while promising directions are suggested for advancing equitable and effective cross-lingual information retrieval. By situating CLIR within the broader landscape of information retrieval and multilingual language processing, this work not only reviews current capabilities but also outlines future directions for building retrieval systems that are robust, inclusive, and adaptable.
\end{abstract}

% keywords can be removed

\setcounter{tocdepth}{2}
% \tableofcontents

\section{Introduction}

% \paragraph{Definition and Scope of CLIR}

Given a query and a set of documents, information retrieval (IR) \cite{lancaster_1979, sparck_1997, baeza-yates_1999} is the task of identifying documents that are relevant to the query. Cross-lingual information retrieval \cite{pevzner1969automatic, salton1972experiments, schauble1998cross,peters2000first} extends this task by enabling queries expressed in one language to retrieve documents written in one or more different languages. Unlike traditional monolingual IR, which assumes a shared language between query and documents, CLIR faces the challenge of bridging language boundaries. This is typically addressed by combining techniques from both information retrieval and multilingual natural language processing (NLP). 

% In practice, CLIR often involves submitting a query in one language to retrieve semantically relevant documents in another, or searching across a multilingual corpus.

% \paragraph{Motivation: Linguistic Inequality and Access}

The emergence of the Internet and search engines in the 1990s revealed striking disparities in linguistic accessibility. At that time, English accounted for almost 80\% of all web content, although it was the native language of only a small share of users \cite{language2015inequality, trotman2000digitaldivide}. This imbalance underscored the need for research into cross-lingual information access. For example, high-resource languages such as English, Spanish, and Chinese developed strong web presence and benefited from early NLP support, while many others, particularly low-resource languages like Swahili or Burmese, lagged behind due to limited digital content and inadequate computational tools. These disparities persist today, with most modern web content and NLP systems still skewed towards a few dominant languages \cite{joshi-etal-2020-state}. The rise of LLMs has only amplified this imbalance, as most high-performance models remain disproportionately trained and optimised for English; English accounts for around 90\% of the training data in most popular models \cite{li2024quantifying}. CLIR offers a compelling response by enabling users to access information written in other languages, thereby helping to democratise knowledge across linguistic boundaries.

Traditional monolingual information retrieval systems are generally organised into a multi-stage pipeline: (i) query expansion, which broadens the query using synonyms, spelling corrections, or related terms to improve recall; (ii) ranking, which performs an efficient first-pass retrieval to select a candidate set of relevant documents; (iii) re-ranking, which applies more computationally intensive models to refine the order of the top documents; and optionally (iv) generation, where an answer or summary is synthesised, often by large language models. %Although ranking is arguably the only essential component of the retrieval pipeline, query expansion, re-ranking and generation are widely used and as such will be covered in this survey.
Cross-lingual retrieval follows the same pipeline while introducing additional complexity,
for example by translating the query and/or the documents,
aligning multilingual embeddings for semantic similarity, or leveraging generative multilingual models to bypass translation altogether.

%The additional complexity of cross-linguality introduces new components to this pipeline, such as translating the query and/or the documents, aligning multilingual embeddings for semantic similarity, or leveraging generative multilingual models to bypass translation altogether.

Recent advances in neural language modelling, particularly in cross-lingual embedding and multilingual pre-training, have enabled more powerful and flexible architectures for CLIR. These systems increasingly use multilingual sentence encoders \cite{vaj2023laser}, dense retrieval \cite{shi-etal-2021-cross, huang-etal-2024-unsupervised, welleretal2025}, and large-scale generative language models \cite{adeyemi-etal-2024-zero} to compare aligned representations across languages. With improved multilingual corpora and evaluation datasets, systems are becoming more effective, scalable, and easier to benchmark. Yet building reliable cross-lingual retrieval remains difficult. Considerable research still treats CLIR as monolingual retrieval plus translation, which oversimplifies the problem and overlooks multilingual challenges. Lexical, syntactic, and semantic differences hinder alignment, and many language pairs lack parallel corpora, especially low-resource languages. Domain mismatch further reduces generalisability. Constructing high-quality, gold-standard judgments is resource-intensive, while annotation transfer introduces noise and inconsistency. Translation and alignment may cause semantic drift, altering meaning and reducing accuracy. To overcome these issues, next-generation systems must go beyond translation-based pipelines and address linguistic, resource, and evaluation mismatches directly. Although there are surveys of information retrieval \cite{hambarde2023information, zhu_2024_ir, li_2025_GenIR} and multilingual NLP \cite{ruder_2019, zhu_2024, huang_2025, xu_2025}, few focus on CLIR, and none fully examine embedding-based retrieval. Early work, such as that by Nie \cite{nie_2010}, focused on translation-based methods, but more recent developments \cite{galuščáková2022} highlight embedding-based retrieval and generative models, which remain in the early stages of adoption.

\begin{figure}
    \centering
    \includegraphics[width=1\linewidth]{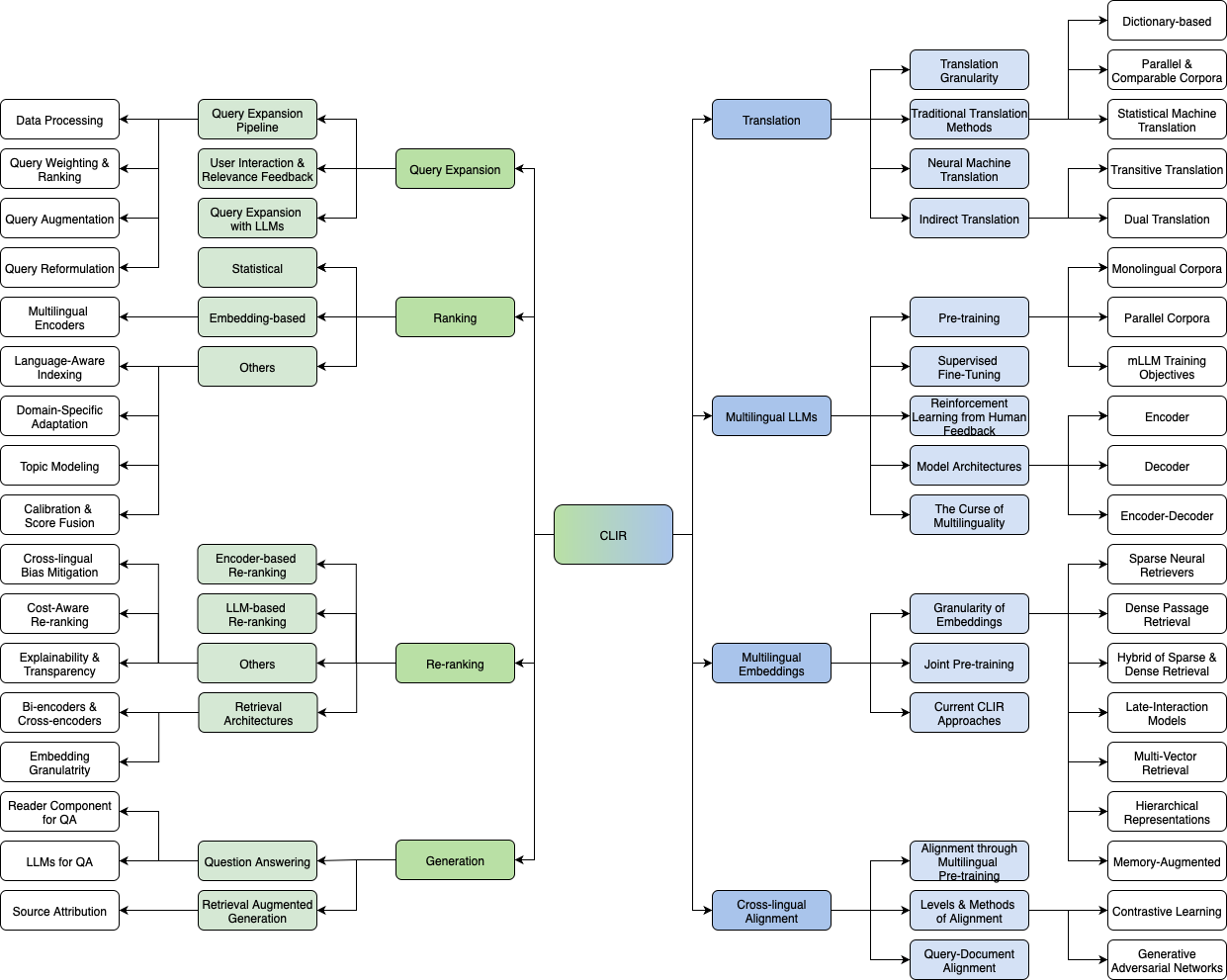}
    \caption{Overview of the survey covering cross-lingual and information retrieval approaches.}
    % We focus on cross-lingual approaches and information retrieval.}
    \label{fig:overview}
\end{figure}

%Cross-lingual retrieval is a distinct research problem that requires attention beyond monolingual retrieval or multilingual NLP. 
This survey provides a comprehensive overview of recent advances across the full pipeline (from query reformulation to re-ranking and answer generation), with a focus on multilingual embedding alignment, contrastive learning, multilingual pre-training strategies, and the integration of generative language models for retrieval and response generation (see Figure \ref{fig:overview}). It presents a unified perspective of current methods, available resources, and the open challenges in cross-lingual system design. Its key contributions are:

\begin{itemize} 
    
    \item \textbf{CLIR techniques.} An analysis of cutting-edge CLIR methods, including embedding alignment, multilingual pre-training, multilingual LLMs, and retrieval architectures, highlighting their strengths and trade-offs.

    \item  \textbf{Multilingual advances.} A review of recent developments in multilingual NLP and their relevance to CLIR showing how progress in cross-lingual models supports multilingual information retrieval. 
        
    \item \textbf{Datasets and evaluation.} A survey of datasets, evaluation protocols, and performance metrics, along with a discussion of current limitations and opportunities for improvement.
    
    \item \textbf{Core challenges.} An analysis of major challenges in CLIR, including linguistic divergence, data scarcity, domain and language generalisation, and fairness, together with their implications for real-world deployment. This identifies obstacles to practical adoption and highlights areas needing further research.
    
\end{itemize}

This work is organised around two central dimensions: (i) how systems implement core components of the retrieval process, and (ii) how they address cross-linguality. Section~\ref{sec:clir architecture} introduces system architectures and integration of cross-lingual representations into scalable retrieval pipelines. Section~\ref{sec:cross-lingual} examines strategies for cross-linguality, including translation-based methods, multilingual LLMs, embeddings, and alignment techniques. Section~\ref{sec:evaluation} reviews evaluation practices, focusing on benchmark datasets, performance metrics, and the need for fair multilingual assessment. Section~\ref{sec:applications} explores real-world applications such as multilingual search, cross-lingual question answering (QA), and domain-specific information access. Section~\ref{sec:challenges} discusses key challenges such as linguistic divergence, resource scarcity, evaluation difficulties, model limitations, and highlights future research directions. Finally, section~\ref{sec:conclusion} concludes with a synthesis of insights and a discussion of the broader impact of cross-lingual retrieval.

\section{CLIR Architecture}
\label{sec:clir architecture}

% Briefly explain the basics of information retrieval.
% And of the overall architecture, of translation terms, etc.

% \textcolor{red}{Should we have a summary table of techniques in each subsection (with e.g. advantages and limitations)? - Eda mentioned it would be better for advantages and limitations to be in the text rather than table }

\begin{figure}[b]
    \centering
    \includegraphics[width=0.9\linewidth]{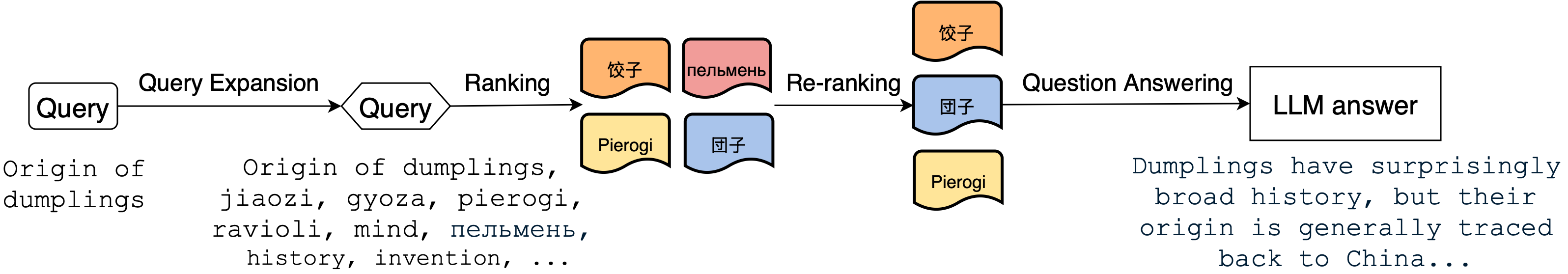}
    \caption{CLIR architecture example. The query "origin of dumplings" passes through expansion, translation and/or embedding before relevant documents are retrieved and re-ranked.}
    \label{fig:clir}
\end{figure}

While the primary challenge in CLIR lies in bridging language gaps, it is equally important to consider how the core architecture of IR systems can be utilised and extended. This section focuses on the foundational stages of the retrieval pipeline: query expansion, ranking, re-ranking, and, optionally, question answering. It highlights recent advances in monolingual IR that have the potential to strengthen CLIR performance rather than addressing how systems handle cross-linguality itself. Figure \ref{fig:clir} illustrates how the query "origin of dumplings" may be processed through the CLIR architecture: beginning with query expansion, moving to initial ranking and re-ranking, followed by question answering, and culminating in a final response generated by a LLM.
%(see section~\ref{sec:cross-lingual}), we examine how innovations in these core components contribute to retrieval quality, and how they might be adapted to multilingual contexts.

\subsection{Query Expansion} 
\label{sec:query-expansion}
% see also Entity Query Feature Expansion (EQFE, 2014), Contextualized Embeddings for Query Expansion (CEQE), LameR

The effectiveness of information retrieval relies on the assumption that the user's initial query accurately reflects their information requirements. However, this assumption is often too strong. In web search, several studies have found that the average length of queries is less than 2.5 words \cite{jansen_2000, spink_2001}. Such short queries provide insufficient context and leave substantial room for ambiguity. Users may also misspell terms, resulting in lower retrieval performance. In the case of cross-lingual retrieval, additional difficulties arise, as there may be multiple possible translations or transliterations of a given word. 

Another issue is the \textit{vocabulary problem} \cite{furnas_1987}: the terms used in a query may not match those used in the documents themselves or terms used to index the documents. As a result, lexical-based retrieval approaches can fail to identify relevant documents. Information retrieval faces the added difficulty of synonyms and word inflections, which can lower recall \cite{carpineto_2012}. Similarly, the concept of polysemy, where a single word may contain multiple meanings, can further reduce retrieval performance. 

Query expansion provides an effective method for addressing these issues, particularly in cases of ambiguity or lower-quality queries. The user's initial query is expanded using relevant terms or synonyms, in so doing improving the retrieval performance and more closely aligning with the user's information needs. For example, Figure \ref{fig:clir} demonstrates how the query "origin of dumplings" may be expanded with terms such as \textit{gyoza} or \textit{history}. Query expansion is especially relevant for what \citet{broder_2002} denotes as informational queries, which are broad in scope and often correspond to many related documents. In contrast, navigational queries typically seek a specific result, and transactional queries involve the intent to perform a particular activity. 

Although query expansion is the more common strategy for overcoming the limitations of short or underspecified queries, recent research has also explored expanding the documents themselves. One example is Nogueira et al.'s \cite{nogueira_2019_doc2query} Doc2query approach, in which a set of possible queries is predicted for each document and appended as pseudo-queries. The expanded document is then indexed and used in retrieval with traditional methods such as BM25 \cite{robertson2009probabilistic}. In practice, some document expansion techniques have proven more effective than query expansion, particularly in sparse retrieval settings \cite{wang_2023_query2doc}.

\subsubsection{Query Expansion Pipeline}

The query expansion pipeline can be divided into four components: preprocessing of the data source, term  weighting and ranking, expansion term selection, and query reformulation~\cite{carpineto_2012, azad_2019}. The main challenges are determining which terms to use, how to weight them, and how to integrate them into the query \cite{nie_2010}. 
%Potential terms are first extracted from the chosen data source, which these are the terms that may then be used to expand the query. As we will see, the use of LLMs for query expansion allows for query reformulation into other formats that do not necessarily follow this pipeline, including the generation of pseudo-documents. 

\paragraph{Data Processing.} In the first stage, potential expansion terms are extracted from the chosen data source. These sources may include thesauri and ontologies such as WordNet \cite{miller_1990_wordnet}, Wikipedia datasets, search and query logs, and word embeddings. Relevance feedback \cite{rocchio_1965, rocchio_1971} is another widely used strategy: feedback may be global, drawing on the entire collection, or local, relying on the set of documents initially retrieved. Xu and Croft \cite{xu_croft_2017} note that local feedback methods are often more effective, since they exploit context-specific evidence rather than general collection statistics. LLMs have also been used to generate pseudo-documents, thereby enriching the document side of retrieval.

%so relying on the documents used in the retrieval process; user search or query logs; and more recently word embeddings to find semantically similar words. Most of these data sources can be processed and terms extracted before the retrieval is carried out, whereas approaches that rely on relevance feedback require an initial retrieval step. This distinction is referred to by Xu and Croft \cite{xu_croft_2017} as techniques that employ \textit{global} as opposed to \textit{local} data, finding local feedback to generally be more effective. 

\paragraph{Query Weighting and Ranking.} Once the relevant data sources have been processed to collect potential expansion terms, these are weighted and ranked to determine the most appropriate ones. Relevant terms are selected, either through lexical approaches to identify potential synonyms, hypernyms, or words identified through statistical or semantic similarity. Approaches differ in how they model the relationship between original query terms and candidate terms \cite{carpineto_2012}. One-to-one associations link each query term directly to an expansion term, often using stemming \cite{porter_1980} or thesauri. By contrast, one-to-many associations identify terms related to multiple query terms simultaneously. For example, through co-occurrence analysis \cite{hsu_2006, hsu_2008}. Other methods apply statistical or model-based weighting techniques, in which probabilistic or language models estimate the strength of association between query and candidate terms \cite{lavrenko_croft_2001, zhai_lafferty_2001}.

%this approach does not handle well relationships between terms in the query, and meanings conveyed through phrases. A one-to-many association approach instead selects words that are related to more than one query terms - for instance, Hsu et al. \cite{hsu_2006, hsu_2008} set a requirement of an association to a minimum of two query terms. Rather than relying on the initial query, other methods extract the top weighted terms from the initially retrieved documents. Finally, another alternative is to use a model-based approach by constructing a statistical model from the query and selecting the expansion terms with the highest probability \cite{lavrenko_croft_2001, zhai_lafferty_2001}.
%\textcolor{red}{--> talk about models for weighting expansion terms?}

\paragraph{Query Augmentation.} Having created a ranked or weighted set of possible expansion terms, the next step is to determine which of those will be used to augment the query. Some have argued that having a smaller number of expansion terms is beneficial as it reduces the noise that can be introduced into the query \cite{salton_buckley_1990}, whereas others have claimed that the quality of the selected terms is more important than how many of them are used \cite{sihvonen_vakkari_2004}. Instead of trying to determine how many terms is the optimum number, others have achieved higher performance by employing more informed selection techniques \cite{cao_nie_2008, carpineto_2002}. Nevertheless, the number of expansion terms proposed has varied widely in the literature, from 5-10 terms \cite{amati_2002, CHANG_2006}, to a few hundred terms \cite{bernardini_2008, WONG_2008, Buckley_1994}, to a third of all candidate terms \cite{robertson_willett_1993}. 

\paragraph{Query Reformulation.} In the final stage, the selected expansion terms are incorporated into the original query to produce an augmented version \cite{salton_buckley_1990}. The objective is to capture the user's information need more precisely and to mitigate the vocabulary problem. 

%of the query expansion process is the reformulation of the initial query into the expanded version. The weights assigned in the previous step are used to generate the augmented query \cite{salton_buckley_1990}. The goal is for this reformulated query to better convey the user's original intent, thus addressing the vocabulary problem. When indexing is used for the documents, query expansion generally involves appending additional terms that may be relevant with varying weights. 

\subsubsection{User Interaction \& Relevance Feedback}

User involvement in query expansion varies by approach. Automatic query expansion \cite{carpineto_2012} requires no input, while manual and interactive methods involve users to prevent concept drift \cite{efthimiadis_1996}. In manual expansion, users directly select terms, often with query log support. Interactive query expansion offers system-suggested terms from which users choose.

Relevance feedback provides another interaction mechanism. Early work by Rocchio \cite{rocchio_1965, rocchio_1971}, based on the SMART retrieval system \cite{salton_1965_smart, salton1965evaluation}, uses user-identified relevant documents to expand queries through local feedback expansion, which proved more effective than global, resource-based methods.

Pseudo-relevance feedback (PRF), introduced by Croft and Harper~\cite{croft_harper_1979}, employs a similar approach but removes user input by assuming the top-ranked documents are relevant. Though generally less robust than explicit feedback, PRF provides advantages as it is fully automated and so is widely adopted. RM3~\cite{abdul-jaleel_2004_rm3}, 
an influential PRF algorithm, builds on Lavrenko and Croft's relevance model \cite{lavrenko_croft_2001} by reintroducing original query terms back into the expansion.

\subsubsection{Query Expansion with LLMs}

The development of neural network-based language models has opened up new possibilities for query expansion \cite{li_2025_qe}. Zhu et al. \cite{zhu_2024_ir} classify LLM-based approaches into three main categories: prompting, supervised fine-tuning and reinforcement learning. Li et al. \cite{li_2025_qe} further consider alignment techniques such as preference optimisation and distillation.  Prompting methods, where models directly generate expanded queries or related documents, include zero-shot, few-shot, and Chain-of-Thought (CoT) techniques. Fine-tuning requires large task-specific datasets, and has so far seen limited use in query expansion \cite{peng_2024}. Reinforcement learning allows models to refine expansions using feedback from retrieval systems. LLMs typically generate reformulated queries (e.g. rephrasings or keyword expansions \cite{li_2024_qe}), concept-based queries, or answer-enriched queries that bridge the semantic gap between short queries and long documents \cite{zhu_2024_ir}.

Most research on LLM-based query expansion has focused on prompting. For instance, Clauveau \cite{claveau_2022} applies zero-shot prompts to generate multiple expansions, concatenated with the original query for retrieval via BM25+ \cite{lv_zhai_2011}. Similarly, Gao et al. \cite{gao_2023_hyde} generate pseudo-documents using zero-shot prompting, later embedded for dense retrieval. Query2doc also produces pseudo-documents, which can then be concatenated with the query for both sparse and dense retrieval \cite{wang_2023_query2doc}. Hypothetical Document Embeddings (HyDE) \cite{gao_2023_hyde} adopts a comparable approach focused on dense retrieval. Few shot-prompting, also known as in-context learning,
has also been used in the Query2doc method to generate pseudo-documents which are then concatenated to the original
query, used for both sparse and dense retrieval \cite{wang_2023_query2doc}. A comparative study found CoT prompting to be the most effective \cite{jagerman_2023}.

Despite these advances, a key challenge lies in mitigating hallucinations (i.e. irrelevant or erroneous expansions that degrade retrieval performance). Abe et al. \cite{abe_2025llmQE} attribute these issues to knowledge gaps and ambiguous prompts, while other work suggests consistency verification to filter meaningless outputs \cite{zhang_2024_QE}. LLMs offer another potential approach to address the vocabulary problem and ambiguity in queries through conversation search, where user interaction can be used to clarify their intent. This helps mitigate concept drift, especially when models generate long-term pseudo-documents.

Research on multilingual query expansion remains limited. Recent work often combines translation and expansion, performed sequentially or jointly \cite{saleh_2019, gaillard_2010}. Implicit expansion, where vector representations are adjusted instead of adding related terms \cite{li_2025_qe}, has shown promise. Query expansion in CLIR can both improve alignment with user intent and reduce translation-induced errors \cite{ballesteros_croft_1997}. Sequential pre-expansion translation has been shown to be effective, though hybrid pre- and post-expansion methods often yield the best performance \cite{ballesteros_croft_1998, mcnamee_mayfield_2002}.

\subsection{Ranking} 
\label{sec:ranking}

The ranking stage retrieves an initial list of documents from the collection and scores them based on relevance to the query. This stage prioritises efficiency and recall, providing a coarse-grained ordering that can later be refined during re-ranking. Ranking methods fall into three categories: (i) traditional statistical approaches; (ii) embedding-based neural retrieval; and (iii) hybrid or enhanced strategies that incorporate language awareness and fine-tuning.

\subsubsection{Traditional Statistical Ranking}
Statistical methods such as Term Frequency-Inverse Document Frequency (TF-IDF) \cite{sparck1972statistical} and BM25~\cite{robertson2009probabilistic} form the historical foundation of CLIR. These approaches rely on exact term matching, making them fast, interpretable, and computationally efficient. BM25, in particular, has become a de facto baseline due to its robustness across domains. However, these models require the query and document to share a language, limiting their direct use in CLIR without translation. Extensions with translation enable their application in cross-lingual contexts, where they remain competitive in low-resource or high-latency settings.

\subsubsection{Neural Embedding-based Ranking}
With the rise of multilingual pre-trained language models, embedding-based retrieval has become dominant in CLIR. These models map queries and documents into a shared vector space, enabling retrieval via similarity metrics such as cosine or dot product rather than lexical overlap.

A common architecture is the bi-encoder, also known as a dual-encoder, in which queries and documents are independently encoded into fixed-size embeddings. The document embeddings can be pre-computed and indexed, making retrieval highly efficient even over large corpora. These and other ranker and re-ranker architectures are described in Section~\ref{subsec:model architecture} as there is significant overlap in the techniques used. Similar models can be used for both ranking and re-ranking, however, ranking requires fast retrieval over extremely large corpora, making only efficient methods applicable.

The performance of multilingual embeddings can sometimes be improved by incorporating language-specific information. This helps filter irrelevant matches and reduce semantic drift in multilingual corpora. One strategy is to augment document or query embeddings with language identifiers or metadata \cite{kuwa2020embedmeta, tashu2024mapping}. By constraining retrieval to the appropriate linguistic space, these methods reduce noise and increase precision. %Another line of work aligns retrieval outputs with the user's preferred or native language \cite{jansen2020predicting, kasambula2022langpref}. Such biasing not only improves relevance but also enhances user satisfaction in practical multilingual systems.

%\paragraph{Domain-Specific Adaptation.} 
General-purpose multilingual models often underperform in specialised domains, where vocabulary and semantic relations differ from open-domain text. Domain-specific adaptation addresses these challenges by tailoring retrieval models to the target field. Methods include training with domain-specific positive and negative pairs to improve semantic alignment~\cite{hu2023language}, continued pre-training on in-domain corpora to expand coverage of specialised vocabulary~\cite{jorgensen_2021_mdapt}, and instruction tuning with domain-relevant tasks to better capture real information needs~\cite{liu2021xlbel}. These strategies enhance retrieval accuracy in areas such as law, medicine, and technology, where precision and contextual understanding are critical.

%\paragraph{Topic Modelling Approaches.}
Before the widespread adoption of neural embeddings, CLIR systems frequently employed topic modelling to capture latent semantics across languages. Latent Dirichlet Allocation (LDA) and multilingual extensions grouped documents into topics, providing a coarse but effective basis for cross-lingual retrieval. Further developments such as Bilingual LDA (BiLDA)~\cite{vulic2013bilda} and Polylingual Topic Models (PLTM)~\cite{mimno2009polylingual} aligned topic spaces across languages using dictionaries or parallel corpora, while supervised models like Polylingual Labelled LDA \cite{posch2015polylingual} improved alignment with labelled data. Although largely supplanted by neural embeddings, topic modelling remains useful in low-resource environments or as a complementary component in hybrid systems.

%\paragraph{Feature-Enhanced Ranking.} 
In many retrieval scenarios, particularly web search and e-commerce, performance depends on features beyond query-document similarity. Query properties such as length, phrasing, and intent are strong predictors of retrieval success \cite{devapujula2019broad}. Structural features, including PageRank, freshness (i.e. a temporal feature that measures how recent a document is relative to the query or current time), and categorical tags, further refine ranking. Behavioural signals such as clicks, ratings, and purchases, serve as implicit relevance feedback, often underperforming purely textual features. Learning-to-rank methods like LambdaMART~\cite{karmaker2017application} incorporate these signals effectively, and normalisation strategies \cite{zhang2022evaluating} help reduce bias, making feature-enhanced ranking especially valuable in multilingual applications.

\subsection{Re-ranking} 
\label{sec:reranking}

After an initial retrieval by a first-stage ranker, re-ranking refines the ordering of candidate documents using more detailed relevance signals. Since it applies only to a limited subset (e.g. top 100 results), computationally intensive models can be used, often incorporating semantic analysis, external signals, or joint ranking strategies. Re-ranking is particularly important in high-stakes or user-facing CLIR systems.

Models differ in how they formulate the relevance estimation objective. The main paradigms are pointwise, pairwise, and listwise, each balancing complexity, interpretability, and optimisation goals. In pointwise approaches, each document is scored independently via regression or classification \cite{cooper1992probabilistic, fuhr1989optimum}. This simple and scalable setup is suited to early-stage applications or when labelled data is scarce, but it may miss fine-grained distinctions. Pairwise models compare document pairs, predicting which is more relevant; methods such as RankNet \cite{burges2005learning} and LambdaMART \cite{burges2010ranknet} remain widely used for the ability to handle noisy or ordinal labels. However, they may not capture global ranking structures as effectively as listwise methods. Listwise approaches optimise over the entire ranked list using loss functions such as ListNet \cite{cao2007learning} and ListMLE \cite{xia2008listwise}, with recent transformer-based extensions further improving alignment with evaluation metrics like Normalised Discounted Cumulative Gain (nDCG)~\cite{jarvelin2002cumulated} or Mean Average Precision (MAP)~\cite{manning2008introduction}. Though more complex, listwise models better capture interdependencies and are particularly effective for long-context or diversity-sensitive ranking tasks \cite{liu2009learning}.

\paragraph{Encoder-based Re-ranking.} Encoder-based methods form the foundation of neural ranking in CLIR, primarily realised as cross-encoders and bi-encoders, explained in detail in Section \ref{subsec:model architecture}. Unlike bi-encoders, which encode queries and documents separately, cross-encoders concatenate them into a single sequence so that the transformer’s attention layers can model fine-grained token-level interactions. For this reason, bi-encoders are often used for initial ranking, whereas cross-encoders are preferred for re-ranking. 
The latter typically yields higher precision but incurs substantial computational costs, limiting scalability to large candidate sets. Multilingual pre-trained models such as XLM‑R \cite{conneau_2020}, mT5 \cite{xue_2021_mt5}, and MiniLM \cite{wang2020minilmdeepselfattentiondistillation} have been used as cross-encoders in CLIR, fine-tuned on datasets like multilingual MS MARCO \cite{bonifacio_2022_mmarco}, XOR‑Retrieve \cite{asai_2021_qa}, or MIRACL \cite{zhang_2023_miracl}, achieving strong language-agnostic ranking. The encoders are often trained with contrastive objectives such as Multiple Negatives Ranking Loss or Margin Ranking Loss, allowing large-scale retrieval with competitive accuracy.

In knowledge-distillation frameworks such as Translate-Distill \cite{yang2024translate_distill}, a cross-encoder acts as a teacher by producing high-quality relevance scores, while a bi-encoder serves as a student, learning to approximate those scores \cite{jeronymo2023neuralmindunicamp2022trecneuclir,yang2024hltcoe}.
This allows the final bi-encoder model to achieve efficient dense retrieval with accuracy close to the more computationally expensive cross-encoder \cite{yang2024translate_distill}.

%\begin{table}[ht]
%\centering
%\small
%\begin{tabular}{@{}p{0.95\linewidth}@{}}
%\toprule
%\textbf{Standard Fine‑tuned Cross‑Encoders} \\
%\textit{Key Idea:} Pretrained multilingual transformer (e.g., mT5, XLM‑R) finetuned on labeled ranking data (e.g., mMARCO, MIRACL, XOR‑Retrieve). \\
%\textit{Pros:} Strong semantic modeling; language-agnostic. \\
%\textit{Cons:} High inference cost; limited to top‑k. \\
%\midrule
%\textbf{Distillation‑Teacher %(Translate‑Distill)~\cite{yang2024translate_distill}} \\
%\textit{Key Idea:} Cross‑encoder teachers trained via MS MARCO translations teach dual‑encoder students. \\
%\textit{Pros:} Maintains cross‑encoder quality in efficient models; avoids full translation train. \\
%\textit{Cons:} Still requires cross‑encoder inference at distillation time. \\
%\midrule
%\textbf{Cross‑Encoder in NeuCLIR} \\
%\textit{Key Idea:} mT5-XXL fine‑tuned cross‑encoder applied to rerank top-100 candidates in zero‑shot multilingual retrieval~\cite{jeronymo2023neuralmindunicamp2022trecneuclir, yang2024hltcoe}. \\
%\textit{Pros:} Robust zero-shot performance across languages; empirical success. \\
%\textit{Cons:} Huge model sizes; latency and compute overhead. \\
%\bottomrule
%\end{tabular}
%\caption{Comparison of cross‑encoders for CLIR re-ranking.}
%\label{tab:crossencoder_paradigms}
%\end{table}

\paragraph{LLM-based Re-ranking.} Language models have recently emerged as powerful re-rankers, leveraging advanced semantic understanding that often surpasses traditional cross-encoders, particularly in zero-shot and few-shot scenarios. By reasoning directly over candidate lists, LLMs can improve retrieval effectiveness without requiring task-specific fine-tuning, though challenges such as inference cost, latency, and prompt sensitivity remain.

\begin{itemize}
    \item \textit{Prompt-based re-ranking.} Prompting LLMs directly enables document scoring (pointwise), pairwise comparisons, or listwise reordering.  Frameworks such as HyDE \cite{gao_2023_hyde}, InPars \cite{bonifacio_2022_inpars}, or RankGPT\cite{sun-etal-2023-chatgpt} adopt this approach. Pairwise ranking prompting has been shown to outperform GPT‑4 pointwise re-ranking in some cases.

    \item \textit{Zero‑shot \& few‑shot listwise re‑ranking.} LLMs can reorder candidates even without specific fine-tuning, often surpassing cross-encoders in zero-shot and few-shot settings. \citet{ma2023lrl} propose a zero-shot listwise re-ranker that achieves strong nDCG gains, particularly in multilingual datasets. Few-shot prompting further improves performance as shown by PaRaDe \cite{drozdov2023paradepassagerankingusing}.

    \item \textit{Document entailment \& instruction‑tuned models.} Instruction-tuned LLMs like Flan‑T5 \cite{chung2022scaling}, Zephyr \cite{tunstall2023zephyrdirectdistillationlm}, and ChatGPT \cite{openai_chatgpt_announcement} demonstrate strong capability in assessing document relevance. For instance, Flan‑T5 XXL outperforms baselines when treated as an entailment verifier. 

    \item \textit{Attention-based ranking (in-context re-ranking).} In-context re-ranking (ICR) leverages transformer attention to rank documents without generating full text. \citet{chen2025icr} show that ICR reduces latency by over 60\% compared to generative prompting while maintaining accuracy.

    \item \textit{Cross-encoder vs LLM re-rankers.} While LLMs like GPT-4 achieve impressive zero-shot performance, cross-encoders remain competitive in domain-matched settings \cite{déjean2024thoroughcomparisoncrossencodersllms}. This suggests that LLMs are promising but not yet universally dominant.
\end{itemize}

\paragraph{Additional Approaches.}

While cross-encoders, bi-encoders, and other neural re-ranking architectures remain central to CLIR, several complementary strategies have been explored. These approaches address challenges such as efficiency, bias, interpretability, and integration with downstream tasks, providing practical extensions beyond the core paradigms.

\begin{itemize}
\item \textit{QA‑oriented re-ranking.} In CLIR pipelines aimed at question answering, retrieval can be optimised for answerability rather than raw relevance. The MIA shared task~\cite{mialowresource2025mia} applied a zero-shot multilingual question-generation model to top‑k passages, scoring by the probability of regenerating the query. This eliminated the need for annotated data and outperformed BM25 by 6–18\% in top‑20 accuracy.

\item \textit{Pre-trained and joint QA models.} Pre-trained extractive QA models act as implicit cross-language re-rankers by extracting answer spans. They assess the likelihood of a passage supporting the correct answer by extracting start/end
positions. This approach has been validated on multilingual QA datasets like MLQA \cite{lewis_2020_mlqa} and XOR-TyDi QA \cite{asai_2021_qa},
where re-ranking based on QA model confidence yields higher answer recall and precision \cite{asai_2021_qa}.

\item \textit{Feature‑enhanced re‑ranking.} Industrial systems often incorporate structural and behavioural features alongside textual similarity. Features such as freshness, domain authority, click-through rates, and user engagement improve ranking utility. Feature-aware models (e.g. LambdaMART \cite{pi2024featurebasedecom}) yield notable gains, especially in low-resource or underrepresented languages where semantic alignment is noisy.

\item \textit{Score calibration and fusion.} High quality re-ranking often requires integrating outputs from multiple retrieval components (BM25, dense encoders, cross-encoders). Reciprocal Rank Fusion \cite{cormack2009rrf} avoids score-scale mismatches and is used in Zilliz~\cite{zilliz_rrf_ranker} and Azure AI~\cite{azure_rrf_hybrid}. Beyond fusion, calibration techniques (e.g. linear interpolation, normalisation) prevent biases from dominant languages or models.

\item \textit{Mitigating cross-lingual bias.} Language-specific scoring biases arise when source-language documents receive inflated relevance scores \citet{zhang2022evaluating}. Mitigation strategies include (i) score normalisation \cite{dietz2023ecir}; (ii) balanced training with models such as LambdaMART~\cite{burges2010ranknet, pi2024featurebasedecom}; and (iii) adaptive thresholds \cite{zhang-etal-2020-2019}. Positional bias is further addressed by the Cascade Model/UBM click models and re-ranking with RandPair or FairPair~\cite{craswell2008cumulative, chapelle2009dynamic, wang2018position, joachims2007position}. For LLM-based re-ranking, prompt-shuffling or permutation improves output stability~\cite{islam2025position,tang2024permutation}. To avoid repeated content or excessive focus on a single topic, methods like Maximal Marginal Relevance selectively promote new information while maintaining relevance. Early work \cite{goldstein1998mmr} demonstrated improved information coverage using this method in both retrieval and summarisation contexts. \citet{lin2023simple} show that explicit novelty scoring yields better user satisfaction in multilingual CLIR systems.

\item \textit{Cost-aware re‑ranking.} To reduce computational cost, cascading approaches filter candidates before applying expensive re-rankers. Bi-encoder cascades cut compute by up to sixfold with minimal performance loss \cite{wang2011cascade, chen2017costaware, honig2023bi}. Adaptive candidate truncation for language model-based re-rankers \cite{meng2024rlt} further optimises the trade-off between efficiency and retrieval quality.

\item \textit{Dynamic and adaptive re‑ranking.} Reinforcement learning and feedback-driven re-rankers dynamically adapt rankings to user interactions. RLIRank~\cite{zhou2021rlirank} uses reinforcement learning with Long Short-Term Memory (LSTM) \cite{hochreiter1997long} based click modelling, outperforming static rankers. Studies \cite{baumgartner2022feedback, nguyen2024rlaf} demonstrate that reinforcement feedback enhances performance on dynamic tasks such as Text REtrieval Conference (TREC) Dynamic Track \cite{yang2016trec}, even with limited supervision. 

\item \textit{Explainability and transparency.} In high-stakes domains (e.g. legal or medical), interpretable ranking is essential. Approaches include extractive explanations (Select‑And‑Rank~\cite{leonhardt2023extractive}), attention-based rationales \cite{jain2019attention, wiegreffe2019attention}, and post-hoc explanation platforms. Frameworks such as Stable and Explainable Attention (SEAT)~\cite{hu2022seat} 
align attention and predictions, while benchmarks \cite{pandian2024explainability, anand2022survey} stress balancing performance with interpretability, accountability, and auditability in CLIR deployment.  

\end{itemize}

\paragraph{Retrieval Architectures.}
\label{subsec:model architecture}

There are many different architectures which can be used for the ranking and re-ranking stages of the CLIR pipeline.
Bi-encoders, the most common architecture for ranking, encode queries and documents into dense vectors using a shared transformer backbone, estimating relevance via similarity functions such as dot product or cosine similarity. Document embeddings are precomputed offline, while query embeddings are generated online and matched using the similarity function. They do not allow token-level interaction but are more efficient since embeddings can be precomputed.  This design allows efficient and scalable retrieval, but struggles with fine-grained text interactions. Cross-encoders, by contrast, concatenate query and document inputs and process them jointly within a transformer model with full cross-attention, capturing detailed query-document interactions and improving ranking accuracy. However, each query-document pair must be processed individually at inference, making this approach computationally expensive and therefore typically reserved for re-ranking. In practice, systems often combine both approaches: bi-encoders for first-stage retrieval over large collections, and cross-encoders for re-ranking a smaller candidate set.

More recent models, such as late interaction methods and sparse neural retrievers, have been designed specifically for information retrieval and often outperform traditional approaches. To balance efficiency and accuracy, late-interaction models (e.g. ColBERT~\cite{khattab2020colbert} and ColBERTv2~\cite{santhanam2022colbertv2}) separately encode queries and documents, then compute token-level similarity via MaxSim \cite{chan-ng-2008-maxsim} (i.e. compute the sum of maximum similarities between each query token and all document tokens), enabling fine-grained matching at reduced cost. Sparse neural retrievers, meanwhile, exploit sparse inverted-index efficiency while incorporating semantic richness. Approaches include term-weighting methods such as DeepCT \cite{dai2019deepct} and expansion models such as SPLADE \cite{formal2021splade, formal2021spladev2} and SpaDE \cite{choi2022spade}, which use transformers to predict term importance or generate additional relevant terms. These maintain compatibility with inverted indices while introducing semantic depth.

Each architecture offers distinct trade-offs. Bi-encoders enable scalability and high efficiency but may miss nuanced semantics. Cross-encoders capture complex interactions but are computationally prohibitive for large-scale retrieval. Late-interaction models offer a middle ground, retaining token-level richness with manageable overhead. Sparse retrievers combine the efficiency of inverted indices with neural modelling, making them well suited for large-scale tasks. The effectiveness of these models also depends on embedding granularity, whether at the word, sentence, passage, or document level. Short queries are often represented well by a single vector, but longer texts demand finer-grained embeddings. Traditional sparse methods like BM25 \cite{robertson1995okapi} rely on term-frequency statistics (e.g. TF–IDF~\cite{sparck1972statistical}), surfacing documents through keyword overlap, but embedding-based models risk collapsing diverse topical content into a single dense vector, overlooking relevant information. To mitigate this, fine-grained approaches have been developed.

Sparse retrieval models such as BM25, SPLADE, SPLADE‑X represent queries and documents as high-dimensional sparse vectors, enabling inverted-index lookup while incorporating learned expressions (e.g. cross-lingual mappings \cite{formal2021splade, nair2022spladex, ctitech2022_SPARSEUDA}) to improve CLIR effectiveness. Dense Passage Retrieval (DPR) embeds text segments (100--300 words) independently and has shown substantial improvements over BM25, though performance depends on segmentation. Hybrid models combine sparse and dense signals \cite{Bruch2023_Hybrid, Weaviate2025_Hybrid}, either through parallel retrieval with later rank fusion (e.g. Reciprocal Rank Fusion \cite{cormack2009rrf, Weaviate2025_Hybrid} or concatenation, which often outperforms either approach alone).

Late-interaction models like ColBERT and ColBERTv2~ \cite{khattab2020colbert, santhanam2022colbertv2} retain token-level embeddings and compute fine-grained similarity at the cost of higher storage requirements. Multi-vector retrieval methods (e.g. ME‑BERT~\cite{luan2020mebert}, COIL~\cite{gao2021coil}) similarly encode documents into multiple vectors for semantic matching, offering strong performance on nuanced queries but demanding greater storage and computation. 

Further refinements include hierarchical representations, such as Dense Hierarchical Retrieval \cite{liu-etal-2021-dense-hierarchical}, which retrieves at the document-level and refines at the passage level, preserving both global and local context. Memory-augmented architectures (e.g. EMAT \cite{wu2022emat}, MoMA \cite{ge2023moma}) store distinct document segments in explicit memory slots, enabling selective attention during retrieval and dynamic external memory access. These enhance performance but introduce added complexity and computational overhead.

In summary, retrieval architectures vary across dual, cross, late-interaction, sparse, hybrid, hierarchical, and memory-augmented models, each offering trade-offs among scalability, semantic depth, precision, and efficiency. The choice of embedding granularity and retrieval mechanism should ultimately align with the task's demands for speed, scale, and ranking accuracy. Table~\ref{tab:granularity_methods} summarises these approaches across embedding granularity, representation style, and retrieval mechanism.

\begin{table}
\centering
\small
\resizebox{\textwidth}{!}{
\begin{tabular}{p{3.5cm} p{2.7cm} p{3.5cm} p{3.5cm} p{4cm}}
\toprule
\textbf{Approach} & \textbf{Granularity} & \textbf{Representation} & \textbf{Retrieval Level} & \textbf{Context Modelling} \\
\midrule
\textbf{Sparse Retrieval} & Term-level & High‑dimensional sparse vectors (e.g. BM25, SPLADE) & Term matching via inverted index & Encoded term importance; no dense semantic context \cite{formal2021splade, nair2022spladex} \\
\midrule
\textbf{Dense Passage Retrieval} & Passage-level (100--300 words) & One embedding per passage & Passage retrieval & Independent encoding of each passage without cross-passage context \\
\midrule
\textbf{Hybrid Sparse + Dense Retrieval} & Term + Document (via dense) & Concatenated or parallel sparse and dense vectors & Retrieval via fusion or unified indexing & Sparse term precision + dense semantic bridging between languages \cite{Bruch2023_Hybrid, Weaviate2025_Hybrid} \\
\midrule
\textbf{Late Interaction Models (e.g. ColBERT)} & Token-level embeddings & Contextualised token embeddings & Token-level MaxSim retrieval & Combines local token interactions with global context via transformer encoders \\
\midrule
\textbf{Multi-Vector Retrieval Models (e.g. ME-BERT, COIL)} & Token- or span-level & Multiple dense vectors per document (token/span) & Token or span matching & Combines fine-grained token matching with optional global encoding \\
\midrule
\textbf{Hierarchical Representations} & Multi-level: Word $\rightarrow$ Sentence $\rightarrow$ Paragraph & Hierarchical combination of embeddings & Document or segment retrieval & Captures local structure and aggregates into global document context \\
\midrule
\textbf{Memory-Augmented Models} & Embeddings for document parts (e.g. passages, sentences) & Memory slots for different parts & Retrieval via memory attention & Query dynamically attends to relevant parts of the document \\
\bottomrule
\addlinespace[1ex]
\end{tabular}
}
\caption{Comparison of document representation and retrieval approaches at different embedding granularities}
\label{tab:granularity_methods}
\end{table}

\subsection{Question Answering}

Question answering systems aim to provide users with direct, contextually appropriate answers rather than requiring them to sift through retrieved documents. Approaches range from factoid-style responses, containing discrete pieces of information, to more complex outputs such as passage extraction or abstractive summaries. This positions QA as a natural progression of information access, aligning more closely with user needs.

Information retrieval systems usually generate a ranked list of documents or re-rank results to directly satisfy a user's query. To bridge the gap between retrieval and direct answering, many systems introduce a ``reader" component \cite{chen_2017_qa, zhu_2024_ir}, often framed as machine reading \cite{izacard_2021, lewis_2020_rag}, or question answering \cite{iida_2019, goodwin_2020, deng_2020}. LLMs such as GenQA \cite{hsu_2021_qa, muller_2022} are increasingly applied to QA tasks. However, they face challenges: hallucinations \cite{jurafsky_2025} (e.g. \citet{dahl_2024} observed at rates of 69\% to 88\% in the legal domain), overconfidence despite uncertainty, and limited access to recent or proprietary data \cite{zhou_2024_calibration}.

\paragraph{Retrieval-augmented QA.} Retrieval augmented generation (RAG) addresses these issues by combining IR with generative models. Retrieved documents ground outputs \cite{jurafsky_2025}, ensuring factuality, timeliness, and transparency. Some systems further enhance reliability by incorporating references and citations \cite{aksitov_2023}. Some researchers view RAG as a complete system that integrates IR and QA, while others conceptualise QA itself as compromising two components: a retriever, which selects relevant information, and a reader, which generates or extracts the answer \cite{chen_2017_qa}. In either view, RAG-based QA represents a subset of broader QA approaches, particularly relevant in open-book scenarios, where grounding in retrieved content enables accurate, context-aware responses. 

In cross-lingual QA, methods mirror monolingual IR/QA but often include translation for sparse retrieval or multilingual embeddings for dense retrieval. Research predominantly involves English plus one other language, utilising English's data abundance to support low-resource settings \cite{lee_2019_qa, asai_2021_qa}. Reader modules are particularly beneficial in CLIR systems, as the retrieved documents may be in a language that the user is unable to understand, so the generation of an answer or summary in the original query language allows for the bridging of this language gap.

\subsection{Current CLIR Approaches}
\label{sec:current_approaches}

CLIR has recently drawn from improvements in representation learning, multilingual modelling, and scalable retrieval architectures. Contemporary systems increasingly integrate these elements into full retrieval pipelines, with methods broadly categorised into sparse retrieval, dense retrieval, late-interaction, hybrid, cross-encoders, and multimodal models. In this section, we highlight representative recent work; for definitions and trade-offs of the retrieval architectures mentioned below, refer back to Section \ref{subsec:model architecture}.

\textit{Sparse retrieval models.} The SPLADE family~\cite{formal2021splade}, SPLADEv2 \cite{formal2021spladev2}, and extensions such as SPLADE-X~\cite{nair2022spladex} and MultiSPLADE~\cite{lassance2023multisplade} exemplify this line of work by generating sparse token-level representations and enabling effective multilingual retrieval.

\textit{Dense retrieval models.} Systems such as LaBSE~\cite{feng2022language}, mSBERT~\cite{reimers2020making}, and mDPR~\cite{zhang2022mr} demonstrate robust multilingual and zero-shot performance across diverse languages.
    
\textit{Late-interaction models.} ColBERT~\cite{khattab2020colbert} and ColBERT v2~\cite{santhanam2022colbertv2} balance fine-grained matching with scalable retrieval, achieving strong effectiveness-efficiency trade-offs.

\textit{Hybrid models.} Examples include pipelines that merge BM25 with neural retrieval \cite{tu-padmanabhan-2022-mia, ogundepo_2022_africlirmatrix}, which show effectiveness in low-resource and typologically diverse settings.

\textit{Cross-encoder \& re-ranking models.} Approaches such as Translate–Distill \cite{yang2024translate_distill}, Multilingual RAG \cite{ranaldi2025multilingual}, and CoConDenser \cite{gao2022cocondense} achieve strong performance, while large-scale resources like CLIRMatrix \cite{sun_2020_clirmatrix} facilitate multilingual evaluation. Recent work, including OPTICAL \cite{huang2023optimal}, mContriever-X \cite{izacard_2022_mcontriever}, and SWIM-X \cite{thakur_2024_swimir}, further extend retrieval capabilities across dozens of languages.

\textit{Multimodal \& speech-based CLIR.} These systems expand retrieval beyond text. Cross-modal pre-training~\cite{fei-etal-2021-cross}, LECCR \cite{wang2024multimodalllmenhancedcrosslingual}, and M‑SpeechCLIP \cite{berry2023mspeechclipleveraginglargescalepretrained} align text with image or speech embeddings, enabling multilingual retrieval across different modalities.

\textit{Other directions.} Additional approaches include unsupervised CLIR~\cite{litschko2018unsupervised}, cross-lingual text encoders~\cite{litschko2022cross}, and task-specific benchmarks such as CrossMath~\cite{gore2024crossmath} and MTD/MLIR~\cite{Yang_2024}. Re-rankers increasingly integrate multilingual supervision and teacher-student learning, while hybrid and generative methods adapt retrieval to noisy or adversarial conditions. Despite these advances, gaps remain in domain adaptation, handling morphologically rich and code-switched languages, and incorporating LLMs as multilingual rankers. Addressing these challenges is essential for developing CLIR systems that are accurate, scalable, and equitable.

\section{Dealing with Cross-Linguality}
\label{sec:cross-lingual}

While some components of the CLIR pipeline, such as cross-lingual query expansion or translation modules, can
be adapted in a modular fashion, approaches specifically developed for cross-lingual retrieval generally yield better performance and more balanced language coverage. Traditional approaches translate queries or documents directly, or use a pivot language, whereas recent methods employ cross-lingual embeddings to map texts into a shared semantic space for more effective comparison. Alignment strategies such as contrastive learning, adversarial alignment, and self-supervised objectives mitigate linguistic and resource disparities, making them critical for effective and inclusive CLIR systems.

\subsection{Translation}
\label{sec:translation}

% \textcolor{blue}{Historically, translation was the main method
% It's still used today in different forms
% Embedding methods have become more popular
% Sometimes both are used in combination (e.g. mLLMs)}

Traditional CLIR approaches are usually divided into two main stages: translation and monolingual IR. Unlike full-text translation, CLIR requires only a representation suitable for the retrieval system, meaning that strict syntactic or grammatical fidelity can be relaxed \cite{nie_2010}. Queries, which are often very short and ambiguous, are pre-processed through tokenisation, stopword removal, and term expansion \cite{zhou_2012, fox_1989}. Whereas most translation applications aim to produce a single, readable output, CLIR can benefit from multiple translation alternatives, which can function as part of the query expansion process. After retrieval and ranking, further translation may sometimes be necessary to ensure the user can interpret the documents, but this step is not always required.

\textbf{Translation Granualrity.} A central design choice concerns translation granularity: whether to translate the query, the document, or both. Techniques range from dictionaries and traditional Statistical Machine Translation to neural and embedding-based approaches, with pivot or dual translation sometimes needed for low-resource languages. Query translation is more widely used in CLIR than document translation, as it is computationally cheaper, avoids large-scale translation, and can be performed at retrieval time, though it introduces risks of ambiguity and misinterpretation \cite{zhou_2012, nie_2010,ture_2014}. Document translation, in contrast, is more resource-intensive but benefits from added context and reduced ambiguity. Moreover, the mistranslation of a single word has less effect on retrieval performance \cite{zhou_2012}. Query translation remains popular due to its efficiency and flexibility, while document translation can be advantageous when all queries are in a single language. Ultimately, query translation may still necessitate subsequent document translation for user access, whereas document translation allows direct examination of retrieved texts.

\subsubsection{Translation Techniques}

Translation can be classified as ``direct" where the source language is translated straight from the target language, or ``indirect", where a pivot language is used to overcome source-target limitations for particular language pairs. Several translation techniques have been developed, including dictionary-based approaches, corpus-driven strategies, statistical models and more recently neural models.

\paragraph{Dictionary-based Methods.}  These rely on bilingual, machine-readable dictionaries (MRDs)~\cite{hull_1996,ballesteros_1996}. For each word in the source language, MRDs contain one or multiple synonymous words and phrases in the target language. For each term in the given query, dictionary-based translation simply finds the word in the dictionary and selects the translation. Ambiguity arises because many words have multiple meanings. One solution is to select the most frequent translation \cite{nie_2010}, while another retains all possible translations in structured query translation \cite{hull_1997, pirkola_1998, darwish_oard_2003}. The latter improves recall \cite{pirkola_1998} but requires weighting schemes to balance translation probabilities \cite{levow_oard_2000, leek_jin_2000, xu_2005}. A key limitation of bilingual dictionaries is their poor handling of proper nouns and out-of-vocabulary (OOV) terms, especially newly-coined technical terms \cite{zhou_2012}.

\paragraph{Parallel and Comparable Corpora.} Parallel corpora are aligned texts in two languages, such as the Hansard Corpus \cite{roukos-graff-melamed-1995-hansard}, EuroParl \cite{koehn-2005-europarl}, or UN documents \cite{ziemski-etal-2016-united}, and are widely used to induce bilingual dictionaries (e.g. \cite{ture_2014, chew_2006}). They support multiple translation options but are costly to collect and often limited in domain. Comparable corpora instead consist of texts that are not translations but share topical or communicative similarity, for example, Wikipedia pages \cite{braschler_2000, moulinier_2003, franz_1999}. These corpora are easier to obtain, yet translation quality is typically lower than with parallel corpora \cite{zhou_2012}.

\paragraph{Statistical Machine Translation (SMT).} SMT was the dominant approach from the 1990s to the early 2010s. It is based on noisy channel models \cite{nie_2010, ture_2014}, particularly IBM models \cite{brown_1990, brown_1993}, which assign probabilities to candidate translations and choose the most likely output. A noisy channel model treats the source language text as a misspelled or distorted version of the target language, where the goal is to recover the most likely original target language text \cite{nie_2010, brown_1990}. SMT proved effective when large parallel corpora were available, but it typically produced a single best translation, reducing ambiguity that could otherwise aid retrieval \cite{ture_2014}.

\paragraph{Neural Machine Translation (NMT).} NMT has overtaken SMT as the preferred paradigm, replacing phrase-based systems with single neural network architectures \cite{stahlberg_2020_nmt, mohamed_2021_nmt, ranathunga_2023_nmt}. Early models handled only one language pair, but subsequent developments expanded to multilingual systems \cite{dabre_2020_mnmt}. NMT produces fluent, context-aware translations, handles OOV terms through subword segmentation, and benefits from contextualised embeddings \cite{McCann_2017, peters_2018}. Encoder-decoder architectures with attention \cite{kalchbrenner_13, cho_2014_nmt, bahdanau_2015_attention}, and later Transformers \cite{vaswani_2017}, enabled more effective long-sequence translation. Subword methods such as byte pair encoding \cite{gage_1994_bpe, sennrich_2016_bpe, provilkov_2020_bpe}, WordPiece \cite{schuster_2012_wordpiece, wu_2016_wordpiece}, SentencePiece \cite{kudo_2018_sentencepiece} are now standard.

NMT has been applied to CLIR through systems such as Translate-Train~\cite{nair_2022, yang_2024_translatetraincolbertxafrican}, Translate-Distill~\cite{yang2024translate_distill}, mDPR~\cite{zhang_2021_mrtydi} and ColBERT-X~\cite{nair_2022}, which optimise retrieval quality by integrating translation with retrieval. Despite improvements, CLIR performance is still hindered by issues relating to short queries, ambiguity, semantic drift, domain adaptation, and high training data requirements \cite{koehn_2017_nmt}. Overall, while NMT represents a substantial improvement over SMT in terms of fluency, adaptability, and context handling, both approaches remain limited by data availability, computational demands, and scalability challenges in low-resource settings.

\subsubsection{Indirect Translation}

When direct translation from source to target language is infeasible, indirect translation offers an effective alternative by exploiting resources available for intermediate languages. Two common approaches are transitive translation and dual translation.

\paragraph{Transitive Translation.} Transitive translation uses a pivot language: the source text is first translated into a high-resource intermediate language, then into the target (e.g. \cite{hiemstra_1999, kishida2003two, kando2005two, kishida2006hybrid, kraaij2004transitive, kishida2005technical, chen_gey_2003, lin_chen_2003, gey_1998}).
Gollins and Sanderson \cite{gollins_sanderson_2001} highlight triangulation, where the use of multiple pivot languages reduces ambiguity compared to a single pivot, which can accumulate errors. Their findings show that triangulation via three intermediate languages outperforms pairwise merging, though subsequent studies note that its benefits are most evident for unstructured queries \cite{lehtokangas_2004}.

\paragraph{Dual Translation.} Dual translation translates both source and target texts into a third language, which can be concrete or abstract (e.g. a semantic space). When concrete languages are used, high-resource ones yield superior translations. Deerwester et al. \cite{deerwester_1990} introduced Latent Semantic Indexing later adapted for CLIR via parallel-corpora \cite{littman_1998, berry_young_1995}. Similarly, Explicit Semantic Analysis (ESA) employs human-readable labels, with \citet{gabrilovich_2007} deriving a machine learning approach using weighted vectors. ESA representations, often built from Wikipedia, are interpretable and have been extended to cross-lingual applications.

% \subsection{Embedding} 
% Cross-lingual embeddings, not just in CLIR. Multilingual models in general. 

% \begin{itemize}
%     \item Methods for creating multilingual embeddings \cite{ruder_2019} - pretraining, fine-tuning
%     \item Doc vs. paragraph vs. query (\cite{ruder_2019} classifies into word vs. sentence vs. document (mention multi-word expression as an open challenge, even in monolingual), also parallel vs. comparable corpora - note first published in 2017) 
%     \item Mapping Function
%     \begin{itemize}
%         \item Between Languages (? - not as current)
%         \item Between Query and Doc embeddings
%     \end{itemize}
%     \item Token imbalance
%     Tokenization issues
%     Differing performance across languages
%     \item Bilingual models vs. multilingual models
% \end{itemize}

\subsection{Multilingual LLMs for CLIR}
\label{sec:mllms}

The emergence of LLMs, particularly those with multilingual capabilities, has transformed CLIR. Traditional CLIR approaches relied on query or document translation, but the development of transformer architectures \cite{vaswani_2017} and large-scale text corpora has enabled LLMs to achieve strong zero-shot performance with notable generalisation potential. This section reviews standard training steps and architectures for multilingual LLMs. While many LLMs can handle multiple languages,  multilingual LLMs are explicitly trained on multilingual corpora. A useful distinction is that if a significant proportion of training data is multilingual, the model can be considered a multilingual LLM \cite{huang_2025}.

Both LLMs and multilingual LLMs follow three main training stages: pre-training on large corpora, fine-tuning for task specialisation, and alignment with human preferences via reinforcement learning from feedback (RLHF). Architecturally, three variants dominate: encoder-only, encoder-decoder, and decoder-only, with the latter being the most common for text generation. Encoder-decoder models may better suit CLIR \cite{reusch_2025}, particularly for question answering, while decoder-only models are preferable for certain specialised tasks \cite{li_2025_GenIR}.

Different training objectives are used to optimise task-specific performance, evaluated through established metrics and benchmarks. Key challenges include the ``curse of multilinguality"~\cite{conneau_2020} discussed below, which highlights trade-offs between performance and multilingual ability. Table \ref{mllm} summarises widely used multilingual LLMs, with further surveys in \cite{ huang_2025, qin_2025, xu_2025, zhu_2024, gurgurov_2024, doddapaneni_2025}.

\begin{table}[htp]
\centering
\small
\begin{tabular}{@{}p{4cm}>{\centering\arraybackslash}p{1.5cm}>{\centering\arraybackslash}p{2.5cm}>{\centering\arraybackslash}p{2cm}p{5cm}@{}}
\toprule
\textbf{Model} & \textbf{Architecture}  & \textbf{Languages} & \textbf{Open-source} & \textbf{Training Data} \\
\midrule
\textbf{mBERT} \cite{devlin_2019_bert} & E & 104 & \ding{51} & Wikipedia \\
\textbf{XLM-R} \cite{conneau_2020} & E & 100 & \ding{51} & CommonCrawl \\
\textbf{mContriever} \cite{izacard_2022_mcontriever} & E & 29 & \ding{51} & CCNet and Wikipedia \\
\textbf{ColBERT-X} \cite{nair_2022} & E  & 7 & \ding{51} & MS MARCO \\
\textbf{Qwen3-Reranker} \cite{qwen3-embedding} & E & 100{+} & \ding{51} & Multilingual query–document pairs \\
\textbf{multilingual‑E5‑large}  \cite{wang2024multilinguale5textembeddings} & E & 100+ & \ding{51} & 1B text pairs and retrieval/TyDi tasks \\
\textbf{XLM‑V} \cite{liang_2023_xlm}& E & 100+ & \ding{51} & CommonCrawl \\
\textbf{Nomic Embed v2} \cite{nussbaum_2025} & E & 100+ & \ding{51} & Data from a variety of sources \\
\textbf{Gemini Embedding} \cite{lee_2025_gemini} & E & 100+ & \ding{55} & Undisclosed \\
\textbf{LaBSE} \cite{feng2020labse} & E & 109+ & \ding{51} & CommonCrawl, Wikipedia and webpage translation pairs \\
\textbf{mT5} \cite{xue_2021_mt5} & E-D & 101 & \ding{51} & CommonCrawl \\
\textbf{mBART} \cite{liu_2020_mbart} & E-D & 25 & \ding{51} & CommonCrawl \\
\textbf{NLLB-200} \cite{nllb_2022} & E-D & 202 & \ding{51} & Parallel data \\
\textbf{mLongT5} \cite{uthus_2023_mlongt5} & E‑D & 101 & \ding{51} & mC4 \\
\textbf{Qwen3-Omni} & E-D & 119 & \ding{51} & Undisclosed \\
\textbf{Aya} \cite{ustun_2024_aya} & E-D & 101 & \ding{51} & xP3x, Aya Dataset, Aya Collection, Data Provenance and ShareGPT-Command \\

\textbf{XGLM} \cite{lin_2022_xlgm} & D & 30 & \ding{51} & CommonCrawl \\
\textbf{LLaMA 3} \cite{llama3_2024} & D & 30 & \ding{51} & Publicly available data \\
\textbf{Mistral} \cite{mistral_2024} & D & Dozens & \ding{51} & Publicly available data \\
\textbf{BLOOM} \cite{bloom_2023} & D & 46 natural, 13 programming & \ding{51} & ROOTS \\
\textbf{Yi-01} \cite{ai_2025_yi} & D & 2 (English and Chinese) & \ding{51} & Publicly available data including CommonCrawl \\
\textbf{GPT-4o} \cite{openai_2024_gpt4o} & D & 50+ & \ding{55} & Publicly available and proprietary data \\
\textbf{Gemini 2.5} \cite{gemini_2025} & D & ND (40+) & \ding{55} & Publicly available data \\
\textbf{Claude 3} \cite{claude_2024} & D & ND & \ding{55} & Publicly available and proprietary data \\
\textbf{PaLM 2} \cite{anil_2023_palm2} & D & ND (100+) & \ding{55} & Data from a variety of sources including Wikipedia, webpages and news articles \\
\textbf{DeepSeek-V3} \cite{deepseek_2025} & D & ND & \ding{51} & Data from a variety of sources  \\
\textbf{Gemma} \cite{gemmateam_2024} & D & 20+ & \ding{51} & Data from a variety of sources \\
\textbf{PolyLM} \cite{wei2023polylm} & D & 18 & \ding{51} & mC4, CC-100, The Pile, GitHub and OPUS \\
\textbf{Nemotron‑4 15B} \cite{parmar2024nemotron415b} & D & 53+ & \ding{51} & Data from a variety of sources (53 languages, 43 code) \\
\bottomrule
\addlinespace[1ex]
\end{tabular}
\caption{Overview of explicitly multilingual LLMs as well as LLMs with multilingual capabilities. Architecture: E = encoder-only, E-D = encoder-decoder, D = decoder-only. ND = not disclosed.}
\label{mllm}
\end{table}

\subsubsection{Training Stages of Multilingual LLMs}
% \begin{figure}
%     \centering
%     \includegraphics[width=0.75\linewidth]{figures/mllm_training.pdf}
%     \caption{mLLM training stages.}
%     \label{fig:mllm_training}
% \end{figure}
As mentioned, the distinction between LLMs and multilingual LLMs is often blurred given that both follow similar training stages, but multilingual LLMs additionally rely on multilingual corpora. Training generally proceeds through pre-training, fine-tuning, and RLHF. \citet{qin_2025} highlight the importance of alignment strategies for multilingual performance, distinguishing parameter-tuning alignment from parameter-frozen alignment, where alignment occurs after post-training via methods such as prompting or code-switching.

Pre-training is based on large-scale multilingual corpora. These usually consist of monolingual texts across many languages (still predominantly English) and a smaller set of parallel corpora, which are less widely available \cite{xu_2025, gurgurov_2024}. Typical sources include  Common Crawl and Wikipedia. The aim is knowledge acquisition and learning universal language structures. Pre-training can start from scratch, with parameters randomly initialised, or follow a continual pre-training approach, where an existing LLM is adapted with multilingual or domain-specific data \cite{zhu_2024, huang_2025}. While continual pre-training is computationally cheaper and allows faster domain adaptation, it risks ``catastrophic forgetting"~\cite{McCloskey_1989} where previously learned knowledge is lost due to distributional shifts between old and new data \cite{shi_2024}. To mitigate this, methods such as replay buffers, parameter freezing, and elastic weight consolidation are used \cite{kirkpatrick_2017}.

Supervised fine-tuning (SFT) adapts pre-trained models to specific tasks using labelled datasets. Unlike pre-training, which focuses on broad knowledge acquisition, fine-tuning specialises models for instruction following or task-specific objectives \cite{zhu_2024}. Increasingly, fine-tuning data combines human- and model-generated content, with effectiveness depending on data quality and diversity. Beyond instruction tuning, multilingual SFT extends to tasks such as CLIR, Named Entity Recognition (NER), Sentiment Analysis and Text Classification \cite{xu_2005}. 

RLHF further aligns multilingual LLM outputs with human preferences. Human annotators either rank outputs or select between alternatives \cite{bai_2022, ouyang_2022}, producing preference data used to train a reward model. The base model is then fine-tuned using algorithms such as Proximal Policy Optimisation~\cite{schulmanetal2017}. RLHF is resource-intensive due to the need for large-scale human annotation, though synthetic data has reduced some costs. Nonetheless, concerns remain over the potential for manipulative behaviours to be learned by models when optimising for human feedback \cite{williams_carroll_2025}.

\subsubsection{Model Architectures of Multilingual LLMs}
\begin{figure}[h]
    \centering
    \includegraphics[width=0.9\linewidth]{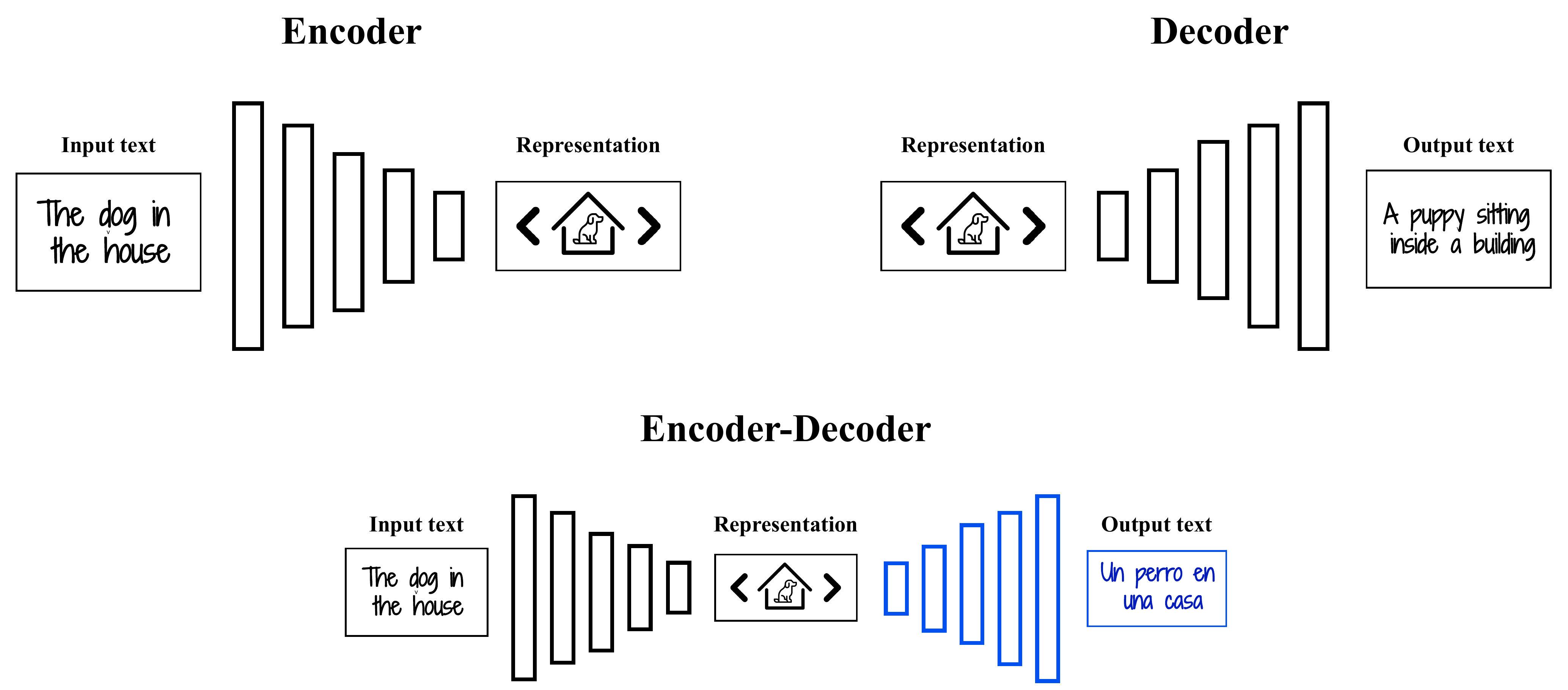}
    \caption{LLM model architectures.}
    \label{fig:mllm_arch}
\end{figure}

Like monolingual LLMs, multilingual LLMs are based on the transformer architecture \cite{vaswani_2017}, which consists of encoder and decoder modules that rely on self-attention. Variants fall into three categories: encoder-only, encoder-decoder, and decoder-only (see Figure \ref{fig:mllm_arch} for an illustration and Table \ref{mllm} for details on the architectures of the listed multilingual LLMs).

\textit{Encoder-only} models (e.g. BERT~\cite{devlin_2019_bert}) tokenise input into embeddings that encode semantic and positional information. These embeddings are processed by stacked encoder layers with multi-head self-attention and feed-forward networks, producing contextualised representations of the input \cite{gurgurov_2024}. Because they attend bidirectionally, encoder-only models excel at transforming data into compressed representations useful for language understanding tasks such as sentiment analysis, NER, and classification in both monolingual and multilingual settings \cite{xu_2025}. However, they are not well suited for generative tasks like next-token prediction or translation. 

BERT was initially released as a monolingual English model, which then led to other language-specific variations (e.g. FlauBERT \cite{le2020flaubert} for French, BERTje \cite{devries2019bertje} for Dutch and AfriBERT \cite{ralethe-2020-adaptation} for Afrikaans), eventually resulting in the multilingual version mBERT that is trained on 104 languages. Other multilingual encoder-only models include XLM-R \cite{conneau_2020} and LaBSE \cite{feng2020labse}.

\textit{Encoder-decoder} models preserve both components of the transformer, allowing the encoder to process input and the decoder to generate output conditioned on the encoded representation \cite{vaswani_2017}. This makes them well-suited to sequence-to-sequence tasks such as summarisation and, in the multilingual domain, machine translation. Examples include mT5 \cite{xue_2021_mt5} and mBART \cite{liu_2020_mbart}.

\textit{Decoder-only} models omit the encoder and generate text autoregressively, producing tokens sequentially while attending only to previous tokens. This unidirectional structure makes them effective for text generation and completion tasks \cite{zhu_2024, xu_2025}. Their popularity has grown significantly with the release of GPT-style models. Widely used multilingual decoder-only models include GPT-4o \cite{openai_2024_gpt4o}, PaLM \cite{chowdhery_2022_palm}, XLGM \cite{lin_2022_xlgm} and BLOOM \cite{bloom_2023}.

\begin{comment}
\begin{figure}
    \centering
    \includegraphics[width=1\linewidth]{figures/enc_dec-2.pdf}
    \caption{mLLM model architectures.}
    \label{fig:mllm_arch}
\end{figure}
\end{comment}

\subsubsection{Multilingual Pre-training Datasets} \label{mllm_data}

Monolingual LLMs are trained on large monolingual corpora. Multilingual LLMs follow a similar approach, extending the corpora to monolingual texts in multiple languages and parallel corpora (collections of translations). Model performance depends on corpus choice: some prioritise higher-resource languages, while others (e.g. IndicBERT \cite{doddapaneni-etal-2023-towards} for 12 Indian languages, or AfriBERTa \cite{ogueji-etal-2021-small} for 11 African languages) target low-resource settings. Broader multilingual coverage requires varied corpora, including mid- and low-resource languages. 

\paragraph{Monolingual Corpora.} Massive monolingual corpora enable learning universal language representations, critical for multilingual models, and reduce reliance on parallel data. Most pre-training data comes from web sources, particularly Common Crawl and Wikipedia. Crawled data, however, often contains harmful or low-quality content; thus, cleaned and filtered datasets are used (e.g. CC-100 for XLM-R \cite{conneau_2020}), and additional training stages like RLHF aim to mitigate undesirable outputs. Despite this, English dominates many training corpora. For example, \citet{xu_2025} report that English comprises about 92.1\% of ChatGPT’s training corpus, leaving relatively little representation for other widely spoken languages. \citet{xu_2025} further highlight that when English is excluded, Indo-European languages (e.g. German, French) still constitute over 50\% of the remaining data in their language-family analysis.

\paragraph{Parallel Corpora} Parallel corpora in multilingual pre-training resemble those in machine translation models, including manually created datasets (e.g. Bible Corpus \cite{chew_2006_bible}, MultiUN \cite{tiedemann_2012}) and machine-generated corpora via multilingual LLM-aided generation.

\subsubsection{Multilingual LLM Training Objectives}

Pre-training objectives for LLMs and multilingual LLMs aim to specialise models for specific tasks. \citet{doddapaneni_2025} categorise them into three types. The first adapts monolingual functions for multilingual use, e.g. Probabilistic Language Modelling \cite{bengio_2003_plm}, Masked Language Modelling \cite{devlin_2019_bert} and Next Sentence Prediction \cite{devlin_2019_bert}. The second leverages parallel corpora at the sentence/document level, including Translation Language Modelling \cite{conneau_lample_2019}, Cross-Attention Masked Language Modelling \cite{ouyang_2021_camlm} and Cross-Lingual Masked Language Modelling \cite{huang_2019_clmlm}. The third exploits other parallel resources such as word alignments, e.g. Cross-Lingual Word Recovery \cite{huang_2019_clmlm}, Alternating Language Model \cite{Yang_2020_alm} and Back Translation Masked Language Modelling \cite{ouyang_2021_camlm}. Table \ref{tab:mllm_obj} presents the most common objective functions used to train multilingual LLMs following this categorisation \cite{doddapaneni_2025}, and Table \ref{table:mlm_objectives_examples} provides illustrative examples of some of these multilingual training objectives using the sentence pair ``A dog in a house" and ``Un perro en una casa".

\begin{table}[h!]
\small
\centering
\begin{tabular}{@{}p{7cm}p{9cm}@{}}
\toprule
\textbf{Training Objective} & \textbf{Description}\\
\midrule
\multicolumn{2}{c}{\textbf{Adapted from monolingual}}\\
\midrule
\textbf{Probabilistic Language Modelling (PLM) }\cite{bengio_2003_plm} & Estimates the probability distribution of sequences of words in a \hspace{1cm} language. \\

\textbf{Masked Language Modelling (MLM)} \cite{devlin_2019_bert} & Inspired by a Cloze task \cite{taylor_1953}. Certain tokens are randomly masked, and the model predicts the masked tokens based on the available context. This objective encourages the model to learn bidirectional representations and dependencies between words in a sentence. \\

\textbf{Next Sentence Prediction (NSP)} \cite{devlin_2019_bert} & Predict whether a given pair of sentences is contiguous or not. Through this objective the model learns to understand coherence and logical flow between sentences. \\

\textbf{Denoising Autoencoder (DAE)} \cite{vincent_2008_dae} & Given a partially corrupted or noisy input, the model is trained to recover the original undistorted input. \\

\textbf{Causal Language Modelling (CLM)} \cite{conneau_lample_2019} & Autoregressive next-token prediction: predict the next token in a sequence of tokens; the model has access to unidirectional context. \\

\textbf{Multilingual Replaced Token Detection (MRTD)} \cite{chi_2022_xlme} & Tokens are replaced in a multilingual sequence, and the model is trained to detect which are the real input tokens from the corrupted sentences. \\

\midrule
\multicolumn{2}{c}{\textbf{Parallel corpora}}\\
\midrule
\textbf{Translation Language Modelling (TLM)} \cite{conneau_lample_2019} & Sentences in different languages are concatenated, and tokens are masked at random. The model then has to predict the masked tokens. \\

\textbf{Cross-Attention Masked Language Modelling (CAMLM)} \cite{ouyang_2021_camlm}  & Using a parallel sentence pair, the model is trained to predict masked tokens in one language using the other language.  \\

\textbf{Cross-Lingual Masked Language Modelling (CLMLM)} \cite{huang_2019_clmlm} & Similar to TLM, but the input is constructed at the document level. Sentences in a cross-lingual document are masked at random, and the model is trained to predict these masked tokens. \\

\textbf{Cross-Lingual Contrastive Learning (XLCO)} \cite{chi_2021_xlco} & Contrastive learning is used: the model learns to bring representations of semantically similar sentences together, and push negative pairs apart. \\

\textbf{Hierarchical Contrastive Learning (HICTL)} \cite{wei_2021_hictl} & Contrastive learning is applied both at sentence and word-level, with the goal for the model to learn language-invariant sentence representations. \\

\textbf{Cross-Lingual Sentence Alignment (CLSA)} \cite{hu_2021_objectives} & Using parallel data, the model is encouraged to align sentence representations across languages. \\

\textbf{Translation Replaced Token Detection (TRTD)} \cite{chi_2022_xlme} & Given translation pairs, the model is trained to detect masked tokens in both languages. \\

\midrule
\multicolumn{2}{c}{\textbf{Parallel resources}}\\
\midrule
\textbf{Cross-Lingual Word Recovery (CLWR)} \cite{huang_2019_clmlm} & Goal is to learn underlying word alignments between two languages by predicting missing words in one language using aligned source sentences. \\

\textbf{Cross-Lingual Paraphrase Classification (CLPC)} \cite{huang_2019_clmlm} & Given two sentences from different languages, classifies whether they have the same meaning. \\

\textbf{Alternating Language Model (ALM)} \cite{Yang_2020_alm} & Using code-switched sentences (alternating languages between phrases), the model is trained to predict masked language modelling. \\ 

\textbf{Denoising Word Alignment (DWA) with Self-Labelling} \cite{chi_2021_dwa} & Two alternating steps: word alignments are first estimated, and the model then predicts masked tokens in parallel sentence pairs.  \\

\textbf{Bidirectional Word Alignment (BWA)} \cite{hu_2021_objectives} & Employs the attention mechanism in the transformer model to align word representations across languages using parallel data. \\

\textbf{Back Translation Masked Language Modeling (BTMLM)} \cite{ouyang_2021_camlm} & Tokens from a source language are predicted (translated) into a target language. Tokens in the source language are then masked, and the target language tokens are used to predict them. \\

\bottomrule
\addlinespace[1ex]
\end{tabular}
\caption{Multilingual LLM training objectives. The objectives are classified into those that are adapted from monolingual training objectives, those that leverage parallel corpora at the sentence or document level, and those that exploit other parallel resources such as word alignments.}
\label{tab:mllm_obj}
\end{table}

\begin{table}[h!]
\centering
\small
\begin{tabular}{p{5.5cm}p{10cm}}
\toprule

\multicolumn{2}{c}{\textbf{Monolingual-Inspired Objectives}} \\
\midrule
\textbf{Masked Language Modelling} & 
EN: A dog in a \textcolor{red}{[MASK]}. \\
& Context: “dog”, “in” → predict “house” \\[1ex]

\textbf{Causal Language Modelling} &
EN: A → dog → in → a → house → [EOS] \\
& Predict next token based on previous tokens only \\[1ex]

\textbf{Multilingual Replaced Token Detection} &
ES: Un perro en \textcolor{red}{una} casa. \\
& Detect whether “una” is real or replaced (binary label) \\
\midrule
\multicolumn{2}{c}{\textbf{Parallel Corpora Objectives}} \\
\midrule
\textbf{Translation Language Modelling} &
EN: A dog in a \textcolor{red}{[MASK]}. \\
% & \hspace{7.7em} $\updownarrow$  \\

& ES: Un \textcolor{red}{[MASK]} en una casa. \\
& Predict “house” and “perro” using cross-lingual context \\[1ex]

\textbf{Cross-Attention MLM} &
EN: A dog in a \textcolor{blue}{house}. \\
& \hspace{8em} $\updownarrow$  \\
& ES: Un perro en una \textcolor{red}{[MASK]}. \\
& Use EN tokens via cross-attention to predict “casa” \\[1ex]
\textbf{Cross-Lingual Contrastive Learning} &
Positive pair: \\
& EN: A dog in a house. $\rightarrow\leftarrow$ ES: Un perro en una casa.  \\
% & ES: Perro en una casa. \\
& \textit{Pull their embeddings closer} \\[0.5ex]
& Negative pair: \\
& EN: A dog in a house. $\leftarrow\rightarrow$ ES: El gato duerme. \\
% &  \\
& \textit{Push these embeddings apart} \\
\midrule
\multicolumn{2}{c}{\textbf{Parallel Resources Objectives}} \\
\midrule
\textbf{Cross-Lingual Word Recovery} &
EN: A dog in a \textcolor{red}{[MASK]}. \\
& \hspace{8em} $\updownarrow$  \\

& Aligned: ES: Un perro en una casa. \\
& Use aligned “casa” to recover “house” (alignment rather than attention, unlike Cross-Attention MLM) \\ [4ex]
\textbf{Alternating Language Modelling} &
Mixed: A perro in una \textcolor{red}{[MASK]}. \\
& Predict “casa” using context in both languages \\ [2ex]
\textbf{Bidirectional Word Alignment} &
EN: A \hspace{0.5em} dog \hspace{0.5em} in \hspace{0.5em} a \hspace{0.5em} house. \\
& \hspace{2em} $\updownarrow$ \hspace{0.7em} $\updownarrow$ \hspace{1.2em} $\updownarrow$ \hspace{1.2em} $\updownarrow$ \hspace{0.8em} $\updownarrow$ \\ 
& ES: Un \hspace{0.7em} perro \hspace{0.7em} en \hspace{1.2em} una \hspace{0.9em} casa. \\ [1ex]
& Model learns token-to-token alignments in both directions \\
\bottomrule
\addlinespace[1ex]
\end{tabular}
\caption{Illustrative examples of multilingual training objectives using the sentence pair \{EN: \textit{A dog in a house}, ES: \textit{Un perro en una casa}\}, following the same classification as Table \ref{tab:mllm_obj}.}
\label{table:mlm_objectives_examples}
\end{table}

\subsubsection{The Curse of Multilinguality}

\citet{conneau_2020} describe the ``curse of multilinguality," where the inclusion of additional languages during pre-training improves performance up to a point, after which both monolingual and cross-lingual performance declines. This reflects a trade-off between expanding language coverage and maintaining per-language capability. Proposed solutions include parameter sharing across pre-training languages such as Cross-lingual Modular \cite{pfeiffer_2022}, or language-specific parameter subsets such as those proposed by \citet{blevins_2024} for Cross-lingual Expert Language Models (X-ELM). Another approach is that employed by \citet{artetxe_2020}, who train with a masked objective in one language before extending to a new language via an embedding matrix that freezes earlier parameters, ensuring stability while enabling transfer to new languages.

\subsection{Embeddings}

Transformer-based models have become central to information retrieval due to their ability to produce contextualised text embeddings. Unlike co-occurrence-based representations, embeddings encode deep semantic information, enabling cross-lingual comparison without explicit translation. Early work relied on static embeddings and alignment techniques, supervised~\cite{mikolov2013} or unsupervised~\cite{lample2018}, which enabled bilingual lexicon induction and translation by mapping monolingual embeddings into a shared space.

The introduction of textual models such as BERT \cite{devlin_2019_bert} marked a major shift. Multilingual variants like mBERT~\cite{devlin_2019_bert} extended pre-training to dozens of languages, showing promising zero-shot transfer, though scaling to many languages often reduced performance due to the aforementioned curse of multilinguality \cite{conneau_2020, chang-etal-2024-multilinguality}. Models such as XLM \cite{conneau_lample_2019} addressed this by incorporating translation-based objectives like Translation Language Modelling (i.e. leveraging sentence-level parallel data to strengthen cross-lingual transfer), while RoBERTa \cite{liu2019robertarobustlyoptimizedbert} introduced architectural improvements, leading to XLM-RoBERTa \cite{conneau_2020} as a strong multilingual baseline.

These models are particularly impactful for low-resource languages, supporting zero-shot \cite{Wang2023SALT} and few-shot \cite{Park2024LoResMT, Asai2023BUFFET} transfer, and enabling applications with  limited data \cite{Bafna2023}. Multilingual embeddings have proven effective for cross-lingual retrieval, surpassing translation-based methods \cite{chiu2025translate_multilingual, litschko2021unsupervised}. By embedding text from multiple languages into a shared space, they allow direct comparison of meaning across languages \cite{vaj2023laser, primer2022eval} without the need for translation. In CLIR, this enables retrieval of semantically relevant documents across languages based on embedding similarity rather than parallel corpora or keyword overlap~\cite{feng2020labse, zhao2020inducing}.

\begin{figure}
    \centering
    \includegraphics[width=0.4\linewidth]{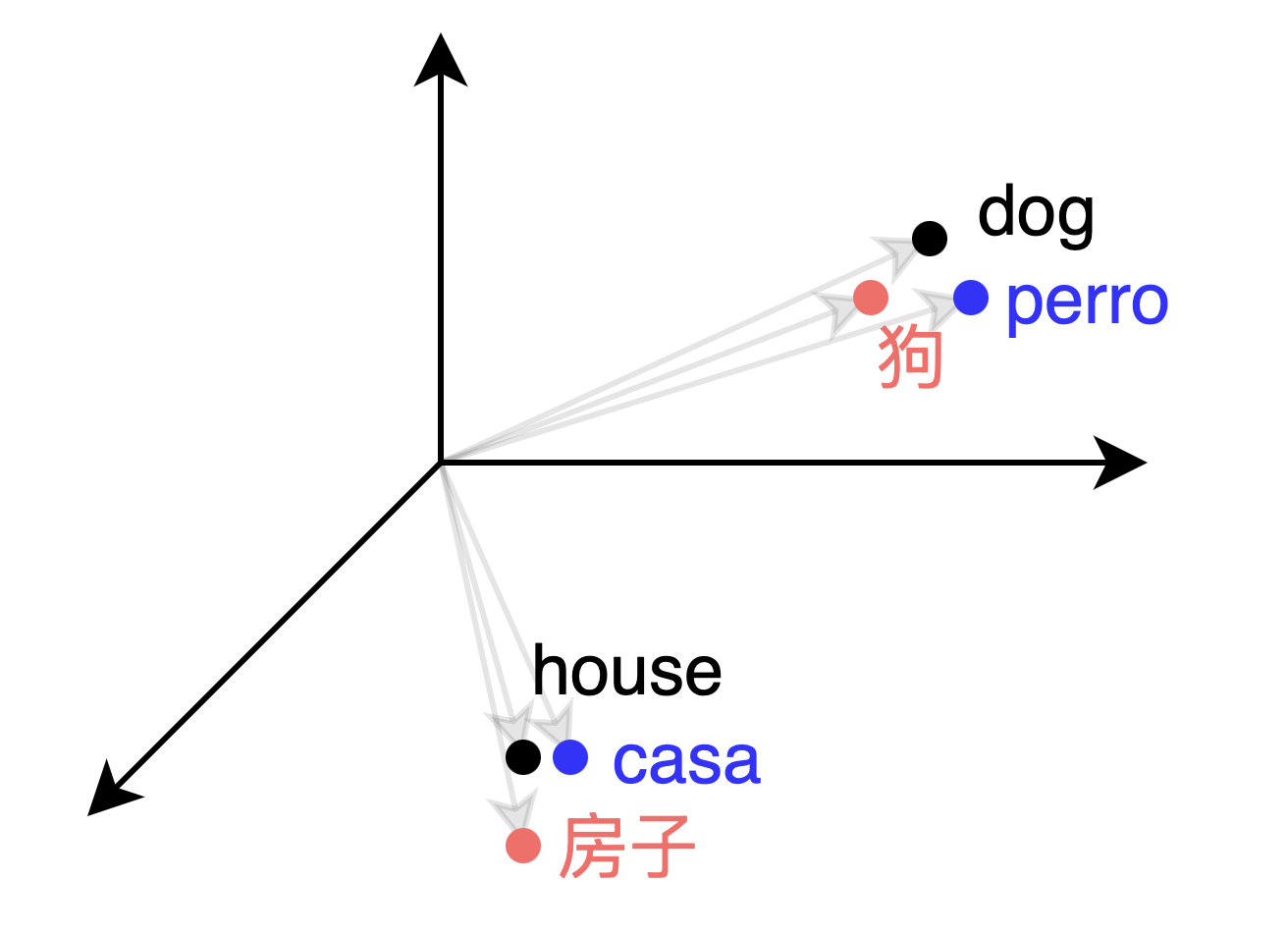}
    % \caption{Simple diagram representation of a multilingual embedding space, showing embeddings corresponding to the same concepts close together, regardless of language}
    \caption{Simplified representation of a multilingual embedding space, highlighting how semantically equivalent concepts are mapped to nearby positions regardless of language.}
    \label{fig:embedding}
\end{figure}
\label{sec:embedding}

\paragraph{Creating Multilingual Embeddings.} The most common approach to constructing multilingual embeddings is \textit{joint pre-training}, where a model is trained on multilingual corpora using Masked Language Modelling \cite{ruder_2019}. This yields language-independent representations by exploiting shared structural and lexical regularities. The quality of embeddings depends on training data, with alignment enhanced via parallel corpora at the word 
\cite{luong_2015}, sentence \cite{gouws_2015}, or domain \cite{vulić2016} level. Large bilingual dictionaries further improve semantic alignment \cite{duong_2016}. The aim is to map semantically equivalent concepts to nearby positions in the embedding space, as visualised in Figure \ref{fig:embedding}.

Before contextual embeddings, cross-lingual retrieval relied on mapping functions aligning monolingual word embeddings. Early work applied linear transformations from bilingual dictionaries \cite{mikolov2013}, while unsupervised methods such as MUSE \cite{conneau2017_wordtranslation} used adversarial learning. Later models refined alignment iteratively \cite{chen-cardie-2018-unsupervised}, enabling applications in lexicon induction and retrieval, but remained limited by domain differences and pairwise mapping requirements \cite{sogaard-etal-2018-limitations}. This motivated the shift toward multilingual pre-training and contextual embedding spaces.
 
Even with shared spaces, mismatches often occur due to style or structural differences, weakening similarity measures. To address this, mapping functions project queries and documents into shared spaces using methods such as student-teacher training \cite{wang-hong-2023}, geometric alignment \cite{kim2023}, or post-hoc projections \cite{tashu_2025}. These lightweight techniques provide efficient alternative to full fine-tuning or translation-based retrieval.

\subsection{Alignment in CLIR} 
\label{sec:alignment}

Effective alignment of embedding spaces is central to CLIR. It ensures semantically equivalent content across languages, whether this be queries, documents, or other distributions, is mapped to comparable vector representations. This enables retrieval without explicit translation, parallel corpora, or shared vocabulary. Cross-lingual alignment positions semantically similar words, phrases, or sentences from different languages close in the embedding space (Figure \ref{fig:embedding}), ideally abstracting away surface-level differences (e.g. syntax, orthography) in favour of shared meaning.

Alignment is particularly valuable as it allows models trained in one language to generalise across others. Even fine-tuning on monolingual query-document pairs can update monolingual embedding spaces, enabling transfer across languages and reducing the need for language-specific supervision, which is especially critical for low-resource settings \cite{reimers2020making, artetxe2018laser, litschko2022cross, zhao2020leveraging}.

However, alignment does not naturally arise from multilingual pre-training alone. Models trained on large multilingual corpora often develop language-specific subspaces. Thus, additional mechanisms are required to ensure semantically equivalent content aligns across languages.

\subsubsection{Levels and Methods of Alignment}

Alignment can be achieved at different granularities. Word-level alignment maps individual lexical tokens but is limited by lexical gaps and cultural specificity \cite{ruder_2019, conneau2017_wordtranslation}. Sentence-level alignment instead captures entire sentences or phrases, preserving semantic meaning while mitigating lexical mismatches. Tools like Awesome-Align \cite{dou2021word} and SimAlign \cite{jalili-sabet-etal-2020-simalign} exploit parallel corpora, while parallel-data objectives such as TLM in XLM \citep{conneau_lample_2019}, where aligned sentences are concatenated and masked so the model learns from both monolingual and cross-lingual context, or LASER, \cite{artetxe2018laser} enhance alignment through encoder-based architectures.

Contrastive learning has emerged as a scalable alternative when parallel data are scarce. It optimises directly over paired inputs: semantically equivalent pairs are pulled closer, while dissimilar ones are pushed apart. Applied to CLIR, this often involves query-document pairs (positive/negative) generated via translation or pseudo-parallel data \cite{gao2022cocondense, lo2023clwic}. Objectives such as triplet loss \cite{schroff2015facenet}, InfoNCE~\cite{oord2018representation}, NT-Xent~\cite{chen2020simple}, and Multiple Negatives Ranking Loss~\cite{henderson2017efficient, reimers2019sentencebert} improve efficiency and scalability, making them well-suited for CLIR. Models like LaBSE~\cite{feng2022language} and mSBERT~\cite{reimers2020making} apply contrastive losses for multilingual sentence embeddings, achieving zero-shot transfer. ALIGN~\cite{jia2021scaling} further scales contrastive learning to massive multilingual datasets, demonstrating generalisation across modalities and languages, while CoConDenser~\cite{gao2022cocondense} combines contrastive pre-training with dense retriever fine-tuning for stronger retrieval benchmarks.

\begin{figure}[h!]
    \centering
    \includegraphics[width=0.65\linewidth]{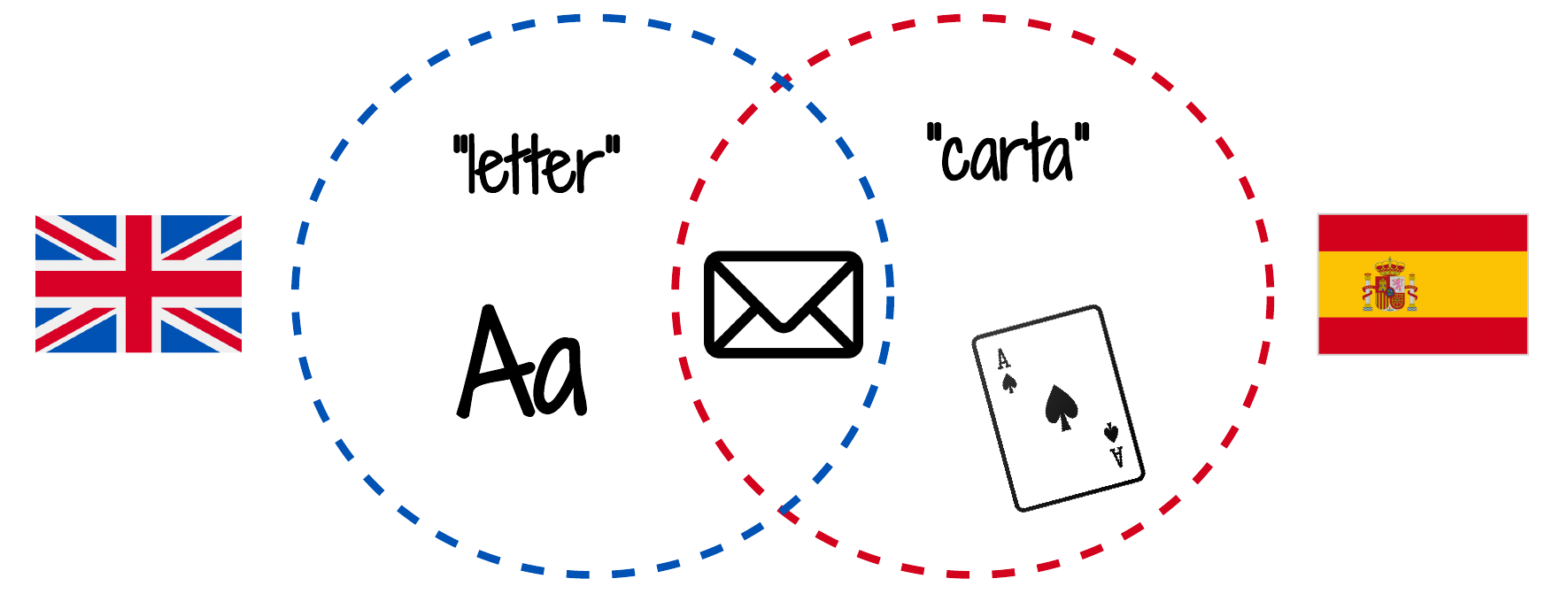}
    \caption{Cross-lingual polysemy mismatch example using the word pair \{EN: \textit{letter}, ES: \textit{carta}\}.}
    \label{fig:polysemy}
\end{figure}

Generative Adversarial Networks (GANs) have also been used to map source embeddings into target-language spaces \cite{fu2020absent}. Early methods like MUSE~\cite{conneau2017_wordtranslation} showed unsupervised adversarial alignment could rival supervised baselines. Yet GAN-based methods remain unstable and less effective in CLIR due to difficulties in handling polysemy (see for example English--Spanish polysemy mismatch illustrated in Figure \ref{fig:polysemy}), compositional semantics, and high-dimensional instability \cite{sogaard-etal-2018-limitations}. These
methods typically assume that the source and target embedding spaces are approximately isomorphic, an assumption
that holds reasonably well for similar languages with comparable corpora and training objectives, but often fails in
practice for distant languages or mismatched domains , leading to degraded alignment quality and limited transfer
effectiveness \cite{sogaard-etal-2018-limitations}.

%\begin{figure}
%    \centering
%    \includegraphics[width=0.5\linewidth]{figures/contrastive_learning_Spanish.png}
%    \caption{Example of English-Spanish positive and negative sentence pairs for contrastive learning.}
%    \label{fig:enter-label}
%\end{figure}

%\begin{figure}
%    \centering
%    \includegraphics[width=0.4\linewidth]{figures/language_agnostic.png}
%    \caption{Illustrative example of a language-agnostic embedding. }
%    \label{fig:enter-label}
%\end{figure}

\subsubsection{Query-Document Alignment} 

Unlike cross-lingual alignment, aligning queries with documents is complicated by their inherent asymmetry. Queries are short, focused, and varied, while documents are longer, diverse, and often cover multiple topics. A single document may serve numerous queries, but a query typically has only a few relevant documents. Consequently, strict embedding alignment fails to capture the necessary contextual nuances. To address this, dual-encoder models use separate query and document encoders while training to align them in a shared space. Dense Passage Retrieval  \cite{karpukhin-etal-2020-dense} exemplifies this by jointly encoding queries and answers, while later models like ANCE~\cite{xiong2021approximate} and TAS-B~\cite{hofstatter2021efficient} improve via dynamic negative sampling. These approaches aim to optimise embeddings for semantic similarity rather than enforcing exact pairwise alignment.

A central issue in query-document alignment is whether to: (i) map queries into the document embedding space (the more common approach) - queries are often expanded or transformed to better resemble documents, thereby reducing style and length mismatches (see Section~\ref{sec:query-expansion}); or (ii) map documents into the query embedding space. The latter is less common, but useful when aiming to directly align document semantics with short, focused queries.

Through LLM-based generation, this idea extends to producing pseudo-documents from queries or pseudo-queries from documents (see Section~\ref{sec:mllms}). In both directions, the objective is to minimise the gap between generated text and its true match. Another promising direction is fine-tuning embeddings of generated text so they align directly with those of their correct counterparts.

\subsubsection{Is Alignment Necessary?}  
While cross-lingual alignment is a desirable property, it is neither always necessary nor sufficient for transfer of model capabilities in tasks like CLIR. Fine-tuning may even weaken initial alignment \cite{chronopoulou2019embarrassingly}, and alignment itself can reinforce cultural or linguistic biases inherited from high-resource languages \cite{lauscher2020limitations}. Instead, CLIR performance often benefits more from representations that preserve semantic distinctions and disambiguate meaning in context. Effective retrieval depends on resolving polysemy, retaining query-relevant information, and ensuring embeddings capture objective-specific semantics, even if perfectly co-located alignment is absent~\cite{philippy-etal-2023-towards, raffel2020exploring}. 

Thus, what ultimately matters for CLIR is not strict geometric uniformity but whether embedding proximity reflects semantic relevance. Approximate, task-sensitive alignment combined with contextual understanding often proves more valuable than perfect alignment.

\section{Evaluation}
\label{sec:evaluation}

Since CLIR systems are composed of multiple components, evaluation can be conducted either component-wise or end-to-end. Full-system evaluation is essential for assessing performance on the overall task of cross-lingual retrieval \cite{thellmannetal2024}, spanning RAG, QA, and domain-specific retrieval. Evaluation employs diverse datasets and metrics, including those from machine translation, monolingual IR, multilingual generation, and embedding alignment. While some datasets provide gold labels for cross-lingual relevance, many rely on proxy or reference-free metrics. Metrics may target lexical overlap, semantic similarity, or ranking quality, depending on the evaluation goal. A range of CLIR benchmarks support this, covering multiple retrieval settings. This section reviews the most relevant datasets and evaluation practices for CLIR, spanning both system-level and component-level assessment.

\subsection{Translation Evaluation}

In CLIR, translation is not assessed for readability or fluency, but for retrieval effectiveness. Thus, the focus shifts from translation accuracy to its impact on retrieval. Traditional metrics like Bilingual Evaluation Understudy (BLEU)\footnote{https://huggingface.co/spaces/evaluate-metric/bleu}, which evaluates n-gram overlap, remain widely used but have limitations, such as poor handling of synonym and word order. Alternatives include the Metric for Evaluation of Translation with Explicit Ordering (METEOR)\footnote{https://huggingface.co/spaces/evaluate-metric/meteor} score, which correlates better with human judgements, and Translation Edit Rate (TER)\footnote{https://huggingface.co/spaces/evaluate-metric/ter}, which measures edit distance to reference translations.

Recent neural metrics leverage pre-trained language models, such as BLEURT (Bilingual Evaluation Understudy with Representations from Transformers)\footnote{https://huggingface.co/spaces/evaluate-metric/bleurt} and COMET (Crosslingual Optimized Metric for Evaluation of Translation)\footnote{https://huggingface.co/spaces/evaluate-metric/comet}. These better capture semantic adequacy and fluency. For retrieval-oriented evaluation, recall is generally prioritised over precision, as the goal is to ensure relevant documents are retrieved, minimising false negatives. Although handling irrelevant documents remains important, precision is usually secondary in CLIR~\cite{galuščáková2022}.

\subsection{Datasets for CLIR}

A major challenge in CLIR is dataset availability, particularly for low-resource languages. Table \ref{tab:datasets} lists multilingual datasets for CLIR and cross-lingual QA. Large-scale CLIR datasets covering diverse topologies include CLIRMatrix~\cite{sun_2020_clirmatrix}, Large-scale CLIR Dataset \cite{sasaki_2018}, SWIM-IR~\cite{thakur_2024_swimir}, XOR-TyDi QA \cite{asai_2021_qa}, LAReQA~\cite{roy_2020_lareqa} and mMARCO~\cite{bonifacio_2022_mmarco}. Other datasets focus on specific languages (e.g. WikiCLIR \cite{schamoni_2014}, NeuMARCO~\cite{lawrie2024overview}, CLIRudit \cite{valentini2025clirudit}), or regions (e.g. AfriCLIRMatrix~\cite{ogundepo_2022_africlirmatrix}, CIRAL~\cite{adeyemi_2024_ciral}). BordIRlines~\cite{li2024bordirlines} targets geopolitical disputes, while MIRACL~\cite{zhang_2023_miracl} and other monolingual IR datasets contain diverse languages but are not CLIR-specific. Cross-lingual resources include Mr. TyDi~\cite{zhang_2021_mrtydi}, XQA~\cite{liu_2019_xqa}, MLQA~\cite{lewis_2020_mlqa}, XQuAD~\cite{artetxe_2020}, TyDi QA~\cite{clark_2020_tydiqa}, MKQA~ \cite{longpre_2021_mkqa}, and XRAG \cite{liu2025xrag}.

Most CLIR datasets are derived from Wikipedia, which offers a broad coverage but limited domain diversity and data variation. As with LLM training data, English dominates, often serving as the query or document language (e.g. AfriCLIRMatrix~\cite{ogundepo_2022_africlirmatrix}, NeuMARCO~\cite{lawrie2024overview}, WikiCLIR~\cite{schamoni_2014}, XOR-TyDi QA~\cite{asai_2021_qa}). This reliance highlights the cross-lingual imbalance, with English functioning as the basis for transfer to lower-resource languages.

Many multilingual datasets extend monolingual English IR/QA corpora, such as TyDi QA ~\cite{clark_2020_tydiqa} and MS MARCO~\cite{bajaj_2018_msmarco}. The latter is notable as it is created using human-generated text, whereas most others rely on automatically created content. Increasingly, datasets covering additional languages use LLM-aided generation and machine translation, producing query-document triplets (query, positive, negative). These methods support query expansion, training augmentation, and evaluation particularly valuable for low-resource languages. Finally, large-scale multilingual parallel corpora (e.g. OPUS\footnote{https://opus.nlpl.eu}) are widely used to train MT systems and LLMs, though they fall outside strict CLIR scope.

\begin{table}[h!]
\centering
\begin{tabular}{@{}p{3.1cm}p{1.5cm}p{7.7cm}p{2.7cm}@{}}
\toprule
\textbf{Dataset} & \textbf{Application} & \textbf{Description} & \textbf{Data source}\\ \midrule

\textbf{CLIRMatrix} \cite{sun_2020_clirmatrix} & CLIR & Two IR datasets: bilingual dataset in 139 languages (BI-139) and multilingual dataset in 8 languages (MULTI-8) & Wikipedia  \\
\textbf{AfriCLIRMatrix} \cite{ogundepo_2022_africlirmatrix} & CLIR & English queries with relevance judgements for documents in 15 African languages & Wikipedia \\
\textbf{Large-Scale CLIR Dataset} \cite{sasaki_2018} & CLIR & English queries, relevant documents in 25 other languages & Wikipedia \\
\textbf{SWIM-IR} \cite{thakur_2024_swimir} & CLIR & Query-passage pairs in 33 languages, queries generated by PaLM-2 from Wikipedia passages using summarise-then-ask prompting & Wikipedia \\
\textbf{CIRAL} \cite{adeyemi_2024_ciral} & CLIR & English queries, documents in four African languages (Hausa, Swahili, Somali and Yoruba) & News articles \\
\textbf{WikiCLIR} \cite{schamoni_2014} & CLIR & German queries and English documents & Wikipedia \\
\textbf{XOR-TyDi QA} \cite{asai_2021_qa} & CLIR & Expanded TyDi QA dataset into 7 typologically diverse languages, documents in English & Wikipedia \\
\textbf{LAReQA} \cite{roy_2020_lareqa} & CLIR & Convert XQuAD and MLQA into answer retrieval tasks (XQuAD-R and MLQA-R) in 11 languages & Wikipedia \\
\textbf{NeuMARCO} \cite{lawrie2024overview}  & CLIR & English queries with documents from MS MARCO translated into Chinese, Persian, and Russian (machine translated) & Human generated \\
\textbf{mMARCO} \cite{bonifacio_2022_mmarco} & CLIR & Multilingual version of MS MARCO in 13 languages (machine translated) & Human generated \\
\textbf{CLIRudit} \cite{valentini2025clirudit} & CLIR & French and English research articles with English queries and French documents & Érudit \\

\textbf{BordIRlines} \cite{li_2025_rag} & CL-RAG & Queries about geopolitical disputes, documents in languages covering all claimant countries for each territory (251 disputes, 720 queries, 49 languages) & Wikipedia  \\
\textbf{MIRACL} \cite{zhang_2023_miracl} & IR & Relevance judgements for documents in 18 languages from 10 language families for monolingual retrieval & Wikipedia \\
\textbf{Mr. TyDi} \cite{zhang_2021_mrtydi} & IR & Question-passage pairs 11 typologically diverse languages for monolingual retrieval & Wikipedia \\

\textbf{XQA} \cite{liu_2019_xqa} & CL-QA & Questions, answers and top 10 retrieved articles in 9 languages & Wikipedia \\
\textbf{MLQA} \cite{lewis_2020_mlqa} & CL-QA & 7 languages (English, Arabic, German, Spanish, Hindi, Vietnamese and Simplified Chinese), each instance parallel between 4 languages on average & Wikipedia \\
\textbf{XQuAD} \cite{artetxe_2020} & CL-QA & Translated subset of SQuAD v1.1 into ten languages (Spanish, German, Greek, Russian, Turkish, Arabic, Vietnamese, Thai, Chinese, and Hindi) & Wikipedia \\
\textbf{TyDi QA} \cite{clark_2020_tydiqa} & CL-QA & Question-answer pairs in 11 typologically diverse languages & Wikipedia \\
\textbf{MKQA} \cite{longpre_2021_mkqa} & CL-QA & Question-answer pairs in 26 typologically diverse languages & Web/human generated \\
\textbf{XRAG} \cite{liu2025xrag} & CL-RAG & Multilingual RAG across 5 languages, supporting both document retrieval and generation tasks; includes natural queries and Wikipedia passages & Wikipedia  \\

%\textbf{CCMatrix} & MT & & \cite{schwenk_2020_ccmatrix} \\
%\textbf{CCAligned} & MT & & \cite{elkishky_2020_ccaligned} \\
%\textbf{} & & & \\

\bottomrule
\addlinespace[1ex]
\end{tabular}
\caption{Multilingual datasets for information retrieval and question answering.}
\label{tab:datasets}
\end{table}

\subsection{CLIR Evaluation Metrics} 
\label{sec:metrics}

Evaluating CLIR systems typically relies on the Cranfield paradigm, which defines a test collection by fixing a document corpus, a set of query topics, and relevance judgments, enabling reproducible, comparable assessment of retrieval models \cite{cleverdon1962report}. Recall often takes precedence over precision or F$_1$ scores, as identifying all relevant documents is more critical than excluding irrelevant ones. Rank-aware measures such as Mean Reciprocal Rank (MRR)~\cite{voorhees1999trec} and nDCG~\cite{jarvelin2002cumulated} build on recall/precision to reward early retrieval of relevant results \cite{jurafsky2025speech}. CLIR adds further evaluation complexities: resource-imbalances and over-looked low-resource languages introduce potential biases, while translation or embedding quality can significantly affect retrieval performance. Although standard IR metrics remain central for end-to-end evaluation, complementary measures are also used, such as translation fluency scores, back-translation retrieval, or embedding alignment, even if they are not typically employed as the final evaluation criteria.

\subsubsection{Retrieval Metrics}

Among retrieval metrics, Hit Ratio@K (Hit@K)~\cite{voorhees1999trec} measures whether at least one relevant document appears in the top \textit{K} results for a query, focusing on binary relevance.  Recall@K~\cite{manning2008introduction} computes the fraction of all relevant documents returned in the top \textit{K}, thus highlighting completeness. MRR~\cite{voorhees1999trec} calculates the inverse rank of the first relevant document for each query and averages across queries, rewarding early retrieval. 
MAP~\cite{manning2008introduction} averages precision scores across all relevant documents and queries, favouring systems that rank relevant documents earlier. Discounted Cumulative Gain (DCG)~\cite{jarvelin2002cumulated} assigns higher weights to relevant documents retrieved earlier, using a logarithmic discount. Its normalised variant, nDCG~\cite{jarvelin2002cumulated}, divides DCG by the ideal DCG, allowing fair comparison across queries with different numbers of relevant documents.

MRR and MAP are especially useful when early retrieval matters, while nDCG is popular in multilingual and graded-relevance contexts (e.g. cross-lingual QA), as it considers both rank position and degree of relevance. In recent years, nDCG and MRR have emerged as preferred metrics for evaluating dense and neural retrieval systems due to their robustness in handling ranking nuances and graded relevance judgements.

%\begin{table}[ht]
%\centering
%\begin{tabular}{@{}p{3cm}p{13cm}@{}}
%\toprule
%\textbf{Metric} & \textbf{Description} \\ \midrule

%\textbf{Hit@K} \cite{voorhees1999trec} & Measures whether at least one relevant document appears in the top \(K\) results for a given query; focuses on binary relevance. \\

%\textbf{Recall@K} \cite{manning2008introduction} & Computes the fraction of all relevant documents returned in the top \(K\) results; highlights completeness. \\

%\textbf{MRR} \cite{voorhees1999trec} & Calculates the inverse rank of the first relevant document for each query and averages over all queries; rewards early precision. \\

%\textbf{MAP} \cite{manning2008introduction} & Averages precision scores across all relevant documents and all queries; favours rankings that surface more relevant items earlier. \\

%\textbf{DCG} \cite{jarvelin2002cumulated} & Measures the utility of a ranking by assigning higher weights to relevant documents retrieved earlier; uses logarithmic discounting. \\

%\textbf{nDCG} \cite{jarvelin2002cumulated} & Normalises DCG by dividing by the ideal DCG, allowing comparison across queries with different numbers of relevant documents. \\

%\bottomrule
%\end{tabular}
%\caption{Comparison of common retrieval evaluation metrics and their core properties.}
%\label{tab:retrieval-metrics}
%\end{table}

\subsubsection{Cross-lingual Performance Metrics}

Measuring the cross-lingual ability of language models introduces challenges as models must handle different scripts, linguistic structures, and cultural contexts. General multilingual benchmarks such as M-RewardBench \cite{gureja2025mrewardbenchevaluatingrewardmodels}, INCLUDE \cite{romanou2024includeevaluatingmultilinguallanguage}, and Global-MMLU \cite{singh_2025} address these aspects while differing in task format but sharing a common evaluation metric of accuracy. M-RewardBench evaluates reward models across 23 languages by testing whether they correctly prefer human-preferred responses in paired comparisons spanning chat, safety, reasoning, and translation. INCLUDE tests knowledge and reasoning using about 197k multiple-choice questions sourced from local exams in 44 languages, scoring models by the fraction of correct answers. Global-MMLU extends the original MMLU into 42 languages with professional translation and annotation, using multiple-choice accuracy to assess model understanding and offering additional analysis on culturally sensitive versus agnostic subsets.

Beyond benchmark accuracy, specialised metrics assess multilingual models at finer levels, with and without gold-standard labels. Translation and generation metrics include BLEU~\cite{papineni2002bleu}, which as mentioned above measures n-gram precision between system output and reference with a brevity penalty; ROUGE~\cite{lin2004rouge}, which emphasises recall of n-grams, subsequences, and skip-bigrams often used in summarisation; METEOR \cite{banerjee2005meteor}, which incorporates stemming, synonym matching, and alignment-based precision and recall for stronger correlation with human judgements; chrF~\cite{popovic2015chrf}, a character level F-score suitable for morphologically rich languages; and TER~\cite{snover2006study}, which computes the number of edits required to change a hypothesis into a reference.

Embedding alignment metrics target semantic coherence across languages. Backretrieval~\cite{fain2021backretrieval} measures alignment quality by checking whether multilingual captions retrieve the same image, while MEXA~\cite{kargaran2024mexa} evaluates how well non-English sentences align with English-centric representations in multilingual models. Learned metrics rely on pre-trained encoders to approximate human judgements. BERTScore~\cite{zhang2019bertscore} computes token-level semantic similarity using contextual embeddings from BERT, COMET~\cite{rei2020cometneuralframeworkmt} trains a regression model on pre-trained encoders to predict human judgement scores, and BLEURT~\cite{sellam2020bleurt} fine-tunes BERT to estimate human-likeness scores based on reference comparisons.

In practice, standalone cross-lingual metrics are rarely applied directly in CLIR evaluation beyond development and diagnostic stages. Most deployed CLIR methods depend on standard retrieval metrics applied to multilingual benchmarks. Translation-quality metrics often inform CLIR evaluation by serving as extrinsic measures of query or document translation quality. Alignment-focused metrics are also used for probing multilingual embedding coherence or representation quality, typically in ablation studies or auxiliary evaluations with pivots such as English. Toolkit-based proxies such as CLIReval~\cite{sun2020clireval} attempt to bridge the gap by adapting machine translation evaluation datasets into retrieval tasks, binding translation-oriented metrics to retrieval performance. Overall, while cross-lingual metrics support model development, alignment analysis, and translation-tuning, CLIR evaluation remains primarily centred on retrieval-focused performance.

\subsubsection{Generation Evaluation Metrics}

For question answering, model performance is often evaluated through human judgments, typically using crowdworkers, though they may lack the expertise to assess factuality and other qualities accurately \cite{gillick_2010, iskender_2020}. A/B testing is common, with annotators comparing answers (e.g. HURDLES \cite{krishna_2021_hurdles}, WEBGPT \cite{nakano_2022_webgpt}), while some studies instead rely on domain experts \cite{xu_2023_QAeval}, though they are harder to obtain. Human annotators typically emphasise attributes such as relevance, factuality, and ease of understanding, which are difficult to capture automatically, meaning there is no single comprehensive metric that does not require human annotators. Existing automatic metrics, often adapted from summarisation tasks rather than designed for QA, include measures like ROUGE \cite{lin2004rouge}, BERTScore \cite{zhang2019bertscore} and BLEURT \cite{sellam2020bleurt}, which require human-written references and are limited for long-form QA due to the diversity of valid answers \cite{krishna_2021_hurdles, wang_2022}. Automatic metrics frequently fail to align with human judgments \cite{xu_2023_QAeval}, tending to capture narrow aspects such as fluency or query relevance, while factuality remains especially difficult, with some approaches borrowing from summarisation faithfulness metrics such as QAFactEval \cite{fabbri_2022_qafacteval}. Additional metrics include Self-BLEU \cite{zhu_2018_selfbleu}, which measures the diversity of generated text, and Perplexity \cite{jelinek_1977_perplexity}, which evaluated linguistic fluency. Particularly pertinent to CLIR, some metrics instead capture the relevance of a question to a given answer (e.g. RankGen \cite{krishna_2022_rankgen}, BARTScore \cite{yuan_2021_bartscore} and Question Likelihood \cite{ponte_1998_likelihood}) - these can be useful for QA in information retrieval.

\section{Applications} 
\label{sec:applications}

\begin{figure}[h!]
    \centering
    \includegraphics[width=0.5\linewidth]{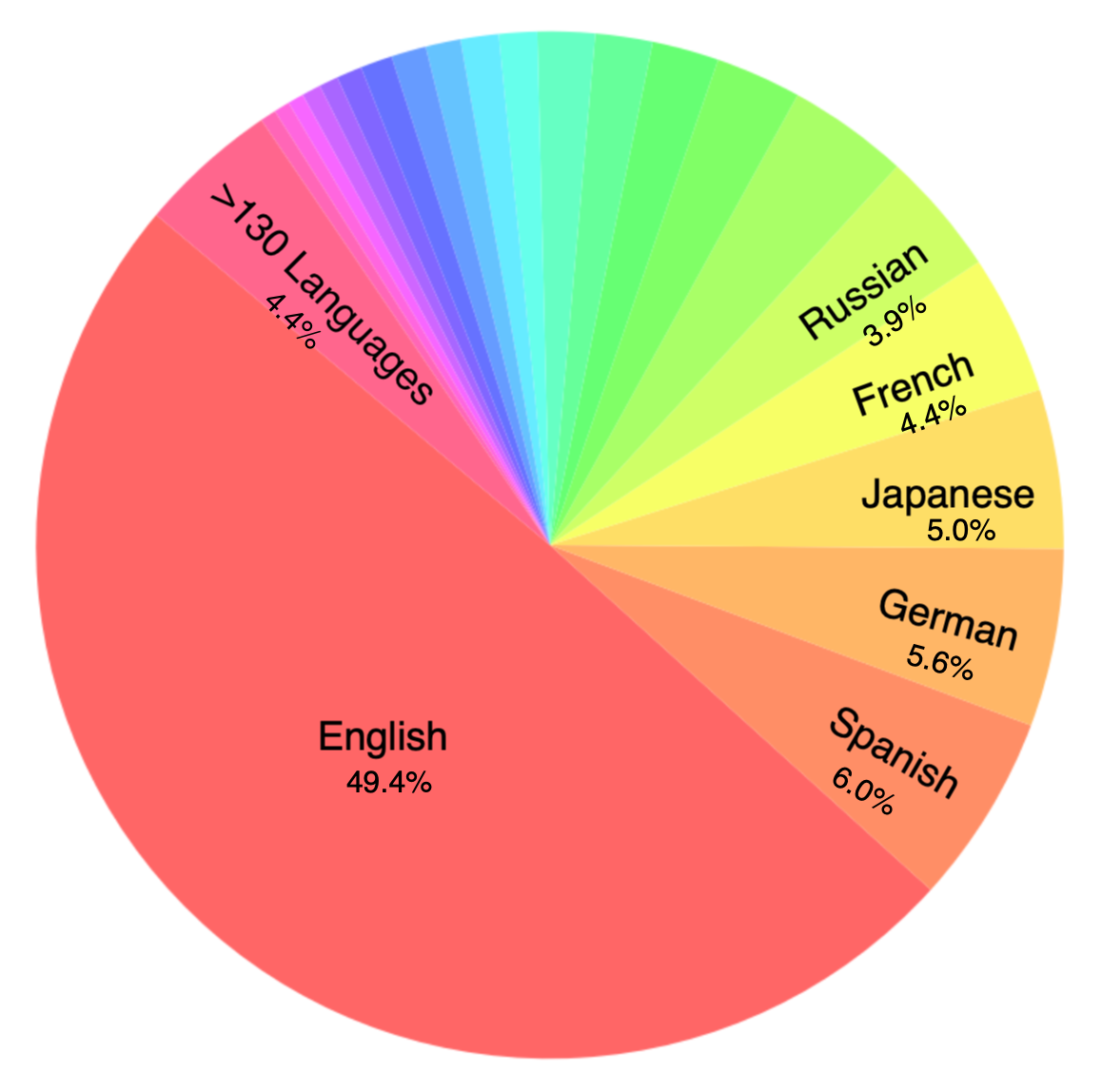}
    \caption{Share of websites by content language as of February 2025. Data represents the most frequently used languages for web content, based on analysis published by Statista \cite{statista_2025}.
}
    \label{fig:lang-share}
\end{figure}

CLIR is essential whenever queries, documents, or both queries and documents appear in multiple languages. Since English dominates online content (English makes up 49.1\% of websites as of February 2025, followed by Spanish (6.0\%), German (5.6\%), and Japanese (5.0\%) \cite{statista_2025}, see Figure \ref{fig:lang-share}), users of low-resource languages face significant barriers \cite{w3techs_content_language}. CLIR addresses this imbalance by enabling access to information beyond a user's native language.

\paragraph{Search Engines.} To provide relevant results, search engines must retrieve content in higher-resource languages and present it in accessible ways \cite{foo2011cross}. Cross-lingual search on Google underperforms compared to monolingual search \cite{foo2011cross}, though syntactic analysis improves results \cite{semmar2006lic2m}. Tools such as Google Translate and AI summaries for search engines highlight the potential of CLIR for translation and summarisation \cite{chen2009cross}.

\paragraph{Specialised Databases.} Professionals in fields such as law and medicine rely on technical resources often available only in English. CLIR enables access without requiring duplicate content in multiple linguistic versions. Examples include English–Persian retrieval~\cite{rahmani2017adapted} and Hindi–English medical retrieval using morphology and query expansion \cite{sharma2021semantic} .

\paragraph{LLMs and Question Answering.} Retrieval-augmented generation systems benefit from CLIR by retrieving relevant information from multilingual sources and generating answers independent the source language. This makes them effective in multilingual environments where queries and evidence span different languages \cite{ranaldi2025multilingual}. At the same time, benchmarks point to challenges in cross-lingual QA and reasoning \cite{li2024bordirlines, liu2025xrag}, underlining areas where further progress is needed.

\paragraph{News, Media and Security.} Journalists, researchers, and businesses use CLIR to monitor global events and gather multilingual insights. Benchmarks such as NeuCLIR~\cite{lawrie2024overview},  xMIND~ \cite{iana2024mind}, and CLSD~\cite{maurer2025clsd} evaluate news retrieval, while MMTweets~\cite{singh2023mmtweets} and MultiClaim~\cite{pikuliak2023multiclaim} assess fact-checking. These functions are equally important in crisis response and security, where timely access to local protocols, reports, and threat indicators supports situational awareness and the detection of terrorism and cyber activity \cite{quelle2025lost, singhal2024translationbias, vitiugin2022cross, sanchez2022cross,lamsal2024se,vitiugin2022crosslang, yin2025disastir, ranade2018using, markov2005identification}

\paragraph{Scientific Research.}
English dominance in academic publishing creates obstacles for non-English-speaking researchers. CLIR enables access to international literature without requiring translation into English, fostering collaborations and reducing duplications. Benchmarks such as CLIRudit~\cite{valentini2025clirudit} and OPTICAL~\cite{huang2023optimal} address retrieval over academic documents, while studies track misinformation diffusion and evaluate claim verification across languages.

\paragraph{E-commerce.} E-commerce platforms apply CLIR to help customers search product catalogues and reviews in their native language. Benchmarks such as CLPR‑9M~\cite{zhu2022clpr9m} and multilingual ranking systems~\cite{zhang2022evaluating} show improvements in retrieval, while datasets like xPQA~\cite{shen2023xpqa} support cross-lingual product question and answering.

Across domains, CLIR reduces language barriers and ensures equitable access to information. It enables accurate retrieval in medicine, law, research, media, security, and commerce, while promoting inclusivity in the digital ecosystem. Timeliness, accuracy, and coverage remain central for credibility \cite{metzger2007credibility}. By moving beyond English dominance, CLIR fosters collaboration and ensures high-quality information is accessible worldwide.

\section{Challenges and Future Directions}
\label{sec:challenges}

\subsection{CLIR Challenges}
CLIR approaches are shaped by advances in multilingual NLP as well as monolingual IR. Challenges from both areas interact and create compound problems. For instance, we have seen that short queries in monolingual IR often contain ambiguity. In CLIR, additional uncertainty arises from translation or embedding, which can shift meaning and lead to the retrieval of irrelevant documents. Multilingual LLMs face further obstacles related to data, linguistic representation, model robustness, and generalisation. Below, we summarise the main challenges.

\paragraph{Cross-Linguality and Language.} Short queries frequently lack context, making them ambiguous and reducing retrieval accuracy. Polysemy and homonymy are especially problematic across languages (as seen above in Figure \ref{fig:polysemy}). Without contextual cues, it is difficult to identify intended meanings, leading to semantic divergence. OOV terms also hinder retrieval, particularly proper nouns or newly-coined technical terms. Large training corpora partly address this, but low-resource languages remain unsupported. Dissimilar character sets add complications, since transliteration is inconsistent \cite{galuščáková2022}. Morphological and syntactic variation across typologically distant languages makes alignment less reliable~\cite{sogaard-etal-2018-limitations, patra-etal-2019-bilingual}. Across languages, morphology is also relevant for the translation direction: translating from a language with a simple morphology to a more morphologically rich one tends to perform poorly compared to the other way around \cite{galuščáková2022}.

% \begin{figure}[h!]
%     \centering
%     \includegraphics[width=0.65\linewidth]{figures/polysemy.pdf}
%     \caption{Cross-lingual polysemy mismatch example using the word pair \{EN: \textit{letter}, ES: \textit{carta}\}.}
%     \label{fig:polysemy}
% \end{figure}

\paragraph{Language Representations.} Languages differ not only in structure and vocabulary but also in semantic priorities. Some concepts are easily expressed in one language but rare or absent in another, especially across semantically distant pairs \cite{sogaard-etal-2018-limitations} (see Table \ref{fig:untranslatable}, which illustrates translation challenges in Japanese and German). These differences hinder the creation of a language-agnostic embedding space. Shared embeddings reduce reliance on translation, but catastrophic forgetting, where a model loses performance on previously learned languages while adapting to new ones, remains a challenge \cite{rust_2021}. Tokenisation also poses difficulties: inadequate segmentation increases OOV issues and fragments representations, especially in under-resourced languages \cite{qin_2025}.

\begin{table}[h!]
\centering
\begin{tabular}{p{5cm} p{10.5cm}}
\toprule
\textbf{Term} & \textbf{Translation Challenge Explanation} \\
\midrule

\textbf{Japanese: \raisebox{-0.2em}{\includegraphics[height=1.2em]{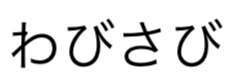}} (Wabi-sabi)} &
\textbf{Literal Components:} \\
& \hspace{1em} • \textit{Wabi (\raisebox{-0.2em}{\includegraphics[height=1.em]{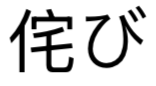}})} — rustic simplicity, quietness, subtle melancholy \\
& \hspace{1em} • \textit{Sabi (\raisebox{-0.2em}{\includegraphics[height=1.em]{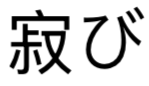}})} — the beauty of ageing, weathering, and impermanence \\
& \textbf{Overall Meaning:} \\
& A worldview centred on the acceptance of transience and imperfection. It celebrates the beauty of things that are humble, weathered, and incomplete - something hard to express succinctly in English. \\
\\
\textbf{German: Waldensamkeit} &
\textbf{Literal Components:} \\
& \hspace{1em} • \textit{Wald} — forest \\
& \hspace{1em} • \textit{Einsamkeit} — solitude, loneliness \\
& \textbf{Overall Meaning:} \\
& A profound, peaceful feeling of solitude and connection with nature experienced while being alone in the forest, more emotional and poetic than a simple “lonely forest” or “solitude in woods”. \\
\bottomrule
\addlinespace[1ex]
\end{tabular}
\caption{Examples of culturally embedded terms that defy literal translation.}
\label{fig:untranslatable}
\end{table}

\paragraph{Data and Resources.} Data imbalance affects every stage of CLIR. High-resource languages dominate pre-training corpora (Figure \ref{fig:lang-share} presents the imbalance of languages across online websites), skewing token representations and alignment. While high-quality parallel corpora are costly to build, machine-generated ones are cheaper but often distort meaning \cite{muennighoff_2023, singh_2025}. Most datasets rely heavily on Wikipedia, enabling strong general-domain performance but weak results in specialised areas such as medicine or law. Dedicated resources for CLIR remain limited.

% \begin{figure}[h!]
%     \centering
%     \includegraphics[width=0.5\linewidth]{figures/pie_chart2.png}
%     \caption{Share of websites by content language as of February 2025. Data represents the most frequently used languages for web content, based on analysis published by Statista \cite{statista_2025}.
% }
%     \label{fig:lang-share}
% \end{figure}

\textbf{Bias.} Multilingual models often perform well on high-resource languages but poorly on low-resource ones, leading to weak embeddings and alignment \cite{levy_2023_bias, cabello_2022, huang2023optimal, litschko2018unsupervised}. Translation-based pipelines mitigate this through pivot languages, but embedding-based CLIR requires uniform representations across languages. Generative models also reflect demographic bias present in training data \cite{naous_2024_bias, touileb_2022}. Bias further affects evaluation, as skewed datasets or metric choices can distort results and give a misleading perception of system effectiveness \cite{Ferrara_2023, cao_2022, sun_2022}.

\paragraph{Evaluation.} Evaluation metrics for CLIR are largely adapted from monolingual IR and often fail to capture cross-lingual challenges. With the rise of LLMs, answer generation becomes central, yet metrics to evaluate it remain insufficient. Hallucinations where models produce inaccurate or nonsensical responses, pose particular risks when users rely on summarises rather than original documents. Trust declines further when fabricated references are included \cite{aksitov_2023}. 

\paragraph{Engineering Challenges.} CLIR involves resource-intensive steps such as translation, cross-lingual embedding, and retrieval. The high inference cost of multilingual LLMs exacerbate this, especially for interactive systems that require speed. Current pipelines combine monolingual and cross-lingual techniques, but subcomponents (e.g. query expansion and ranking) are often designed independently, complicating integration.

CLIR faces persistent challenges particularly due to the combination of applying monolingual information retrieval methods, which assume a uniform and balanced document corpus, to multilingual settings, where data, representations, and performance are often uneven, especially in low-resource languages.
Addressing these interconnected issues is essential for progress. 

\subsection{Future Directions in CLIR}
Cross-lingual information retrieval has seen significant progress in recent years with the rise of multilingual generative models and embedding-based retrieval techniques. This section covers some promising research directions:

\begin{itemize}

\item \textit{Language-agnostic representations.} Advances in contrastive pre-training and multilingual embedding alignment are promising \cite{feng2020labse, gao2022cocondense}, but models still underperform in low-resource and typologically diverse languages \cite{sogaard-etal-2018-limitations}. Future work should explore training objectives that enforce semantic consistency across languages at the sentence and document level.

\item \textit{Low-resource languages.} Addressing data imbalance remains essential \cite{li2024quantifying}. Expanding training and pseudo-parallel corpora across diverse languages would enhance model performance. LLMs also show potential for generating training data, whether for queries, documents, or translation \cite{muennighoff_2023}. 

%\item \textit{End-to-end CLIR.} Traditional CLIR has relied on pipelines of query expansion, ranking and re-ranking. Multilingual generative models can reduce on such pipelines by directly supporting dense retrieval and question answering \cite{feng2020labse, izacard_2022_mcontriever, Zhang_2023_mdpr}.

\item \textit{Multimodal CLIR.} Incorporating images, captions, and speech queries broadens CLIR applications. Expanding to cross-lingual multimodal inputs requires new pre-training strategies, benchmarks and evaluation metrics \cite{samuel_2025,arora_2020}.

%\item \textit{Microtexts.} Short and informal texts, such as social media content, introduce challenges including misspelling, code-switching (switching between languages), and abbreviations. Early progress has been made with dedicated test collections \cite{galuščáková2022m,lawrie_2023}.

\item \textit{Graph-based retrieval.} For complex knowledge-grounded CLIR, graph-based approaches are promising \cite{peng_2024_grag, hu_2024_grag, Procko_2024_grag}. RAG frameworks could be extended to multilingual knowledge graphs for factual grounding and disambiguation.

\item \textit{Misinformation.} As CLIR systems increasingly incorporate generative models, risks of hallucination and inaccurate outputs remain a concern \cite{jurafsky_2025, abe_2025llmQE}. Future research should focus on integrating robust fact-checking and hallucination-aware retrieval mechanisms \cite{aksitov_2023}, as well as developing models that can provide transparent references and assess the reliability of information sources across languages.

\item \textit{Disambiguation.} CLIR ambiguity arises from
short queries and polysemy, which are especially problematic in multilingual contexts. Future systems could allow clarification of user intent through conversational search and build on query expansion methods \cite{rocchio_1965, rocchio_1971, efthimiadis_1996}. 

\item \textit{Benchmarks and metrics.} Continued development of multilingual and multi-domain benchmarks is essential, building on datasets such as MIRACL~\cite{zhang_2023_miracl}. Metrics are required to evaluate semantic drift, retrieval robustness, and QA performance.

\item \textit{Bias and representation robustness.} CLIR systems are susceptible to bias (e.g. stereotypical, cultural, linguistic) due to imbalances in training data, leading to disparities in retrieval performance. Recent work proposes methods such as OPTICAL (Optimal Transport Distillation) to transfer alignment from high-resource to low-resource languages \cite{huang2023optimal} and adversarial training~\cite{litschko2018unsupervised} to reduce embedding biases and improve zero-shot performance \cite{pengetal2025}. Another line of research addresses tokenisation challenges. Universal tokenisers trained across diverse languages have shown improved adaptability to unseen languages \cite{abagyan2025tokenizerruleallemergent}, while cross-lingual vocabulary transfer methods such as trans-tokenisation \cite{remy2024transtokenization} initialise embeddings using semantically related tokens from high-resource languages. These strategies highlight the need for continued exploration of bias reduction and robust representation of learning across diverse linguistic contexts, especially beyond European languages. 

\end{itemize}

Taken together, these directions highlight that although significant progress has been achieved in CLIR, substantial challenges remain. Addressing resource imbalance, bias, misinformation, and disambiguation, while extending systems to multimodal and graph-based contexts, will be key to advancing robust and equitable CLIR. 

\section{Conclusion}
\label{sec:conclusion}

% \paragraph{The Importance of CLIR.}
Cross-lingual information retrieval lies at the intersection of two rapidly growing fields: multilingual representation learning and information retrieval. As NLP technologies become increasingly integrated into everyday tools and services, the ability to access information across language boundaries has become more critical than ever. This survey has presented a comprehensive overview of CLIR systems, covering advances in representation learning, embedding alignment, retrieval techniques, evaluation methods, and system architectures. The discussion underscores that effective CLIR requires more than the adaptation of monolingual approaches or the direct reuse of multilingual models. Progress depends on innovations that address linguistic mismatch, resource imbalance, and the development of reliable evaluation frameworks. At the same time, CLIR plays a vital role in democratising access to knowledge. Since the majority of digital resources are concentrated in a small number of high-resource languages, equitable access relies on systems that are accurate, robust, and language-agnostic, with the potential to expand opportunities in science, education, healthcare, commerce, and many other domains.

Future development must confront persistent challenges. The scarcity of high-quality multilingual data, particularly for low-resource languages, remains a major bottleneck, making inclusive training corpora and stronger evaluation benchmarks essential. Sustained collaboration between academic and industrial communities will be crucial, along with attention to fairness, bias mitigation, and linguistic diversity in the system design. This survey provides both a foundation and a reference point for advancing the field. By uniting progress in multilingual modelling with innovations in information retrieval, CLIR has the capacity to transform global access to knowledge. Continued efforts will be necessary to ensure that these technologies serve all languages equitably, creating a future where language is no longer a barrier to information.

\bibliographystyle{unsrt}  
\bibliography{references}  %%% Remove comment to use the external .bib file (using bibtex).

\begin{thebibliography}{429}
\providecommand{\natexlab}[1]{#1}
\providecommand{\url}[1]{\texttt{#1}}
\expandafter\ifx\csname urlstyle\endcsname\relax
  \providecommand{\doi}[1]{doi: #1}\else
  \providecommand{\doi}{doi: \begingroup \urlstyle{rm}\Url}\fi

\bibitem[Lancaster(1979)]{lancaster_1979}
F.~Wilfrid Lancaster.
\newblock \emph{Information Retrieval Systems: Characteristics, Testing and Evaluation}.
\newblock Wiley, New York, 1979.

\bibitem[Sparck~Jones and Willett(1997)]{sparck_1997}
Karen Sparck~Jones and Peter Willett, editors.
\newblock \emph{Readings in information retrieval}.
\newblock Morgan Kaufmann Publishers Inc., San Francisco, CA, USA, 1997.
\newblock ISBN 1558604545.

\bibitem[Baeza-Yates and Ribeiro-Neto(1999)]{baeza-yates_1999}
Ricardo~A. Baeza-Yates and Berthier Ribeiro-Neto.
\newblock \emph{Modern Information Retrieval}.
\newblock Addison-Wesley Longman Publishing Co., Inc., USA, 1999.
\newblock ISBN 020139829X.

\bibitem[Pevzner(1969)]{pevzner1969automatic}
BR~Pevzner.
\newblock Automatic translation of english text to the language of the pusto-nepusto-2 system.
\newblock \emph{Automatic Documentation and Mathematical Linguistics}, 3\penalty0 (4):\penalty0 40--48, 1969.

\bibitem[Salton(1972)]{salton1972experiments}
Gerard Salton.
\newblock Experiments in multi-lingual information retrieval.
\newblock Technical report, Cornell University, 1972.

\bibitem[Sch{\"a}uble and Sheridan(1998)]{schauble1998cross}
Peter Sch{\"a}uble and P{\'a}raic Sheridan.
\newblock Cross-language information retrieval (clir) track overview.
\newblock \emph{NIST SPECIAL PUBLICATION SP}, pages 31--44, 1998.

\bibitem[PETERS(2000)]{peters2000first}
CAROL PETERS.
\newblock First results of the clef 2000 cross-language text retrieval system evaluation campaign.
\newblock In \emph{Working Notes for the CLEF 2000 Workshop. Lisboa}, 2000.

\bibitem[{Language Magazine}(2015)]{language2015inequality}
{Language Magazine}.
\newblock Information inequality and the languages of the internet, 2015.
\newblock URL \url{https://languagemagazine.com/2015/05/29/information-inequality-and-the-languages-of-the-internet/}.
\newblock Accessed: 2025-07-03.

\bibitem[Trotman(2000)]{trotman2000digitaldivide}
Andrew Trotman.
\newblock Education equity and the digital divide.
\newblock \emph{ResearchGate}, 2000.
\newblock URL \url{https://www.researchgate.net/publication/228616386_Education_equity_and_the_digital_divide}.
\newblock Accessed: 2025-07-03.

\bibitem[Joshi et~al.(2020)Joshi, Santy, Budhiraja, Bali, and Choudhury]{joshi-etal-2020-state}
Pratik Joshi, Sebastin Santy, Amar Budhiraja, Kalika Bali, and Monojit Choudhury.
\newblock The state and fate of linguistic diversity and inclusion in the {NLP} world.
\newblock In Dan Jurafsky, Joyce Chai, Natalie Schluter, and Joel Tetreault, editors, \emph{Proceedings of the 58th Annual Meeting of the Association for Computational Linguistics}, pages 6282--6293, Online, July 2020. Association for Computational Linguistics.
\newblock \doi{10.18653/v1/2020.acl-main.560}.
\newblock URL \url{https://aclanthology.org/2020.acl-main.560/}.

\bibitem[Li et~al.(2024{\natexlab{a}})Li, Shi, Liu, Yang, Payani, Liu, and Du]{li2024quantifying}
Zihao Li, Yucheng Shi, Zirui Liu, Fan Yang, Ali Payani, Ninghao Liu, and Mengnan Du.
\newblock Language ranker: A metric for quantifying llm performance across high and low-resource languages, 2024{\natexlab{a}}.
\newblock URL \url{https://arxiv.org/abs/2404.11553}.

\bibitem[Vaj(2023)]{vaj2023laser}
Tiya Vaj.
\newblock Laser (language‑agnostic sentence representations).
\newblock Online article, 2023.

\bibitem[Shi et~al.(2021)Shi, Zhang, Bai, and Lin]{shi-etal-2021-cross}
Peng Shi, Rui Zhang, He~Bai, and Jimmy Lin.
\newblock Cross-lingual training of dense retrievers for document retrieval.
\newblock In \emph{Proceedings of the 1st Workshop on Multilingual Representation Learning}, pages 251--253, November 2021.
\newblock \doi{10.18653/v1/2021.mrl-1.24}.
\newblock URL \url{https://aclanthology.org/2021.mrl-1.24}.

\bibitem[Huang et~al.(2024)Huang, Li, Hsu, Hsu, and Chen]{huang-etal-2024-unsupervised}
Chao-Wei Huang, Chen-An Li, Tsu-Yuan Hsu, Chen-Yu Hsu, and Yun-Nung Chen.
\newblock Unsupervised multilingual dense retrieval via generative pseudo labeling.
\newblock In Yvette Graham and Matthew Purver, editors, \emph{Findings of the Association for Computational Linguistics: EACL 2024}, pages 736--746, St. Julian{'}s, Malta, March 2024. Association for Computational Linguistics.
\newblock URL \url{https://aclanthology.org/2024.findings-eacl.49/}.

\bibitem[Weller et~al.(2025)Weller, Boratko, Naim, and Lee]{welleretal2025}
Orion Weller, Michael Boratko, Iftekhar Naim, and Jinhyuk Lee.
\newblock On the theoretical limitations of embedding-based retrieval, 2025.
\newblock URL \url{https://arxiv.org/abs/2508.21038}.

\bibitem[Adeyemi et~al.(2024{\natexlab{a}})Adeyemi, Oladipo, Pradeep, and Lin]{adeyemi-etal-2024-zero}
Mofetoluwa Adeyemi, Akintunde Oladipo, Ronak Pradeep, and Jimmy Lin.
\newblock Zero-shot cross-lingual reranking with large language models for low-resource languages.
\newblock In Lun-Wei Ku, Andre Martins, and Vivek Srikumar, editors, \emph{Proceedings of the 62nd Annual Meeting of the Association for Computational Linguistics (Volume 2: Short Papers)}, pages 650--656, Bangkok, Thailand, August 2024{\natexlab{a}}. Association for Computational Linguistics.
\newblock \doi{10.18653/v1/2024.acl-short.59}.
\newblock URL \url{https://aclanthology.org/2024.acl-short.59/}.

\bibitem[Hambarde and Proenca(2023)]{hambarde2023information}
Kailash~A Hambarde and Hugo Proenca.
\newblock Information retrieval: recent advances and beyond.
\newblock \emph{IEEE Access}, 11:\penalty0 76581--76604, 2023.

\bibitem[Zhu et~al.(2024{\natexlab{a}})Zhu, Yuan, Wang, Liu, Liu, Deng, Chen, Liu, Dou, and Wen]{zhu_2024_ir}
Yutao Zhu, Huaying Yuan, Shuting Wang, Jiongnan Liu, Wenhan Liu, Chenlong Deng, Haonan Chen, Zheng Liu, Zhicheng Dou, and Ji-Rong Wen.
\newblock Large language models for information retrieval: A survey, 2024{\natexlab{a}}.
\newblock URL \url{https://arxiv.org/abs/2308.07107}.

\bibitem[Li et~al.(2025{\natexlab{a}})Li, Jin, Zhou, Zhang, Zhang, Zhu, and Dou]{li_2025_GenIR}
Xiaoxi Li, Jiajie Jin, Yujia Zhou, Yuyao Zhang, Peitian Zhang, Yutao Zhu, and Zhicheng Dou.
\newblock From matching to generation: A survey on generative information retrieval.
\newblock \emph{ACM Trans. Inf. Syst.}, 43\penalty0 (3), May 2025{\natexlab{a}}.
\newblock ISSN 1046-8188.
\newblock \doi{10.1145/3722552}.
\newblock URL \url{https://doi.org/10.1145/3722552}.

\bibitem[Ruder et~al.(2019)Ruder, Vulić, and Søgaard]{ruder_2019}
Sebastian Ruder, Ivan Vulić, and Anders Søgaard.
\newblock A survey of cross-lingual word embedding models.
\newblock \emph{Journal of Artificial Intelligence Research}, 65:\penalty0 569–631, August 2019.
\newblock ISSN 1076-9757.
\newblock \doi{10.1613/jair.1.11640}.
\newblock URL \url{http://dx.doi.org/10.1613/jair.1.11640}.

\bibitem[Zhu et~al.(2024{\natexlab{b}})Zhu, Supryadi, Xu, Sun, Pan, Cui, Du, Jin, Branco, and Xiong]{zhu_2024}
Shaolin Zhu, Supryadi, Shaoyang Xu, Haoran Sun, Leiyu Pan, Menglong Cui, Jiangcun Du, Renren Jin, António Branco, and Deyi Xiong.
\newblock Multilingual large language models: A systematic survey, 2024{\natexlab{b}}.
\newblock URL \url{https://arxiv.org/abs/2411.11072}.

\bibitem[Huang et~al.(2025)Huang, Mo, Zhang, Li, Li, Zhang, Yi, Mao, Liu, Xu, Xu, Nie, and Liu]{huang_2025}
Kaiyu Huang, Fengran Mo, Xinyu Zhang, Hongliang Li, You Li, Yuanchi Zhang, Weijian Yi, Yulong Mao, Jinchen Liu, Yuzhuang Xu, Jinan Xu, Jian-Yun Nie, and Yang Liu.
\newblock A survey on large language models with multilingualism: Recent advances and new frontiers, 2025.
\newblock URL \url{https://arxiv.org/abs/2405.10936}.

\bibitem[Xu et~al.(2025)Xu, Hu, Zhao, Qiu, Xu, Ye, and Gu]{xu_2025}
Yuemei Xu, Ling Hu, Jiayi Zhao, Zihan Qiu, Kexin Xu, Yuqi Ye, and Hanwen Gu.
\newblock A survey on multilingual large language models: corpora, alignment, and bias.
\newblock \emph{Frontiers of Computer Science}, 19\penalty0 (11), April 2025.
\newblock ISSN 2095-2236.
\newblock \doi{10.1007/s11704-024-40579-4}.
\newblock URL \url{http://dx.doi.org/10.1007/s11704-024-40579-4}.

\bibitem[Nie(2010)]{nie_2010}
Jian-Yun Nie.
\newblock \emph{Cross-Language Information Retrieval}.
\newblock Springer Cham, 2010.

\bibitem[Galuščáková et~al.(2022)Galuščáková, Oard, and Nair]{galuščáková2022}
Petra Galuščáková, Douglas~W. Oard, and Suraj Nair.
\newblock Cross-language information retrieval, 2022.
\newblock URL \url{https://arxiv.org/abs/2111.05988}.

\bibitem[Jansen et~al.(2000)Jansen, Spink, and Saracevic]{jansen_2000}
Bernard~J. Jansen, Amanda Spink, and Tefko Saracevic.
\newblock Real life, real users, and real needs: a study and analysis of user queries on the web.
\newblock \emph{Information Processing \& Management}, 36\penalty0 (2):\penalty0 207--227, 2000.
\newblock ISSN 0306-4573.
\newblock \doi{https://doi.org/10.1016/S0306-4573(99)00056-4}.
\newblock URL \url{https://www.sciencedirect.com/science/article/pii/S0306457399000564}.

\bibitem[Spink et~al.(2001)Spink, Wolfram, Jansen, and Saracevic]{spink_2001}
Amanda Spink, Dietmar Wolfram, Major B.~J. Jansen, and Tefko Saracevic.
\newblock Searching the web: The public and their queries.
\newblock \emph{Journal of the American Society for Information Science and Technology}, 52\penalty0 (3):\penalty0 226--234, 2001.
\newblock \doi{https://doi.org/10.1002/1097-4571(2000)9999:9999<::AID-ASI1591>3.0.CO;2-R}.
\newblock URL \url{https://asistdl.onlinelibrary.wiley.com/doi/abs/10.1002/1097-4571%282000%299999%3A9999%3C%3A%3AAID-ASI1591%3E3.0.CO%3B2-R}.

\bibitem[Furnas et~al.(1987)Furnas, Landauer, Gomez, and Dumais]{furnas_1987}
G.~W. Furnas, T.~K. Landauer, L.~M. Gomez, and S.~T. Dumais.
\newblock The vocabulary problem in human-system communication.
\newblock \emph{Commun. ACM}, 30\penalty0 (11):\penalty0 964–971, November 1987.
\newblock ISSN 0001-0782.
\newblock \doi{10.1145/32206.32212}.
\newblock URL \url{https://doi.org/10.1145/32206.32212}.

\bibitem[Carpineto and Romano(2012)]{carpineto_2012}
Claudio Carpineto and Giovanni Romano.
\newblock A survey of automatic query expansion in information retrieval.
\newblock \emph{ACM Comput. Surv.}, 44\penalty0 (1), January 2012.
\newblock ISSN 0360-0300.
\newblock \doi{10.1145/2071389.2071390}.
\newblock URL \url{https://doi.org/10.1145/2071389.2071390}.

\bibitem[Broder(2002)]{broder_2002}
Andrei Broder.
\newblock A taxonomy of web search.
\newblock \emph{SIGIR Forum}, 36\penalty0 (2):\penalty0 3–10, September 2002.
\newblock ISSN 0163-5840.
\newblock \doi{10.1145/792550.792552}.
\newblock URL \url{https://doi.org/10.1145/792550.792552}.

\bibitem[Nogueira et~al.(2019)Nogueira, Yang, Lin, and Cho]{nogueira_2019_doc2query}
Rodrigo Nogueira, Wei Yang, Jimmy Lin, and Kyunghyun Cho.
\newblock Document expansion by query prediction, 2019.
\newblock URL \url{https://arxiv.org/abs/1904.08375}.

\bibitem[Robertson(2009)]{robertson2009probabilistic}
Stephen~E. Robertson.
\newblock The probabilistic relevance framework: Bm25 and beyond.
\newblock \emph{Foundations and Trends® in Information Retrieval}, 3\penalty0 (4):\penalty0 333--389, 2009.

\bibitem[Wang et~al.(2023{\natexlab{a}})Wang, Yang, and Wei]{wang_2023_query2doc}
Liang Wang, Nan Yang, and Furu Wei.
\newblock Query2doc: Query expansion with large language models, 2023{\natexlab{a}}.
\newblock URL \url{https://arxiv.org/abs/2303.07678}.

\bibitem[Azad and Deepak(2019)]{azad_2019}
Hiteshwar~Kumar Azad and Akshay Deepak.
\newblock Query expansion techniques for information retrieval: A survey.
\newblock \emph{Information Processing \& Management}, 56\penalty0 (5):\penalty0 1698--1735, 2019.
\newblock ISSN 0306-4573.
\newblock \doi{https://doi.org/10.1016/j.ipm.2019.05.009}.
\newblock URL \url{https://www.sciencedirect.com/science/article/pii/S0306457318305466}.

\bibitem[Miller et~al.(1990)Miller, Beckwith, Fellbaum, Gross, and Miller]{miller_1990_wordnet}
George~A. Miller, Richard Beckwith, Christiane Fellbaum, Derek Gross, and Katherine~J. Miller.
\newblock Introduction to wordnet: An on-line lexical database*.
\newblock \emph{International Journal of Lexicography}, 3\penalty0 (4):\penalty0 235--244, 12 1990.
\newblock ISSN 0950-3846.
\newblock \doi{10.1093/ijl/3.4.235}.
\newblock URL \url{https://doi.org/10.1093/ijl/3.4.235}.

\bibitem[Rocchio and Salton(1965)]{rocchio_1965}
J.~J. Rocchio and G.~Salton.
\newblock Information search optimization and interactive retrieval techniques.
\newblock In \emph{Proceedings of the November 30--December 1, 1965, Fall Joint Computer Conference, Part I}, AFIPS '65 (Fall, part I), page 293–305, New York, NY, USA, 1965. Association for Computing Machinery.
\newblock ISBN 9781450378857.
\newblock \doi{10.1145/1463891.1463926}.
\newblock URL \url{https://doi.org/10.1145/1463891.1463926}.

\bibitem[Rocchio(1971)]{rocchio_1971}
J.~J. Rocchio.
\newblock Relevance feedback in information retrieval.
\newblock In G.~Salton, editor, \emph{The Smart retrieval system - experiments in automatic document processing}, pages 313--323. Englewood Cliffs, NJ: Prentice-Hall, 1971.

\bibitem[Xu and Croft(2017)]{xu_croft_2017}
Jinxi Xu and W.~Bruce Croft.
\newblock Quary expansion using local and global document analysis.
\newblock \emph{SIGIR Forum}, 51\penalty0 (2):\penalty0 168–175, August 2017.
\newblock ISSN 0163-5840.
\newblock \doi{10.1145/3130348.3130364}.
\newblock URL \url{https://doi.org/10.1145/3130348.3130364}.

\bibitem[Porter(1980)]{porter_1980}
M.~F. Porter.
\newblock An algorithm for suffix stripping.
\newblock \emph{Program: electronic library and information systems}, 14\penalty0 (3):\penalty0 130--137, 1980.

\bibitem[Hsu et~al.(2006)Hsu, Tsai, and Chen]{hsu_2006}
Ming-Hung Hsu, Ming-Feng Tsai, and Hsin-Hsi Chen.
\newblock Query expansion with conceptnet and wordnet: An intrinsic comparison.
\newblock In Hwee~Tou Ng, Mun-Kew Leong, Min-Yen Kan, and Donghong Ji, editors, \emph{Information Retrieval Technology}, pages 1--13, Berlin, Heidelberg, 2006. Springer Berlin Heidelberg.
\newblock ISBN 978-3-540-46237-8.

\bibitem[Hsu et~al.(2008)Hsu, Tsai, and Chen]{hsu_2008}
Ming-Hung Hsu, Ming-Feng Tsai, and Hsin-Hsi Chen.
\newblock Combining wordnet and conceptnet for automatic query expansion: A learning approach.
\newblock In Hang Li, Ting Liu, Wei-Ying Ma, Tetsuya Sakai, Kam-Fai Wong, and Guodong Zhou, editors, \emph{Information Retrieval Technology}, pages 213--224, Berlin, Heidelberg, 2008. Springer Berlin Heidelberg.
\newblock ISBN 978-3-540-68636-1.

\bibitem[Lavrenko and Croft(2001)]{lavrenko_croft_2001}
Victor Lavrenko and W.~Bruce Croft.
\newblock Relevance based language models.
\newblock In \emph{Proceedings of the 24th Annual International ACM SIGIR Conference on Research and Development in Information Retrieval}, SIGIR '01, page 120–127, New York, NY, USA, 2001. Association for Computing Machinery.
\newblock ISBN 1581133316.
\newblock \doi{10.1145/383952.383972}.
\newblock URL \url{https://doi.org/10.1145/383952.383972}.

\bibitem[Zhai and Lafferty(2001)]{zhai_lafferty_2001}
Chengxiang Zhai and John Lafferty.
\newblock Model-based feedback in the language modeling approach to information retrieval.
\newblock In \emph{Proceedings of the Tenth International Conference on Information and Knowledge Management}, CIKM '01, page 403–410, New York, NY, USA, 2001. Association for Computing Machinery.
\newblock ISBN 1581134363.
\newblock \doi{10.1145/502585.502654}.
\newblock URL \url{https://doi.org/10.1145/502585.502654}.

\bibitem[Salton and Buckley(1990)]{salton_buckley_1990}
Gerard Salton and Chris Buckley.
\newblock Improving retrieval performance by relevance feedback.
\newblock \emph{Journal of the American Society for Information Science}, 41\penalty0 (4):\penalty0 288--297, 1990.
\newblock \doi{https://doi.org/10.1002/(SICI)1097-4571(199006)41:4<288::AID-ASI8>3.0.CO;2-H}.
\newblock URL \url{https://asistdl.onlinelibrary.wiley.com/doi/abs/10.1002/%28SICI%291097-4571%28199006%2941%3A4%3C288%3A%3AAID-ASI8%3E3.0.CO%3B2-H}.

\bibitem[Sihvonen and Vakkari(2004)]{sihvonen_vakkari_2004}
Anne Sihvonen and Pertti Vakkari.
\newblock Subject knowledge improves interactive query expansion assisted by a thesaurus.
\newblock \emph{Journal of Documentation}, 60\penalty0 (6):\penalty0 673--690, 2004.
\newblock ISSN 0022-0418.
\newblock \doi{10.1108/00220410410568151}.

\bibitem[Cao et~al.(2008)Cao, Nie, Gao, and Robertson]{cao_nie_2008}
Guihong Cao, Jian-Yun Nie, Jianfeng Gao, and Stephen Robertson.
\newblock Selecting good expansion terms for pseudo-relevance feedback.
\newblock In \emph{Proceedings of the 31st Annual International ACM SIGIR Conference on Research and Development in Information Retrieval}, SIGIR '08, page 243–250, New York, NY, USA, 2008. Association for Computing Machinery.
\newblock ISBN 9781605581644.
\newblock \doi{10.1145/1390334.1390377}.
\newblock URL \url{https://doi.org/10.1145/1390334.1390377}.

\bibitem[Carpineto et~al.(2002)Carpineto, Romano, and Giannini]{carpineto_2002}
Claudio Carpineto, Giovanni Romano, and Vittorio Giannini.
\newblock Improving retrieval feedback with multiple term-ranking function combination.
\newblock \emph{ACM Trans. Inf. Syst.}, 20\penalty0 (3):\penalty0 259–290, July 2002.
\newblock ISSN 1046-8188.
\newblock \doi{10.1145/568727.568728}.
\newblock URL \url{https://doi.org/10.1145/568727.568728}.

\bibitem[Amati and Van~Rijsbergen(2002)]{amati_2002}
Gianni Amati and Cornelis~Joost Van~Rijsbergen.
\newblock Probabilistic models of information retrieval based on measuring the divergence from randomness.
\newblock \emph{ACM Trans. Inf. Syst.}, 20\penalty0 (4):\penalty0 357–389, October 2002.
\newblock ISSN 1046-8188.
\newblock \doi{10.1145/582415.582416}.
\newblock URL \url{https://doi.org/10.1145/582415.582416}.

\bibitem[Chang et~al.(2006)Chang, Ounis, and Kim]{CHANG_2006}
Youjin Chang, Iadh Ounis, and Minkoo Kim.
\newblock Query reformulation using automatically generated query concepts from a document space.
\newblock \emph{Information Processing \& Management}, 42\penalty0 (2):\penalty0 453--468, 2006.
\newblock ISSN 0306-4573.
\newblock \doi{https://doi.org/10.1016/j.ipm.2005.03.025}.
\newblock URL \url{https://www.sciencedirect.com/science/article/pii/S0306457305000567}.

\bibitem[Bernardini and Carpineto(2008)]{bernardini_2008}
Andrea Bernardini and Claudio Carpineto.
\newblock {FUB} at {TREC} 2008 relevance feedback track: Extending rocchio with distributional term analysis.
\newblock In Ellen~M. Voorhees and Lori~P. Buckland, editors, \emph{Proceedings of The Seventeenth Text REtrieval Conference, {TREC} 2008, Gaithersburg, Maryland, USA, November 18-21, 2008}, volume 500-277 of \emph{{NIST} Special Publication}. National Institute of Standards and Technology {(NIST)}, 2008.
\newblock URL \url{http://trec.nist.gov/pubs/trec17/papers/fondazione.rf.rev.pdf}.

\bibitem[Wong et~al.(2008)Wong, Luk, Leong, Ho, and Lee]{WONG_2008}
W.S. Wong, R.W.P. Luk, H.V. Leong, K.S. Ho, and D.L. Lee.
\newblock Re-examining the effects of adding relevance information in a relevance feedback environment.
\newblock \emph{Information Processing \& Management}, 44\penalty0 (3):\penalty0 1086--1116, 2008.
\newblock ISSN 0306-4573.
\newblock \doi{https://doi.org/10.1016/j.ipm.2007.12.002}.
\newblock URL \url{https://www.sciencedirect.com/science/article/pii/S0306457307002191}.

\bibitem[Buckley et~al.(1994)Buckley, Salton, Allan, and Singhal]{Buckley_1994}
Chris Buckley, Gerard Salton, James Allan, and Amit Singhal.
\newblock Automatic query expansion using smart: Trec 3.
\newblock In \emph{Text Retrieval Conference}, 1994.
\newblock URL \url{https://api.semanticscholar.org/CorpusID:14683127}.

\bibitem[Robertson and Willett(1993)]{robertson_willett_1993}
Alexander~M. Robertson and Peter Willett.
\newblock A comparison of spelling-correction methods for the identification of word forms in historical text databases*.
\newblock \emph{Literary and Linguistic Computing}, 8\penalty0 (3):\penalty0 143--152, 01 1993.
\newblock ISSN 0268-1145.
\newblock \doi{10.1093/llc/8.3.143}.
\newblock URL \url{https://doi.org/10.1093/llc/8.3.143}.

\bibitem[Efthimiadis(1996)]{efthimiadis_1996}
E.N. Efthimiadis.
\newblock Query expansion.
\newblock In M.E. Williams, editor, \emph{Annual review of information science and technology}, pages 121--187. Information Today, Medford, NJ, 1996.

\bibitem[Salton and Lesk(1965)]{salton_1965_smart}
G.~Salton and M.~E. Lesk.
\newblock The smart automatic document retrieval systems—an illustration.
\newblock \emph{Commun. ACM}, 8\penalty0 (6):\penalty0 391–398, June 1965.
\newblock ISSN 0001-0782.
\newblock \doi{10.1145/364955.364990}.
\newblock URL \url{https://doi.org/10.1145/364955.364990}.

\bibitem[Salton(1965)]{salton1965evaluation}
Gerard Salton.
\newblock The evaluation of automatic retrieval procedures—selected test results using the smart system.
\newblock \emph{American Documentation}, 16\penalty0 (3):\penalty0 209--222, 1965.

\bibitem[Croft and Harper(1979)]{croft_harper_1979}
W.~B. Croft and D.~J. Harper.
\newblock Using probabilistic models of document retrieval without relevance information.
\newblock \emph{Journal of Documentation}, 35\penalty0 (4):\penalty0 285--295, 1979.

\bibitem[Abdul-Jaleel et~al.(2004)Abdul-Jaleel, Allan, Croft, Diaz, Larkey, Li, Smucker, and Wade]{abdul-jaleel_2004_rm3}
Nasreen Abdul-Jaleel, James Allan, W.~Bruce Croft, Fernando Diaz, Leah Larkey, Xiaoyan Li, Mark~D. Smucker, and Courtney Wade.
\newblock Umass at trec 2004: Novelty and hard.
\newblock In \emph{Proceedings of the Thirteenth Text REtrieval Conference (TREC-13)}, 2004.

\bibitem[Li et~al.(2025{\natexlab{b}})Li, Lv, Zou, Chen, Zhang, An, Nie, and Zhou]{li_2025_qe}
Minghan Li, Xinxuan Lv, Junjie Zou, Tongna Chen, Chao Zhang, Suchao An, Ercong Nie, and Guodong Zhou.
\newblock Query expansion in the age of pre-trained and large language models: A comprehensive survey, 2025{\natexlab{b}}.
\newblock URL \url{https://arxiv.org/abs/2509.07794}.

\bibitem[Peng et~al.(2024{\natexlab{a}})Peng, Li, Jiang, Wang, Ou, Zeng, Xu, Xu, and Chen]{peng_2024}
Wenjun Peng, Guiyang Li, Yue Jiang, Zilong Wang, Dan Ou, Xiaoyi Zeng, Derong Xu, Tong Xu, and Enhong Chen.
\newblock Large language model based long-tail query rewriting in taobao search, 2024{\natexlab{a}}.
\newblock URL \url{https://arxiv.org/abs/2311.03758}.

\bibitem[Li et~al.(2024{\natexlab{b}})Li, Zhuang, Hui, Qin, Lin, Jagerman, Wang, and Bendersky]{li_2024_qe}
Minghan Li, Honglei Zhuang, Kai Hui, Zhen Qin, Jimmy Lin, Rolf Jagerman, Xuanhui Wang, and Michael Bendersky.
\newblock Can query expansion improve generalization of strong cross-encoder rankers?
\newblock In \emph{Proceedings of the 47th International ACM SIGIR Conference on Research and Development in Information Retrieval}, SIGIR '24, page 2321–2326, New York, NY, USA, 2024{\natexlab{b}}. Association for Computing Machinery.
\newblock ISBN 9798400704314.
\newblock \doi{10.1145/3626772.3657979}.
\newblock URL \url{https://doi.org/10.1145/3626772.3657979}.

\bibitem[Claveau(2022)]{claveau_2022}
Vincent Claveau.
\newblock Neural text generation for query expansion in information retrieval.
\newblock In \emph{IEEE/WIC/ACM International Conference on Web Intelligence and Intelligent Agent Technology}, WI-IAT '21, page 202–209, New York, NY, USA, 2022. Association for Computing Machinery.
\newblock ISBN 9781450391153.
\newblock \doi{10.1145/3486622.3493957}.
\newblock URL \url{https://doi.org/10.1145/3486622.3493957}.

\bibitem[Lv and Zhai(2011)]{lv_zhai_2011}
Yuanhua Lv and ChengXiang Zhai.
\newblock Lower-bounding term frequency normalization.
\newblock In \emph{Proceedings of the 20th ACM International Conference on Information and Knowledge Management}, CIKM '11, page 7–16, New York, NY, USA, 2011. Association for Computing Machinery.
\newblock ISBN 9781450307178.
\newblock \doi{10.1145/2063576.2063584}.
\newblock URL \url{https://doi.org/10.1145/2063576.2063584}.

\bibitem[Gao et~al.(2023)Gao, Ma, Lin, and Callan]{gao_2023_hyde}
Luyu Gao, Xueguang Ma, Jimmy Lin, and Jamie Callan.
\newblock Precise zero-shot dense retrieval without relevance labels.
\newblock In Anna Rogers, Jordan Boyd-Graber, and Naoaki Okazaki, editors, \emph{Proceedings of the 61st Annual Meeting of the Association for Computational Linguistics (Volume 1: Long Papers)}, pages 1762--1777, Toronto, Canada, July 2023. Association for Computational Linguistics.
\newblock \doi{10.18653/v1/2023.acl-long.99}.
\newblock URL \url{https://aclanthology.org/2023.acl-long.99/}.

\bibitem[Jagerman et~al.(2023)Jagerman, Zhuang, Qin, Wang, and Bendersky]{jagerman_2023}
Rolf Jagerman, Honglei Zhuang, Zhen Qin, Xuanhui Wang, and Michael Bendersky.
\newblock Query expansion by prompting large language models, 2023.
\newblock URL \url{https://arxiv.org/abs/2305.03653}.

\bibitem[Abe et~al.(2025)Abe, Takeoka, Kato, and Oyamada]{abe_2025llmQE}
Kenya Abe, Kunihiro Takeoka, Makoto~P. Kato, and Masafumi Oyamada.
\newblock Llm-based query expansion fails for unfamiliar and ambiguous queries, 2025.
\newblock URL \url{https://arxiv.org/abs/2505.12694}.

\bibitem[Zhang et~al.(2024)Zhang, Liu, Wang, and Lian]{zhang_2024_QE}
Wenjing Zhang, Zhaoxiang Liu, Kai Wang, and Shiguo Lian.
\newblock Query expansion and verification with large language model for information retrieval.
\newblock In De-Shuang Huang, Zhanjun Si, and Chuanlei Zhang, editors, \emph{Advanced Intelligent Computing Technology and Applications}, pages 341--351, Singapore, 2024. Springer Nature Singapore.
\newblock ISBN 978-981-97-5672-8.

\bibitem[Saleh and Pecina(2019)]{saleh_2019}
Shadi Saleh and Pavel Pecina.
\newblock Term selection for query expansion in medical cross-lingual information retrieval.
\newblock In Leif Azzopardi, Benno Stein, Norbert Fuhr, Philipp Mayr, Claudia Hauff, and Djoerd Hiemstra, editors, \emph{Advances in Information Retrieval}, pages 507--522, Cham, 2019. Springer International Publishing.
\newblock ISBN 978-3-030-15712-8.

\bibitem[Gaillard et~al.(2010)Gaillard, Bouraoui, Guimier~de Neef, and Boualem]{gaillard_2010}
Benoît Gaillard, Jean-Léon Bouraoui, Emilie Guimier~de Neef, and Malek Boualem.
\newblock Query expansion for cross language information retrieval improvement.
\newblock In \emph{2010 Fourth International Conference on Research Challenges in Information Science (RCIS)}, pages 337--342, 2010.
\newblock \doi{10.1109/RCIS.2010.5507393}.

\bibitem[Ballesteros and Croft(1997)]{ballesteros_croft_1997}
Lisa Ballesteros and W.~Bruce Croft.
\newblock Phrasal translation and query expansion techniques for cross-language information retrieval.
\newblock In \emph{Proceedings of the 20th Annual International ACM SIGIR Conference on Research and Development in Information Retrieval}, SIGIR '97, page 84–91, New York, NY, USA, 1997. Association for Computing Machinery.
\newblock ISBN 0897918363.
\newblock \doi{10.1145/258525.258540}.
\newblock URL \url{https://doi.org/10.1145/258525.258540}.

\bibitem[Ballesteros and Croft(1998)]{ballesteros_croft_1998}
Lisa Ballesteros and W.~Bruce Croft.
\newblock Resolving ambiguity for cross-language retrieval.
\newblock In \emph{Proceedings of the 21st Annual International ACM SIGIR Conference on Research and Development in Information Retrieval}, SIGIR '98, page 64–71, New York, NY, USA, 1998. Association for Computing Machinery.
\newblock ISBN 1581130155.
\newblock \doi{10.1145/290941.290958}.
\newblock URL \url{https://doi.org/10.1145/290941.290958}.

\bibitem[McNamee and Mayfield(2002)]{mcnamee_mayfield_2002}
Paul McNamee and James Mayfield.
\newblock Comparing cross-language query expansion techniques by degrading translation resources.
\newblock In \emph{Proceedings of the 25th Annual International ACM SIGIR Conference on Research and Development in Information Retrieval}, SIGIR '02, page 159–166, New York, NY, USA, 2002. Association for Computing Machinery.
\newblock ISBN 1581135610.
\newblock \doi{10.1145/564376.564406}.
\newblock URL \url{https://doi.org/10.1145/564376.564406}.

\bibitem[Sp{\"a}rck~Jones(1972)]{sparck1972statistical}
Karen Sp{\"a}rck~Jones.
\newblock A statistical interpretation of term specificity and its application in retrieval.
\newblock \emph{Journal of Documentation}, 28\penalty0 (1):\penalty0 11--21, 1972.
\newblock \doi{10.1108/eb026526}.

\bibitem[Kuwa et~al.(2020)Kuwa, Schamoni, and Riezler]{kuwa2020embedmeta}
Toshitaka Kuwa, Shigehiko Schamoni, and Stefan Riezler.
\newblock Embedding meta‑textual information for improved learning to rank.
\newblock In \emph{Proceedings of the 28th International Conference on Computational Linguistics (COLING 2020)}, pages 5558--5568, 2020.

\bibitem[Tashu et~al.(2024)Tashu, Kontos, Sabatelli, and Valdenegro-Toro]{tashu2024mapping}
Tsegaye~Misikir Tashu, Eduard‑Raul Kontos, Matthia Sabatelli, and Matias Valdenegro-Toro.
\newblock Mapping transformer leveraged embeddings for cross-lingual document representation.
\newblock \emph{arXiv preprint arXiv:2401.06583}, 2024.

\bibitem[Hu et~al.(2023)Hu, Chen, Qi, Kong, Liu, Wang, and Huang]{hu2023language}
Xiyang Hu, Xinchi Chen, Peng Qi, Deguang Kong, Kunlun Liu, William~Yang Wang, and Zhiheng Huang.
\newblock Language agnostic multilingual information retrieval with contrastive learning.
\newblock In \emph{Findings of the Association for Computational Linguistics: ACL 2023}, pages 9133--9146, 2023.
\newblock \doi{10.18653/v1/2023.findings-acl.581}.
\newblock URL \url{https://aclanthology.org/2023.findings-acl.581/}.

\bibitem[K{\ae}r~J{\o}rgensen et~al.(2021)K{\ae}r~J{\o}rgensen, Hartmann, Dai, and Elliott]{jorgensen_2021_mdapt}
Rasmus K{\ae}r~J{\o}rgensen, Mareike Hartmann, Xiang Dai, and Desmond Elliott.
\newblock {mDAPT}: Multilingual domain adaptive pretraining in a single model.
\newblock In \emph{Findings of the Association for Computational Linguistics: EMNLP 2021}, pages 3404--3418, 2021.
\newblock \doi{10.18653/v1/2021.findings-emnlp.290}.
\newblock URL \url{https://aclanthology.org/2021.findings-emnlp.290/}.

\bibitem[Liu et~al.(2021{\natexlab{a}})Liu, Vulić, Korhonen, and Collier]{liu2021xlbel}
Fangyu Liu, Ivan Vulić, Anna Korhonen, and Nigel Collier.
\newblock Learning domain-specialised representations for cross-lingual biomedical entity linking, 2021{\natexlab{a}}.
\newblock URL \url{https://arxiv.org/abs/2105.14398}.

\bibitem[Vuli{\'c} et~al.(2013)Vuli{\'c}, De~Smet, and Moens]{vulic2013bilda}
Ivan Vuli{\'c}, Wim De~Smet, and Marie-Francine Moens.
\newblock Cross-language information retrieval models based on latent topic models trained on document-aligned comparable corpora.
\newblock \emph{Information Retrieval Journal}, 16\penalty0 (3):\penalty0 331--368, 2013.
\newblock \doi{10.1007/s10791-012-9200-5}.

\bibitem[Mimno et~al.(2009)Mimno, Wallach, Naradowsky, Smith, and McCallum]{mimno2009polylingual}
David Mimno, Hanna Wallach, Jason Naradowsky, David~A. Smith, and Andrew McCallum.
\newblock Polylingual topic models.
\newblock In \emph{Proceedings of EMNLP 2009}, pages 880--889, 2009.

\bibitem[Posch et~al.(2015)Posch, Bleier, Schaer, and Strohmaier]{posch2015polylingual}
Lisa Posch, Arnim Bleier, Philipp Schaer, and Markus Strohmaier.
\newblock The polylingual labeled topic model.
\newblock In \emph{Proceedings of COLING 2014}, pages 2982--2992, 2015.

\bibitem[Devapujula et~al.(2019)Devapujula, Arora, and Borar]{devapujula2019broad}
Siddhartha Devapujula, Sagar Arora, and Sumit Borar.
\newblock Learning to rank broad and narrow queries in e-commerce.
\newblock \emph{arXiv preprint arXiv:1907.01549}, 2019.
\newblock URL \url{https://arxiv.org/abs/1907.01549}.

\bibitem[Karmaker~Santu et~al.(2017)Karmaker~Santu, Sondhi, and Zhai]{karmaker2017application}
Shubhra~Kanti Karmaker~Santu, Parikshit Sondhi, and ChengXiang Zhai.
\newblock On application of learning to rank for e-commerce search.
\newblock In \emph{Proceedings of SIGIR 2017 e-Commerce Workshop}, 2017.
\newblock URL \url{https://arxiv.org/abs/1903.04263}.

\bibitem[Zhang et~al.(2022{\natexlab{a}})Zhang, Tan, and Misra]{zhang2022evaluating}
Bryan Zhang, Liling Tan, and Amita Misra.
\newblock Evaluating machine translation in cross‑lingual e‑commerce search.
\newblock In \emph{Proceedings of the 15th Biennial Conference of the Association for Machine Translation in the Americas}, pages 322--334, Orlando, USA, 2022{\natexlab{a}}. AMTA.

\bibitem[Cooper et~al.(1992)Cooper, Gey, and Dabney]{cooper1992probabilistic}
William~S Cooper, Fredric~C Gey, and Daniel~P Dabney.
\newblock Probabilistic retrieval based on staged logistic regression.
\newblock In \emph{Proceedings of the 15th annual international ACM SIGIR conference on Research and development in information retrieval}, pages 198--210, 1992.

\bibitem[Fuhr(1989)]{fuhr1989optimum}
Norbert Fuhr.
\newblock Optimum polynomial retrieval functions based on the probability ranking principle.
\newblock \emph{ACM Transactions on Information Systems (TOIS)}, 7\penalty0 (3):\penalty0 183--204, 1989.

\bibitem[Burges et~al.(2005)Burges, Shaked, Renshaw, Lazier, Deeds, Hamilton, and Hullender]{burges2005learning}
Chris Burges, Tal Shaked, Erin Renshaw, Ari Lazier, Matt Deeds, Nicole Hamilton, and Greg Hullender.
\newblock Learning to rank using gradient descent.
\newblock In \emph{Proceedings of the 22nd international conference on Machine learning}, pages 89--96, 2005.

\bibitem[Burges(2010)]{burges2010ranknet}
Chris J.~C. Burges.
\newblock From ranknet to lambdarank to lambdamart: An overview.
\newblock Technical report msr-tr-2010-82, Microsoft Research, 2010.
\newblock URL \url{http://research.microsoft.com/en-us/um/people/cburges/tech_reports/MSR-TR-2010-82.pdf}.

\bibitem[Cao et~al.(2007)Cao, Qin, Liu, Tsai, and Li]{cao2007learning}
Zhe Cao, Tao Qin, Tie-Yan Liu, Ming-Feng Tsai, and Hang Li.
\newblock Learning to rank: From pairwise approach to listwise approach.
\newblock In \emph{Proceedings of the 24th International Conference on Machine Learning (ICML)}, pages 129--136, 2007.
\newblock \doi{10.1145/1273496.1273513}.

\bibitem[Xia et~al.(2008)Xia, Liu, Wang, Zhang, and Li]{xia2008listwise}
Fen Xia, Tie-Yan Liu, Jue Wang, Wensheng Zhang, and Hang Li.
\newblock Listwise approach to learning to rank: Theory and algorithm.
\newblock In \emph{Proceedings of the 25th International Conference on Machine Learning (ICML)}, pages 1192--1199, 2008.
\newblock \doi{10.1145/1390156.1390306}.

\bibitem[J{\"a}rvelin and Kek{\"a}l{\"a}inen(2002)]{jarvelin2002cumulated}
Kalervo J{\"a}rvelin and Jaana Kek{\"a}l{\"a}inen.
\newblock Cumulated gain-based evaluation of ir techniques.
\newblock \emph{ACM Transactions on Information Systems}, 20\penalty0 (4):\penalty0 422--446, 2002.

\bibitem[Manning et~al.(2008)Manning, Raghavan, and Schütze]{manning2008introduction}
D.~Manning, Christopher\, Prabhakar Raghavan, and Hinrich Schütze.
\newblock \emph{Introduction to Information Retrieval}.
\newblock Cambridge University Press, 2008.

\bibitem[Liu(2009)]{liu2009learning}
Tie-Yan Liu.
\newblock Learning to rank for information retrieval.
\newblock \emph{Foundations and Trends® in Information Retrieval}, 3\penalty0 (3):\penalty0 225--331, 2009.
\newblock \doi{10.1561/1500000016}.

\bibitem[Conneau et~al.(2020)Conneau, Khandelwal, Goyal, Chaudhary, Wenzek, Guzmán, Grave, Ott, Zettlemoyer, and Stoyanov]{conneau_2020}
Alexis Conneau, Kartikay Khandelwal, Naman Goyal, Vishrav Chaudhary, Guillaume Wenzek, Francisco Guzmán, Edouard Grave, Myle Ott, Luke Zettlemoyer, and Veselin Stoyanov.
\newblock Unsupervised cross-lingual representation learning at scale.
\newblock In \emph{Proceedings of the 58th Annual Meeting of the Association for Computational Linguistics}, pages 8440--8451. Association for Computational Linguistics, 2020.

\bibitem[Xue et~al.(2021)Xue, Constant, Roberts, Kale, Al-Rfou, Siddhant, Barua, and Raffel]{xue_2021_mt5}
Linting Xue, Noah Constant, Adam Roberts, Mihir Kale, Rami Al-Rfou, Aditya Siddhant, Aditya Barua, and Colin Raffel.
\newblock mt5: A massively multilingual pre-trained text-to-text transformer, 2021.
\newblock URL \url{https://arxiv.org/abs/2010.11934}.

\bibitem[Wang et~al.(2020)Wang, Wei, Dong, Bao, Yang, and Zhou]{wang2020minilmdeepselfattentiondistillation}
Wenhui Wang, Furu Wei, Li~Dong, Hangbo Bao, Nan Yang, and Ming Zhou.
\newblock Minilm: Deep self-attention distillation for task-agnostic compression of pre-trained transformers, 2020.
\newblock URL \url{https://arxiv.org/abs/2002.10957}.

\bibitem[Bonifacio et~al.(2022{\natexlab{a}})Bonifacio, Jeronymo, Abonizio, Campiotti, Fadaee, Lotufo, and Nogueira]{bonifacio_2022_mmarco}
Luiz Bonifacio, Vitor Jeronymo, Hugo~Queiroz Abonizio, Israel Campiotti, Marzieh Fadaee, Roberto Lotufo, and Rodrigo Nogueira.
\newblock mmarco: A multilingual version of the ms marco passage ranking dataset, 2022{\natexlab{a}}.
\newblock URL \url{https://arxiv.org/abs/2108.13897}.

\bibitem[Asai et~al.(2021)Asai, Kasai, Clark, Lee, Choi, and Hajishirzi]{asai_2021_qa}
Akari Asai, Jungo Kasai, Jonathan~H. Clark, Kenton Lee, Eunsol Choi, and Hannaneh Hajishirzi.
\newblock Xor qa: Cross-lingual open-retrieval question answering, 2021.
\newblock URL \url{https://arxiv.org/abs/2010.11856}.

\bibitem[Zhang et~al.(2023)Zhang, Thakur, Ogundepo, Kamalloo, Alfonso-Hermelo, Li, Liu, Rezagholizadeh, and Lin]{zhang_2023_miracl}
Xinyu Zhang, Nandan Thakur, Odunayo Ogundepo, Ehsan Kamalloo, David Alfonso-Hermelo, Xiaoguang Li, Qun Liu, Mehdi Rezagholizadeh, and Jimmy Lin.
\newblock {MIRACL}: A multilingual retrieval dataset covering 18 diverse languages.
\newblock \emph{Transactions of the Association for Computational Linguistics}, 11:\penalty0 1114--1131, 2023.
\newblock \doi{10.1162/tacl_a_00595}.
\newblock URL \url{https://aclanthology.org/2023.tacl-1.63/}.

\bibitem[Yang et~al.(2024{\natexlab{a}})Yang, Lawrie, Mayfield, Oard, and Miller]{yang2024translate_distill}
Eugene Yang, Dawn~J. Lawrie, James Mayfield, Douglas~W. Oard, and Scott Miller.
\newblock Translate‑distill: Learning cross‑language dense retrieval by translation and distillation.
\newblock In \emph{Advances in Information Retrieval: 46th European Conference on IR (ECIR)}, pages 50--65, 2024{\natexlab{a}}.
\newblock \doi{10.1007/978-3-031-56060-6_4}.

\bibitem[Jeronymo et~al.(2023)Jeronymo, Lotufo, and Nogueira]{jeronymo2023neuralmindunicamp2022trecneuclir}
Vitor Jeronymo, Roberto Lotufo, and Rodrigo Nogueira.
\newblock Neuralmind-unicamp at 2022 trec neuclir: Large boring rerankers for cross-lingual retrieval, 2023.
\newblock URL \url{https://arxiv.org/abs/2303.16145}.

\bibitem[Yang et~al.(2024{\natexlab{b}})Yang, Lawrie, and Mayfield]{yang2024hltcoe}
Eugene Yang, Dawn Lawrie, and James Mayfield.
\newblock Hltcoe at trec 2023 neuclir track.
\newblock \emph{arXiv preprint arXiv:2404.08118}, 2024{\natexlab{b}}.

\bibitem[Bonifacio et~al.(2022{\natexlab{b}})Bonifacio, Abonizio, Fadaee, and Nogueira]{bonifacio_2022_inpars}
Luiz Bonifacio, Hugo Abonizio, Marzieh Fadaee, and Rodrigo Nogueira.
\newblock Inpars: Unsupervised dataset generation for information retrieval.
\newblock In \emph{Proceedings of the 45th International ACM SIGIR Conference on Research and Development in Information Retrieval}, SIGIR '22, page 2387–2392, New York, NY, USA, 2022{\natexlab{b}}. Association for Computing Machinery.
\newblock ISBN 9781450387323.
\newblock \doi{10.1145/3477495.3531863}.

\bibitem[Sun et~al.(2023)Sun, Yan, Ma, Wang, Ren, Chen, Yin, and Ren]{sun-etal-2023-chatgpt}
Weiwei Sun, Lingyong Yan, Xinyu Ma, Shuaiqiang Wang, Pengjie Ren, Zhumin Chen, Dawei Yin, and Zhaochun Ren.
\newblock Is {C}hat{GPT} good at search? investigating large language models as re-ranking agents.
\newblock In Houda Bouamor, Juan Pino, and Kalika Bali, editors, \emph{Proceedings of the 2023 Conference on Empirical Methods in Natural Language Processing}, pages 14918--14937, Singapore, December 2023. Association for Computational Linguistics.
\newblock \doi{10.18653/v1/2023.emnlp-main.923}.
\newblock URL \url{https://aclanthology.org/2023.emnlp-main.923/}.

\bibitem[Ma et~al.(2023)Ma, Zhang, Pradeep, and Lin]{ma2023lrl}
Xueguang Ma, Xinyu Zhang, Ronak Pradeep, and Jimmy Lin.
\newblock Zero-shot listwise document reranking with a large language model.
\newblock \emph{arXiv preprint arXiv:2305.02156}, 2023.

\bibitem[Drozdov et~al.(2023)Drozdov, Zhuang, Dai, Qin, Rahimi, Wang, Alon, Iyyer, McCallum, Metzler, and Hui]{drozdov2023paradepassagerankingusing}
Andrew Drozdov, Honglei Zhuang, Zhuyun Dai, Zhen Qin, Razieh Rahimi, Xuanhui Wang, Dana Alon, Mohit Iyyer, Andrew McCallum, Donald Metzler, and Kai Hui.
\newblock Parade: Passage ranking using demonstrations with large language models, 2023.
\newblock URL \url{https://arxiv.org/abs/2310.14408}.

\bibitem[Chung et~al.(2022)Chung, Hou, Longpre, Zoph, Tay, Fedus, Li, Wang, Dehghani, Brahma, Webson, Gu, Dai, Suzgun, Chen, Chowdhery, Narang, Mishra, Yu, Zhao, Huang, Dai, Yu, Petrov, Chi, Dean, Devlin, Roberts, Zhou, Le, and Wei]{chung2022scaling}
Hyung~Won Chung, Le~Hou, Shayne Longpre, Barret Zoph, Yi~Tay, William Fedus, Eric Li, Xuezhi Wang, Mostafa Dehghani, Siddhartha Brahma, Albert Webson, Shixiang~Shane Gu, Zhuyun Dai, Mirac Suzgun, Xinyun Chen, Aakanksha Chowdhery, Sharan Narang, Gaurav Mishra, Adams Yu, Vincent Zhao, Yanping Huang, Andrew Dai, Hongkun Yu, Slav Petrov, Ed~H. Chi, Jeff Dean, Jacob Devlin, Adam Roberts, Denny Zhou, Quoc~V. Le, and Jason Wei.
\newblock Scaling instruction-finetuned language models, 2022.
\newblock URL \url{https://arxiv.org/abs/2210.11416}.

\bibitem[Tunstall et~al.(2023)Tunstall, Beeching, Lambert, Rajani, Rasul, Belkada, Huang, von Werra, Fourrier, Habib, Sarrazin, Sanseviero, Rush, and Wolf]{tunstall2023zephyrdirectdistillationlm}
Lewis Tunstall, Edward Beeching, Nathan Lambert, Nazneen Rajani, Kashif Rasul, Younes Belkada, Shengyi Huang, Leandro von Werra, Clémentine Fourrier, Nathan Habib, Nathan Sarrazin, Omar Sanseviero, Alexander~M. Rush, and Thomas Wolf.
\newblock Zephyr: Direct distillation of lm alignment, 2023.
\newblock URL \url{https://arxiv.org/abs/2310.16944}.

\bibitem[{OpenAI}(2022)]{openai_chatgpt_announcement}
{OpenAI}.
\newblock Introducing chatgpt.
\newblock \url{https://openai.com/blog/chatgpt}, 2022.
\newblock Blog post announcing ChatGPT by OpenAI.

\bibitem[Chen et~al.(2025)Chen, Jimenez~Gutierrez, and Su]{chen2025icr}
Shijie Chen, Bernal Jimenez~Gutierrez, and Yu~Su.
\newblock Attention in large language models yields efficient zero‑shot re‑rankers.
\newblock In \emph{ICLR 2025 Workshop}, 2025.

\bibitem[Déjean et~al.(2024)Déjean, Clinchant, and Formal]{déjean2024thoroughcomparisoncrossencodersllms}
Hervé Déjean, Stéphane Clinchant, and Thibault Formal.
\newblock A thorough comparison of cross-encoders and llms for reranking splade, 2024.
\newblock URL \url{https://arxiv.org/abs/2403.10407}.

\bibitem[Anonymous(2025)]{mialowresource2025mia}
Anonymous.
\newblock Multilingual open qa on the mia shared task.
\newblock \emph{arXiv preprint arXiv:2501.04153}, 2025.

\bibitem[Lewis et~al.(2020{\natexlab{a}})Lewis, Oguz, Rinott, Riedel, and Schwenk]{lewis_2020_mlqa}
Patrick Lewis, Barlas Oguz, Ruty Rinott, Sebastian Riedel, and Holger Schwenk.
\newblock {MLQA}: Evaluating cross-lingual extractive question answering.
\newblock In Dan Jurafsky, Joyce Chai, Natalie Schluter, and Joel Tetreault, editors, \emph{Proceedings of the 58th Annual Meeting of the Association for Computational Linguistics}, pages 7315--7330, Online, July 2020{\natexlab{a}}. Association for Computational Linguistics.
\newblock \doi{10.18653/v1/2020.acl-main.653}.
\newblock URL \url{https://aclanthology.org/2020.acl-main.653/}.

\bibitem[Karmaker~Santu et~al.(2019)Karmaker~Santu, Sondhi, and Zhai]{pi2024featurebasedecom}
Shubhra~Kanti Karmaker~Santu, Parikshit Sondhi, and ChengXiang Zhai.
\newblock On application of learning to rank for e-commerce search, 2019.

\bibitem[Cormack et~al.(2009)Cormack, Clarke, and Büttcher]{cormack2009rrf}
Gordon~V. Cormack, Charles L.~A. Clarke, and Stefan Büttcher.
\newblock Reciprocal rank fusion outperforms condorcet and individual rank learning methods.
\newblock In \emph{SIGIR ’09}, 2009.

\bibitem[zil(2025)]{zilliz_rrf_ranker}
Rrf ranker — zilliz cloud developer hub.
\newblock \url{https://docs.zilliz.com/docs/reranking-rrf}, 2025.
\newblock Describes use of Reciprocal Rank Fusion for multilingual and hybrid search.

\bibitem[azu(2025)]{azure_rrf_hybrid}
Hybrid search scoring via reciprocal rank fusion — azure ai search.
\newblock \url{https://learn.microsoft.com/en-us/azure/search/hybrid-search-ranking}, 2025.
\newblock Explains RRF for combining vector and lexical search results.

\bibitem[Dietz et~al.(2023)Dietz, Bast, Chatterjee, Dalton, Meij, and de~Vries]{dietz2023ecir}
Laura Dietz, Hannah Bast, Shubham Chatterjee, Jeff Dalton, Edgar Meij, and Arjen de~Vries.
\newblock Ecir 23 tutorial: Neuro-symbolic approaches for information retrieval.
\newblock In \emph{European Conference on Information Retrieval}, pages 324--330. Springer, 2023.

\bibitem[Zhang et~al.(2020)Zhang, Karakos, Hartmann, Srivastava, Tarlin, Akodes, Gouda, Bathool, Zhao, Jiang, Schwartz, and Makhoul]{zhang-etal-2020-2019}
Le~Zhang, Damianos Karakos, William Hartmann, Manaj Srivastava, Lee Tarlin, David Akodes, Sanjay~Krishna Gouda, Numra Bathool, Lingjun Zhao, Zhuolin Jiang, Richard Schwartz, and John Makhoul.
\newblock The 2019 {BBN} cross-lingual information retrieval system.
\newblock In Kathy McKeown, Douglas~W. Oard, Elizabeth, and Richard Schwartz, editors, \emph{Proceedings of the workshop on Cross-Language Search and Summarization of Text and Speech (CLSSTS2020)}, pages 44--51, Marseille, France, May 2020. European Language Resources Association.
\newblock ISBN 979-10-95546-55-9.
\newblock URL \url{https://aclanthology.org/2020.clssts-1.8/}.

\bibitem[Craswell et~al.(2008)Craswell, Zoeter, and Taylor]{craswell2008cumulative}
Nick Craswell, Onno Zoeter, and Michael Taylor.
\newblock An experimental comparison of click position‐bias models.
\newblock In \emph{WSDM '08}, pages 87--94, 2008.

\bibitem[Chapelle and Zhang(2009)]{chapelle2009dynamic}
Olivier Chapelle and Yi~Zhang.
\newblock A dynamic bayesian network click model for web search ranking.
\newblock In \emph{WWW '09}, pages 1--10, 2009.

\bibitem[Wang et~al.(2018)Wang, Golbandi, Bendersky, Metzler, and Najork]{wang2018position}
Xuanhui Wang, Nadav Golbandi, Michael Bendersky, Donald Metzler, and Marc Najork.
\newblock Position bias estimation for unbiased learning to rank in personal search.
\newblock In \emph{Proceedings of the eleventh ACM international conference on web search and data mining}, pages 610--618, 2018.

\bibitem[Joachims et~al.(2007)Joachims, Granka, Pan, Hembrooke, Radlinski, and Gay]{joachims2007position}
Thorsten Joachims, Laura Granka, Bing Pan, Helene Hembrooke, Filip Radlinski, and Geri Gay.
\newblock Evaluating the accuracy of implicit feedback from clicks and query reformulations in web search.
\newblock \emph{ACM Trans. Inf. Syst.}, 25\penalty0 (2):\penalty0 7–es, April 2007.
\newblock ISSN 1046-8188.
\newblock \doi{10.1145/1229179.1229181}.
\newblock URL \url{https://doi.org/10.1145/1229179.1229181}.

\bibitem[Islam and Faruk(2025)]{islam2025position}
Md~Aminul Islam and Ahmed~Sayeed Faruk.
\newblock Prompt‑based llms for position bias‑aware reranking in personalized recommendations.
\newblock In \emph{arXiv preprint arXiv:2505.04948}, 2025.

\bibitem[Tang et~al.(2024)Tang, Zhang, Ma, Lin, and Ture]{tang2024permutation}
Raphael Tang, Xinyu Zhang, Xueguang Ma, Jimmy Lin, and Ferhan Ture.
\newblock Found in the middle: Permutation self‑consistency improves listwise ranking in large language models.
\newblock In \emph{NAACL 2024 Long Papers}, 2024.
\newblock \doi{10.18653/v1/2024.naacl-long.129}.

\bibitem[Goldstein and Carbonell(1998)]{goldstein1998mmr}
Jade Goldstein and Jaime Carbonell.
\newblock Using mmr, diversity-based reranking for reordering query results.
\newblock In \emph{Proceedings of the 21st Annual International ACM SIGIR Conference on Research and Development in Information Retrieval}, pages 335--336, 1998.
\newblock \doi{10.1145/290941.291025}.

\bibitem[Lin et~al.(2023)Lin, Alfonso-Hermelo, Jeronymo, Kamalloo, Lassance, Nogueira, Ogundepo, Rezagholizadeh, Thakur, Yang, and Zhang]{lin2023simple}
Jimmy Lin, David Alfonso-Hermelo, Vitor Jeronymo, Ehsan Kamalloo, Carlos Lassance, Rodrigo Nogueira, Odunayo Ogundepo, Mehdi Rezagholizadeh, Nandan Thakur, Jheng-Hong Yang, and Xinyu Zhang.
\newblock Simple yet effective neural ranking and reranking baselines for cross-lingual information retrieval, 2023.
\newblock URL \url{https://arxiv.org/abs/2304.01019}.

\bibitem[Wang et~al.(2011)Wang, Lin, and Metzler]{wang2011cascade}
Lidan Wang, Jimmy~J. Lin, and Donald Metzler.
\newblock A cascade ranking model for efficient ranked retrieval.
\newblock In \emph{SIGIR '11}, pages 105--114, 2011.

\bibitem[Chen et~al.(2017{\natexlab{a}})Chen, Gallagher, Blanco, and Culpepper]{chen2017costaware}
Ruey-Cheng Chen, Luke Gallagher, Roi Blanco, and J.\~Shane Culpepper.
\newblock Efficient cost-aware cascade ranking in multi-stage retrieval.
\newblock In \emph{SIGIR '17}, pages 445--454, 2017{\natexlab{a}}.

\bibitem[Honig et~al.(2023)Honig, Ackermann, and Chi]{honig2023bi}
Robert Honig, Jan Ackermann, and Mingyuan Chi.
\newblock Bi-encoder cascades for efficient image search.
\newblock In \emph{ICCV Workshop RCV '23}, 2023.

\bibitem[Meng et~al.(2024)Meng, Arabzadeh, Askari, Aliannejadi, and de~Rijke]{meng2024rlt}
Chuan Meng, Negar Arabzadeh, Arian Askari, Mohammad Aliannejadi, and Maarten de~Rijke.
\newblock Ranked list truncation for large language model-based re-ranking.
\newblock In \emph{SIGIR '24}, 2024.

\bibitem[Zhou and Agichtein(2021)]{zhou2021rlirank}
Jianghong Zhou and Eugene Agichtein.
\newblock Rlirank: Learning to rank with reinforcement learning for dynamic search.
\newblock \emph{arXiv preprint arXiv:2105.10124}, 2021.

\bibitem[Hochreiter and Schmidhuber(1997)]{hochreiter1997long}
Sepp Hochreiter and J{\"u}rgen Schmidhuber.
\newblock Long short-term memory.
\newblock \emph{Neural computation}, 9\penalty0 (8):\penalty0 1735--1780, 1997.

\bibitem[Baumgärtner et~al.(2022)Baumgärtner, Ribeiro, Reimers, and Gurevych]{baumgartner2022feedback}
Tim Baumgärtner, Leonardo F.~R. Ribeiro, Nils Reimers, and Iryna Gurevych.
\newblock Incorporating relevance feedback for information‑seeking retrieval using few‑shot document re‑ranking.
\newblock In \emph{EMNLP Findings 2022}, 2022.

\bibitem[Nguyen et~al.(2024)Nguyen, Nguyen, KC, Zhang, and Vu]{nguyen2024rlaf}
Minh Nguyen, Toan~Quoc Nguyen, Kishan KC, Zeyu Zhang, and Thuy Vu.
\newblock Reinforcement learning from answer reranking feedback for retrieval-augmented answer generation.
\newblock In \emph{Proceedings of INTERSPEECH}, 2024.

\bibitem[Yang and Soboroff(2016)]{yang2016trec}
Grace~Hui Yang and Ian Soboroff.
\newblock Trec 2016 dynamic domain track overview.
\newblock In \emph{TREC}, 2016.

\bibitem[Leonhardt et~al.(2023)Leonhardt, Rudra, and Anand]{leonhardt2023extractive}
Jurek Leonhardt, Koustav Rudra, and Avishek Anand.
\newblock Extractive explanations for interpretable text ranking.
\newblock \emph{ACM Transactions on Information Systems}, 41\penalty0 (4):\penalty0 1--27, 2023.

\bibitem[Jain and Wallace(2019)]{jain2019attention}
Sarthak Jain and Byron~C. Wallace.
\newblock Attention is not explanation.
\newblock In \emph{NAACL-HLT '19}, pages 3543--3556, 2019.

\bibitem[Wiegreffe and Pinter(2019)]{wiegreffe2019attention}
Sarah Wiegreffe and Yuval Pinter.
\newblock Attention is not not explanation.
\newblock In \emph{NAACL-HLT '19}, pages 854--863, 2019.

\bibitem[Hu et~al.(2022)Hu, Liu, Liu, Huai, Sun, and Wang]{hu2022seat}
Lijie Hu, Yixin Liu, Ninghao Liu, Mengdi Huai, Lichao Sun, and Di~Wang.
\newblock Seat: Stable and explainable attention, 2022.
\newblock URL \url{https://arxiv.org/abs/2211.13290}.

\bibitem[Pandian et~al.(2024)Pandian, Ganguly, and MacAvaney]{pandian2024explainability}
Saran Pandian, Debasis Ganguly, and Sean MacAvaney.
\newblock Evaluating the explainability of neural rankers, 2024.
\newblock URL \url{https://arxiv.org/abs/2403.01981}.

\bibitem[Anand et~al.(2022)Anand, Lyu, Idahl, Wang, Wallat, and Zhang]{anand2022survey}
Avishek Anand, Lijun Lyu, Maximilian Idahl, Yumeng Wang, Jonas Wallat, and Zijian Zhang.
\newblock Explainable information retrieval: A survey, 2022.
\newblock URL \url{https://arxiv.org/abs/2211.02405}.

\bibitem[Khattab and Zaharia(2020)]{khattab2020colbert}
Omar Khattab and Matei Zaharia.
\newblock Colbert: Efficient and effective passage search via contextualized late interaction over bert.
\newblock In \emph{SIGIR}, 2020.

\bibitem[Santhanam et~al.(2022)Santhanam, Khattab, Rekatsinas, and Zaharia]{santhanam2022colbertv2}
Keshav Santhanam, Omar Khattab, Theodoros Rekatsinas, and Matei Zaharia.
\newblock Colbertv2: Effective and efficient retrieval via lightweight late interaction.
\newblock In \emph{NAACL}, 2022.

\bibitem[Chan and Ng(2008)]{chan-ng-2008-maxsim}
Yee~Seng Chan and Hwee~Tou Ng.
\newblock {MAXSIM}: A maximum similarity metric for machine translation evaluation.
\newblock In Johanna~D. Moore, Simone Teufel, James Allan, and Sadaoki Furui, editors, \emph{Proceedings of ACL-08: HLT}, pages 55--62, Columbus, Ohio, June 2008. Association for Computational Linguistics.
\newblock URL \url{https://aclanthology.org/P08-1007/}.

\bibitem[Dai and Callan(2019)]{dai2019deepct}
Zhuyun Dai and Jamie Callan.
\newblock Context-aware term weighting for first-stage passage retrieval.
\newblock In \emph{SIGIR}, pages 985--988, 2019.
\newblock \doi{10.1145/3397271.3401204}.

\bibitem[Formal et~al.(2021{\natexlab{a}})Formal, Piwowarski, and Clinchant]{formal2021splade}
Thibault Formal, Benjamin Piwowarski, and Stéphane Clinchant.
\newblock Splade: Sparse lexical and expansion model for first stage ranking.
\newblock In \emph{SIGIR}, 2021{\natexlab{a}}.
\newblock arXiv preprint arXiv:2107.05720.

\bibitem[Formal et~al.(2021{\natexlab{b}})Formal, Lassance, Piwowarski, and Clinchant]{formal2021spladev2}
Thibault Formal, Carlos Lassance, Benjamin Piwowarski, and Stéphane Clinchant.
\newblock Splade v2: Sparse lexical and expansion model for information retrieval.
\newblock In \emph{SIGIR}, 2021{\natexlab{b}}.
\newblock arXiv preprint arXiv:2109.10086.

\bibitem[Choi et~al.(2022)Choi, Lee, Choi, Ko, Song, and Lee]{choi2022spade}
Eunseong Choi, Sunkyung Lee, Minjin Choi, Hyeseon Ko, Young‑In Song, and Jongwuk Lee.
\newblock Spade: Improving sparse representations using a dual document encoder for first‑stage retrieval.
\newblock In \emph{arXiv preprint arXiv:2209.05917}, 2022.

\bibitem[Robertson et~al.(1995)Robertson, Walker, Jones, Hancock-Beaulieu, and Gatford]{robertson1995okapi}
Stephen~E Robertson, Steve Walker, Susan Jones, Micheline~M Hancock-Beaulieu, and Mike Gatford.
\newblock Okapi at trec-3.
\newblock In \emph{Proceedings of the Third Text REtrieval Conference (TREC-3)}, pages 109--126. NIST, 1995.

\bibitem[Nair et~al.(2022{\natexlab{a}})Nair, Yang, Lawrie, Mayfield, and Oard]{nair2022spladex}
Suraj Nair, Eugene Yang, Dawn Lawrie, James Mayfield, and Douglas Oard.
\newblock Learning a sparse representation model for neural clir.
\newblock In \emph{Proceedings of DESIRES 2022: Design of Experimental Search \& Information REtrieval Systems}, 2022{\natexlab{a}}.

\bibitem[Iida and Okazaki(2022)]{ctitech2022_SPARSEUDA}
Hiroki Iida and Naoaki Okazaki.
\newblock Unsupervised domain adaptation for sparse retrieval by filling vocabulary and word frequency gaps.
\newblock \emph{arXiv preprint arXiv:2211.03988}, 2022.

\bibitem[Bruch et~al.(2023)Bruch, Gai, and Ingber]{Bruch2023_Hybrid}
Sebastian Bruch, Siyu Gai, and Amir Ingber.
\newblock An analysis of fusion functions for hybrid retrieval.
\newblock \emph{ACM Transactions on Information Systems}, 42\penalty0 (1):\penalty0 1--35, 2023.

\bibitem[Team(2025{\natexlab{a}})]{Weaviate2025_Hybrid}
Weaviate Team.
\newblock Hybrid search explained, 2025{\natexlab{a}}.
\newblock Weaviate blog.

\bibitem[Luan et~al.(2020)Luan, Eisenstein, Toutanova, and Collins]{luan2020mebert}
Yiqun Luan, Jacob Eisenstein, Kristina Toutanova, and Michael Collins.
\newblock Sparse, dense, and attentional representations for text retrieval.
\newblock In \emph{TACL}, 2020.

\bibitem[Gao et~al.(2021)Gao, Dai, and Callan]{gao2021coil}
Luyu Gao, Zhuyun Dai, and Jamie Callan.
\newblock {COIL}: Revisit exact lexical match in information retrieval with contextualized inverted list.
\newblock In \emph{NAACL}, 2021.

\bibitem[Liu et~al.(2021{\natexlab{b}})Liu, Hashimoto, Zhou, Yavuz, Xiong, and Yu]{liu-etal-2021-dense-hierarchical}
Ye~Liu, Kazuma Hashimoto, Yingbo Zhou, Semih Yavuz, Caiming Xiong, and Philip Yu.
\newblock Dense hierarchical retrieval for open-domain question answering.
\newblock In Marie-Francine Moens, Xuanjing Huang, Lucia Specia, and Scott Wen-tau Yih, editors, \emph{Findings of the Association for Computational Linguistics: EMNLP 2021}, pages 188--200, Punta Cana, Dominican Republic, November 2021{\natexlab{b}}. Association for Computational Linguistics.
\newblock \doi{10.18653/v1/2021.findings-emnlp.19}.
\newblock URL \url{https://aclanthology.org/2021.findings-emnlp.19/}.

\bibitem[Wu et~al.(2022)Wu, Zhao, Hu, Minervini, Stenetorp, and Riedel]{wu2022emat}
Yuxiang Wu, Yu~Zhao, Baotian Hu, Pasquale Minervini, Pontus Stenetorp, and Sebastian Riedel.
\newblock An efficient memory-augmented transformer for knowledge-intensive nlp tasks.
\newblock In \emph{EMNLP}, 2022.

\bibitem[Ge et~al.(2023)Ge, Xiong, Rosset, Overwijk, Han, and Bennett]{ge2023moma}
Suyu Ge, Chenyan Xiong, Corby Rosset, Arnold Overwijk, Jiawei Han, and Paul Bennett.
\newblock Augmenting zero-shot dense retrievers with plug-in mixture-of-memories (moma).
\newblock \emph{arXiv preprint arXiv:2302.03754}, 2023.

\bibitem[Chen et~al.(2017{\natexlab{b}})Chen, Fisch, Weston, and Bordes]{chen_2017_qa}
Danqi Chen, Adam Fisch, Jason Weston, and Antoine Bordes.
\newblock Reading {W}ikipedia to answer open-domain questions.
\newblock In Regina Barzilay and Min-Yen Kan, editors, \emph{Proceedings of the 55th Annual Meeting of the Association for Computational Linguistics (Volume 1: Long Papers)}, pages 1870--1879, Vancouver, Canada, July 2017{\natexlab{b}}. Association for Computational Linguistics.
\newblock \doi{10.18653/v1/P17-1171}.
\newblock URL \url{https://aclanthology.org/P17-1171/}.

\bibitem[Izacard and Grave(2021)]{izacard_2021}
Gautier Izacard and Edouard Grave.
\newblock Leveraging passage retrieval with generative models for open domain question answering.
\newblock In Paola Merlo, Jorg Tiedemann, and Reut Tsarfaty, editors, \emph{Proceedings of the 16th Conference of the European Chapter of the Association for Computational Linguistics: Main Volume}, pages 874--880, Online, April 2021. Association for Computational Linguistics.
\newblock \doi{10.18653/v1/2021.eacl-main.74}.
\newblock URL \url{https://aclanthology.org/2021.eacl-main.74/}.

\bibitem[Lewis et~al.(2020{\natexlab{b}})Lewis, Perez, Piktus, Petroni, Karpukhin, Goyal, K\"{u}ttler, Lewis, Yih, Rockt\"{a}schel, Riedel, and Kiela]{lewis_2020_rag}
Patrick Lewis, Ethan Perez, Aleksandra Piktus, Fabio Petroni, Vladimir Karpukhin, Naman Goyal, Heinrich K\"{u}ttler, Mike Lewis, Wen-tau Yih, Tim Rockt\"{a}schel, Sebastian Riedel, and Douwe Kiela.
\newblock Retrieval-augmented generation for knowledge-intensive nlp tasks.
\newblock In H.~Larochelle, M.~Ranzato, R.~Hadsell, M.F. Balcan, and H.~Lin, editors, \emph{Advances in Neural Information Processing Systems}, volume~33, pages 9459--9474. Curran Associates, Inc., 2020{\natexlab{b}}.
\newblock URL \url{https://proceedings.neurips.cc/paper_files/paper/2020/file/6b493230205f780e1bc26945df7481e5-Paper.pdf}.

\bibitem[Iida et~al.(2019)Iida, Kruengkrai, Ishida, Torisawa, Oh, and Kloetzer]{iida_2019}
Ryu Iida, Canasai Kruengkrai, Ryo Ishida, Kentaro Torisawa, Jong-Hoon Oh, and Julien Kloetzer.
\newblock Exploiting background knowledge in compact answer generation for why-questions.
\newblock \emph{Proceedings of the AAAI Conference on Artificial Intelligence}, 33\penalty0 (01):\penalty0 142--151, Jul. 2019.
\newblock \doi{10.1609/aaai.v33i01.3301142}.
\newblock URL \url{https://ojs.aaai.org/index.php/AAAI/article/view/3779}.

\bibitem[Goodwin et~al.(2020)Goodwin, Savery, and Demner-Fushman]{goodwin_2020}
Travis Goodwin, Max Savery, and Dina Demner-Fushman.
\newblock Towards {Z}ero-{S}hot {C}onditional {S}ummarization with {A}daptive {M}ulti-{T}ask {F}ine-{T}uning.
\newblock In Trevor Cohn, Yulan He, and Yang Liu, editors, \emph{Findings of the Association for Computational Linguistics: EMNLP 2020}, pages 3215--3226, Online, November 2020. Association for Computational Linguistics.
\newblock \doi{10.18653/v1/2020.findings-emnlp.289}.
\newblock URL \url{https://aclanthology.org/2020.findings-emnlp.289/}.

\bibitem[Deng et~al.(2020)Deng, Lam, Xie, Chen, Li, Yang, and Shen]{deng_2020}
Yang Deng, Wai Lam, Yuexiang Xie, Daoyuan Chen, Yaliang Li, Min Yang, and Ying Shen.
\newblock Joint learning of answer selection and answer summary generation in community question answering.
\newblock \emph{Proceedings of the AAAI Conference on Artificial Intelligence}, 34\penalty0 (05):\penalty0 7651--7658, Apr. 2020.
\newblock \doi{10.1609/aaai.v34i05.6266}.
\newblock URL \url{https://ojs.aaai.org/index.php/AAAI/article/view/6266}.

\bibitem[Hsu et~al.(2021)Hsu, Lind, Soldaini, and Moschitti]{hsu_2021_qa}
Chao-Chun Hsu, Eric Lind, Luca Soldaini, and Alessandro Moschitti.
\newblock Answer generation for retrieval-based question answering systems.
\newblock In Chengqing Zong, Fei Xia, Wenjie Li, and Roberto Navigli, editors, \emph{Findings of the Association for Computational Linguistics: ACL-IJCNLP 2021}, pages 4276--4282, Online, August 2021. Association for Computational Linguistics.
\newblock \doi{10.18653/v1/2021.findings-acl.374}.
\newblock URL \url{https://aclanthology.org/2021.findings-acl.374/}.

\bibitem[Muller et~al.(2022)Muller, Soldaini, Koncel-Kedziorski, Lind, and Moschitti]{muller_2022}
Benjamin Muller, Luca Soldaini, Rik Koncel-Kedziorski, Eric Lind, and Alessandro Moschitti.
\newblock Cross-lingual open-domain question answering with answer sentence generation, 2022.
\newblock URL \url{https://arxiv.org/abs/2110.07150}.

\bibitem[Jurafsky and Martin(2025{\natexlab{a}})]{jurafsky_2025}
Daniel Jurafsky and James~H. Martin.
\newblock \emph{Speech and Language Processing: An Introduction to Natural Language Processing, Computational Linguistics, and Speech Recognition with Language Models}, chapter Question Answering, Information Retrieval, and Retrieval-Augmented Generation.
\newblock Forthcoming, 3rd edition, 2025{\natexlab{a}}.

\bibitem[Dahl et~al.(2024)Dahl, Magesh, Suzgun, and Ho]{dahl_2024}
Matthew Dahl, Varun Magesh, Mirac Suzgun, and Daniel~E Ho.
\newblock Large legal fictions: Profiling legal hallucinations in large language models.
\newblock \emph{Journal of Legal Analysis}, 16\penalty0 (1):\penalty0 64--93, 06 2024.
\newblock ISSN 2161-7201.
\newblock \doi{10.1093/jla/laae003}.
\newblock URL \url{https://doi.org/10.1093/jla/laae003}.

\bibitem[Zhou et~al.(2024)Zhou, Hwang, Ren, and Sap]{zhou_2024_calibration}
Kaitlyn Zhou, Jena~D. Hwang, Xiang Ren, and Maarten Sap.
\newblock Relying on the unreliable: The impact of language models' reluctance to express uncertainty, 2024.
\newblock URL \url{https://arxiv.org/abs/2401.06730}.

\bibitem[Aksitov et~al.(2023)Aksitov, Chang, Reitter, Shakeri, and Sung]{aksitov_2023}
Renat Aksitov, Chung-Ching Chang, David Reitter, Siamak Shakeri, and Yunhsuan Sung.
\newblock Characterizing attribution and fluency tradeoffs for retrieval-augmented large language models, 2023.
\newblock URL \url{https://arxiv.org/abs/2302.05578}.

\bibitem[Lee and Lee(2019)]{lee_2019_qa}
Chia-Hsuan Lee and Hung-Yi Lee.
\newblock Cross-lingual transfer learning for question answering, 2019.
\newblock URL \url{https://arxiv.org/abs/1907.06042}.

\bibitem[Lassance(2023)]{lassance2023multisplade}
Carlos Lassance.
\newblock Extending english ir methods to multi-lingual ir.
\newblock arXiv preprint arXiv:2302.14723, 2023.

\bibitem[Feng et~al.(2022)Feng, Yang, Cer, Arivazhagan, and Wang]{feng2022language}
Fangxiaoyu Feng, Yinfei Yang, Daniel Cer, Naveen Arivazhagan, and Wei Wang.
\newblock Language-agnostic bert sentence embedding.
\newblock In \emph{ACL}, 2022.

\bibitem[Reimers and Gurevych(2020)]{reimers2020making}
Nils Reimers and Iryna Gurevych.
\newblock Making monolingual sentence embeddings multilingual using knowledge distillation.
\newblock In \emph{Proceedings of the 2020 Conference on Empirical Methods in Natural Language Processing (EMNLP)}, pages 4513--4525. ACL, 2020.

\bibitem[Zhang et~al.(2022{\natexlab{b}})Zhang, Bajaj, Ma, Asai, Xiong, Callan, and Hajishirzi]{zhang2022mr}
Yi~Zhang, Payal Bajaj, Xiao Ma, Akari Asai, Chenyan Xiong, Jamie Callan, and Hannaneh Hajishirzi.
\newblock Mr. tydi: A multi-lingual benchmark for dense retrieval.
\newblock In \emph{Proceedings of the 45th International ACM SIGIR Conference on Research and Development in Information Retrieval}, 2022{\natexlab{b}}.

\bibitem[Tu and Padmanabhan(2022)]{tu-padmanabhan-2022-mia}
Zhucheng Tu and Sarguna~Janani Padmanabhan.
\newblock {MIA} 2022 shared task submission: Leveraging entity representations, dense‑sparse hybrids, and fusion‑in‑decoder for cross‑lingual question answering.
\newblock In \emph{Proceedings of the Workshop on Multilingual Information Access (MIA)}, pages 100--107, Seattle, USA, July 2022. Association for Computational Linguistics.
\newblock \doi{10.18653/v1/2022.mia-1.10}.
\newblock URL \url{https://aclanthology.org/2022.mia-1.10/}.

\bibitem[Ogundepo et~al.(2022)Ogundepo, Zhang, Sun, Duh, and Lin]{ogundepo_2022_africlirmatrix}
Odunayo Ogundepo, Xinyu Zhang, Shuo Sun, Kevin Duh, and Jimmy Lin.
\newblock {A}fri{CLIRM}atrix: Enabling cross-lingual information retrieval for {A}frican languages.
\newblock In Yoav Goldberg, Zornitsa Kozareva, and Yue Zhang, editors, \emph{Proceedings of the 2022 Conference on Empirical Methods in Natural Language Processing}, pages 8721--8728, Abu Dhabi, United Arab Emirates, December 2022. Association for Computational Linguistics.
\newblock \doi{10.18653/v1/2022.emnlp-main.597}.
\newblock URL \url{https://aclanthology.org/2022.emnlp-main.597/}.

\bibitem[Ranaldi et~al.(2025)Ranaldi, Haddow, and Birch]{ranaldi2025multilingual}
Leonardo Ranaldi, Barry Haddow, and Alexandra Birch.
\newblock Multilingual retrieval‑augmented generation for knowledge‑intensive task.
\newblock \emph{arXiv preprint arXiv:2504.03616}, 2025.
\newblock Published April 2025.

\bibitem[Gao et~al.(2022)Gao, Zhong, Callan, Xiong, and Yih]{gao2022cocondense}
Luyu Gao, Haoyu Zhong, Jamie Callan, Longqing Xiong, and Wen-tau Yih.
\newblock Cocondenser: Contrastive pretraining with hard negatives for dense passage retrieval.
\newblock In \emph{EMNLP}, 2022.

\bibitem[Sun and Duh(2020)]{sun_2020_clirmatrix}
Shuo Sun and Kevin Duh.
\newblock {CLIRM}atrix: A massively large collection of bilingual and multilingual datasets for cross-lingual information retrieval.
\newblock In Bonnie Webber, Trevor Cohn, Yulan He, and Yang Liu, editors, \emph{Proceedings of the 2020 Conference on Empirical Methods in Natural Language Processing (EMNLP)}, pages 4160--4170, Online, November 2020. Association for Computational Linguistics.
\newblock \doi{10.18653/v1/2020.emnlp-main.340}.
\newblock URL \url{https://aclanthology.org/2020.emnlp-main.340/}.

\bibitem[Huang et~al.(2023)Huang, Yu, and Allan]{huang2023optimal}
Zhiqi Huang, Puxuan Yu, and James Allan.
\newblock Improving cross‐lingual information retrieval on low‑resource languages via optimal transport distillation.
\newblock In \emph{Proceedings of the Sixteenth ACM International Conference on Web Search and Data Mining (WSDM)}, pages 1048--1056, 2023.
\newblock \doi{10.1145/3539597.3570468}.

\bibitem[Izacard et~al.(2022)Izacard, Caron, Hosseini, Riedel, Bojanowski, Joulin, and Grave]{izacard_2022_mcontriever}
Gautier Izacard, Mathilde Caron, Lucas Hosseini, Sebastian Riedel, Piotr Bojanowski, Armand Joulin, and Edouard Grave.
\newblock Unsupervised dense information retrieval with contrastive learning, 2022.
\newblock URL \url{https://arxiv.org/abs/2112.09118}.

\bibitem[Thakur et~al.(2024)Thakur, Ni, Ábrego, Wieting, Lin, and Cer]{thakur_2024_swimir}
Nandan Thakur, Jianmo Ni, Gustavo~Hernández Ábrego, John Wieting, Jimmy Lin, and Daniel Cer.
\newblock Leveraging llms for synthesizing training data across many languages in multilingual dense retrieval, 2024.
\newblock URL \url{https://arxiv.org/abs/2311.05800}.

\bibitem[Fei et~al.(2021)Fei, Yu, and Li]{fei-etal-2021-cross}
Hongliang Fei, Tan Yu, and Ping Li.
\newblock Cross-lingual cross-modal pretraining for multimodal retrieval.
\newblock In Kristina Toutanova, Anna Rumshisky, Luke Zettlemoyer, Dilek Hakkani-Tur, Iz~Beltagy, Steven Bethard, Ryan Cotterell, Tanmoy Chakraborty, and Yichao Zhou, editors, \emph{Proceedings of the 2021 Conference of the North American Chapter of the Association for Computational Linguistics: Human Language Technologies}, pages 3644--3650, Online, June 2021. Association for Computational Linguistics.
\newblock \doi{10.18653/v1/2021.naacl-main.285}.
\newblock URL \url{https://aclanthology.org/2021.naacl-main.285/}.

\bibitem[Wang et~al.(2024{\natexlab{a}})Wang, Wang, Zhou, Wang, Li, Hua, and Tang]{wang2024multimodalllmenhancedcrosslingual}
Yabing Wang, Le~Wang, Qiang Zhou, Zhibin Wang, Hao Li, Gang Hua, and Wei Tang.
\newblock Multimodal llm enhanced cross-lingual cross-modal retrieval, 2024{\natexlab{a}}.
\newblock URL \url{https://arxiv.org/abs/2409.19961}.

\bibitem[Berry et~al.(2023)Berry, Shih, Wang, Chang, yi~Lee, and Harwath]{berry2023mspeechclipleveraginglargescalepretrained}
Layne Berry, Yi-Jen Shih, Hsuan-Fu Wang, Heng-Jui Chang, Hung yi~Lee, and David Harwath.
\newblock M-speechclip: Leveraging large-scale, pre-trained models for multilingual speech to image retrieval, 2023.
\newblock URL \url{https://arxiv.org/abs/2211.01180}.

\bibitem[Litschko et~al.(2018)Litschko, Glavaš, Ponzetto, and Vulić]{litschko2018unsupervised}
Robert Litschko, Goran Glavaš, Simone~Paolo Ponzetto, and Ivan Vulić.
\newblock Unsupervised cross-lingual information retrieval using monolingual data only.
\newblock \emph{arXiv preprint arXiv:1805.00879}, May 2018.

\bibitem[Litschko et~al.(2022)Litschko, Vulić, Ponzetto, and Glavaš]{litschko2022cross}
Robert Litschko, Ivan Vulić, Simone~Paolo Ponzetto, and Goran Glavaš.
\newblock On cross-lingual retrieval with multilingual text encoders.
\newblock \emph{Information Retrieval Journal}, 25:\penalty0 149--183, 2022.

\bibitem[Gore et~al.(2024)Gore, Polletta, and Mansouri]{gore2024crossmath}
James Gore, Joseph Polletta, and Behrooz Mansouri.
\newblock Crossmath: Towards cross-lingual math information retrieval.
\newblock In \emph{Proceedings of the 2024 ACM SIGIR International Conference on Theory of Information Retrieval}, pages 101--105, 2024.

\bibitem[Yang et~al.(2024{\natexlab{c}})Yang, Lawrie, and Mayfield]{Yang_2024}
Eugene Yang, Dawn Lawrie, and James Mayfield.
\newblock Distillation for multilingual information retrieval.
\newblock In \emph{Proceedings of the 47th International ACM SIGIR Conference on Research and Development in Information Retrieval}, SIGIR 2024, page 2368–2373. ACM, July 2024{\natexlab{c}}.
\newblock \doi{10.1145/3626772.3657955}.
\newblock URL \url{http://dx.doi.org/10.1145/3626772.3657955}.

\bibitem[Zhou et~al.(2012)Zhou, Truran, Brailsford, Wade, and Ashman]{zhou_2012}
Dong Zhou, Mark Truran, Tim Brailsford, Vincent Wade, and Helen Ashman.
\newblock Translation techniques in cross-language information retrieval.
\newblock \emph{ACM Computing Surveys}, 45\penalty0 (1), 2012.
\newblock ISSN 0360-0300.
\newblock \doi{10.1145/2379776.2379777}.
\newblock URL \url{https://doi.org/10.1145/2379776.2379777}.

\bibitem[Fox(1989)]{fox_1989}
Christopher Fox.
\newblock A stop list for general text.
\newblock \emph{SIGIR Forum}, 24\penalty0 (1–2):\penalty0 19–21, September 1989.
\newblock ISSN 0163-5840.
\newblock \doi{10.1145/378881.378888}.
\newblock URL \url{https://doi.org/10.1145/378881.378888}.

\bibitem[Ture and Lin(2014)]{ture_2014}
Ferhan Ture and Jimmy Lin.
\newblock Exploiting representations from statistical machine translation for cross-language information retrieval.
\newblock \emph{ACM Trans. Inf. Syst.}, 32\penalty0 (4), October 2014.
\newblock ISSN 1046-8188.
\newblock \doi{10.1145/2644807}.
\newblock URL \url{https://doi.org/10.1145/2644807}.

\bibitem[Hull and Grefenstette(1996)]{hull_1996}
David~A. Hull and Gregory Grefenstette.
\newblock Querying across languages: a dictionary-based approach to multilingual information retrieval.
\newblock In \emph{Proceedings of the 19th Annual International ACM SIGIR Conference on Research and Development in Information Retrieval}, SIGIR '96, page 49–57, New York, NY, USA, 1996. Association for Computing Machinery.
\newblock ISBN 0897917928.
\newblock \doi{10.1145/243199.243212}.
\newblock URL \url{https://doi.org/10.1145/243199.243212}.

\bibitem[Ballesteros and Croft(1996)]{ballesteros_1996}
Lisa Ballesteros and W.~Bruce Croft.
\newblock Dictionary methods for cross-lingual information retrieval.
\newblock In \emph{Proceedings of the 7th International Conference on Database and Expert Systems Applications}, DEXA '96, page 791–801, Berlin, Heidelberg, 1996. Springer-Verlag.
\newblock ISBN 354061656X.

\bibitem[Hull(1997)]{hull_1997}
David~A. Hull.
\newblock Using structured queries for disambiguation in cross-language information retrieval.
\newblock In \emph{Proceedings of the AAAI Spring Symposium on Cross-Language Text and Speech Retrieval}, pages 84--98, Stanford, CA, 1997. AAAI Press.

\bibitem[Pirkola(1998)]{pirkola_1998}
Ari Pirkola.
\newblock The effects of query structure and dictionary setups in dictionary-based cross-language information retrieval.
\newblock In \emph{Proceedings of the 21st Annual International ACM SIGIR Conference on Research and Development in Information Retrieval}, SIGIR '98, page 55–63, New York, NY, USA, 1998. Association for Computing Machinery.
\newblock ISBN 1581130155.
\newblock \doi{10.1145/290941.290957}.
\newblock URL \url{https://doi.org/10.1145/290941.290957}.

\bibitem[Darwish and Oard(2003)]{darwish_oard_2003}
Kareem Darwish and Douglas~W. Oard.
\newblock Probabilistic structured query methods.
\newblock In \emph{Proceedings of the 26th Annual International ACM SIGIR Conference on Research and Development in Informaion Retrieval}, SIGIR '03, page 338–344, New York, NY, USA, 2003. Association for Computing Machinery.
\newblock ISBN 1581136463.
\newblock \doi{10.1145/860435.860497}.
\newblock URL \url{https://doi.org/10.1145/860435.860497}.

\bibitem[Levow and Oard(2000)]{levow_oard_2000}
Gina-Anne Levow and Douglas~W. Oard.
\newblock Translingual topic tracking with prise.
\newblock In \emph{Working Notes of the 3rd Topic Detection and Tracking Workshop (TDT-3)}. National Institutes of Standards and Technology, 2000.

\bibitem[Leek et~al.(2000)Leek, Jin, Sista, and Schwartz]{leek_jin_2000}
Tim Leek, Hubert Jin, Sreenivasa Sista, and Richard Schwartz.
\newblock The bbn cross-lingual topic detection and tracking system.
\newblock In \emph{Working Notes of the 3rd Topic Detection and Tracking Workshop (TDT-3)}. National Institutes of Standards and Technology, 2000.

\bibitem[Xu and Weischedel(2005)]{xu_2005}
Jinxi Xu and Ralph Weischedel.
\newblock Empirical studies on the impact of lexical resources on clir performance.
\newblock \emph{Information Processing and Management}, 41\penalty0 (3):\penalty0 475–487, May 2005.
\newblock ISSN 0306-4573.
\newblock \doi{10.1016/j.ipm.2004.06.009}.
\newblock URL \url{https://doi.org/10.1016/j.ipm.2004.06.009}.

\bibitem[Roukos et~al.(1995)Roukos, Graff, and Melamed]{roukos-graff-melamed-1995-hansard}
Salim Roukos, David Graff, and Dan Melamed.
\newblock Hansard french/english (ldc95t20).
\newblock Linguistic Data Consortium, Web Download, 1995.
\newblock URL \url{https://catalog.ldc.upenn.edu/LDC95T20}.
\newblock LDC95T20.

\bibitem[Koehn(2005)]{koehn-2005-europarl}
Philipp Koehn.
\newblock {E}uroparl: A parallel corpus for statistical machine translation.
\newblock In \emph{Proceedings of Machine Translation Summit X: Papers}, pages 79--86, Phuket, Thailand, September 13-15 2005.
\newblock URL \url{https://aclanthology.org/2005.mtsummit-papers.11/}.

\bibitem[Ziemski et~al.(2016)Ziemski, Junczys-Dowmunt, and Pouliquen]{ziemski-etal-2016-united}
Micha{\l} Ziemski, Marcin Junczys-Dowmunt, and Bruno Pouliquen.
\newblock The united nations parallel corpus v1.0.
\newblock In \emph{Proceedings of the Tenth International Conference on Language Resources and Evaluation (LREC’16)}, pages 3530--3534, Portoro{\v{z}}, Slovenia, May 2016. European Language Resources Association (ELRA).
\newblock URL \url{https://aclanthology.org/L16-1561/}.

\bibitem[Chew et~al.(2006{\natexlab{a}})Chew, Verzi, Bauer, and McClain]{chew_2006}
Peter~A. Chew, Steve~J. Verzi, Travis~L. Bauer, and Jonathan~T. McClain.
\newblock Evaluation of the {B}ible as a resource for cross-language information retrieval.
\newblock In Andreas Witt, Gilles S{\'e}rasset, Susan Armstrong, Jim Breen, Ulrich Heid, and Felix Sasaki, editors, \emph{Proceedings of the Workshop on Multilingual Language Resources and Interoperability}, pages 68--74, Sydney, Australia, July 2006{\natexlab{a}}. Association for Computational Linguistics.
\newblock URL \url{https://aclanthology.org/W06-1009/}.

\bibitem[Braschler and Schäuble(2000)]{braschler_2000}
Martin Braschler and Peter Schäuble.
\newblock Using corpus-based approaches in a system for multilingual information retrieval.
\newblock \emph{Information Retrieval}, 3\penalty0 (3):\penalty0 273–284, October 2000.
\newblock ISSN 1386-4564.
\newblock \doi{10.1023/A:1026525127581}.
\newblock URL \url{https://doi.org/10.1023/A:1026525127581}.

\bibitem[Moulinier and Molina-Salgado(2003)]{moulinier_2003}
Isabelle Moulinier and Hugo Molina-Salgado.
\newblock Thomson legal and regulatory experiments for clef 2002.
\newblock In Carol Peters, Martin Braschler, Julio Gonzalo, and Michael Kluck, editors, \emph{Advances in Cross-Language Information Retrieval}, pages 155--163, Berlin, Heidelberg, 2003. Springer Berlin Heidelberg.
\newblock ISBN 978-3-540-45237-9.

\bibitem[Franz et~al.(1999)Franz, Mccarley, and Roukos]{franz_1999}
Martin Franz, J.~Mccarley, and Salim Roukos.
\newblock Ad hoc and multilingual information retrieval at ibm.
\newblock In \emph{Proceedings of trec Conference}, 1999.

\bibitem[Brown et~al.(1990)Brown, Cocke, Pietra, Pietra, Jelinek, Lafferty, Mercer, and Roossin]{brown_1990}
Peter~F. Brown, John Cocke, Stephen A.~Della Pietra, Vincent J.~Della Pietra, Fredrick Jelinek, John~D. Lafferty, Robert~L. Mercer, and Paul~S. Roossin.
\newblock A statistical approach to machine translation.
\newblock \emph{Computational Linguistics}, 16\penalty0 (2):\penalty0 79–85, 1990.
\newblock ISSN 0891-2017.

\bibitem[Brown et~al.(1993)Brown, Pietra, Pietra, and Mercer]{brown_1993}
Peter~F. Brown, Vincent J.~Della Pietra, Stephen A.~Della Pietra, and Robert~L. Mercer.
\newblock The mathematics of statistical machine translation: parameter estimation.
\newblock \emph{Computational Linguistics}, 19\penalty0 (2):\penalty0 263–311, 1993.
\newblock ISSN 0891-2017.

\bibitem[Stahlberg(2020)]{stahlberg_2020_nmt}
Felix Stahlberg.
\newblock Neural machine translation: A review and survey, 2020.
\newblock URL \url{https://arxiv.org/abs/1912.02047}.

\bibitem[Mohamed et~al.(2021)Mohamed, Elsayed, Hassan, and Abdou]{mohamed_2021_nmt}
Shereen~A. Mohamed, Ashraf~A. Elsayed, Y.~F. Hassan, and Mohamed~A. Abdou.
\newblock Neural machine translation: past, present, and future.
\newblock \emph{Neural Comput. Appl.}, 33\penalty0 (23):\penalty0 15919–15931, December 2021.
\newblock ISSN 0941-0643.
\newblock \doi{10.1007/s00521-021-06268-0}.
\newblock URL \url{https://doi.org/10.1007/s00521-021-06268-0}.

\bibitem[Ranathunga et~al.(2023)Ranathunga, Lee, Prifti~Skenduli, Shekhar, Alam, and Kaur]{ranathunga_2023_nmt}
Surangika Ranathunga, En-Shiun~Annie Lee, Marjana Prifti~Skenduli, Ravi Shekhar, Mehreen Alam, and Rishemjit Kaur.
\newblock Neural machine translation for low-resource languages: A survey.
\newblock \emph{ACM Comput. Surv.}, 55\penalty0 (11), February 2023.
\newblock ISSN 0360-0300.
\newblock \doi{10.1145/3567592}.
\newblock URL \url{https://doi.org/10.1145/3567592}.

\bibitem[Dabre et~al.(2020)Dabre, Chu, and Kunchukuttan]{dabre_2020_mnmt}
Raj Dabre, Chenhui Chu, and Anoop Kunchukuttan.
\newblock A survey of multilingual neural machine translation.
\newblock \emph{ACM Comput. Surv.}, 53\penalty0 (5), September 2020.
\newblock ISSN 0360-0300.
\newblock \doi{10.1145/3406095}.
\newblock URL \url{https://doi.org/10.1145/3406095}.

\bibitem[McCann et~al.(2017)McCann, Bradbury, Xiong, and Socher]{McCann_2017}
Bryan McCann, James Bradbury, Caiming Xiong, and Richard Socher.
\newblock Learned in translation: Contextualized word vectors.
\newblock In I.~Guyon, U.~Von Luxburg, S.~Bengio, H.~Wallach, R.~Fergus, S.~Vishwanathan, and R.~Garnett, editors, \emph{Advances in Neural Information Processing Systems}, volume~30. Curran Associates, Inc., 2017.
\newblock URL \url{https://proceedings.neurips.cc/paper_files/paper/2017/file/20c86a628232a67e7bd46f76fba7ce12-Paper.pdf}.

\bibitem[Peters et~al.(2018)Peters, Neumann, Zettlemoyer, and tau Yih]{peters_2018}
Matthew~E. Peters, Mark Neumann, Luke Zettlemoyer, and Wen tau Yih.
\newblock Dissecting contextual word embeddings: Architecture and representation, 2018.
\newblock URL \url{https://arxiv.org/abs/1808.08949}.

\bibitem[Kalchbrenner and Blunsom(2013)]{kalchbrenner_13}
Nal Kalchbrenner and Phil Blunsom.
\newblock Recurrent continuous translation models.
\newblock In \emph{Proceedings of the 2013 Conference on Empirical Methods in Natural Language Processing}, Seattle, October 2013. Association for Computational Linguistics.

\bibitem[Cho et~al.(2014)Cho, van Merri{\"e}nboer, Bahdanau, and Bengio]{cho_2014_nmt}
Kyunghyun Cho, Bart van Merri{\"e}nboer, Dzmitry Bahdanau, and Yoshua Bengio.
\newblock On the properties of neural machine translation: Encoder{--}decoder approaches.
\newblock In Dekai Wu, Marine Carpuat, Xavier Carreras, and Eva~Maria Vecchi, editors, \emph{Proceedings of {SSST}-8, Eighth Workshop on Syntax, Semantics and Structure in Statistical Translation}, pages 103--111, Doha, Qatar, October 2014. Association for Computational Linguistics.
\newblock \doi{10.3115/v1/W14-4012}.
\newblock URL \url{https://aclanthology.org/W14-4012/}.

\bibitem[Bahdanau et~al.(2015)Bahdanau, Cho, and Bengio]{bahdanau_2015_attention}
Dzmitry Bahdanau, Kyunghyun Cho, and Yoshua Bengio.
\newblock Neural machine translation by jointly learning to align and translate.
\newblock In Yoshua Bengio and Yann LeCun, editors, \emph{3rd International Conference on Learning Representations, {ICLR} 2015}, 2015.
\newblock URL \url{http://arxiv.org/abs/1409.0473}.

\bibitem[Vaswani et~al.(2017)Vaswani, Shazeer, Parmar, Uszkoreit, Jones, Gomez, Kaiser, and Polosukhin]{vaswani_2017}
Ashish Vaswani, Noam Shazeer, Niki Parmar, Jakob Uszkoreit, Llion Jones, Aidan~N Gomez, \L~ukasz Kaiser, and Illia Polosukhin.
\newblock Attention is all you need.
\newblock In I.~Guyon, U.~Von Luxburg, S.~Bengio, H.~Wallach, R.~Fergus, S.~Vishwanathan, and R.~Garnett, editors, \emph{Advances in Neural Information Processing Systems}, volume~30. Curran Associates, Inc., 2017.
\newblock URL \url{https://proceedings.neurips.cc/paper_files/paper/2017/file/3f5ee243547dee91fbd053c1c4a845aa-Paper.pdf}.

\bibitem[Gage(1994)]{gage_1994_bpe}
Philip Gage.
\newblock A new algorithm for data compression.
\newblock \emph{C Users J.}, 12\penalty0 (2):\penalty0 23–38, February 1994.
\newblock ISSN 0898-9788.

\bibitem[Sennrich et~al.(2016)Sennrich, Haddow, and Birch]{sennrich_2016_bpe}
Rico Sennrich, Barry Haddow, and Alexandra Birch.
\newblock Neural machine translation of rare words with subword units.
\newblock In Katrin Erk and Noah~A. Smith, editors, \emph{Proceedings of the 54th Annual Meeting of the Association for Computational Linguistics (Volume 1: Long Papers)}, pages 1715--1725, Berlin, Germany, August 2016. Association for Computational Linguistics.
\newblock \doi{10.18653/v1/P16-1162}.
\newblock URL \url{https://aclanthology.org/P16-1162/}.

\bibitem[Provilkov et~al.(2020)Provilkov, Emelianenko, and Voita]{provilkov_2020_bpe}
Ivan Provilkov, Dmitrii Emelianenko, and Elena Voita.
\newblock Bpe-dropout: Simple and effective subword regularization, 2020.
\newblock URL \url{https://arxiv.org/abs/1910.13267}.

\bibitem[Schuster and Nakajima(2012)]{schuster_2012_wordpiece}
Mike Schuster and Kaisuke Nakajima.
\newblock Japanese and korean voice search.
\newblock In \emph{2012 IEEE International Conference on Acoustics, Speech and Signal Processing (ICASSP)}, pages 5149--5152, 2012.
\newblock \doi{10.1109/ICASSP.2012.6289079}.

\bibitem[Wu et~al.(2016)Wu, Schuster, Chen, Le, Norouzi, Macherey, Krikun, Cao, Gao, Macherey, Klingner, Shah, Johnson, Liu, Łukasz Kaiser, Gouws, Kato, Kudo, Kazawa, Stevens, Kurian, Patil, Wang, Young, Smith, Riesa, Rudnick, Vinyals, Corrado, Hughes, and Dean]{wu_2016_wordpiece}
Yonghui Wu, Mike Schuster, Zhifeng Chen, Quoc~V. Le, Mohammad Norouzi, Wolfgang Macherey, Maxim Krikun, Yuan Cao, Qin Gao, Klaus Macherey, Jeff Klingner, Apurva Shah, Melvin Johnson, Xiaobing Liu, Łukasz Kaiser, Stephan Gouws, Yoshikiyo Kato, Taku Kudo, Hideto Kazawa, Keith Stevens, George Kurian, Nishant Patil, Wei Wang, Cliff Young, Jason Smith, Jason Riesa, Alex Rudnick, Oriol Vinyals, Greg Corrado, Macduff Hughes, and Jeffrey Dean.
\newblock Google's neural machine translation system: Bridging the gap between human and machine translation, 2016.
\newblock URL \url{https://arxiv.org/abs/1609.08144}.

\bibitem[Kudo and Richardson(2018)]{kudo_2018_sentencepiece}
Taku Kudo and John Richardson.
\newblock {S}entence{P}iece: A simple and language independent subword tokenizer and detokenizer for neural text processing.
\newblock In Eduardo Blanco and Wei Lu, editors, \emph{Proceedings of the 2018 Conference on Empirical Methods in Natural Language Processing: System Demonstrations}, pages 66--71, Brussels, Belgium, November 2018. Association for Computational Linguistics.
\newblock \doi{10.18653/v1/D18-2012}.
\newblock URL \url{https://aclanthology.org/D18-2012/}.

\bibitem[Nair et~al.(2022{\natexlab{b}})Nair, Yang, Lawrie, Duh, McNamee, Murray, Mayfield, and Oard]{nair_2022}
Suraj Nair, Eugene Yang, Dawn Lawrie, Kevin Duh, Paul McNamee, Kenton Murray, James Mayfield, and Douglas~W. Oard.
\newblock Transfer learning approaches for building cross-language dense retrieval models, 2022{\natexlab{b}}.
\newblock URL \url{https://arxiv.org/abs/2201.08471}.

\bibitem[Yang et~al.(2024{\natexlab{d}})Yang, Lawrie, McNamee, and Mayfield]{yang_2024_translatetraincolbertxafrican}
Eugene Yang, Dawn~J. Lawrie, Paul McNamee, and James Mayfield.
\newblock Extending translate-train for colbert-x to african language clir, 2024{\natexlab{d}}.
\newblock URL \url{https://arxiv.org/abs/2404.08134}.

\bibitem[Zhang et~al.(2021)Zhang, Ma, Shi, and Lin]{zhang_2021_mrtydi}
Xinyu Zhang, Xueguang Ma, Peng Shi, and Jimmy Lin.
\newblock Mr. {T}y{D}i: A multi-lingual benchmark for dense retrieval.
\newblock In Duygu Ataman, Alexandra Birch, Alexis Conneau, Orhan Firat, Sebastian Ruder, and Gozde~Gul Sahin, editors, \emph{Proceedings of the 1st Workshop on Multilingual Representation Learning}, pages 127--137, Punta Cana, Dominican Republic, November 2021. Association for Computational Linguistics.
\newblock \doi{10.18653/v1/2021.mrl-1.12}.
\newblock URL \url{https://aclanthology.org/2021.mrl-1.12/}.

\bibitem[Koehn and Knowles(2017)]{koehn_2017_nmt}
Philipp Koehn and Rebecca Knowles.
\newblock Six challenges for neural machine translation.
\newblock In Thang Luong, Alexandra Birch, Graham Neubig, and Andrew Finch, editors, \emph{Proceedings of the First Workshop on Neural Machine Translation}, pages 28--39, Vancouver, August 2017. Association for Computational Linguistics.
\newblock \doi{10.18653/v1/W17-3204}.
\newblock URL \url{https://aclanthology.org/W17-3204/}.

\bibitem[Hiemstra and Kraaij(1999)]{hiemstra_1999}
Djoerd Hiemstra and Wessel Kraaij.
\newblock Twenty-one at trec-7: ad-hoc and cross-language track.
\newblock In E.M Voorhees and D.K. Harman, editors, \emph{Proceedings of the seventh Text Retrieval Conference (TREC)}, NIST Special Publications, pages 227--238, United States, 1999. National Institute of Standards and Technology.
\newblock Seventh Text REtrieval Conference, TREC-7 1998 ; Conference date: 09-11-1999 Through 11-11-1999.

\bibitem[Kishida and Kando(2003)]{kishida2003two}
Kazuaki Kishida and Noriko Kando.
\newblock Two-stage refinement of query translation in a pivot language approach to cross-lingual information retrieval: An experiment at clef 2003.
\newblock In \emph{Workshop of the Cross-Language Evaluation Forum for European Languages}, pages 253--262. Springer, 2003.

\bibitem[Kando et~al.(2005)Kando, Chen, and Kishida]{kando2005two}
Noriko Kando, Kuang-hua Chen, and Kazuaki Kishida.
\newblock Two-stage refinement of transitive query translation with english disambiguation for cross-language information retrieval: An experiment at clef 2004.
\newblock \emph{(No Title)}, 2005.

\bibitem[Kishida and Kando(2006)]{kishida2006hybrid}
Kazuaki Kishida and Noriko Kando.
\newblock A hybrid approach to query and document translation using a pivot language for cross-language information retrieval.
\newblock In Carol Peters, Fredric~C. Gey, Julio Gonzalo, Henning M{\"u}ller, Gareth J.~F. Jones, Michael Kluck, Bernardo Magnini, and Maarten de~Rijke, editors, \emph{Accessing Multilingual Information Repositories}, pages 93--101, Berlin, Heidelberg, 2006. Springer Berlin Heidelberg.
\newblock ISBN 978-3-540-45700-8.

\bibitem[Kraaij and de~Jong(2004)]{kraaij2004transitive}
Wessel Kraaij and Franciska~MG de~Jong.
\newblock Transitive probabilistic clir models.
\newblock In \emph{7th International Conference on Computer-Assisted Information Retrieval, RIAO 2004:(Recherche d'Information et ses Applications)}, pages 69--81. Centre de Hautes Etudes Internationales d'Informatique Documentaire (CID), 2004.

\bibitem[Kishida(2005)]{kishida2005technical}
Kazuaki Kishida.
\newblock Technical issues of cross-language information retrieval: a review.
\newblock \emph{Information processing \& management}, 41\penalty0 (3):\penalty0 433--455, 2005.

\bibitem[Chen and Gey(2003)]{chen_gey_2003}
Aitao Chen and Fredric~C. Gey.
\newblock Experiments on cross-language and patent retrieval at ntcir-3 workshop.
\newblock In \emph{Proceedings of the Third NTCIR Workshop}, 2003.

\bibitem[Lin and Chen(2003)]{lin_chen_2003}
Wen-Cheng Lin and Hsin-Hsi Chen.
\newblock Description of ntu approach to ntcir3 multilingual information retrieval.
\newblock In \emph{Proceedings of the Third NTCIR Workshop}, 2003.

\bibitem[Gey et~al.(1998)Gey, Jiang, Chen, and Larson]{gey_1998}
Fredric~C. Gey, Hailing Jiang, Aitao Chen, and Ray~R. Larson.
\newblock Manual queries and machine translation in cross-language retrieval and interactive retrieval with cheshire ii at trec-7.
\newblock In \emph{Proceedings of the seventh Text Retrieval Conference (TREC)}, TREC'98, pages 463--476, 1998.

\bibitem[Gollins and Sanderson(2001)]{gollins_sanderson_2001}
Tim Gollins and Mark Sanderson.
\newblock Improving cross language retrieval with triangulated translation.
\newblock In \emph{Proceedings of the 24th Annual International ACM SIGIR Conference on Research and Development in Information Retrieval}, SIGIR '01, page 90–95, New York, NY, USA, 2001. Association for Computing Machinery.
\newblock ISBN 1581133316.
\newblock \doi{10.1145/383952.383965}.
\newblock URL \url{https://doi.org/10.1145/383952.383965}.

\bibitem[Lehtokangas et~al.(2004)Lehtokangas, Airio, and Järvelin]{lehtokangas_2004}
Raija Lehtokangas, Eija Airio, and Kalervo Järvelin.
\newblock Transitive dictionary translation challenges direct dictionary translation in clir.
\newblock \emph{Information Processing \& Management}, 40\penalty0 (6):\penalty0 973--988, 2004.
\newblock ISSN 0306-4573.
\newblock \doi{https://doi.org/10.1016/j.ipm.2003.10.005}.
\newblock URL \url{https://www.sciencedirect.com/science/article/pii/S0306457303000864}.

\bibitem[Deerwester et~al.(1990)Deerwester, Dumais, Furnas, Landauer, and Harshman]{deerwester_1990}
Scott Deerwester, Susan~T. Dumais, George~W. Furnas, Thomas~K. Landauer, and Richard Harshman.
\newblock Indexing by latent semantic analysis.
\newblock \emph{Journal of the American Society for Information Science}, 41\penalty0 (6):\penalty0 391--407, 1990.
\newblock \doi{https://doi.org/10.1002/(SICI)1097-4571(199009)41:6<391::AID-ASI1>3.0.CO;2-9}.
\newblock URL \url{https://asistdl.onlinelibrary.wiley.com/doi/abs/10.1002/%28SICI%291097-4571%28199009%2941%3A6%3C391%3A%3AAID-ASI1%3E3.0.CO%3B2-9}.

\bibitem[Littman et~al.(1998)Littman, Dumais, and Landauer]{littman_1998}
Michael~L. Littman, Susan~T. Dumais, and Thomas~K. Landauer.
\newblock \emph{Automatic Cross-Language Information Retrieval Using Latent Semantic Indexing}, pages 51--62.
\newblock Springer US, Boston, MA, 1998.
\newblock ISBN 978-1-4615-5661-9.
\newblock \doi{10.1007/978-1-4615-5661-9_5}.
\newblock URL \url{https://doi.org/10.1007/978-1-4615-5661-9_5}.

\bibitem[Berry and Young(1995)]{berry_young_1995}
Michael~W. Berry and Paul~G. Young.
\newblock Using latent semantic indexing for multilanguage information retrieval.
\newblock \emph{Computers and the Humanities}, 29\penalty0 (6):\penalty0 413--429, 1995.
\newblock ISSN 00104817.
\newblock URL \url{http://www.jstor.org/stable/30200366}.

\bibitem[Gabrilovich and Markovitch(2007)]{gabrilovich_2007}
Evgeniy Gabrilovich and Shaul Markovitch.
\newblock Computing semantic relatedness using wikipedia-based explicit semantic analysis.
\newblock In \emph{Proceedings of the 20th International Joint Conference on Artifical Intelligence}, IJCAI'07, page 1606–1611, San Francisco, CA, USA, 2007. Morgan Kaufmann Publishers Inc.

\bibitem[Reusch and Belinkov(2025)]{reusch_2025}
Anja Reusch and Yonatan Belinkov.
\newblock Reverse-engineering the retrieval process in genir models, 2025.
\newblock URL \url{https://arxiv.org/abs/2503.19715}.

\bibitem[Qin et~al.(2025)Qin, Chen, Zhou, Chen, Li, Liao, Li, Che, and Yu]{qin_2025}
Libo Qin, Qiguang Chen, Yuhang Zhou, Zhi Chen, Yinghui Li, Lizi Liao, Min Li, Wanxiang Che, and Philip~S. Yu.
\newblock A survey of multilingual large language models.
\newblock \emph{Patterns}, 6\penalty0 (1):\penalty0 101118, 2025.
\newblock ISSN 2666-3899.
\newblock \doi{https://doi.org/10.1016/j.patter.2024.101118}.
\newblock URL \url{https://www.sciencedirect.com/science/article/pii/S2666389924002903}.

\bibitem[Gurgurov et~al.(2024)Gurgurov, Bäumel, and Anikina]{gurgurov_2024}
Daniil Gurgurov, Tanja Bäumel, and Tatiana Anikina.
\newblock Multilingual large language models and curse of multilinguality, 2024.
\newblock URL \url{https://arxiv.org/abs/2406.10602}.

\bibitem[Doddapaneni et~al.(2025)Doddapaneni, Ramesh, Khapra, Kunchukuttan, and Kumar]{doddapaneni_2025}
Sumanth Doddapaneni, Gowtham Ramesh, Mitesh Khapra, Anoop Kunchukuttan, and Pratyush Kumar.
\newblock A primer on pretrained multilingual language models.
\newblock \emph{ACM Comput. Surv.}, 57\penalty0 (9), May 2025.
\newblock ISSN 0360-0300.
\newblock \doi{10.1145/3727339}.
\newblock URL \url{https://doi.org/10.1145/3727339}.

\bibitem[Devlin et~al.(2019)Devlin, Chang, Lee, and Toutanova]{devlin_2019_bert}
Jacob Devlin, Ming-Wei Chang, Kenton Lee, and Kristina Toutanova.
\newblock {BERT}: Pre-training of deep bidirectional transformers for language understanding.
\newblock In Jill Burstein, Christy Doran, and Thamar Solorio, editors, \emph{Proceedings of the 2019 Conference of the North {A}merican Chapter of the Association for Computational Linguistics: Human Language Technologies, Volume 1 (Long and Short Papers)}, pages 4171--4186, Minneapolis, Minnesota, June 2019. Association for Computational Linguistics.
\newblock \doi{10.18653/v1/N19-1423}.
\newblock URL \url{https://aclanthology.org/N19-1423/}.

\bibitem[Team(2025{\natexlab{b}})]{qwen3-embedding}
Qwen Team.
\newblock Qwen3-embedding (including qwen3-reranker-8b).
\newblock \url{https://qwenlm.github.io/blog/qwen3/}, May 2025{\natexlab{b}}.
\newblock URL \url{https://qwenlm.github.io/blog/qwen3/}.
\newblock Includes Qwen3-Reranker-8B reranker model.

\bibitem[Wang et~al.(2024{\natexlab{b}})Wang, Yang, Huang, Yang, Majumder, and Wei]{wang2024multilinguale5textembeddings}
Liang Wang, Nan Yang, Xiaolong Huang, Linjun Yang, Rangan Majumder, and Furu Wei.
\newblock Multilingual e5 text embeddings: A technical report, 2024{\natexlab{b}}.
\newblock URL \url{https://arxiv.org/abs/2402.05672}.

\bibitem[Liang et~al.(2023)Liang, Gonen, Mao, Hou, Goyal, Ghazvininejad, Zettlemoyer, and Khabsa]{liang_2023_xlm}
Davis Liang, Hila Gonen, Yuning Mao, Rui Hou, Naman Goyal, Marjan Ghazvininejad, Luke Zettlemoyer, and Madian Khabsa.
\newblock {XLM}-{V}: Overcoming the vocabulary bottleneck in multilingual masked language models.
\newblock In Houda Bouamor, Juan Pino, and Kalika Bali, editors, \emph{Proceedings of the 2023 Conference on Empirical Methods in Natural Language Processing}, pages 13142--13152, Singapore, December 2023. Association for Computational Linguistics.
\newblock \doi{10.18653/v1/2023.emnlp-main.813}.
\newblock URL \url{https://aclanthology.org/2023.emnlp-main.813/}.

\bibitem[Nussbaum and Duderstadt(2025)]{nussbaum_2025}
Zach Nussbaum and Brandon Duderstadt.
\newblock Training sparse mixture of experts text embedding models, 2025.
\newblock URL \url{https://arxiv.org/abs/2502.07972}.

\bibitem[Lee et~al.(2025)Lee, Chen, Dua, Cer, Shanbhogue, Naim, Ábrego, Li, Chen, Vera, Ren, Zhang, Salz, Boratko, Han, Chen, Huang, Rao, Suganthan, Han, Doumanoglou, Gupta, Moiseev, Yip, Jain, Baumgartner, Shahi, Gomez, Mariserla, Choi, Shah, Goenka, Chen, Xia, Chen, Duddu, Chen, Walker, Zhou, Ghiya, Gleicher, Gill, Dong, Seyedhosseini, Sung, Hoffmann, and Duerig]{lee_2025_gemini}
Jinhyuk Lee, Feiyang Chen, Sahil Dua, Daniel Cer, Madhuri Shanbhogue, Iftekhar Naim, Gustavo~Hernández Ábrego, Zhe Li, Kaifeng Chen, Henrique~Schechter Vera, Xiaoqi Ren, Shanfeng Zhang, Daniel Salz, Michael Boratko, Jay Han, Blair Chen, Shuo Huang, Vikram Rao, Paul Suganthan, Feng Han, Andreas Doumanoglou, Nithi Gupta, Fedor Moiseev, Cathy Yip, Aashi Jain, Simon Baumgartner, Shahrokh Shahi, Frank~Palma Gomez, Sandeep Mariserla, Min Choi, Parashar Shah, Sonam Goenka, Ke~Chen, Ye~Xia, Koert Chen, Sai Meher~Karthik Duddu, Yichang Chen, Trevor Walker, Wenlei Zhou, Rakesh Ghiya, Zach Gleicher, Karan Gill, Zhe Dong, Mojtaba Seyedhosseini, Yunhsuan Sung, Raphael Hoffmann, and Tom Duerig.
\newblock Gemini embedding: Generalizable embeddings from gemini, 2025.
\newblock URL \url{https://arxiv.org/abs/2503.07891}.

\bibitem[Feng et~al.(2020)Feng, Yang, Cer, Arivazhagan, and Wang]{feng2020labse}
Fangxiaoyu Feng, Yinfei Yang, Daniel Cer, Naveen Arivazhagan, and Wei Wang.
\newblock Language‑agnostic bert sentence embedding.
\newblock 2020.

\bibitem[Liu et~al.(2020)Liu, Gu, Goyal, Li, Edunov, Ghazvininejad, Lewis, and Zettlemoyer]{liu_2020_mbart}
Yinhan Liu, Jiatao Gu, Naman Goyal, Xian Li, Sergey Edunov, Marjan Ghazvininejad, Mike Lewis, and Luke Zettlemoyer.
\newblock Multilingual denoising pre-training for neural machine translation, 2020.
\newblock URL \url{https://arxiv.org/abs/2001.08210}.

\bibitem[Team et~al.(2022)Team, Costa-jussà, Cross, Çelebi, Elbayad, Heafield, Heffernan, Kalbassi, Lam, Licht, Maillard, Sun, Wang, Wenzek, Youngblood, Akula, Barrault, Gonzalez, Hansanti, Hoffman, Jarrett, Sadagopan, Rowe, Spruit, Tran, Andrews, Ayan, Bhosale, Edunov, Fan, Gao, Goswami, Guzmán, Koehn, Mourachko, Ropers, Saleem, Schwenk, and Wang]{nllb_2022}
NLLB Team, Marta~R. Costa-jussà, James Cross, Onur Çelebi, Maha Elbayad, Kenneth Heafield, Kevin Heffernan, Elahe Kalbassi, Janice Lam, Daniel Licht, Jean Maillard, Anna Sun, Skyler Wang, Guillaume Wenzek, Al~Youngblood, Bapi Akula, Loic Barrault, Gabriel~Mejia Gonzalez, Prangthip Hansanti, John Hoffman, Semarley Jarrett, Kaushik~Ram Sadagopan, Dirk Rowe, Shannon Spruit, Chau Tran, Pierre Andrews, Necip~Fazil Ayan, Shruti Bhosale, Sergey Edunov, Angela Fan, Cynthia Gao, Vedanuj Goswami, Francisco Guzmán, Philipp Koehn, Alexandre Mourachko, Christophe Ropers, Safiyyah Saleem, Holger Schwenk, and Jeff Wang.
\newblock No language left behind: Scaling human-centered machine translation, 2022.
\newblock URL \url{https://arxiv.org/abs/2207.04672}.

\bibitem[Uthus et~al.(2023)Uthus, Ontañón, Ainslie, and Guo]{uthus_2023_mlongt5}
David Uthus, Santiago Ontañón, Joshua Ainslie, and Mandy Guo.
\newblock mlongt5: A multilingual and efficient text-to-text transformer for longer sequences, 2023.
\newblock URL \url{https://arxiv.org/abs/2305.11129}.

\bibitem[Üstün et~al.(2024)Üstün, Aryabumi, Yong, Ko, D'souza, Onilude, Bhandari, Singh, Ooi, Kayid, Vargus, Blunsom, Longpre, Muennighoff, Fadaee, Kreutzer, and Hooker]{ustun_2024_aya}
Ahmet Üstün, Viraat Aryabumi, Zheng-Xin Yong, Wei-Yin Ko, Daniel D'souza, Gbemileke Onilude, Neel Bhandari, Shivalika Singh, Hui-Lee Ooi, Amr Kayid, Freddie Vargus, Phil Blunsom, Shayne Longpre, Niklas Muennighoff, Marzieh Fadaee, Julia Kreutzer, and Sara Hooker.
\newblock Aya model: An instruction finetuned open-access multilingual language model, 2024.
\newblock URL \url{https://arxiv.org/abs/2402.07827}.

\bibitem[Lin et~al.(2022)Lin, Mihaylov, Artetxe, Wang, Chen, Simig, Ott, Goyal, Bhosale, Du, Pasunuru, Shleifer, Koura, Chaudhary, O'Horo, Wang, Zettlemoyer, Kozareva, Diab, Stoyanov, and Li]{lin_2022_xlgm}
Xi~Victoria Lin, Todor Mihaylov, Mikel Artetxe, Tianlu Wang, Shuohui Chen, Daniel Simig, Myle Ott, Naman Goyal, Shruti Bhosale, Jingfei Du, Ramakanth Pasunuru, Sam Shleifer, Punit~Singh Koura, Vishrav Chaudhary, Brian O'Horo, Jeff Wang, Luke Zettlemoyer, Zornitsa Kozareva, Mona Diab, Veselin Stoyanov, and Xian Li.
\newblock Few-shot learning with multilingual language models, 2022.
\newblock URL \url{https://arxiv.org/abs/2112.10668}.

\bibitem[Meta(2024)]{llama3_2024}
Meta.
\newblock Introducing meta llama 3: The most capable openly available llm to date, 2024.
\newblock URL \url{https://ai.meta.com/blog/meta-llama-3/}.

\bibitem[AI(2024)]{mistral_2024}
Mistral AI.
\newblock Model card for mistral-large-instruct-2407, 2024.
\newblock URL \url{https://huggingface.co/mistralai/Mistral-Large-Instruct-2407}.

\bibitem[Workshop et~al.(2023)Workshop, :, Scao, Fan, Akiki, Pavlick, Ilić, Hesslow, Castagné, Luccioni, Yvon, Gallé, Tow, Rush, Biderman, Webson, Ammanamanchi, Wang, Sagot, Muennighoff, del Moral, Ruwase, Bawden, Bekman, McMillan-Major, Beltagy, Nguyen, Saulnier, Tan, Suarez, Sanh, Laurençon, Jernite, Launay, Mitchell, Raffel, Gokaslan, Simhi, Soroa, Aji, Alfassy, Rogers, Nitzav, Xu, Mou, Emezue, Klamm, Leong, van Strien, Adelani, Radev, Ponferrada, Levkovizh, Kim, Natan, Toni, Dupont, Kruszewski, Pistilli, Elsahar, Benyamina, Tran, Yu, Abdulmumin, Johnson, Gonzalez-Dios, de~la Rosa, Chim, Dodge, Zhu, Chang, Frohberg, Tobing, Bhattacharjee, Almubarak, Chen, Lo, Werra, Weber, Phan, allal, Tanguy, Dey, Muñoz, Masoud, Grandury, Šaško, Huang, Coavoux, Singh, Jiang, Vu, Jauhar, Ghaleb, Subramani, Kassner, Khamis, Nguyen, Espejel, de~Gibert, Villegas, Henderson, Colombo, Amuok, Lhoest, Harliman, Bommasani, López, Ribeiro, Osei, Pyysalo, Nagel, Bose, Muhammad, Sharma, Longpre, Nikpoor, Silberberg, Pai,
  Zink, Torrent, Schick, Thrush, Danchev, Nikoulina, Laippala, Lepercq, Prabhu, Alyafeai, Talat, Raja, Heinzerling, Si, Taşar, Salesky, Mielke, Lee, Sharma, Santilli, Chaffin, Stiegler, Datta, Szczechla, Chhablani, Wang, Pandey, Strobelt, Fries, Rozen, Gao, Sutawika, Bari, Al-shaibani, Manica, Nayak, Teehan, Albanie, Shen, Ben-David, Bach, Kim, Bers, Fevry, Neeraj, Thakker, Raunak, Tang, Yong, Sun, Brody, Uri, Tojarieh, Roberts, Chung, Tae, Phang, Press, Li, Narayanan, Bourfoune, Casper, Rasley, Ryabinin, Mishra, Zhang, Shoeybi, Peyrounette, Patry, Tazi, Sanseviero, von Platen, Cornette, Lavallée, Lacroix, Rajbhandari, Gandhi, Smith, Requena, Patil, Dettmers, Baruwa, Singh, Cheveleva, Ligozat, Subramonian, Névéol, Lovering, Garrette, Tunuguntla, Reiter, Taktasheva, Voloshina, Bogdanov, Winata, Schoelkopf, Kalo, Novikova, Forde, Clive, Kasai, Kawamura, Hazan, Carpuat, Clinciu, Kim, Cheng, Serikov, Antverg, van~der Wal, Zhang, Zhang, Gehrmann, Mirkin, Pais, Shavrina, Scialom, Yun, Limisiewicz, Rieser,
  Protasov, Mikhailov, Pruksachatkun, Belinkov, Bamberger, Kasner, Rueda, Pestana, Feizpour, Khan, Faranak, Santos, Hevia, Unldreaj, Aghagol, Abdollahi, Tammour, HajiHosseini, Behroozi, Ajibade, Saxena, Ferrandis, McDuff, Contractor, Lansky, David, Kiela, Nguyen, Tan, Baylor, Ozoani, Mirza, Ononiwu, Rezanejad, Jones, Bhattacharya, Solaiman, Sedenko, Nejadgholi, Passmore, Seltzer, Sanz, Dutra, Samagaio, Elbadri, Mieskes, Gerchick, Akinlolu, McKenna, Qiu, Ghauri, Burynok, Abrar, Rajani, Elkott, Fahmy, Samuel, An, Kromann, Hao, Alizadeh, Shubber, Wang, Roy, Viguier, Le, Oyebade, Le, Yang, Nguyen, Kashyap, Palasciano, Callahan, Shukla, Miranda-Escalada, Singh, Beilharz, Wang, Brito, Zhou, Jain, Xu, Fourrier, Periñán, Molano, Yu, Manjavacas, Barth, Fuhrimann, Altay, Bayrak, Burns, Vrabec, Bello, Dash, Kang, Giorgi, Golde, Posada, Sivaraman, Bulchandani, Liu, Shinzato, de~Bykhovetz, Takeuchi, Pàmies, Castillo, Nezhurina, Sänger, Samwald, Cullan, Weinberg, Wolf, Mihaljcic, Liu, Freidank, Kang, Seelam, Dahlberg,
  Broad, Muellner, Fung, Haller, Chandrasekhar, Eisenberg, Martin, Canalli, Su, Su, Cahyawijaya, Garda, Deshmukh, Mishra, Kiblawi, Ott, Sang-aroonsiri, Kumar, Schweter, Bharati, Laud, Gigant, Kainuma, Kusa, Labrak, Bajaj, Venkatraman, Xu, Xu, Xu, Tan, Xie, Ye, Bras, Belkada, and Wolf]{bloom_2023}
BigScience Workshop, :, Teven~Le Scao, Angela Fan, Christopher Akiki, Ellie Pavlick, Suzana Ilić, Daniel Hesslow, Roman Castagné, Alexandra~Sasha Luccioni, François Yvon, Matthias Gallé, Jonathan Tow, Alexander~M. Rush, Stella Biderman, Albert Webson, Pawan~Sasanka Ammanamanchi, Thomas Wang, Benoît Sagot, Niklas Muennighoff, Albert~Villanova del Moral, Olatunji Ruwase, Rachel Bawden, Stas Bekman, Angelina McMillan-Major, Iz~Beltagy, Huu Nguyen, Lucile Saulnier, Samson Tan, Pedro~Ortiz Suarez, Victor Sanh, Hugo Laurençon, Yacine Jernite, Julien Launay, Margaret Mitchell, Colin Raffel, Aaron Gokaslan, Adi Simhi, Aitor Soroa, Alham~Fikri Aji, Amit Alfassy, Anna Rogers, Ariel~Kreisberg Nitzav, Canwen Xu, Chenghao Mou, Chris Emezue, Christopher Klamm, Colin Leong, Daniel van Strien, David~Ifeoluwa Adelani, Dragomir Radev, Eduardo~González Ponferrada, Efrat Levkovizh, Ethan Kim, Eyal~Bar Natan, Francesco~De Toni, Gérard Dupont, Germán Kruszewski, Giada Pistilli, Hady Elsahar, Hamza Benyamina, Hieu Tran,
  Ian Yu, Idris Abdulmumin, Isaac Johnson, Itziar Gonzalez-Dios, Javier de~la Rosa, Jenny Chim, Jesse Dodge, Jian Zhu, Jonathan Chang, Jörg Frohberg, Joseph Tobing, Joydeep Bhattacharjee, Khalid Almubarak, Kimbo Chen, Kyle Lo, Leandro~Von Werra, Leon Weber, Long Phan, Loubna~Ben allal, Ludovic Tanguy, Manan Dey, Manuel~Romero Muñoz, Maraim Masoud, María Grandury, Mario Šaško, Max Huang, Maximin Coavoux, Mayank Singh, Mike Tian-Jian Jiang, Minh~Chien Vu, Mohammad~A. Jauhar, Mustafa Ghaleb, Nishant Subramani, Nora Kassner, Nurulaqilla Khamis, Olivier Nguyen, Omar Espejel, Ona de~Gibert, Paulo Villegas, Peter Henderson, Pierre Colombo, Priscilla Amuok, Quentin Lhoest, Rheza Harliman, Rishi Bommasani, Roberto~Luis López, Rui Ribeiro, Salomey Osei, Sampo Pyysalo, Sebastian Nagel, Shamik Bose, Shamsuddeen~Hassan Muhammad, Shanya Sharma, Shayne Longpre, Somaieh Nikpoor, Stanislav Silberberg, Suhas Pai, Sydney Zink, Tiago~Timponi Torrent, Timo Schick, Tristan Thrush, Valentin Danchev, Vassilina Nikoulina,
  Veronika Laippala, Violette Lepercq, Vrinda Prabhu, Zaid Alyafeai, Zeerak Talat, Arun Raja, Benjamin Heinzerling, Chenglei Si, Davut~Emre Taşar, Elizabeth Salesky, Sabrina~J. Mielke, Wilson~Y. Lee, Abheesht Sharma, Andrea Santilli, Antoine Chaffin, Arnaud Stiegler, Debajyoti Datta, Eliza Szczechla, Gunjan Chhablani, Han Wang, Harshit Pandey, Hendrik Strobelt, Jason~Alan Fries, Jos Rozen, Leo Gao, Lintang Sutawika, M~Saiful Bari, Maged~S. Al-shaibani, Matteo Manica, Nihal Nayak, Ryan Teehan, Samuel Albanie, Sheng Shen, Srulik Ben-David, Stephen~H. Bach, Taewoon Kim, Tali Bers, Thibault Fevry, Trishala Neeraj, Urmish Thakker, Vikas Raunak, Xiangru Tang, Zheng-Xin Yong, Zhiqing Sun, Shaked Brody, Yallow Uri, Hadar Tojarieh, Adam Roberts, Hyung~Won Chung, Jaesung Tae, Jason Phang, Ofir Press, Conglong Li, Deepak Narayanan, Hatim Bourfoune, Jared Casper, Jeff Rasley, Max Ryabinin, Mayank Mishra, Minjia Zhang, Mohammad Shoeybi, Myriam Peyrounette, Nicolas Patry, Nouamane Tazi, Omar Sanseviero, Patrick von
  Platen, Pierre Cornette, Pierre~François Lavallée, Rémi Lacroix, Samyam Rajbhandari, Sanchit Gandhi, Shaden Smith, Stéphane Requena, Suraj Patil, Tim Dettmers, Ahmed Baruwa, Amanpreet Singh, Anastasia Cheveleva, Anne-Laure Ligozat, Arjun Subramonian, Aurélie Névéol, Charles Lovering, Dan Garrette, Deepak Tunuguntla, Ehud Reiter, Ekaterina Taktasheva, Ekaterina Voloshina, Eli Bogdanov, Genta~Indra Winata, Hailey Schoelkopf, Jan-Christoph Kalo, Jekaterina Novikova, Jessica~Zosa Forde, Jordan Clive, Jungo Kasai, Ken Kawamura, Liam Hazan, Marine Carpuat, Miruna Clinciu, Najoung Kim, Newton Cheng, Oleg Serikov, Omer Antverg, Oskar van~der Wal, Rui Zhang, Ruochen Zhang, Sebastian Gehrmann, Shachar Mirkin, Shani Pais, Tatiana Shavrina, Thomas Scialom, Tian Yun, Tomasz Limisiewicz, Verena Rieser, Vitaly Protasov, Vladislav Mikhailov, Yada Pruksachatkun, Yonatan Belinkov, Zachary Bamberger, Zdeněk Kasner, Alice Rueda, Amanda Pestana, Amir Feizpour, Ammar Khan, Amy Faranak, Ana Santos, Anthony Hevia, Antigona
  Unldreaj, Arash Aghagol, Arezoo Abdollahi, Aycha Tammour, Azadeh HajiHosseini, Bahareh Behroozi, Benjamin Ajibade, Bharat Saxena, Carlos~Muñoz Ferrandis, Daniel McDuff, Danish Contractor, David Lansky, Davis David, Douwe Kiela, Duong~A. Nguyen, Edward Tan, Emi Baylor, Ezinwanne Ozoani, Fatima Mirza, Frankline Ononiwu, Habib Rezanejad, Hessie Jones, Indrani Bhattacharya, Irene Solaiman, Irina Sedenko, Isar Nejadgholi, Jesse Passmore, Josh Seltzer, Julio~Bonis Sanz, Livia Dutra, Mairon Samagaio, Maraim Elbadri, Margot Mieskes, Marissa Gerchick, Martha Akinlolu, Michael McKenna, Mike Qiu, Muhammed Ghauri, Mykola Burynok, Nafis Abrar, Nazneen Rajani, Nour Elkott, Nour Fahmy, Olanrewaju Samuel, Ran An, Rasmus Kromann, Ryan Hao, Samira Alizadeh, Sarmad Shubber, Silas Wang, Sourav Roy, Sylvain Viguier, Thanh Le, Tobi Oyebade, Trieu Le, Yoyo Yang, Zach Nguyen, Abhinav~Ramesh Kashyap, Alfredo Palasciano, Alison Callahan, Anima Shukla, Antonio Miranda-Escalada, Ayush Singh, Benjamin Beilharz, Bo~Wang, Caio Brito,
  Chenxi Zhou, Chirag Jain, Chuxin Xu, Clémentine Fourrier, Daniel~León Periñán, Daniel Molano, Dian Yu, Enrique Manjavacas, Fabio Barth, Florian Fuhrimann, Gabriel Altay, Giyaseddin Bayrak, Gully Burns, Helena~U. Vrabec, Imane Bello, Ishani Dash, Jihyun Kang, John Giorgi, Jonas Golde, Jose~David Posada, Karthik~Rangasai Sivaraman, Lokesh Bulchandani, Lu~Liu, Luisa Shinzato, Madeleine~Hahn de~Bykhovetz, Maiko Takeuchi, Marc Pàmies, Maria~A Castillo, Marianna Nezhurina, Mario Sänger, Matthias Samwald, Michael Cullan, Michael Weinberg, Michiel~De Wolf, Mina Mihaljcic, Minna Liu, Moritz Freidank, Myungsun Kang, Natasha Seelam, Nathan Dahlberg, Nicholas~Michio Broad, Nikolaus Muellner, Pascale Fung, Patrick Haller, Ramya Chandrasekhar, Renata Eisenberg, Robert Martin, Rodrigo Canalli, Rosaline Su, Ruisi Su, Samuel Cahyawijaya, Samuele Garda, Shlok~S Deshmukh, Shubhanshu Mishra, Sid Kiblawi, Simon Ott, Sinee Sang-aroonsiri, Srishti Kumar, Stefan Schweter, Sushil Bharati, Tanmay Laud, Théo Gigant, Tomoya
  Kainuma, Wojciech Kusa, Yanis Labrak, Yash~Shailesh Bajaj, Yash Venkatraman, Yifan Xu, Yingxin Xu, Yu~Xu, Zhe Tan, Zhongli Xie, Zifan Ye, Mathilde Bras, Younes Belkada, and Thomas Wolf.
\newblock Bloom: A 176b-parameter open-access multilingual language model, 2023.
\newblock URL \url{https://arxiv.org/abs/2211.05100}.

\bibitem[AI et~al.(2025)AI, :, Young, Chen, Li, Huang, Zhang, Zhang, Wang, Li, Zhu, Chen, Chang, Yu, Liu, Liu, Yue, Yang, Yang, Xie, Huang, Hu, Ren, Niu, Nie, Li, Xu, Liu, Wang, Cai, Gu, Liu, and Dai]{ai_2025_yi}
01. AI, :, Alex Young, Bei Chen, Chao Li, Chengen Huang, Ge~Zhang, Guanwei Zhang, Guoyin Wang, Heng Li, Jiangcheng Zhu, Jianqun Chen, Jing Chang, Kaidong Yu, Peng Liu, Qiang Liu, Shawn Yue, Senbin Yang, Shiming Yang, Wen Xie, Wenhao Huang, Xiaohui Hu, Xiaoyi Ren, Xinyao Niu, Pengcheng Nie, Yanpeng Li, Yuchi Xu, Yudong Liu, Yue Wang, Yuxuan Cai, Zhenyu Gu, Zhiyuan Liu, and Zonghong Dai.
\newblock Yi: Open foundation models by 01.ai, 2025.
\newblock URL \url{https://arxiv.org/abs/2403.04652}.

\bibitem[OpenAI et~al.(2024)OpenAI, :, Hurst, Lerer, Goucher, Perelman, Ramesh, Clark, Ostrow, Welihinda, Hayes, Radford, Mądry, Baker-Whitcomb, Beutel, Borzunov, Carney, Chow, Kirillov, Nichol, Paino, Renzin, Passos, Kirillov, Christakis, Conneau, Kamali, Jabri, Moyer, Tam, Crookes, Tootoochian, Tootoonchian, Kumar, Vallone, Karpathy, Braunstein, Cann, Codispoti, Galu, Kondrich, Tulloch, Mishchenko, Baek, Jiang, Pelisse, Woodford, Gosalia, Dhar, Pantuliano, Nayak, Oliver, Zoph, Ghorbani, Leimberger, Rossen, Sokolowsky, Wang, Zweig, Hoover, Samic, McGrew, Spero, Giertler, Cheng, Lightcap, Walkin, Quinn, Guarraci, Hsu, Kellogg, Eastman, Lugaresi, Wainwright, Bassin, Hudson, Chu, Nelson, Li, Shern, Conger, Barette, Voss, Ding, Lu, Zhang, Beaumont, Hallacy, Koch, Gibson, Kim, Choi, McLeavey, Hesse, Fischer, Winter, Czarnecki, Jarvis, Wei, Koumouzelis, Sherburn, Kappler, Levin, Levy, Carr, Farhi, Mely, Robinson, Sasaki, Jin, Valladares, Tsipras, Li, Nguyen, Findlay, Oiwoh, Wong, Asdar, Proehl, Yang, Antonow,
  Kramer, Peterson, Sigler, Wallace, Brevdo, Mays, Khorasani, Such, Raso, Zhang, von Lohmann, Sulit, Goh, Oden, Salmon, Starace, Brockman, Salman, Bao, Hu, Wong, Wang, Schmidt, Whitney, Jun, Kirchner, de~Oliveira~Pinto, Ren, Chang, Chung, Kivlichan, O'Connell, O'Connell, Osband, Silber, Sohl, Okuyucu, Lan, Kostrikov, Sutskever, Kanitscheider, Gulrajani, Coxon, Menick, Pachocki, Aung, Betker, Crooks, Lennon, Kiros, Leike, Park, Kwon, Phang, Teplitz, Wei, Wolfe, Chen, Harris, Varavva, Lee, Shieh, Lin, Yu, Weng, Tang, Yu, Jang, Candela, Beutler, Landers, Parish, Heidecke, Schulman, Lachman, McKay, Uesato, Ward, Kim, Huizinga, Sitkin, Kraaijeveld, Gross, Kaplan, Snyder, Achiam, Jiao, Lee, Zhuang, Harriman, Fricke, Hayashi, Singhal, Shi, Karthik, Wood, Rimbach, Hsu, Nguyen, Gu-Lemberg, Button, Liu, Howe, Muthukumar, Luther, Ahmad, Kai, Itow, Workman, Pathak, Chen, Jing, Guy, Fedus, Zhou, Mamitsuka, Weng, McCallum, Held, Ouyang, Feuvrier, Zhang, Kondraciuk, Kaiser, Hewitt, Metz, Doshi, Aflak, Simens, Boyd,
  Thompson, Dukhan, Chen, Gray, Hudnall, Zhang, Aljubeh, Litwin, Zeng, Johnson, Shetty, Gupta, Shah, Yatbaz, Yang, Zhong, Glaese, Chen, Janner, Lampe, Petrov, Wu, Wang, Fradin, Pokrass, Castro, de~Castro, Pavlov, Brundage, Wang, Khan, Murati, Bavarian, Lin, Yesildal, Soto, Gimelshein, Cone, Staudacher, Summers, LaFontaine, Chowdhury, Ryder, Stathas, Turley, Tezak, Felix, Kudige, Keskar, Deutsch, Bundick, Puckett, Nachum, Okelola, Boiko, Murk, Jaffe, Watkins, Godement, Campbell-Moore, Chao, McMillan, Belov, Su, Bak, Bakkum, Deng, Dolan, Hoeschele, Welinder, Tillet, Pronin, Tillet, Dhariwal, Yuan, Dias, Lim, Arora, Troll, Lin, Lopes, Puri, Miyara, Leike, Gaubert, Zamani, Wang, Donnelly, Honsby, Smith, Sahai, Ramchandani, Huet, Carmichael, Zellers, Chen, Chen, Nigmatullin, Cheu, Jain, Altman, Schoenholz, Toizer, Miserendino, Agarwal, Culver, Ethersmith, Gray, Grove, Metzger, Hermani, Jain, Zhao, Wu, Jomoto, Wu, Shuaiqi, Xia, Phene, Papay, Narayanan, Coffey, Lee, Hall, Balaji, Broda, Stramer, Xu, Gogineni,
  Christianson, Sanders, Patwardhan, Cunninghman, Degry, Dimson, Raoux, Shadwell, Zheng, Underwood, Markov, Sherbakov, Rubin, Stasi, Kaftan, Heywood, Peterson, Walters, Eloundou, Qi, Moeller, Monaco, Kuo, Fomenko, Chang, Zheng, Zhou, Manassra, Sheu, Zaremba, Patil, Qian, Kim, Cheng, Zhang, He, Zhang, Jin, Dai, and Malkov]{openai_2024_gpt4o}
OpenAI, :, Aaron Hurst, Adam Lerer, Adam~P. Goucher, Adam Perelman, Aditya Ramesh, Aidan Clark, AJ~Ostrow, Akila Welihinda, Alan Hayes, Alec Radford, Aleksander Mądry, Alex Baker-Whitcomb, Alex Beutel, Alex Borzunov, Alex Carney, Alex Chow, Alex Kirillov, Alex Nichol, Alex Paino, Alex Renzin, Alex~Tachard Passos, Alexander Kirillov, Alexi Christakis, Alexis Conneau, Ali Kamali, Allan Jabri, Allison Moyer, Allison Tam, Amadou Crookes, Amin Tootoochian, Amin Tootoonchian, Ananya Kumar, Andrea Vallone, Andrej Karpathy, Andrew Braunstein, Andrew Cann, Andrew Codispoti, Andrew Galu, Andrew Kondrich, Andrew Tulloch, Andrey Mishchenko, Angela Baek, Angela Jiang, Antoine Pelisse, Antonia Woodford, Anuj Gosalia, Arka Dhar, Ashley Pantuliano, Avi Nayak, Avital Oliver, Barret Zoph, Behrooz Ghorbani, Ben Leimberger, Ben Rossen, Ben Sokolowsky, Ben Wang, Benjamin Zweig, Beth Hoover, Blake Samic, Bob McGrew, Bobby Spero, Bogo Giertler, Bowen Cheng, Brad Lightcap, Brandon Walkin, Brendan Quinn, Brian Guarraci, Brian Hsu,
  Bright Kellogg, Brydon Eastman, Camillo Lugaresi, Carroll Wainwright, Cary Bassin, Cary Hudson, Casey Chu, Chad Nelson, Chak Li, Chan~Jun Shern, Channing Conger, Charlotte Barette, Chelsea Voss, Chen Ding, Cheng Lu, Chong Zhang, Chris Beaumont, Chris Hallacy, Chris Koch, Christian Gibson, Christina Kim, Christine Choi, Christine McLeavey, Christopher Hesse, Claudia Fischer, Clemens Winter, Coley Czarnecki, Colin Jarvis, Colin Wei, Constantin Koumouzelis, Dane Sherburn, Daniel Kappler, Daniel Levin, Daniel Levy, David Carr, David Farhi, David Mely, David Robinson, David Sasaki, Denny Jin, Dev Valladares, Dimitris Tsipras, Doug Li, Duc~Phong Nguyen, Duncan Findlay, Edede Oiwoh, Edmund Wong, Ehsan Asdar, Elizabeth Proehl, Elizabeth Yang, Eric Antonow, Eric Kramer, Eric Peterson, Eric Sigler, Eric Wallace, Eugene Brevdo, Evan Mays, Farzad Khorasani, Felipe~Petroski Such, Filippo Raso, Francis Zhang, Fred von Lohmann, Freddie Sulit, Gabriel Goh, Gene Oden, Geoff Salmon, Giulio Starace, Greg Brockman, Hadi
  Salman, Haiming Bao, Haitang Hu, Hannah Wong, Haoyu Wang, Heather Schmidt, Heather Whitney, Heewoo Jun, Hendrik Kirchner, Henrique~Ponde de~Oliveira~Pinto, Hongyu Ren, Huiwen Chang, Hyung~Won Chung, Ian Kivlichan, Ian O'Connell, Ian O'Connell, Ian Osband, Ian Silber, Ian Sohl, Ibrahim Okuyucu, Ikai Lan, Ilya Kostrikov, Ilya Sutskever, Ingmar Kanitscheider, Ishaan Gulrajani, Jacob Coxon, Jacob Menick, Jakub Pachocki, James Aung, James Betker, James Crooks, James Lennon, Jamie Kiros, Jan Leike, Jane Park, Jason Kwon, Jason Phang, Jason Teplitz, Jason Wei, Jason Wolfe, Jay Chen, Jeff Harris, Jenia Varavva, Jessica~Gan Lee, Jessica Shieh, Ji~Lin, Jiahui Yu, Jiayi Weng, Jie Tang, Jieqi Yu, Joanne Jang, Joaquin~Quinonero Candela, Joe Beutler, Joe Landers, Joel Parish, Johannes Heidecke, John Schulman, Jonathan Lachman, Jonathan McKay, Jonathan Uesato, Jonathan Ward, Jong~Wook Kim, Joost Huizinga, Jordan Sitkin, Jos Kraaijeveld, Josh Gross, Josh Kaplan, Josh Snyder, Joshua Achiam, Joy Jiao, Joyce Lee, Juntang
  Zhuang, Justyn Harriman, Kai Fricke, Kai Hayashi, Karan Singhal, Katy Shi, Kavin Karthik, Kayla Wood, Kendra Rimbach, Kenny Hsu, Kenny Nguyen, Keren Gu-Lemberg, Kevin Button, Kevin Liu, Kiel Howe, Krithika Muthukumar, Kyle Luther, Lama Ahmad, Larry Kai, Lauren Itow, Lauren Workman, Leher Pathak, Leo Chen, Li~Jing, Lia Guy, Liam Fedus, Liang Zhou, Lien Mamitsuka, Lilian Weng, Lindsay McCallum, Lindsey Held, Long Ouyang, Louis Feuvrier, Lu~Zhang, Lukas Kondraciuk, Lukasz Kaiser, Luke Hewitt, Luke Metz, Lyric Doshi, Mada Aflak, Maddie Simens, Madelaine Boyd, Madeleine Thompson, Marat Dukhan, Mark Chen, Mark Gray, Mark Hudnall, Marvin Zhang, Marwan Aljubeh, Mateusz Litwin, Matthew Zeng, Max Johnson, Maya Shetty, Mayank Gupta, Meghan Shah, Mehmet Yatbaz, Meng~Jia Yang, Mengchao Zhong, Mia Glaese, Mianna Chen, Michael Janner, Michael Lampe, Michael Petrov, Michael Wu, Michele Wang, Michelle Fradin, Michelle Pokrass, Miguel Castro, Miguel Oom~Temudo de~Castro, Mikhail Pavlov, Miles Brundage, Miles Wang, Minal
  Khan, Mira Murati, Mo~Bavarian, Molly Lin, Murat Yesildal, Nacho Soto, Natalia Gimelshein, Natalie Cone, Natalie Staudacher, Natalie Summers, Natan LaFontaine, Neil Chowdhury, Nick Ryder, Nick Stathas, Nick Turley, Nik Tezak, Niko Felix, Nithanth Kudige, Nitish Keskar, Noah Deutsch, Noel Bundick, Nora Puckett, Ofir Nachum, Ola Okelola, Oleg Boiko, Oleg Murk, Oliver Jaffe, Olivia Watkins, Olivier Godement, Owen Campbell-Moore, Patrick Chao, Paul McMillan, Pavel Belov, Peng Su, Peter Bak, Peter Bakkum, Peter Deng, Peter Dolan, Peter Hoeschele, Peter Welinder, Phil Tillet, Philip Pronin, Philippe Tillet, Prafulla Dhariwal, Qiming Yuan, Rachel Dias, Rachel Lim, Rahul Arora, Rajan Troll, Randall Lin, Rapha~Gontijo Lopes, Raul Puri, Reah Miyara, Reimar Leike, Renaud Gaubert, Reza Zamani, Ricky Wang, Rob Donnelly, Rob Honsby, Rocky Smith, Rohan Sahai, Rohit Ramchandani, Romain Huet, Rory Carmichael, Rowan Zellers, Roy Chen, Ruby Chen, Ruslan Nigmatullin, Ryan Cheu, Saachi Jain, Sam Altman, Sam Schoenholz, Sam
  Toizer, Samuel Miserendino, Sandhini Agarwal, Sara Culver, Scott Ethersmith, Scott Gray, Sean Grove, Sean Metzger, Shamez Hermani, Shantanu Jain, Shengjia Zhao, Sherwin Wu, Shino Jomoto, Shirong Wu, Shuaiqi, Xia, Sonia Phene, Spencer Papay, Srinivas Narayanan, Steve Coffey, Steve Lee, Stewart Hall, Suchir Balaji, Tal Broda, Tal Stramer, Tao Xu, Tarun Gogineni, Taya Christianson, Ted Sanders, Tejal Patwardhan, Thomas Cunninghman, Thomas Degry, Thomas Dimson, Thomas Raoux, Thomas Shadwell, Tianhao Zheng, Todd Underwood, Todor Markov, Toki Sherbakov, Tom Rubin, Tom Stasi, Tomer Kaftan, Tristan Heywood, Troy Peterson, Tyce Walters, Tyna Eloundou, Valerie Qi, Veit Moeller, Vinnie Monaco, Vishal Kuo, Vlad Fomenko, Wayne Chang, Weiyi Zheng, Wenda Zhou, Wesam Manassra, Will Sheu, Wojciech Zaremba, Yash Patil, Yilei Qian, Yongjik Kim, Youlong Cheng, Yu~Zhang, Yuchen He, Yuchen Zhang, Yujia Jin, Yunxing Dai, and Yury Malkov.
\newblock Gpt-4o system card, 2024.
\newblock URL \url{https://arxiv.org/abs/2410.21276}.

\bibitem[DeepMind(2025)]{gemini_2025}
Google DeepMind.
\newblock Gemini 2.5 pro model card.
\newblock Technical report, 2025.
\newblock URL \url{https://storage.googleapis.com/model-cards/documents/gemini-2.5-pro.pdf}.

\bibitem[Anthropic(2024)]{claude_2024}
Anthropic.
\newblock The claude 3 model family: Opus, sonnet, haiku.
\newblock Technical report, 2024.
\newblock URL \url{https://assets.anthropic.com/m/61e7d27f8c8f5919/original/Claude-3-Model-Card.pdf}.

\bibitem[Anil et~al.(2023)Anil, Dai, Firat, Johnson, Lepikhin, Passos, Shakeri, Taropa, Bailey, Chen, Chu, Clark, Shafey, Huang, Meier-Hellstern, Mishra, Moreira, Omernick, Robinson, Ruder, Tay, Xiao, Xu, Zhang, Abrego, Ahn, Austin, Barham, Botha, Bradbury, Brahma, Brooks, Catasta, Cheng, Cherry, Choquette-Choo, Chowdhery, Crepy, Dave, Dehghani, Dev, Devlin, Díaz, Du, Dyer, Feinberg, Feng, Fienber, Freitag, Garcia, Gehrmann, Gonzalez, Gur-Ari, Hand, Hashemi, Hou, Howland, Hu, Hui, Hurwitz, Isard, Ittycheriah, Jagielski, Jia, Kenealy, Krikun, Kudugunta, Lan, Lee, Lee, Li, Li, Li, Li, Li, Lim, Lin, Liu, Liu, Maggioni, Mahendru, Maynez, Misra, Moussalem, Nado, Nham, Ni, Nystrom, Parrish, Pellat, Polacek, Polozov, Pope, Qiao, Reif, Richter, Riley, Ros, Roy, Saeta, Samuel, Shelby, Slone, Smilkov, So, Sohn, Tokumine, Valter, Vasudevan, Vodrahalli, Wang, Wang, Wang, Wang, Wieting, Wu, Xu, Xu, Xue, Yin, Yu, Zhang, Zheng, Zheng, Zhou, Zhou, Petrov, and Wu]{anil_2023_palm2}
Rohan Anil, Andrew~M. Dai, Orhan Firat, Melvin Johnson, Dmitry Lepikhin, Alexandre Passos, Siamak Shakeri, Emanuel Taropa, Paige Bailey, Zhifeng Chen, Eric Chu, Jonathan~H. Clark, Laurent~El Shafey, Yanping Huang, Kathy Meier-Hellstern, Gaurav Mishra, Erica Moreira, Mark Omernick, Kevin Robinson, Sebastian Ruder, Yi~Tay, Kefan Xiao, Yuanzhong Xu, Yujing Zhang, Gustavo~Hernandez Abrego, Junwhan Ahn, Jacob Austin, Paul Barham, Jan Botha, James Bradbury, Siddhartha Brahma, Kevin Brooks, Michele Catasta, Yong Cheng, Colin Cherry, Christopher~A. Choquette-Choo, Aakanksha Chowdhery, Clément Crepy, Shachi Dave, Mostafa Dehghani, Sunipa Dev, Jacob Devlin, Mark Díaz, Nan Du, Ethan Dyer, Vlad Feinberg, Fangxiaoyu Feng, Vlad Fienber, Markus Freitag, Xavier Garcia, Sebastian Gehrmann, Lucas Gonzalez, Guy Gur-Ari, Steven Hand, Hadi Hashemi, Le~Hou, Joshua Howland, Andrea Hu, Jeffrey Hui, Jeremy Hurwitz, Michael Isard, Abe Ittycheriah, Matthew Jagielski, Wenhao Jia, Kathleen Kenealy, Maxim Krikun, Sneha Kudugunta, Chang
  Lan, Katherine Lee, Benjamin Lee, Eric Li, Music Li, Wei Li, YaGuang Li, Jian Li, Hyeontaek Lim, Hanzhao Lin, Zhongtao Liu, Frederick Liu, Marcello Maggioni, Aroma Mahendru, Joshua Maynez, Vedant Misra, Maysam Moussalem, Zachary Nado, John Nham, Eric Ni, Andrew Nystrom, Alicia Parrish, Marie Pellat, Martin Polacek, Alex Polozov, Reiner Pope, Siyuan Qiao, Emily Reif, Bryan Richter, Parker Riley, Alex~Castro Ros, Aurko Roy, Brennan Saeta, Rajkumar Samuel, Renee Shelby, Ambrose Slone, Daniel Smilkov, David~R. So, Daniel Sohn, Simon Tokumine, Dasha Valter, Vijay Vasudevan, Kiran Vodrahalli, Xuezhi Wang, Pidong Wang, Zirui Wang, Tao Wang, John Wieting, Yuhuai Wu, Kelvin Xu, Yunhan Xu, Linting Xue, Pengcheng Yin, Jiahui Yu, Qiao Zhang, Steven Zheng, Ce~Zheng, Weikang Zhou, Denny Zhou, Slav Petrov, and Yonghui Wu.
\newblock Palm 2 technical report, 2023.
\newblock URL \url{https://arxiv.org/abs/2305.10403}.

\bibitem[DeepSeek-AI et~al.(2025)DeepSeek-AI, Liu, Feng, Xue, Wang, Wu, Lu, Zhao, Deng, Zhang, Ruan, Dai, Guo, Yang, Chen, Ji, Li, Lin, Dai, Luo, Hao, Chen, Li, Zhang, Bao, Xu, Wang, Zhang, Ding, Xin, Gao, Li, Qu, Cai, Liang, Guo, Ni, Li, Wang, Chen, Chen, Yuan, Qiu, Li, Song, Dong, Hu, Gao, Guan, Huang, Yu, Wang, Zhang, Xu, Xia, Zhao, Wang, Zhang, Li, Wang, Zhang, Zhang, Tang, Li, Tian, Huang, Wang, Zhang, Wang, Zhu, Chen, Du, Chen, Jin, Ge, Zhang, Pan, Wang, Xu, Zhang, Chen, Li, Lu, Zhou, Chen, Wu, Ye, Ye, Ma, Wang, Zhou, Yu, Zhou, Pan, Wang, Yun, Pei, Sun, Xiao, Zeng, Zhao, An, Liu, Liang, Gao, Yu, Zhang, Li, Jin, Wang, Bi, Liu, Wang, Shen, Chen, Zhang, Chen, Nie, Sun, Wang, Cheng, Liu, Xie, Liu, Yu, Song, Shan, Zhou, Yang, Li, Su, Lin, Li, Wang, Wei, Zhu, Zhang, Xu, Xu, Huang, Li, Zhao, Sun, Li, Wang, Yu, Zheng, Zhang, Shi, Xiong, He, Tang, Piao, Wang, Tan, Ma, Liu, Guo, Wu, Ou, Zhu, Wang, Gong, Zou, He, Zha, Xiong, Ma, Yan, Luo, You, Liu, Zhou, Wu, Ren, Ren, Sha, Fu, Xu, Huang, Zhang, Xie, Zhang, Hao,
  Gou, Ma, Yan, Shao, Xu, Wu, Zhang, Li, Gu, Zhu, Liu, Li, Xie, Song, Gao, and Pan]{deepseek_2025}
DeepSeek-AI, Aixin Liu, Bei Feng, Bing Xue, Bingxuan Wang, Bochao Wu, Chengda Lu, Chenggang Zhao, Chengqi Deng, Chenyu Zhang, Chong Ruan, Damai Dai, Daya Guo, Dejian Yang, Deli Chen, Dongjie Ji, Erhang Li, Fangyun Lin, Fucong Dai, Fuli Luo, Guangbo Hao, Guanting Chen, Guowei Li, H.~Zhang, Han Bao, Hanwei Xu, Haocheng Wang, Haowei Zhang, Honghui Ding, Huajian Xin, Huazuo Gao, Hui Li, Hui Qu, J.~L. Cai, Jian Liang, Jianzhong Guo, Jiaqi Ni, Jiashi Li, Jiawei Wang, Jin Chen, Jingchang Chen, Jingyang Yuan, Junjie Qiu, Junlong Li, Junxiao Song, Kai Dong, Kai Hu, Kaige Gao, Kang Guan, Kexin Huang, Kuai Yu, Lean Wang, Lecong Zhang, Lei Xu, Leyi Xia, Liang Zhao, Litong Wang, Liyue Zhang, Meng Li, Miaojun Wang, Mingchuan Zhang, Minghua Zhang, Minghui Tang, Mingming Li, Ning Tian, Panpan Huang, Peiyi Wang, Peng Zhang, Qiancheng Wang, Qihao Zhu, Qinyu Chen, Qiushi Du, R.~J. Chen, R.~L. Jin, Ruiqi Ge, Ruisong Zhang, Ruizhe Pan, Runji Wang, Runxin Xu, Ruoyu Zhang, Ruyi Chen, S.~S. Li, Shanghao Lu, Shangyan Zhou, Shanhuang
  Chen, Shaoqing Wu, Shengfeng Ye, Shengfeng Ye, Shirong Ma, Shiyu Wang, Shuang Zhou, Shuiping Yu, Shunfeng Zhou, Shuting Pan, T.~Wang, Tao Yun, Tian Pei, Tianyu Sun, W.~L. Xiao, Wangding Zeng, Wanjia Zhao, Wei An, Wen Liu, Wenfeng Liang, Wenjun Gao, Wenqin Yu, Wentao Zhang, X.~Q. Li, Xiangyue Jin, Xianzu Wang, Xiao Bi, Xiaodong Liu, Xiaohan Wang, Xiaojin Shen, Xiaokang Chen, Xiaokang Zhang, Xiaosha Chen, Xiaotao Nie, Xiaowen Sun, Xiaoxiang Wang, Xin Cheng, Xin Liu, Xin Xie, Xingchao Liu, Xingkai Yu, Xinnan Song, Xinxia Shan, Xinyi Zhou, Xinyu Yang, Xinyuan Li, Xuecheng Su, Xuheng Lin, Y.~K. Li, Y.~Q. Wang, Y.~X. Wei, Y.~X. Zhu, Yang Zhang, Yanhong Xu, Yanhong Xu, Yanping Huang, Yao Li, Yao Zhao, Yaofeng Sun, Yaohui Li, Yaohui Wang, Yi~Yu, Yi~Zheng, Yichao Zhang, Yifan Shi, Yiliang Xiong, Ying He, Ying Tang, Yishi Piao, Yisong Wang, Yixuan Tan, Yiyang Ma, Yiyuan Liu, Yongqiang Guo, Yu~Wu, Yuan Ou, Yuchen Zhu, Yuduan Wang, Yue Gong, Yuheng Zou, Yujia He, Yukun Zha, Yunfan Xiong, Yunxian Ma, Yuting Yan, Yuxiang
  Luo, Yuxiang You, Yuxuan Liu, Yuyang Zhou, Z.~F. Wu, Z.~Z. Ren, Zehui Ren, Zhangli Sha, Zhe Fu, Zhean Xu, Zhen Huang, Zhen Zhang, Zhenda Xie, Zhengyan Zhang, Zhewen Hao, Zhibin Gou, Zhicheng Ma, Zhigang Yan, Zhihong Shao, Zhipeng Xu, Zhiyu Wu, Zhongyu Zhang, Zhuoshu Li, Zihui Gu, Zijia Zhu, Zijun Liu, Zilin Li, Ziwei Xie, Ziyang Song, Ziyi Gao, and Zizheng Pan.
\newblock Deepseek-v3 technical report, 2025.
\newblock URL \url{https://arxiv.org/abs/2412.19437}.

\bibitem[Team et~al.(2024)Team, Mesnard, Hardin, Dadashi, Bhupatiraju, Pathak, Sifre, Rivière, Kale, Love, Tafti, Hussenot, Sessa, Chowdhery, Roberts, Barua, Botev, Castro-Ros, Slone, Héliou, Tacchetti, Bulanova, Paterson, Tsai, Shahriari, Lan, Choquette-Choo, Crepy, Cer, Ippolito, Reid, Buchatskaya, Ni, Noland, Yan, Tucker, Muraru, Rozhdestvenskiy, Michalewski, Tenney, Grishchenko, Austin, Keeling, Labanowski, Lespiau, Stanway, Brennan, Chen, Ferret, Chiu, Mao-Jones, Lee, Yu, Millican, Sjoesund, Lee, Dixon, Reid, Mikuła, Wirth, Sharman, Chinaev, Thain, Bachem, Chang, Wahltinez, Bailey, Michel, Yotov, Chaabouni, Comanescu, Jana, Anil, McIlroy, Liu, Mullins, Smith, Borgeaud, Girgin, Douglas, Pandya, Shakeri, De, Klimenko, Hennigan, Feinberg, Stokowiec, hui Chen, Ahmed, Gong, Warkentin, Peran, Giang, Farabet, Vinyals, Dean, Kavukcuoglu, Hassabis, Ghahramani, Eck, Barral, Pereira, Collins, Joulin, Fiedel, Senter, Andreev, and Kenealy]{gemmateam_2024}
Gemma Team, Thomas Mesnard, Cassidy Hardin, Robert Dadashi, Surya Bhupatiraju, Shreya Pathak, Laurent Sifre, Morgane Rivière, Mihir~Sanjay Kale, Juliette Love, Pouya Tafti, Léonard Hussenot, Pier~Giuseppe Sessa, Aakanksha Chowdhery, Adam Roberts, Aditya Barua, Alex Botev, Alex Castro-Ros, Ambrose Slone, Amélie Héliou, Andrea Tacchetti, Anna Bulanova, Antonia Paterson, Beth Tsai, Bobak Shahriari, Charline~Le Lan, Christopher~A. Choquette-Choo, Clément Crepy, Daniel Cer, Daphne Ippolito, David Reid, Elena Buchatskaya, Eric Ni, Eric Noland, Geng Yan, George Tucker, George-Christian Muraru, Grigory Rozhdestvenskiy, Henryk Michalewski, Ian Tenney, Ivan Grishchenko, Jacob Austin, James Keeling, Jane Labanowski, Jean-Baptiste Lespiau, Jeff Stanway, Jenny Brennan, Jeremy Chen, Johan Ferret, Justin Chiu, Justin Mao-Jones, Katherine Lee, Kathy Yu, Katie Millican, Lars~Lowe Sjoesund, Lisa Lee, Lucas Dixon, Machel Reid, Maciej Mikuła, Mateo Wirth, Michael Sharman, Nikolai Chinaev, Nithum Thain, Olivier Bachem,
  Oscar Chang, Oscar Wahltinez, Paige Bailey, Paul Michel, Petko Yotov, Rahma Chaabouni, Ramona Comanescu, Reena Jana, Rohan Anil, Ross McIlroy, Ruibo Liu, Ryan Mullins, Samuel~L Smith, Sebastian Borgeaud, Sertan Girgin, Sholto Douglas, Shree Pandya, Siamak Shakeri, Soham De, Ted Klimenko, Tom Hennigan, Vlad Feinberg, Wojciech Stokowiec, Yu~hui Chen, Zafarali Ahmed, Zhitao Gong, Tris Warkentin, Ludovic Peran, Minh Giang, Clément Farabet, Oriol Vinyals, Jeff Dean, Koray Kavukcuoglu, Demis Hassabis, Zoubin Ghahramani, Douglas Eck, Joelle Barral, Fernando Pereira, Eli Collins, Armand Joulin, Noah Fiedel, Evan Senter, Alek Andreev, and Kathleen Kenealy.
\newblock Gemma: Open models based on gemini research and technology, 2024.
\newblock URL \url{https://arxiv.org/abs/2403.08295}.

\bibitem[Wei et~al.(2023)Wei, Wei, Lin, Li, Zhang, Ren, Li, Wan, Cao, Xie, Hu, Li, Hui, Yu, Liu, Yang, Huang, and Xie]{wei2023polylm}
Xiangpeng Wei, Haoran Wei, Huan Lin, Tianhao Li, Pei Zhang, Xingzhang Ren, Mei Li, Yu~Wan, Zhiwei Cao, Binbin Xie, Tianxiang Hu, Shangjie Li, Binyuan Hui, Bowen Yu, Dayiheng Liu, Baosong Yang, Fei Huang, and Jun Xie.
\newblock Polylm: An open source polyglot large language model, 2023.
\newblock URL \url{https://arxiv.org/abs/2307.06018}.

\bibitem[Parmar et~al.(2024)Parmar, Prabhumoye, Jennings, Patwary, Subramanian, Su, Zhu, Narayanan, Jhunjhunwala, Dattagupta, Jawa, Liu, Mahabaleshwarkar, Nitski, Brundyn, Maki, Martinez, You, Kamalu, LeGresley, Fridman, Casper, Aithal, Kuchaiev, Shoeybi, Cohen, and Catanzaro]{parmar2024nemotron415b}
Jupinder Parmar, Shrimai Prabhumoye, Joseph Jennings, Mostofa Patwary, Sandeep Subramanian, Dan Su, Chen Zhu, Deepak Narayanan, Aastha Jhunjhunwala, Ayush Dattagupta, Vibhu Jawa, Jiwei Liu, Ameya Mahabaleshwarkar, Osvald Nitski, Annika Brundyn, James Maki, Miguel Martinez, Jiaxuan You, John Kamalu, Patrick LeGresley, Denys Fridman, Jared Casper, Ashwath Aithal, Oleksii Kuchaiev, Mohammad Shoeybi, Jonathan Cohen, and Bryan Catanzaro.
\newblock Nemotron-4 15b technical report, 2024.
\newblock URL \url{https://arxiv.org/abs/2402.16819}.

\bibitem[McCloskey and Cohen(1989)]{McCloskey_1989}
Michael McCloskey and Neal~J. Cohen.
\newblock Catastrophic interference in connectionist networks: The sequential learning problem.
\newblock volume~24 of \emph{Psychology of Learning and Motivation}, pages 109--165. Academic Press, 1989.
\newblock \doi{https://doi.org/10.1016/S0079-7421(08)60536-8}.
\newblock URL \url{https://www.sciencedirect.com/science/article/pii/S0079742108605368}.

\bibitem[Shi et~al.(2024)Shi, Xu, Wang, Qin, Wang, Wang, Wang, Ebrahimi, and Wang]{shi_2024}
Haizhou Shi, Zihao Xu, Hengyi Wang, Weiyi Qin, Wenyuan Wang, Yibin Wang, Zifeng Wang, Sayna Ebrahimi, and Hao Wang.
\newblock Continual learning of large language models: A comprehensive survey, 2024.
\newblock URL \url{https://arxiv.org/abs/2404.16789}.

\bibitem[Kirkpatrick et~al.(2017)Kirkpatrick, Pascanu, Rabinowitz, Veness, Desjardins, Rusu, Milan, Quan, Ramalho, Grabska-Barwinska, Hassabis, Clopath, Kumaran, and Hadsell]{kirkpatrick_2017}
James Kirkpatrick, Razvan Pascanu, Neil Rabinowitz, Joel Veness, Guillaume Desjardins, Andrei~A. Rusu, Kieran Milan, John Quan, Tiago Ramalho, Agnieszka Grabska-Barwinska, Demis Hassabis, Claudia Clopath, Dharshan Kumaran, and Raia Hadsell.
\newblock Overcoming catastrophic forgetting in neural networks.
\newblock \emph{Proceedings of the National Academy of Sciences}, 114\penalty0 (13):\penalty0 3521--3526, 2017.
\newblock \doi{10.1073/pnas.1611835114}.
\newblock URL \url{https://www.pnas.org/doi/abs/10.1073/pnas.1611835114}.

\bibitem[Bai et~al.(2022)Bai, Jones, Ndousse, Askell, Chen, DasSarma, Drain, Fort, Ganguli, Henighan, Joseph, Kadavath, Kernion, Conerly, El-Showk, Elhage, Hatfield-Dodds, Hernandez, Hume, Johnston, Kravec, Lovitt, Nanda, Olsson, Amodei, Brown, Clark, McCandlish, Olah, Mann, and Kaplan]{bai_2022}
Yuntao Bai, Andy Jones, Kamal Ndousse, Amanda Askell, Anna Chen, Nova DasSarma, Dawn Drain, Stanislav Fort, Deep Ganguli, Tom Henighan, Nicholas Joseph, Saurav Kadavath, Jackson Kernion, Tom Conerly, Sheer El-Showk, Nelson Elhage, Zac Hatfield-Dodds, Danny Hernandez, Tristan Hume, Scott Johnston, Shauna Kravec, Liane Lovitt, Neel Nanda, Catherine Olsson, Dario Amodei, Tom Brown, Jack Clark, Sam McCandlish, Chris Olah, Ben Mann, and Jared Kaplan.
\newblock Training a helpful and harmless assistant with reinforcement learning from human feedback, 2022.
\newblock URL \url{https://arxiv.org/abs/2204.05862}.

\bibitem[Ouyang et~al.(2022)Ouyang, Wu, Jiang, Almeida, Wainwright, Mishkin, Zhang, Agarwal, Slama, Ray, Schulman, Hilton, Kelton, Miller, Simens, Askell, Welinder, Christiano, Leike, and Lowe]{ouyang_2022}
Long Ouyang, Jeffrey Wu, Xu~Jiang, Diogo Almeida, Carroll Wainwright, Pamela Mishkin, Chong Zhang, Sandhini Agarwal, Katarina Slama, Alex Ray, John Schulman, Jacob Hilton, Fraser Kelton, Luke Miller, Maddie Simens, Amanda Askell, Peter Welinder, Paul~F Christiano, Jan Leike, and Ryan Lowe.
\newblock Training language models to follow instructions with human feedback.
\newblock In S.~Koyejo, S.~Mohamed, A.~Agarwal, D.~Belgrave, K.~Cho, and A.~Oh, editors, \emph{Advances in Neural Information Processing Systems}, volume~35, pages 27730--27744. Curran Associates, Inc., 2022.
\newblock URL \url{https://proceedings.neurips.cc/paper_files/paper/2022/file/b1efde53be364a73914f58805a001731-Paper-Conference.pdf}.

\bibitem[Schulman et~al.(2017)Schulman, Wolski, Dhariwal, Radford, and Klimov]{schulmanetal2017}
John Schulman, Filip Wolski, Prafulla Dhariwal, Alec Radford, and Oleg Klimov.
\newblock Proximal policy optimization algorithms, 2017.
\newblock URL \url{https://arxiv.org/abs/1707.06347}.

\bibitem[Williams et~al.(2025)Williams, Carroll, Narang, Weisser, Murphy, and Dragan]{williams_carroll_2025}
Marcus Williams, Micah Carroll, Adhyyan Narang, Constantin Weisser, Brendan Murphy, and Anca Dragan.
\newblock On targeted manipulation and deception when optimizing llms for user feedback, 2025.
\newblock URL \url{https://arxiv.org/abs/2411.02306}.

\bibitem[Le et~al.(2020)Le, Vial, Frej, Segonne, Coavoux, Lecouteux, Allauzen, Crabbé, Besacier, and Schwab]{le2020flaubert}
Hang Le, Loïc Vial, Jibril Frej, Vincent Segonne, Maximin Coavoux, Benjamin Lecouteux, Alexandre Allauzen, Benoît Crabbé, Laurent Besacier, and Didier Schwab.
\newblock Flaubert: Unsupervised language model pre-training for french, 2020.
\newblock URL \url{https://arxiv.org/abs/1912.05372}.

\bibitem[de~Vries et~al.(2019)de~Vries, van Cranenburgh, Bisazza, Caselli, van Noord, and Nissim]{devries2019bertje}
Wietse de~Vries, Andreas van Cranenburgh, Arianna Bisazza, Tommaso Caselli, Gertjan van Noord, and Malvina Nissim.
\newblock Bertje: A dutch bert model, 2019.
\newblock URL \url{https://arxiv.org/abs/1912.09582}.

\bibitem[Ralethe(2020)]{ralethe-2020-adaptation}
Sello Ralethe.
\newblock Adaptation of deep bidirectional transformers for {A}frikaans language.
\newblock In Nicoletta Calzolari, Fr{\'e}d{\'e}ric B{\'e}chet, Philippe Blache, Khalid Choukri, Christopher Cieri, Thierry Declerck, Sara Goggi, Hitoshi Isahara, Bente Maegaard, Joseph Mariani, H{\'e}l{\`e}ne Mazo, Asuncion Moreno, Jan Odijk, and Stelios Piperidis, editors, \emph{Proceedings of the Twelfth Language Resources and Evaluation Conference}, pages 2475--2478, Marseille, France, May 2020. European Language Resources Association.
\newblock ISBN 979-10-95546-34-4.
\newblock URL \url{https://aclanthology.org/2020.lrec-1.301/}.

\bibitem[Chowdhery et~al.(2022)Chowdhery, Narang, Devlin, Bosma, Mishra, Roberts, Barham, Chung, Sutton, Gehrmann, Schuh, Shi, Tsvyashchenko, Maynez, Rao, Barnes, Tay, Shazeer, Prabhakaran, Reif, Du, Hutchinson, Pope, Bradbury, Austin, Isard, Gur-Ari, Yin, Duke, Levskaya, Ghemawat, Dev, Michalewski, Garcia, Misra, Robinson, Fedus, Zhou, Ippolito, Luan, Lim, Zoph, Spiridonov, Sepassi, Dohan, Agrawal, Omernick, Dai, Pillai, Pellat, Lewkowycz, Moreira, Child, Polozov, Lee, Zhou, Wang, Saeta, Diaz, Firat, Catasta, Wei, Meier-Hellstern, Eck, Dean, Petrov, and Fiedel]{chowdhery_2022_palm}
Aakanksha Chowdhery, Sharan Narang, Jacob Devlin, Maarten Bosma, Gaurav Mishra, Adam Roberts, Paul Barham, Hyung~Won Chung, Charles Sutton, Sebastian Gehrmann, Parker Schuh, Kensen Shi, Sasha Tsvyashchenko, Joshua Maynez, Abhishek Rao, Parker Barnes, Yi~Tay, Noam Shazeer, Vinodkumar Prabhakaran, Emily Reif, Nan Du, Ben Hutchinson, Reiner Pope, James Bradbury, Jacob Austin, Michael Isard, Guy Gur-Ari, Pengcheng Yin, Toju Duke, Anselm Levskaya, Sanjay Ghemawat, Sunipa Dev, Henryk Michalewski, Xavier Garcia, Vedant Misra, Kevin Robinson, Liam Fedus, Denny Zhou, Daphne Ippolito, David Luan, Hyeontaek Lim, Barret Zoph, Alexander Spiridonov, Ryan Sepassi, David Dohan, Shivani Agrawal, Mark Omernick, Andrew~M. Dai, Thanumalayan~Sankaranarayana Pillai, Marie Pellat, Aitor Lewkowycz, Erica Moreira, Rewon Child, Oleksandr Polozov, Katherine Lee, Zongwei Zhou, Xuezhi Wang, Brennan Saeta, Mark Diaz, Orhan Firat, Michele Catasta, Jason Wei, Kathy Meier-Hellstern, Douglas Eck, Jeff Dean, Slav Petrov, and Noah Fiedel.
\newblock Palm: Scaling language modeling with pathways, 2022.
\newblock URL \url{https://arxiv.org/abs/2204.02311}.

\bibitem[Doddapaneni et~al.(2023)Doddapaneni, Aralikatte, Ramesh, Goyal, Khapra, Kunchukuttan, and Kumar]{doddapaneni-etal-2023-towards}
Sumanth Doddapaneni, Rahul Aralikatte, Gowtham Ramesh, Shreya Goyal, Mitesh~M. Khapra, Anoop Kunchukuttan, and Pratyush Kumar.
\newblock Towards leaving no {I}ndic language behind: Building monolingual corpora, benchmark and models for {I}ndic languages.
\newblock In Anna Rogers, Jordan Boyd-Graber, and Naoaki Okazaki, editors, \emph{Proceedings of the 61st Annual Meeting of the Association for Computational Linguistics (Volume 1: Long Papers)}, pages 12402--12426, Toronto, Canada, July 2023. Association for Computational Linguistics.
\newblock \doi{10.18653/v1/2023.acl-long.693}.
\newblock URL \url{https://aclanthology.org/2023.acl-long.693/}.

\bibitem[Ogueji et~al.(2021)Ogueji, Zhu, and Lin]{ogueji-etal-2021-small}
Kelechi Ogueji, Yuxin Zhu, and Jimmy Lin.
\newblock Small data? no problem! exploring the viability of pretrained multilingual language models for low-resourced languages.
\newblock In Duygu Ataman, Alexandra Birch, Alexis Conneau, Orhan Firat, Sebastian Ruder, and Gozde~Gul Sahin, editors, \emph{Proceedings of the 1st Workshop on Multilingual Representation Learning}, pages 116--126, Punta Cana, Dominican Republic, November 2021. Association for Computational Linguistics.
\newblock \doi{10.18653/v1/2021.mrl-1.11}.
\newblock URL \url{https://aclanthology.org/2021.mrl-1.11/}.

\bibitem[Chew et~al.(2006{\natexlab{b}})Chew, Verzi, Bauer, and McClain]{chew_2006_bible}
Peter~A. Chew, Steve~J. Verzi, Travis~L. Bauer, and Jonathan~T. McClain.
\newblock Evaluation of the bible as a resource for cross-language information retrieval.
\newblock In \emph{Proceedings of the Workshop on Multilingual Language Resources and Interoperability}, MLRI '06, page 68–74, USA, 2006{\natexlab{b}}. Association for Computational Linguistics.
\newblock ISBN 1932432825.

\bibitem[Tiedemann(2012)]{tiedemann_2012}
Jörg Tiedemann.
\newblock Parallel data, tools and interfaces in opus.
\newblock In Nicoletta Calzolari~(Conference Chair), Khalid Choukri, Thierry Declerck, Mehmet~Ugur Dogan, Bente Maegaard, Joseph Mariani, Jan Odijk, and Stelios Piperidis, editors, \emph{Proceedings of the Eight International Conference on Language Resources and Evaluation (LREC'12)}, Istanbul, Turkey, may 2012. European Language Resources Association (ELRA).
\newblock ISBN 978-2-9517408-7-7.

\bibitem[Bengio et~al.(2003)Bengio, Ducharme, Vincent, and Janvin]{bengio_2003_plm}
Yoshua Bengio, R\'{e}jean Ducharme, Pascal Vincent, and Christian Janvin.
\newblock A neural probabilistic language model.
\newblock \emph{J. Mach. Learn. Res.}, 3\penalty0 (null):\penalty0 1137–1155, March 2003.
\newblock ISSN 1532-4435.

\bibitem[Conneau and Lample(2019)]{conneau_lample_2019}
Alexis Conneau and Guillaume Lample.
\newblock \emph{Cross-lingual language model pretraining}.
\newblock Curran Associates Inc., Red Hook, NY, USA, 2019.

\bibitem[Ouyang et~al.(2021)Ouyang, Wang, Pang, Sun, Tian, Wu, and Wang]{ouyang_2021_camlm}
Xuan Ouyang, Shuohuan Wang, Chao Pang, Yu~Sun, Hao Tian, Hua Wu, and Haifeng Wang.
\newblock Ernie-m: Enhanced multilingual representation by aligning cross-lingual semantics with monolingual corpora, 2021.
\newblock URL \url{https://arxiv.org/abs/2012.15674}.

\bibitem[Huang et~al.(2019)Huang, Liang, Duan, Gong, Shou, Jiang, and Zhou]{huang_2019_clmlm}
Haoyang Huang, Yaobo Liang, Nan Duan, Ming Gong, Linjun Shou, Daxin Jiang, and Ming Zhou.
\newblock {U}nicoder: A universal language encoder by pre-training with multiple cross-lingual tasks.
\newblock In Kentaro Inui, Jing Jiang, Vincent Ng, and Xiaojun Wan, editors, \emph{Proceedings of the 2019 Conference on Empirical Methods in Natural Language Processing and the 9th International Joint Conference on Natural Language Processing (EMNLP-IJCNLP)}, pages 2485--2494, Hong Kong, China, November 2019. Association for Computational Linguistics.
\newblock \doi{10.18653/v1/D19-1252}.
\newblock URL \url{https://aclanthology.org/D19-1252/}.

\bibitem[Yang et~al.(2020)Yang, Ma, Zhang, Wu, Li, and Zhou]{Yang_2020_alm}
Jian Yang, Shuming Ma, Dongdong Zhang, ShuangZhi Wu, Zhoujun Li, and Ming Zhou.
\newblock Alternating language modeling for cross-lingual pre-training.
\newblock \emph{Proceedings of the AAAI Conference on Artificial Intelligence}, 34\penalty0 (05):\penalty0 9386--9393, Apr. 2020.
\newblock \doi{10.1609/aaai.v34i05.6480}.
\newblock URL \url{https://ojs.aaai.org/index.php/AAAI/article/view/6480}.

\bibitem[Taylor(1953)]{taylor_1953}
W.~L. Taylor.
\newblock "cloze procedure": a new tool for measuring readability.
\newblock \emph{Journalism Quarterly}, pages 415--433, 1953.

\bibitem[Vincent et~al.(2008)Vincent, Larochelle, Bengio, and Manzagol]{vincent_2008_dae}
Pascal Vincent, Hugo Larochelle, Yoshua Bengio, and Pierre-Antoine Manzagol.
\newblock Extracting and composing robust features with denoising autoencoders.
\newblock In \emph{Proceedings of the 25th International Conference on Machine Learning}, ICML '08, page 1096–1103, New York, NY, USA, 2008. Association for Computing Machinery.
\newblock ISBN 9781605582054.
\newblock \doi{10.1145/1390156.1390294}.
\newblock URL \url{https://doi.org/10.1145/1390156.1390294}.

\bibitem[Chi et~al.(2022)Chi, Huang, Dong, Ma, Zheng, Singhal, Bajaj, Song, Mao, Huang, and Wei]{chi_2022_xlme}
Zewen Chi, Shaohan Huang, Li~Dong, Shuming Ma, Bo~Zheng, Saksham Singhal, Payal Bajaj, Xia Song, Xian-Ling Mao, Heyan Huang, and Furu Wei.
\newblock Xlm-e: Cross-lingual language model pre-training via electra, 2022.
\newblock URL \url{https://arxiv.org/abs/2106.16138}.

\bibitem[Chi et~al.(2021{\natexlab{a}})Chi, Dong, Wei, Yang, Singhal, Wang, Song, Mao, Huang, and Zhou]{chi_2021_xlco}
Zewen Chi, Li~Dong, Furu Wei, Nan Yang, Saksham Singhal, Wenhui Wang, Xia Song, Xian-Ling Mao, Heyan Huang, and Ming Zhou.
\newblock {I}nfo{XLM}: An information-theoretic framework for cross-lingual language model pre-training.
\newblock In Kristina Toutanova, Anna Rumshisky, Luke Zettlemoyer, Dilek Hakkani-Tur, Iz~Beltagy, Steven Bethard, Ryan Cotterell, Tanmoy Chakraborty, and Yichao Zhou, editors, \emph{Proceedings of the 2021 Conference of the North American Chapter of the Association for Computational Linguistics: Human Language Technologies}, pages 3576--3588, Online, June 2021{\natexlab{a}}. Association for Computational Linguistics.
\newblock \doi{10.18653/v1/2021.naacl-main.280}.
\newblock URL \url{https://aclanthology.org/2021.naacl-main.280/}.

\bibitem[Wei et~al.(2021)Wei, Weng, Hu, Xing, Yu, and Luo]{wei_2021_hictl}
Xiangpeng Wei, Rongxiang Weng, Yue Hu, Luxi Xing, Heng Yu, and Weihua Luo.
\newblock On learning universal representations across languages, 2021.
\newblock URL \url{https://arxiv.org/abs/2007.15960}.

\bibitem[Hu et~al.(2021)Hu, Johnson, Firat, Siddhant, and Neubig]{hu_2021_objectives}
Junjie Hu, Melvin Johnson, Orhan Firat, Aditya Siddhant, and Graham Neubig.
\newblock Explicit alignment objectives for multilingual bidirectional encoders, 2021.
\newblock URL \url{https://arxiv.org/abs/2010.07972}.

\bibitem[Chi et~al.(2021{\natexlab{b}})Chi, Dong, Zheng, Huang, Mao, Huang, and Wei]{chi_2021_dwa}
Zewen Chi, Li~Dong, Bo~Zheng, Shaohan Huang, Xian-Ling Mao, Heyan Huang, and Furu Wei.
\newblock Improving pretrained cross-lingual language models via self-labeled word alignment.
\newblock In Chengqing Zong, Fei Xia, Wenjie Li, and Roberto Navigli, editors, \emph{Proceedings of the 59th Annual Meeting of the Association for Computational Linguistics and the 11th International Joint Conference on Natural Language Processing (Volume 1: Long Papers)}, pages 3418--3430, Online, August 2021{\natexlab{b}}. Association for Computational Linguistics.
\newblock \doi{10.18653/v1/2021.acl-long.265}.
\newblock URL \url{https://aclanthology.org/2021.acl-long.265/}.

\bibitem[Pfeiffer et~al.(2022)Pfeiffer, Goyal, Lin, Li, Cross, Riedel, and Artetxe]{pfeiffer_2022}
Jonas Pfeiffer, Naman Goyal, Xi~Victoria Lin, Xian Li, James Cross, Sebastian Riedel, and Mikel Artetxe.
\newblock Lifting the curse of multilinguality by pre-training modular transformers, 2022.
\newblock URL \url{https://arxiv.org/abs/2205.06266}.

\bibitem[Blevins et~al.(2024)Blevins, Limisiewicz, Gururangan, Li, Gonen, Smith, and Zettlemoyer]{blevins_2024}
Terra Blevins, Tomasz Limisiewicz, Suchin Gururangan, Margaret Li, Hila Gonen, Noah~A. Smith, and Luke Zettlemoyer.
\newblock Breaking the curse of multilinguality with cross-lingual expert language models, 2024.
\newblock URL \url{https://arxiv.org/abs/2401.10440}.

\bibitem[Artetxe et~al.(2020)Artetxe, Ruder, and Yogatama]{artetxe_2020}
Mikel Artetxe, Sebastian Ruder, and Dani Yogatama.
\newblock On the cross-lingual transferability of monolingual representations.
\newblock In Dan Jurafsky, Joyce Chai, Natalie Schluter, and Joel Tetreault, editors, \emph{Proceedings of the 58th Annual Meeting of the Association for Computational Linguistics}, pages 4623--4637, Online, July 2020. Association for Computational Linguistics.
\newblock \doi{10.18653/v1/2020.acl-main.421}.
\newblock URL \url{https://aclanthology.org/2020.acl-main.421/}.

\bibitem[Mikolov et~al.(2013)Mikolov, Le, and Sutskever]{mikolov2013}
Tomas Mikolov, Quoc~V. Le, and Ilya Sutskever.
\newblock Exploiting similarities among languages for machine translation, 2013.
\newblock URL \url{https://arxiv.org/abs/1309.4168}.

\bibitem[Lample et~al.(2018)Lample, Conneau, Denoyer, and Ranzato]{lample2018}
Guillaume Lample, Alexis Conneau, Ludovic Denoyer, and Marc'Aurelio Ranzato.
\newblock Unsupervised machine translation using monolingual corpora only, 2018.
\newblock URL \url{https://arxiv.org/abs/1711.00043}.

\bibitem[Chang et~al.(2024)Chang, Arnett, Tu, and Bergen]{chang-etal-2024-multilinguality}
Tyler~A. Chang, Catherine Arnett, Zhuowen Tu, and Ben Bergen.
\newblock When is multilinguality a curse? language modeling for 250 high- and low-resource languages.
\newblock In Yaser Al-Onaizan, Mohit Bansal, and Yun-Nung Chen, editors, \emph{Proceedings of the 2024 Conference on Empirical Methods in Natural Language Processing}, pages 4074--4096, Miami, Florida, USA, November 2024. Association for Computational Linguistics.
\newblock \doi{10.18653/v1/2024.emnlp-main.236}.
\newblock URL \url{https://aclanthology.org/2024.emnlp-main.236/}.

\bibitem[Liu et~al.(2019{\natexlab{a}})Liu, Ott, Goyal, Du, Joshi, Chen, Levy, Lewis, Zettlemoyer, and Stoyanov]{liu2019robertarobustlyoptimizedbert}
Yinhan Liu, Myle Ott, Naman Goyal, Jingfei Du, Mandar Joshi, Danqi Chen, Omer Levy, Mike Lewis, Luke Zettlemoyer, and Veselin Stoyanov.
\newblock Roberta: A robustly optimized bert pretraining approach, 2019{\natexlab{a}}.
\newblock URL \url{https://arxiv.org/abs/1907.11692}.

\bibitem[Wang et~al.(2023{\natexlab{b}})Wang, Huang, Chang, and Chen]{Wang2023SALT}
Fei Wang, Kuan‑Hao Huang, Kai‑Wei Chang, and Muhao Chen.
\newblock Self‑augmentation improves zero‑shot cross‑lingual transfer.
\newblock \emph{arXiv preprint arXiv:2309.10891}, 2023{\natexlab{b}}.

\bibitem[Park et~al.(2024)Park, Hwang, Lee, Ojha, Liu, Vylomova, Pirinen, Abbott, Washington, Oco, Malykh, Logacheva, and Zhao]{Park2024LoResMT}
Gyutae Park, Seojin Hwang, Hwanhee Lee, Atul~Kr. Ojha, Chao‑hong Liu, Ekaterina Vylomova, Flammie Pirinen, Jade Abbott, Jonathan Washington, Nathaniel Oco, Valentin Malykh, Varvara Logacheva, and Xiaobing Zhao.
\newblock Low‑resource cross‑lingual summarization through few‑shot learning with large language models.
\newblock In \emph{Proceedings of LoResMT 2024}, 2024.

\bibitem[Asai et~al.(2024)Asai, Kudugunta, Yu, Blevins, Gonen, Reid, Tsvetkov, Ruder, and Hajishirzi]{Asai2023BUFFET}
Akari Asai, Sneha Kudugunta, Xinyan Yu, Terra Blevins, Hila Gonen, Machel Reid, Yulia Tsvetkov, Sebastian Ruder, and Hannaneh Hajishirzi.
\newblock Buffet: Benchmarking large language models for few‑shot cross‑lingual transfer.
\newblock In \emph{NAACL Long Papers}, 2024.

\bibitem[Bafna et~al.(2023)Bafna, España‑Bonet, van Genabith, Sagot, and Bawden]{Bafna2023}
Niyati Bafna, Cristina España‑Bonet, Josef van Genabith, Beno{\^i}t Sagot, and Rachel Bawden.
\newblock Cross-lingual strategies for low-resource language modeling: A study on five indic dialects.
\newblock In \emph{Proceedings of CORIA‑TALN 2023}, 2023.

\bibitem[Chiu and Bai(2025)]{chiu2025translate_multilingual}
Yi‑Ting Chiu and Zong‑Han Bai.
\newblock Translation or multilingual retrieval? evaluating cross‑lingual search strategies for traditional chinese financial documents.
\newblock In \emph{FinTech in AI CUP Special Session}, 2025.

\bibitem[Litschko et~al.(2021)Litschko, Vulić, Ponzetto, and Glavaš]{litschko2021unsupervised}
Robert Litschko, Ivan Vulić, Simone~Paolo Ponzetto, and Goran Glavaš.
\newblock Evaluating multilingual text encoders for unsupervised cross‑lingual retrieval.
\newblock \emph{arXiv preprint arXiv:2101.08370}, 2021.

\bibitem[AI(2022)]{primer2022eval}
Primer AI.
\newblock Language agnostic multilingual sentence embedding models for semantic search.
\newblock 2022.

\bibitem[Zhao et~al.(2020{\natexlab{a}})Zhao, Eger, Bjerva, and Augenstein]{zhao2020inducing}
Wei Zhao, Steffen Eger, Johannes Bjerva, and Isabelle Augenstein.
\newblock Inducing language‑agnostic multilingual representations.
\newblock \emph{arXiv preprint arXiv:2008.09112}, 2020{\natexlab{a}}.

\bibitem[Luong et~al.(2015)Luong, Pham, and Manning]{luong_2015}
Thang Luong, Hieu Pham, and Christopher~D. Manning.
\newblock Bilingual word representations with monolingual quality in mind.
\newblock In Phil Blunsom, Shay Cohen, Paramveer Dhillon, and Percy Liang, editors, \emph{Proceedings of the 1st Workshop on Vector Space Modeling for Natural Language Processing}, pages 151--159, Denver, Colorado, June 2015. Association for Computational Linguistics.
\newblock \doi{10.3115/v1/W15-1521}.
\newblock URL \url{https://aclanthology.org/W15-1521/}.

\bibitem[Gouws et~al.(2015)Gouws, Bengio, and Corrado]{gouws_2015}
Stephan Gouws, Yoshua Bengio, and Greg Corrado.
\newblock Bilbowa: Fast bilingual distributed representations without word alignments.
\newblock In Francis Bach and David Blei, editors, \emph{Proceedings of the 32nd International Conference on Machine Learning}, volume~37 of \emph{Proceedings of Machine Learning Research}, pages 748--756, Lille, France, 07--09 Jul 2015. PMLR.
\newblock URL \url{https://proceedings.mlr.press/v37/gouws15.html}.

\bibitem[Vulić and Moens(2016)]{vulić2016}
Ivan Vulić and Marie-Francine Moens.
\newblock Bilingual distributed word representations from document-aligned comparable data, 2016.
\newblock URL \url{https://arxiv.org/abs/1509.07308}.

\bibitem[Duong et~al.(2016)Duong, Kanayama, Ma, Bird, and Cohn]{duong_2016}
Long Duong, Hiroshi Kanayama, Tengfei Ma, Steven Bird, and Trevor Cohn.
\newblock Learning crosslingual word embeddings without bilingual corpora.
\newblock In Jian Su, Kevin Duh, and Xavier Carreras, editors, \emph{Proceedings of the 2016 Conference on Empirical Methods in Natural Language Processing}, pages 1285--1295, Austin, Texas, November 2016. Association for Computational Linguistics.
\newblock \doi{10.18653/v1/D16-1136}.
\newblock URL \url{https://aclanthology.org/D16-1136/}.

\bibitem[Conneau et~al.(2017)Conneau, Lample, Ranzato, Denoyer, and J{\'e}gou]{conneau2017_wordtranslation}
Alexis Conneau, Guillaume Lample, Marc’aurelio Ranzato, Ludovic Denoyer, and Herv{\'e} J{\'e}gou.
\newblock Word translation without parallel data.
\newblock \emph{arXiv preprint arXiv:1710.04087}, 2017.

\bibitem[Chen and Cardie(2018)]{chen-cardie-2018-unsupervised}
Xilun Chen and Claire Cardie.
\newblock Unsupervised multilingual word embeddings.
\newblock In Ellen Riloff, David Chiang, Julia Hockenmaier, and Jun{'}ichi Tsujii, editors, \emph{Proceedings of the 2018 Conference on Empirical Methods in Natural Language Processing}, pages 261--270, Brussels, Belgium, October-November 2018. Association for Computational Linguistics.
\newblock \doi{10.18653/v1/D18-1024}.
\newblock URL \url{https://aclanthology.org/D18-1024/}.

\bibitem[S{\o}gaard et~al.(2018)S{\o}gaard, Ruder, and Vuli{\'c}]{sogaard-etal-2018-limitations}
Anders S{\o}gaard, Sebastian Ruder, and Ivan Vuli{\'c}.
\newblock On the limitations of unsupervised bilingual dictionary induction.
\newblock In Iryna Gurevych and Yusuke Miyao, editors, \emph{Proceedings of the 56th Annual Meeting of the Association for Computational Linguistics (Volume 1: Long Papers)}, pages 778--788, Melbourne, Australia, July 2018. Association for Computational Linguistics.
\newblock \doi{10.18653/v1/P18-1072}.
\newblock URL \url{https://aclanthology.org/P18-1072/}.

\bibitem[Wang and Hong(2023)]{wang-hong-2023}
Yuxuan Wang and Lyu Hong.
\newblock Query encoder distillation via embedding alignment is a strong baseline method to boost dense retriever online efficiency.
\newblock In Nafise Sadat~Moosavi, Iryna Gurevych, Yufang Hou, Gyuwan Kim, Young~Jin Kim, Tal Schuster, and Ameeta Agrawal, editors, \emph{Proceedings of the Fourth Workshop on Simple and Efficient Natural Language Processing (SustaiNLP)}, pages 290--298, Toronto, Canada (Hybrid), July 2023. Association for Computational Linguistics.
\newblock \doi{10.18653/v1/2023.sustainlp-1.23}.
\newblock URL \url{https://aclanthology.org/2023.sustainlp-1.23/}.

\bibitem[Kim et~al.(2023)Kim, Rawat, Zaheer, Jayasumana, Sadhanala, Jitkrittum, Menon, Fergus, and Kumar]{kim2023}
Seungyeon Kim, Ankit~Singh Rawat, Manzil Zaheer, Sadeep Jayasumana, Veeranjaneyulu Sadhanala, Wittawat Jitkrittum, Aditya~Krishna Menon, Rob Fergus, and Sanjiv Kumar.
\newblock Embeddistill: A geometric knowledge distillation for information retrieval, 2023.
\newblock URL \url{https://arxiv.org/abs/2301.12005}.

\bibitem[Tashu et~al.(2025)Tashu, Kontos, Sabatelli, and Valdenegro-Toro]{tashu_2025}
Tsegaye~Misikir Tashu, Eduard-Raul Kontos, Matthia Sabatelli, and Matias Valdenegro-Toro.
\newblock Cross-lingual document recommendations with transformer-based representations: Evaluating multilingual models and mapping techniques.
\newblock In \emph{Proceedings of the Second Workshop on Scaling Up Multilingual {\&} Multi-Cultural Evaluation}, pages 39--47, Abu Dhabi, January 2025. Association for Computational Linguistics.
\newblock URL \url{https://aclanthology.org/2025.sumeval-2.4/}.

\bibitem[Artetxe and Schwenk(2019)]{artetxe2018laser}
Mikel Artetxe and Holger Schwenk.
\newblock Massively multilingual sentence embeddings for zero‑shot cross‑lingual transfer and beyond.
\newblock In \emph{TACL}, 2019.

\bibitem[Zhao et~al.(2020{\natexlab{b}})Zhao, Callan, and Gao]{zhao2020leveraging}
Ruqing Zhao, Jamie Callan, and Luyu Gao.
\newblock Leveraging sentence-aligned parallel data as supervision signal for neural clir.
\newblock In \emph{SIGIR}, 2020{\natexlab{b}}.

\bibitem[Dou and Neubig(2021)]{dou2021word}
Zi-Yi Dou and Graham Neubig.
\newblock Word alignment by fine-tuning embeddings on parallel corpora.
\newblock In \emph{EACL}, 2021.

\bibitem[Jalili~Sabet et~al.(2020)Jalili~Sabet, Dufter, Yvon, and Sch{\"u}tze]{jalili-sabet-etal-2020-simalign}
Masoud Jalili~Sabet, Philipp Dufter, Fran{\c{c}}ois Yvon, and Hinrich Sch{\"u}tze.
\newblock {S}im{A}lign: High quality word alignments without parallel training data using static and contextualized embeddings.
\newblock In Trevor Cohn, Yulan He, and Yang Liu, editors, \emph{Findings of the Association for Computational Linguistics: EMNLP 2020}, pages 1627--1643, Online, November 2020. Association for Computational Linguistics.
\newblock \doi{10.18653/v1/2020.findings-emnlp.147}.
\newblock URL \url{https://aclanthology.org/2020.findings-emnlp.147/}.

\bibitem[Lo et~al.(2023)Lo, Lee, Chen, Kwee, and Lim]{lo2023clwic}
Pei-Chi Lo, Yang-Yin Lee, Hsien-Hao Chen, Agus~Trisnajaya Kwee, and Ee-Peng Lim.
\newblock Contrastive learning approach to word-in-context task for low-resource languages.
\newblock In \emph{Proceedings of the 37th AAAI Workshop on Knowledge Augmented Methods for NLP}, AAAI 2023 Workshops, pages 1--8, Washington, DC, 2023.
\newblock URL \url{https://ink.library.smu.edu.sg/sis_research/8327}.

\bibitem[Schroff et~al.(2015)Schroff, Kalenichenko, and Philbin]{schroff2015facenet}
Florian Schroff, Dmitry Kalenichenko, and James Philbin.
\newblock Facenet: A unified embedding for face recognition and clustering.
\newblock In \emph{Proceedings of the IEEE Conference on Computer Vision and Pattern Recognition (CVPR)}, pages 815--823, 2015.

\bibitem[van~den Oord et~al.(2018)van~den Oord, Li, and Vinyals]{oord2018representation}
Aaron van~den Oord, Yazhe Li, and Oriol Vinyals.
\newblock Representation learning with contrastive predictive coding.
\newblock \emph{arXiv preprint arXiv:1807.03748}, 2018.

\bibitem[Chen et~al.(2020)Chen, Kornblith, Norouzi, and Hinton]{chen2020simple}
Ting Chen, Simon Kornblith, Mohammad Norouzi, and Geoffrey Hinton.
\newblock A simple framework for contrastive learning of visual representations.
\newblock In \emph{Proceedings of the 37th International Conference on Machine Learning (ICML)}, pages 1597--1607, 2020.

\bibitem[Henderson et~al.(2017)Henderson, Al-Rfou, Strope, Sung, Luk{\'a}cs, Guo, Kumar, Miklos, and Kurzweil]{henderson2017efficient}
Matthew Henderson, Rami Al-Rfou, Brian Strope, Yun-Hsuan Sung, L{\'a}szl{\'o} Luk{\'a}cs, Ruiqi Guo, Sanjiv Kumar, Balint Miklos, and Ray Kurzweil.
\newblock Efficient natural language response suggestion for smart reply.
\newblock \emph{arXiv preprint arXiv:1705.00652}, 2017.

\bibitem[Reimers and Gurevych(2019)]{reimers2019sentencebert}
Nils Reimers and Iryna Gurevych.
\newblock Sentence-bert: Sentence embeddings using siamese bert-networks.
\newblock In \emph{Proceedings of the 2019 Conference on Empirical Methods in Natural Language Processing and the 9th International Joint Conference on Natural Language Processing (EMNLP-IJCNLP)}, pages 3982--3992. Association for Computational Linguistics, 2019.

\bibitem[Jia et~al.(2021)Jia, Yang, Xia, Chen, Parekh, Pham, Le, Sung, Li, and Duerig]{jia2021scaling}
Chao Jia, Yinfei Yang, Ye~Xia, Yi-Ting Chen, Zarana Parekh, Hieu Pham, Quoc~V. Le, Yunhsuan Sung, Zhen Li, and Tom Duerig.
\newblock Scaling up visual and vision-language representation learning with noisy text supervision, 2021.
\newblock URL \url{https://arxiv.org/abs/2102.05918}.

\bibitem[Fu et~al.(2020)Fu, Xian, Geng, Ge, Wang, Dong, and de~Melo]{fu2020absent}
Zuohui Fu, Yikun Xian, Shijie Geng, Yingqiang Ge, Yuting Wang, Xin Dong, and Gerard de~Melo.
\newblock Absent: Cross-lingual sentence representation mapping with bidirectional gans.
\newblock arXiv preprint arXiv:2001.11121, 2020.

\bibitem[Karpukhin et~al.(2020)Karpukhin, Oguz, Min, Lewis, Wu, Edunov, Chen, and Yih]{karpukhin-etal-2020-dense}
Vladimir Karpukhin, Barlas Oguz, Sewon Min, Patrick Lewis, Ledell Wu, Sergey Edunov, Danqi Chen, and Wen-tau Yih.
\newblock Dense passage retrieval for open-domain question answering.
\newblock In Bonnie Webber, Trevor Cohn, Yulan He, and Yang Liu, editors, \emph{Proceedings of the 2020 Conference on Empirical Methods in Natural Language Processing (EMNLP)}, pages 6769--6781, Online, November 2020. Association for Computational Linguistics.
\newblock \doi{10.18653/v1/2020.emnlp-main.550}.
\newblock URL \url{https://aclanthology.org/2020.emnlp-main.550/}.

\bibitem[Xiong et~al.(2021)Xiong, Xiong, Li, Tang, Liu, Bennett, Ahmed, and Overwijk]{xiong2021approximate}
Lee Xiong, Chenyan Xiong, Ye~Li, Kwok‑Fung Tang, Jialin Liu, Paul~N. Bennett, Junaid Ahmed, and Arnold Overwijk.
\newblock Approximate nearest neighbor negative contrastive learning for dense text retrieval.
\newblock In \emph{International Conference on Learning Representations (ICLR)}. OpenReview.net, 2021.
\newblock URL \url{https://openreview.net/forum?id=zeFrfgyZln}.

\bibitem[Hofst{\"a}tter et~al.(2021)Hofst{\"a}tter, Lin, Yang, Lin, and Hanbury]{hofstatter2021efficient}
Sebastian Hofst{\"a}tter, Sheng‑Chieh Lin, Jheng‑Hong Yang, Jimmy Lin, and Allan Hanbury.
\newblock Efficiently teaching an effective dense retriever with balanced topic aware sampling.
\newblock In \emph{Proceedings of the 44th International ACM SIGIR Conference on Research and Development in Information Retrieval (SIGIR)}, 2021.
\newblock URL \url{https://arxiv.org/abs/2104.06967}.

\bibitem[Chronopoulou et~al.(2019)Chronopoulou, Baziotis, and Potamianos]{chronopoulou2019embarrassingly}
Alexandra Chronopoulou, Christos Baziotis, and Alexandros Potamianos.
\newblock An embarrassingly simple approach for transfer learning from pretrained language models.
\newblock In \emph{Proceedings of NAACL-HLT}, 2019.

\bibitem[Lauscher et~al.(2020)Lauscher, Ravishankar, Vuli{\'c}, and Glava{\v{s}}]{lauscher2020limitations}
Anne Lauscher, Vinit Ravishankar, Ivan Vuli{\'c}, and Goran Glava{\v{s}}.
\newblock From zero to hero: On the limitations of zero-shot cross-lingual transfer with multilingual transformers.
\newblock In \emph{Findings of EMNLP}, pages 1290--1302, 2020.

\bibitem[Philippy et~al.(2023)Philippy, Guo, and Haddadan]{philippy-etal-2023-towards}
Fred Philippy, Siwen Guo, and Shohreh Haddadan.
\newblock Towards a common understanding of contributing factors for cross-lingual transfer in multilingual language models: A review.
\newblock In Anna Rogers, Jordan Boyd-Graber, and Naoaki Okazaki, editors, \emph{Proceedings of the 61st Annual Meeting of the Association for Computational Linguistics (Volume 1: Long Papers)}, pages 5877--5891, Toronto, Canada, July 2023. Association for Computational Linguistics.
\newblock \doi{10.18653/v1/2023.acl-long.323}.
\newblock URL \url{https://aclanthology.org/2023.acl-long.323/}.

\bibitem[Raffel et~al.(2020)Raffel, Shazeer, Roberts, Lee, Narang, Matena, Zhou, Li, and Liu]{raffel2020exploring}
Colin Raffel, Noam Shazeer, Adam Roberts, Katherine Lee, Sharan Narang, Michael Matena, Yanqi Zhou, Wei Li, and Peter~J. Liu.
\newblock Exploring the limits of transfer learning with a unified text-to-text transformer.
\newblock \emph{Journal of Machine Learning Research}, 21\penalty0 (140):\penalty0 1--67, 2020.

\bibitem[Thellmann et~al.(2024)Thellmann, Stadler, Fromm, Buschhoff, Jude, Barth, Leveling, Flores-Herr, Köhler, Jäkel, and Ali]{thellmannetal2024}
Klaudia Thellmann, Bernhard Stadler, Michael Fromm, Jasper~Schulze Buschhoff, Alex Jude, Fabio Barth, Johannes Leveling, Nicolas Flores-Herr, Joachim Köhler, René Jäkel, and Mehdi Ali.
\newblock Towards multilingual llm evaluation for european languages, 2024.
\newblock URL \url{https://arxiv.org/abs/2410.08928}.

\bibitem[Sasaki et~al.(2018)Sasaki, Sun, Schamoni, Duh, and Inui]{sasaki_2018}
Shota Sasaki, Shuo Sun, Shigehiko Schamoni, Kevin Duh, and Kentaro Inui.
\newblock Cross-lingual learning-to-rank with shared representations.
\newblock In Marilyn Walker, Heng Ji, and Amanda Stent, editors, \emph{Proceedings of the 2018 Conference of the North {A}merican Chapter of the Association for Computational Linguistics: Human Language Technologies, Volume 2 (Short Papers)}, pages 458--463, New Orleans, Louisiana, June 2018. Association for Computational Linguistics.
\newblock \doi{10.18653/v1/N18-2073}.
\newblock URL \url{https://aclanthology.org/N18-2073/}.

\bibitem[Roy et~al.(2020)Roy, Constant, Al-Rfou, Barua, Phillips, and Yang]{roy_2020_lareqa}
Uma Roy, Noah Constant, Rami Al-Rfou, Aditya Barua, Aaron Phillips, and Yinfei Yang.
\newblock Lareqa: Language-agnostic answer retrieval from a multilingual pool, 2020.
\newblock URL \url{https://arxiv.org/abs/2004.05484}.

\bibitem[Schamoni et~al.(2014)Schamoni, Hieber, Sokolov, and Riezler]{schamoni_2014}
Shigehiko Schamoni, Felix Hieber, Artem Sokolov, and Stefan Riezler.
\newblock Learning translational and knowledge-based similarities from relevance rankings for cross-language retrieval.
\newblock In \emph{Proceedings of the 52nd Annual Meeting of the Association for Computational Linguistics}, 2014.
\newblock URL \url{https://www.cl.uni-heidelberg.de/~riezler/publications/papers/ACL2014short.pdf}.

\bibitem[Lawrie et~al.(2024)Lawrie, MacAvaney, Mayfield, McNamee, Oard, Soldaini, and Yang]{lawrie2024overview}
Dawn Lawrie, Sean MacAvaney, James Mayfield, Paul McNamee, Douglas~W. Oard, Luca Soldaini, and Eugene Yang.
\newblock Overview of the trec 2023 neuclir track, 2024.
\newblock URL \url{https://arxiv.org/abs/2404.08071}.

\bibitem[Valentini et~al.(2025)Valentini, Kozlowski, and Larivi{\`e}re]{valentini2025clirudit}
Francisco Valentini, Diego Kozlowski, and Vincent Larivi{\`e}re.
\newblock Clirudit: Cross‐lingual information retrieval of scientific documents.
\newblock \emph{arXiv preprint arXiv:2504.16264}, 2025.
\newblock URL \url{https://arxiv.org/abs/2504.16264}.
\newblock Submitted April 2025.

\bibitem[Adeyemi et~al.(2024{\natexlab{b}})Adeyemi, Oladipo, Zhang, Alfonso-Hermelo, Rezagholizadeh, Chen, Omotayo, Abdulmumin, Etori, Musa, Fanijo, Awoyomi, Salahudeen, Mohammed, Abolade, Lawan, Sabo~Abubakar, Nasir~Iro, Imam~Abubakar, Mohamed, Mohamed, Ajayi, and Lin]{adeyemi_2024_ciral}
Mofetoluwa Adeyemi, Akintunde Oladipo, Xinyu Zhang, David Alfonso-Hermelo, Mehdi Rezagholizadeh, Boxing Chen, Abdul-Hakeem Omotayo, Idris Abdulmumin, Naome~A. Etori, Toyib~Babatunde Musa, Samuel Fanijo, Oluwabusayo~Olufunke Awoyomi, Saheed~Abdullahi Salahudeen, Labaran~Adamu Mohammed, Daud~Olamide Abolade, Falalu~Ibrahim Lawan, Maryam Sabo~Abubakar, Ruqayya Nasir~Iro, Amina Imam~Abubakar, Shafie~Abdi Mohamed, Hanad~Mohamud Mohamed, Tunde~Oluwaseyi Ajayi, and Jimmy Lin.
\newblock Ciral: A test collection for clir evaluations in african languages.
\newblock SIGIR '24, page 293–302, New York, NY, USA, 2024{\natexlab{b}}. Association for Computing Machinery.
\newblock ISBN 9798400704314.
\newblock \doi{10.1145/3626772.3657884}.
\newblock URL \url{https://doi.org/10.1145/3626772.3657884}.

\bibitem[Li et~al.(2024{\natexlab{c}})Li, Haider, Luo, Agashe, and Callison‑Burch]{li2024bordirlines}
Bryan Li, Samar Haider, Fiona Luo, Adwait Agashe, and Chris Callison‑Burch.
\newblock Bordirlines: A dataset for evaluating cross‑lingual retrieval augmented generation.
\newblock Proceedings of the First Workshop on Advancing NLP for Wikipedia, Nov 2024{\natexlab{c}}.

\bibitem[Liu et~al.(2019{\natexlab{b}})Liu, Lin, Liu, and Sun]{liu_2019_xqa}
Jiahua Liu, Yankai Lin, Zhiyuan Liu, and Maosong Sun.
\newblock {XQA}: A cross-lingual open-domain question answering dataset.
\newblock In Anna Korhonen, David Traum, and Llu{\'i}s M{\`a}rquez, editors, \emph{Proceedings of the 57th Annual Meeting of the Association for Computational Linguistics}, pages 2358--2368, Florence, Italy, July 2019{\natexlab{b}}. Association for Computational Linguistics.
\newblock \doi{10.18653/v1/P19-1227}.
\newblock URL \url{https://aclanthology.org/P19-1227/}.

\bibitem[Clark et~al.(2020)Clark, Choi, Collins, Garrette, Kwiatkowski, Nikolaev, and Palomaki]{clark_2020_tydiqa}
Jonathan~H. Clark, Eunsol Choi, Michael Collins, Dan Garrette, Tom Kwiatkowski, Vitaly Nikolaev, and Jennimaria Palomaki.
\newblock Tydi qa: A benchmark for information-seeking question answering in typologically diverse languages.
\newblock \emph{Transactions of the Association for Computational Linguistics}, 8:\penalty0 454--470, 07 2020.
\newblock ISSN 2307-387X.
\newblock \doi{10.1162/tacl_a_00317}.
\newblock URL \url{https://doi.org/10.1162/tacl\_a\_00317}.

\bibitem[Longpre et~al.(2021)Longpre, Lu, and Daiber]{longpre_2021_mkqa}
Shayne Longpre, Yi~Lu, and Joachim Daiber.
\newblock Mkqa: A linguistically diverse benchmark for multilingual open domain question answering, 2021.
\newblock URL \url{https://arxiv.org/abs/2007.15207}.

\bibitem[Liu et~al.(2025)Liu, Trenous, Ribeiro, Byrne, and Hieber]{liu2025xrag}
Wei Liu, Sony Trenous, Leonardo F.~R. Ribeiro, Bill Byrne, and Felix Hieber.
\newblock Xrag: Cross‑lingual retrieval‑augmented generation.
\newblock \emph{arXiv preprint arXiv:2505.10089}, 2025.
\newblock Published May 2025.

\bibitem[Bajaj et~al.(2018)Bajaj, Campos, Craswell, Deng, Gao, Liu, Majumder, McNamara, Mitra, Nguyen, Rosenberg, Song, Stoica, Tiwary, and Wang]{bajaj_2018_msmarco}
Payal Bajaj, Daniel Campos, Nick Craswell, Li~Deng, Jianfeng Gao, Xiaodong Liu, Rangan Majumder, Andrew McNamara, Bhaskar Mitra, Tri Nguyen, Mir Rosenberg, Xia Song, Alina Stoica, Saurabh Tiwary, and Tong Wang.
\newblock Ms marco: A human generated machine reading comprehension dataset, 2018.
\newblock URL \url{https://arxiv.org/abs/1611.09268}.

\bibitem[Li et~al.(2025{\natexlab{c}})Li, Luo, Haider, Agashe, Li, Liu, Miao, Ramakrishnan, Yuan, and Callison-Burch]{li_2025_rag}
Bryan Li, Fiona Luo, Samar Haider, Adwait Agashe, Tammy Li, Runqi Liu, Muqing Miao, Shriya Ramakrishnan, Yuan Yuan, and Chris Callison-Burch.
\newblock Multilingual retrieval augmented generation for culturally-sensitive tasks: A benchmark for cross-lingual robustness, 2025{\natexlab{c}}.
\newblock URL \url{https://arxiv.org/abs/2410.01171}.

\bibitem[Cleverdon(1962)]{cleverdon1962report}
Cyril~W. Cleverdon.
\newblock Report on the testing and analysis of an investigation into the comparative efficiency of indexing systems.
\newblock Technical report, College of Aeronautics, Cranfield, England, 1962.
\newblock Also known as the “Cranfield Report”; established the foundation for test‐collection based IR evaluation.

\bibitem[Voorhees(1999)]{voorhees1999trec}
Ellen~M. Voorhees.
\newblock {The TREC-8 Question Answering Track Evaluation}.
\newblock In \emph{Proceedings of the Eighth Text REtrieval Conference (TREC-8)}, 1999.

\bibitem[Jurafsky and Martin(2025{\natexlab{b}})]{jurafsky2025speech}
Daniel Jurafsky and James~H. Martin.
\newblock \emph{Speech and Language Processing: An Introduction to Natural Language Processing, Computational Linguistics, and Speech Recognition}.
\newblock Draft publicly released, 3rd draft edition, 2025{\natexlab{b}}.
\newblock Draft edition (January 12\, 2025), available at \url{https://web.stanford.edu/~jurafsky/slp3/}.

\bibitem[Gureja et~al.(2025)Gureja, Miranda, Islam, Maheshwary, Sharma, Winata, Lambert, Ruder, Hooker, and Fadaee]{gureja2025mrewardbenchevaluatingrewardmodels}
Srishti Gureja, Lester James~V. Miranda, Shayekh~Bin Islam, Rishabh Maheshwary, Drishti Sharma, Gusti Winata, Nathan Lambert, Sebastian Ruder, Sara Hooker, and Marzieh Fadaee.
\newblock M-rewardbench: Evaluating reward models in multilingual settings, 2025.
\newblock URL \url{https://arxiv.org/abs/2410.15522}.

\bibitem[Romanou et~al.(2024)Romanou, Foroutan, Sotnikova, Chen, Nelaturu, Singh, Maheshwary, Altomare, Haggag, A, Amayuelas, Amirudin, Aryabumi, Boiko, Chang, Chim, Cohen, Dalmia, Diress, Duwal, Dzenhaliou, Florez, Farestam, Imperial, Islam, Isotalo, Jabbarishiviari, Karlsson, Khalilov, Klamm, Koto, Krzemiński, de~Melo, Montariol, Nan, Niklaus, Novikova, Ceron, Paul, Ploeger, Purbey, Rajwal, Ravi, Rydell, Santhosh, Sharma, Skenduli, Moakhar, Moakhar, Tamir, Tarun, Wasi, Weerasinghe, Yilmaz, Zhang, Schlag, Fadaee, Hooker, and Bosselut]{romanou2024includeevaluatingmultilinguallanguage}
Angelika Romanou, Negar Foroutan, Anna Sotnikova, Zeming Chen, Sree~Harsha Nelaturu, Shivalika Singh, Rishabh Maheshwary, Micol Altomare, Mohamed~A. Haggag, Snegha A, Alfonso Amayuelas, Azril~Hafizi Amirudin, Viraat Aryabumi, Danylo Boiko, Michael Chang, Jenny Chim, Gal Cohen, Aditya~Kumar Dalmia, Abraham Diress, Sharad Duwal, Daniil Dzenhaliou, Daniel Fernando~Erazo Florez, Fabian Farestam, Joseph~Marvin Imperial, Shayekh~Bin Islam, Perttu Isotalo, Maral Jabbarishiviari, Börje~F. Karlsson, Eldar Khalilov, Christopher Klamm, Fajri Koto, Dominik Krzemiński, Gabriel~Adriano de~Melo, Syrielle Montariol, Yiyang Nan, Joel Niklaus, Jekaterina Novikova, Johan Samir~Obando Ceron, Debjit Paul, Esther Ploeger, Jebish Purbey, Swati Rajwal, Selvan~Sunitha Ravi, Sara Rydell, Roshan Santhosh, Drishti Sharma, Marjana~Prifti Skenduli, Arshia~Soltani Moakhar, Bardia~Soltani Moakhar, Ran Tamir, Ayush~Kumar Tarun, Azmine~Toushik Wasi, Thenuka~Ovin Weerasinghe, Serhan Yilmaz, Mike Zhang, Imanol Schlag, Marzieh Fadaee, Sara
  Hooker, and Antoine Bosselut.
\newblock Include: Evaluating multilingual language understanding with regional knowledge, 2024.
\newblock URL \url{https://arxiv.org/abs/2411.19799}.

\bibitem[Singh et~al.(2025)Singh, Romanou, Fourrier, Adelani, Ngui, Vila-Suero, Limkonchotiwat, Marchisio, Leong, Susanto, Ng, Longpre, Ko, Ruder, Smith, Bosselut, Oh, Martins, Choshen, Ippolito, Ferrante, Fadaee, Ermis, and Hooker]{singh_2025}
Shivalika Singh, Angelika Romanou, Clémentine Fourrier, David~I. Adelani, Jian~Gang Ngui, Daniel Vila-Suero, Peerat Limkonchotiwat, Kelly Marchisio, Wei~Qi Leong, Yosephine Susanto, Raymond Ng, Shayne Longpre, Wei-Yin Ko, Sebastian Ruder, Madeline Smith, Antoine Bosselut, Alice Oh, Andre F.~T. Martins, Leshem Choshen, Daphne Ippolito, Enzo Ferrante, Marzieh Fadaee, Beyza Ermis, and Sara Hooker.
\newblock Global mmlu: Understanding and addressing cultural and linguistic biases in multilingual evaluation, 2025.
\newblock URL \url{https://arxiv.org/abs/2412.03304}.

\bibitem[Papineni et~al.(2002)Papineni, Roukos, Ward, and Zhu]{papineni2002bleu}
Kishore Papineni, Salim Roukos, Todd Ward, and Wei-Jing Zhu.
\newblock Bleu: a method for automatic evaluation of machine translation.
\newblock In \emph{Proceedings of ACL}, pages 311--318, 2002.

\bibitem[Lin(2004)]{lin2004rouge}
Chin-Yew Lin.
\newblock Rouge: A package for automatic evaluation of summaries.
\newblock In \emph{Proceedings of Workshop on Text Summarization Branches Out}, pages 74--81, 2004.

\bibitem[Banerjee and Lavie(2005)]{banerjee2005meteor}
Satanjeev Banerjee and Alon Lavie.
\newblock Meteor: An automatic metric for mt evaluation with improved correlation with human judgments.
\newblock In \emph{Proceedings of ACL Workshop on Intrinsic and Extrinsic Evaluation Measures for MT and/or Summarization}, pages 65--72, 2005.

\bibitem[Popović(2015)]{popovic2015chrf}
Maja Popović.
\newblock chrf: Character n-gram f-score for automatic mt evaluation.
\newblock In \emph{Proceedings of WMT}, pages 392--395, 2015.

\bibitem[Snover et~al.(2006)Snover, Dorr, Schwartz, Micciulla, and Makhoul]{snover2006study}
Matthew Snover, Bonnie Dorr, Richard Schwartz, Linnea Micciulla, and John Makhoul.
\newblock A study of translation edit rate with targeted human annotation.
\newblock In \emph{Proceedings of AMTA}, pages 223--231, 2006.

\bibitem[Fain et~al.(2021)Fain, Twomey, and Bollegala]{fain2021backretrieval}
Mikhail Fain, Niall Twomey, and Danushka Bollegala.
\newblock Backretrieval: An image‑pivoted evaluation metric for cross‑lingual text representations without parallel corpora.
\newblock In \emph{Proceedings of SIGIR}, pages 1244--1248, 2021.

\bibitem[Kargaran et~al.(2024)Kargaran, Modarressi, Nikeghbal, Diesner, Yvon, and Schütze]{kargaran2024mexa}
Amir~Hossein Kargaran, Ali Modarressi, Nafiseh Nikeghbal, Jana Diesner, François Yvon, and Hinrich Schütze.
\newblock Mexa: Multilingual evaluation of english‑centric llms via cross‑lingual alignment.
\newblock In \emph{Findings of ACL}, pages 123--137, 2024.

\bibitem[Zhang et~al.(2019)Zhang, Kishore, Wu, Weinberger, and Artzi]{zhang2019bertscore}
Tianyi Zhang, Varsha Kishore, Felix Wu, Kilian Q Weinberger, and Yoav Artzi.
\newblock Bertscore: Evaluating text generation with bert.
\newblock In \emph{Proceedings of ICLR}, 2019.

\bibitem[Rei et~al.(2020)Rei, Stewart, Farinha, and Lavie]{rei2020cometneuralframeworkmt}
Ricardo Rei, Craig Stewart, Ana~C Farinha, and Alon Lavie.
\newblock Comet: A neural framework for mt evaluation, 2020.
\newblock URL \url{https://arxiv.org/abs/2009.09025}.

\bibitem[Sellam et~al.(2020)Sellam, Das, and Parikh]{sellam2020bleurt}
Omer Sellam, Noah Das, and Ankur Parikh.
\newblock Bleurt: Learning robust metrics for text generation.
\newblock In \emph{Proceedings of ACL}, 2020.

\bibitem[Sun et~al.(2020)Sun, Sia, and Duh]{sun2020clireval}
Shuo Sun, Suzanna Sia, and Kevin Duh.
\newblock Clireval: Evaluating machine translation as a cross‑lingual information retrieval task.
\newblock In \emph{Proceedings of the 58th Annual Meeting of the Association for Computational Linguistics: System Demonstrations}, 2020.

\bibitem[Gillick and Liu(2010)]{gillick_2010}
Dan Gillick and Yang Liu.
\newblock Non-expert evaluation of summarization systems is risky.
\newblock In Chris Callison-Burch and Mark Dredze, editors, \emph{Proceedings of the {NAACL} {HLT} 2010 Workshop on Creating Speech and Language Data with {A}mazon{'}s Mechanical Turk}, pages 148--151, Los Angeles, June 2010. Association for Computational Linguistics.
\newblock URL \url{https://aclanthology.org/W10-0722/}.

\bibitem[Iskender et~al.(2020)Iskender, Polzehl, and M{\"o}ller]{iskender_2020}
Neslihan Iskender, Tim Polzehl, and Sebastian M{\"o}ller.
\newblock Best practices for crowd-based evaluation of {G}erman summarization: Comparing crowd, expert and automatic evaluation.
\newblock In Steffen Eger, Yang Gao, Maxime Peyrard, Wei Zhao, and Eduard Hovy, editors, \emph{Proceedings of the First Workshop on Evaluation and Comparison of NLP Systems}, pages 164--175, Online, November 2020. Association for Computational Linguistics.
\newblock \doi{10.18653/v1/2020.eval4nlp-1.16}.
\newblock URL \url{https://aclanthology.org/2020.eval4nlp-1.16/}.

\bibitem[Krishna et~al.(2021)Krishna, Roy, and Iyyer]{krishna_2021_hurdles}
Kalpesh Krishna, Aurko Roy, and Mohit Iyyer.
\newblock Hurdles to progress in long-form question answering, 2021.
\newblock URL \url{https://arxiv.org/abs/2103.06332}.

\bibitem[Nakano et~al.(2022)Nakano, Hilton, Balaji, Wu, Ouyang, Kim, Hesse, Jain, Kosaraju, Saunders, Jiang, Cobbe, Eloundou, Krueger, Button, Knight, Chess, and Schulman]{nakano_2022_webgpt}
Reiichiro Nakano, Jacob Hilton, Suchir Balaji, Jeff Wu, Long Ouyang, Christina Kim, Christopher Hesse, Shantanu Jain, Vineet Kosaraju, William Saunders, Xu~Jiang, Karl Cobbe, Tyna Eloundou, Gretchen Krueger, Kevin Button, Matthew Knight, Benjamin Chess, and John Schulman.
\newblock Webgpt: Browser-assisted question-answering with human feedback, 2022.
\newblock URL \url{https://arxiv.org/abs/2112.09332}.

\bibitem[Xu et~al.(2023)Xu, Song, Iyyer, and Choi]{xu_2023_QAeval}
Fangyuan Xu, Yixiao Song, Mohit Iyyer, and Eunsol Choi.
\newblock A critical evaluation of evaluations for long-form question answering, 2023.
\newblock URL \url{https://arxiv.org/abs/2305.18201}.

\bibitem[Wang et~al.(2022)Wang, Xu, Thompson, Choi, and Iyyer]{wang_2022}
Shufan Wang, Fangyuan Xu, Laure Thompson, Eunsol Choi, and Mohit Iyyer.
\newblock Modeling exemplification in long-form question answering via retrieval.
\newblock In Marine Carpuat, Marie-Catherine de~Marneffe, and Ivan~Vladimir Meza~Ruiz, editors, \emph{Proceedings of the 2022 Conference of the North American Chapter of the Association for Computational Linguistics: Human Language Technologies}, pages 2079--2092, Seattle, United States, July 2022. Association for Computational Linguistics.
\newblock \doi{10.18653/v1/2022.naacl-main.151}.
\newblock URL \url{https://aclanthology.org/2022.naacl-main.151/}.

\bibitem[Fabbri et~al.(2022)Fabbri, Wu, Liu, and Xiong]{fabbri_2022_qafacteval}
Alexander Fabbri, Chien-Sheng Wu, Wenhao Liu, and Caiming Xiong.
\newblock {QAF}act{E}val: Improved {QA}-based factual consistency evaluation for summarization.
\newblock In Marine Carpuat, Marie-Catherine de~Marneffe, and Ivan~Vladimir Meza~Ruiz, editors, \emph{Proceedings of the 2022 Conference of the North American Chapter of the Association for Computational Linguistics: Human Language Technologies}, pages 2587--2601, Seattle, United States, July 2022. Association for Computational Linguistics.
\newblock \doi{10.18653/v1/2022.naacl-main.187}.
\newblock URL \url{https://aclanthology.org/2022.naacl-main.187/}.

\bibitem[Zhu et~al.(2018)Zhu, Lu, Zheng, Guo, Zhang, Wang, and Yu]{zhu_2018_selfbleu}
Yaoming Zhu, Sidi Lu, Lei Zheng, Jiaxian Guo, Weinan Zhang, Jun Wang, and Yong Yu.
\newblock Texygen: A benchmarking platform for text generation models.
\newblock In \emph{The 41st International ACM SIGIR Conference on Research \& Development in Information Retrieval}, SIGIR '18, page 1097–1100, New York, NY, USA, 2018. Association for Computing Machinery.
\newblock ISBN 9781450356572.
\newblock \doi{10.1145/3209978.3210080}.
\newblock URL \url{https://doi.org/10.1145/3209978.3210080}.

\bibitem[Jelinek et~al.(2005)Jelinek, Mercer, Bahl, and Baker]{jelinek_1977_perplexity}
F.~Jelinek, R.~L. Mercer, L.~R. Bahl, and J.~K. Baker.
\newblock Perplexity—a measure of the difficulty of speech recognition tasks.
\newblock \emph{The Journal of the Acoustical Society of America}, 62\penalty0 (S1):\penalty0 S63--S63, 08 2005.
\newblock ISSN 0001-4966.
\newblock \doi{10.1121/1.2016299}.
\newblock URL \url{https://doi.org/10.1121/1.2016299}.

\bibitem[Krishna et~al.(2022)Krishna, Chang, Wieting, and Iyyer]{krishna_2022_rankgen}
Kalpesh Krishna, Yapei Chang, John Wieting, and Mohit Iyyer.
\newblock Rankgen: Improving text generation with large ranking models, 2022.
\newblock URL \url{https://arxiv.org/abs/2205.09726}.

\bibitem[Yuan et~al.(2021)Yuan, Neubig, and Liu]{yuan_2021_bartscore}
Weizhe Yuan, Graham Neubig, and Pengfei Liu.
\newblock Bartscore: evaluating generated text as text generation.
\newblock In \emph{Proceedings of the 35th International Conference on Neural Information Processing Systems}, NIPS '21, Red Hook, NY, USA, 2021. Curran Associates Inc.
\newblock ISBN 9781713845393.

\bibitem[Ponte and Croft(1998)]{ponte_1998_likelihood}
Jay~M. Ponte and W.~Bruce Croft.
\newblock A language modeling approach to information retrieval.
\newblock In \emph{Proceedings of the 21st Annual International ACM SIGIR Conference on Research and Development in Information Retrieval}, SIGIR '98, page 275–281, New York, NY, USA, 1998. Association for Computing Machinery.
\newblock ISBN 1581130155.
\newblock \doi{10.1145/290941.291008}.
\newblock URL \url{https://doi.org/10.1145/290941.291008}.

\bibitem[Statista(2025)]{statista_2025}
Statista.
\newblock Languages most frequently used for web content as of february 2025, by share of websites, 2025.
\newblock URL \url{https://www.statista.com/statistics/262946/most-common-languages-on-the-internet/}.

\bibitem[W3Techs(2025)]{w3techs_content_language}
W3Techs.
\newblock Usage statistics of content languages for websites.
\newblock \url{https://w3techs.com/technologies/overview/content_language}, 2025.
\newblock Accessed June 12, 2025.

\bibitem[Foo(2011)]{foo2011cross}
Schubert Foo.
\newblock Retrieval effectiveness of cross language information retrieval search engines.
\newblock In Chunxiao Xing, Fabio Crestani, and Andreas Rauber, editors, \emph{Digital Libraries: For Cultural Heritage, Knowledge Dissemination, and Future Creation}, pages 296--306, Berlin, Heidelberg, 2011. Springer Berlin Heidelberg.
\newblock ISBN 978-3-642-24826-9.

\bibitem[Semmar et~al.(2006)Semmar, Laib, and Fluhr]{semmar2006lic2m}
Nasredine Semmar, Meriama Laib, and Christian Fluhr.
\newblock A deep linguistic analysis for cross-language information retrieval.
\newblock In \emph{LREC}, pages 2507--2510, 2006.

\bibitem[Chen and Bao(2009)]{chen2009cross}
Jiangping Chen and Yu~Bao.
\newblock Cross-language search: The case of google language tools.
\newblock \emph{First Monday}, 14\penalty0 (3), Feb. 2009.
\newblock \doi{10.5210/fm.v14i3.2335}.
\newblock URL \url{https://firstmonday.org/ojs/index.php/fm/article/view/2335}.

\bibitem[Rahmani and Fumani(2017)]{rahmani2017adapted}
Amin Rahmani and Mohammad Reza Falahati~Qadimi Fumani.
\newblock Adapting google translate for english–persian cross‑lingual information retrieval in medical domain.
\newblock In \emph{International Conference on Information Technology}, 2017.

\bibitem[Sharma et~al.(2023)Sharma, Mittal, and Vidyarthi]{sharma2021semantic}
Vijay~Kumar Sharma, Namita Mittal, and Ankit Vidyarthi.
\newblock Semantic morphological variant selection and translation disambiguation for cross‑lingual information retrieval.
\newblock \emph{Multimedia Tools and Applications}, 2023.

\bibitem[Iana et~al.(2024)Iana, Glava{\v s}, and Paulheim]{iana2024mind}
Andreea Iana, Goran Glava{\v s}, and Heiko Paulheim.
\newblock Mind your language: A multilingual dataset for cross-lingual news recommendation.
\newblock \emph{arXiv preprint arXiv:2403.17876}, 2024.

\bibitem[Maurer and Voigt(2025)]{maurer2025clsd}
Fritz Maurer and Sebastian Voigt.
\newblock Examining multilingual embedding models cross-lingually through adversarial german–french news search.
\newblock \emph{arXiv preprint arXiv:2502.08638}, 2025.

\bibitem[Singh et~al.(2023)Singh, Scarton, Song, and Bontcheva]{singh2023mmtweets}
Iknoor Singh, Carolina Scarton, Xingyi Song, and Kalina Bontcheva.
\newblock Breaking language barriers with mmtweets: Advancing cross-lingual debunked narrative retrieval for fact-checking.
\newblock \emph{arXiv preprint arXiv:2308.05680}, 2023.
\newblock Introduces a cross-lingual tweet–debunk retrieval dataset.

\bibitem[Pikuliak et~al.(2023)Pikuliak, Srba, Moro, Hromadka, Smolen, Melisek, Vykopal, Simko, Podrouzek, and Bielikova]{pikuliak2023multiclaim}
Matúš Pikuliak, Ivan Srba, Robert Moro, Timo Hromadka, Timotej Smolen, Martin Melisek, Ivan Vykopal, Jakub Simko, Juraj Podrouzek, and Maria Bielikova.
\newblock Multilingual previously fact-checked claim retrieval.
\newblock In \emph{EMNLP 2023}, pages 16477--16500, 2023.
\newblock \doi{10.18653/v1/2023.emnlp-main.1027}.
\newblock URL \url{https://aclanthology.org/2023.emnlp-main.1027}.

\bibitem[Quelle et~al.(2025)Quelle, Cheng, Bovet, and Hale]{quelle2025lost}
Dorian Quelle, Calvin~Yixiang Cheng, Alexandre Bovet, and Scott~A. Hale.
\newblock Lost in translation: Using global fact‑checks to measure multilingual misinformation prevalence, spread, and evolution.
\newblock \emph{EPJ Data Science}, 14\penalty0 (1):\penalty0 22, 2025.
\newblock \doi{10.1140/epjds/s13688-025-00520-6}.
\newblock Analyzes 264K fact-checks in 95 languages with multilingual embeddings.

\bibitem[Singhal et~al.(2024)Singhal, Shao, Sun, Ding, Lu, and Zhu]{singhal2024translationbias}
Aryan Singhal, Veronica Shao, Gary Sun, Ryan Ding, Jonathan Lu, and Kevin Zhu.
\newblock A comparative study of translation bias and accuracy in multilingual large language models for cross-language claim verification.
\newblock \emph{arXiv preprint arXiv:2410.10303}, 2024.
\newblock Systematically evaluates translation bias across 15 languages using XFACT.

\bibitem[Vitiugin and Castillo(2022{\natexlab{a}})]{vitiugin2022cross}
Fedor Vitiugin and Carlos Castillo.
\newblock Cross‑lingual query‑based summarization of crisis‑related social media: An abstractive approach using transformers.
\newblock In \emph{Proceedings of the 33rd ACM Conference on Hypertext and Social Media (HT ’22)}, pages 21--31, 2022{\natexlab{a}}.
\newblock \doi{10.1145/3511095.3531279}.
\newblock Evaluated on five disasters in 10 languages:contentReference[oaicite:1]{index=1}.

\bibitem[S{\'a}nchez et~al.(2022)S{\'a}nchez, Sarmiento, Abeliuk, P{\'e}rez, and Poblete]{sanchez2022cross}
Cinthia S{\'a}nchez, Hernan Sarmiento, Andr{\'e}s Abeliuk, Jorge P{\'e}rez, and Barbara Poblete.
\newblock Cross‑lingual and cross‑domain crisis classification for low‑resource scenarios.
\newblock \emph{arXiv preprint arXiv:2209.02139}, 2022.
\newblock Achieved ~80% F1 transferring from English to Spanish/Italian:contentReference[oaicite:2]{index=2}.

\bibitem[Lamsal et~al.(2024)Lamsal, Rodriguez~Read, and Karunasekera]{lamsal2024se}
Rabindra Lamsal, Maria Rodriguez~Read, and Shanika Karunasekera.
\newblock Semantically enriched cross‑lingual sentence embeddings for crisis-related social media texts.
\newblock \emph{arXiv preprint arXiv:2403.16614}, 2024.
\newblock Multilingual encoder across 50 languages improving crisis embedding tasks:contentReference[oaicite:3]{index=3}.

\bibitem[Vitiugin and Castillo(2022{\natexlab{b}})]{vitiugin2022crosslang}
Fedor Vitiugin and Carlos Castillo.
\newblock Cross‑language classification of crisis‑related tweets.
\newblock \emph{ICWSM Workshop}, 2022{\natexlab{b}}.
\newblock XLM‑R fine‑tuned English→Arabic crisis tweet classification :contentReference[oaicite:5]{index=5}.

\bibitem[Yin et~al.(2025)Yin, Dong, Liu, Huang, Xiao, Liu, Mostafavi, and Caverlee]{yin2025disastir}
Kai Yin, Xiangjue Dong, Chengkai Liu, Lipai Huang, Yiming Xiao, Zhewei Liu, Ali Mostafavi, and James Caverlee.
\newblock Disastir: A comprehensive information retrieval benchmark for disaster management.
\newblock \emph{arXiv preprint arXiv:2505.15856}, 2025.
\newblock Introduced 301 crisis IR tasks across 16 languages :contentReference[oaicite:4]{index=4}.

\bibitem[Ranade et~al.(2018)Ranade, Mittal, Joshi, and Pande]{ranade2018using}
Priyanka Ranade, Sudip Mittal, Anupam Joshi, and Karuna Pande.
\newblock Using deep neural networks to translate multilingual threat intelligence.
\newblock In \emph{Proceedings of the IEEE Intelligence and Security Informatics Conference (ISI)}, November 2018.
\newblock Neural pipeline translating Russian-language threat intelligence into English.

\bibitem[Markov and Last(2005)]{markov2005identification}
Alex Markov and Mark Last.
\newblock Identification of terrorist web sites with cross‑lingual classification tools.
\newblock In \emph{Fighting Terror in Cyberspace}, volume~65 of \emph{Studies in Computational Intelligence}, page 117–? Springer, 2005.
\newblock Early cross‑lingual classification for security monitoring.

\bibitem[Zhu et~al.(2022)Zhu, Lv, Yang, Zhang, Xu, Xu, Feng, Zhang, Da, Zeng, and Chen]{zhu2022clpr9m}
Wenya Zhu, Xiaoyu Lv, Baosong Yang, Yinghua Zhang, Yong Xu, Linlong Xu, Yinfu Feng, Haibo Zhang, Qing Da, Anxiang Zeng, and Ronghua Chen.
\newblock Cross‐lingual product retrieval in e‐commerce search (clpr‐9m).
\newblock In \emph{Advances in Information Retrieval}, volume 13238 of \emph{Lecture Notes in Computer Science}, pages 441--457. Springer, 2022.

\bibitem[Shen et~al.(2023)Shen, Asai, Byrne, and de~Gispert]{shen2023xpqa}
Xiaoyu Shen, Akari Asai, Bill Byrne, and Adrià de~Gispert.
\newblock xpqa: Cross‑lingual product question answering across 12 languages.
\newblock \emph{arXiv preprint arXiv:2305.09249}, 2023.

\bibitem[Metzger(2007)]{metzger2007credibility}
Miriam~J. Metzger.
\newblock Making sense of credibility on the web: Models for evaluating online information and recommendations for future research.
\newblock \emph{Journal of the American Society for Information Science and Technology}, 58\penalty0 (13):\penalty0 2078--2091, 2007.
\newblock \doi{10.1002/asi.20672}.

\bibitem[Patra et~al.(2019)Patra, Moniz, Garg, Gormley, and Neubig]{patra-etal-2019-bilingual}
Barun Patra, Joel Ruben~Antony Moniz, Sarthak Garg, Matthew~R. Gormley, and Graham Neubig.
\newblock Bilingual lexicon induction with semi-supervision in non-isometric embedding spaces.
\newblock In Anna Korhonen, David Traum, and Llu{\'i}s M{\`a}rquez, editors, \emph{Proceedings of the 57th Annual Meeting of the Association for Computational Linguistics}, pages 184--193, Florence, Italy, July 2019. Association for Computational Linguistics.
\newblock \doi{10.18653/v1/P19-1018}.
\newblock URL \url{https://aclanthology.org/P19-1018/}.

\bibitem[Rust et~al.(2021)Rust, Pfeiffer, Vuli{\'c}, Ruder, and Gurevych]{rust_2021}
Phillip Rust, Jonas Pfeiffer, Ivan Vuli{\'c}, Sebastian Ruder, and Iryna Gurevych.
\newblock How good is your tokenizer? on the monolingual performance of multilingual language models.
\newblock In Chengqing Zong, Fei Xia, Wenjie Li, and Roberto Navigli, editors, \emph{Proceedings of the 59th Annual Meeting of the Association for Computational Linguistics and the 11th International Joint Conference on Natural Language Processing (Volume 1: Long Papers)}, pages 3118--3135, Online, August 2021. Association for Computational Linguistics.
\newblock \doi{10.18653/v1/2021.acl-long.243}.
\newblock URL \url{https://aclanthology.org/2021.acl-long.243/}.

\bibitem[Muennighoff et~al.(2023)Muennighoff, Wang, Sutawika, Roberts, Biderman, Le~Scao, Bari, Shen, Yong, Schoelkopf, Tang, Radev, Aji, Almubarak, Albanie, Alyafeai, Webson, Raff, and Raffel]{muennighoff_2023}
Niklas Muennighoff, Thomas Wang, Lintang Sutawika, Adam Roberts, Stella Biderman, Teven Le~Scao, M~Saiful Bari, Sheng Shen, Zheng~Xin Yong, Hailey Schoelkopf, Xiangru Tang, Dragomir Radev, Alham~Fikri Aji, Khalid Almubarak, Samuel Albanie, Zaid Alyafeai, Albert Webson, Edward Raff, and Colin Raffel.
\newblock Crosslingual generalization through multitask finetuning.
\newblock In Anna Rogers, Jordan Boyd-Graber, and Naoaki Okazaki, editors, \emph{Proceedings of the 61st Annual Meeting of the Association for Computational Linguistics (Volume 1: Long Papers)}, pages 15991--16111, Toronto, Canada, July 2023. Association for Computational Linguistics.
\newblock \doi{10.18653/v1/2023.acl-long.891}.
\newblock URL \url{https://aclanthology.org/2023.acl-long.891/}.

\bibitem[Levy et~al.(2023)Levy, John, Liu, Vyas, Ma, Fujinuma, Ballesteros, Castelli, and Roth]{levy_2023_bias}
Sharon Levy, Neha John, Ling Liu, Yogarshi Vyas, Jie Ma, Yoshinari Fujinuma, Miguel Ballesteros, Vittorio Castelli, and Dan Roth.
\newblock Comparing biases and the impact of multilingual training across multiple languages.
\newblock In Houda Bouamor, Juan Pino, and Kalika Bali, editors, \emph{Proceedings of the 2023 Conference on Empirical Methods in Natural Language Processing}, pages 10260--10280, Singapore, December 2023. Association for Computational Linguistics.
\newblock \doi{10.18653/v1/2023.emnlp-main.634}.
\newblock URL \url{https://aclanthology.org/2023.emnlp-main.634/}.

\bibitem[Cabello~Piqueras and S{\o}gaard(2022)]{cabello_2022}
Laura Cabello~Piqueras and Anders S{\o}gaard.
\newblock Are pretrained multilingual models equally fair across languages?
\newblock In Nicoletta Calzolari, Chu-Ren Huang, Hansaem Kim, James Pustejovsky, Leo Wanner, Key-Sun Choi, Pum-Mo Ryu, Hsin-Hsi Chen, Lucia Donatelli, Heng Ji, Sadao Kurohashi, Patrizia Paggio, Nianwen Xue, Seokhwan Kim, Younggyun Hahm, Zhong He, Tony~Kyungil Lee, Enrico Santus, Francis Bond, and Seung-Hoon Na, editors, \emph{Proceedings of the 29th International Conference on Computational Linguistics}, pages 3597--3605, Gyeongju, Republic of Korea, October 2022. International Committee on Computational Linguistics.
\newblock URL \url{https://aclanthology.org/2022.coling-1.318/}.

\bibitem[Naous et~al.(2024)Naous, Ryan, Ritter, and Xu]{naous_2024_bias}
Tarek Naous, Michael~J. Ryan, Alan Ritter, and Wei Xu.
\newblock Having beer after prayer? measuring cultural bias in large language models, 2024.
\newblock URL \url{https://arxiv.org/abs/2305.14456}.

\bibitem[Touileb et~al.(2022)Touileb, {\O}vrelid, and Velldal]{touileb_2022}
Samia Touileb, Lilja {\O}vrelid, and Erik Velldal.
\newblock Occupational biases in {N}orwegian and multilingual language models.
\newblock In Christian Hardmeier, Christine Basta, Marta~R. Costa-juss{\`a}, Gabriel Stanovsky, and Hila Gonen, editors, \emph{Proceedings of the 4th Workshop on Gender Bias in Natural Language Processing (GeBNLP)}, pages 200--211, Seattle, Washington, July 2022. Association for Computational Linguistics.
\newblock \doi{10.18653/v1/2022.gebnlp-1.21}.
\newblock URL \url{https://aclanthology.org/2022.gebnlp-1.21/}.

\bibitem[Ferrara(2023)]{Ferrara_2023}
Emilio Ferrara.
\newblock Should chatgpt be biased? challenges and risks of bias in large language models.
\newblock \emph{First Monday}, November 2023.
\newblock ISSN 1396-0466.
\newblock \doi{10.5210/fm.v28i11.13346}.
\newblock URL \url{http://dx.doi.org/10.5210/fm.v28i11.13346}.

\bibitem[Cao et~al.(2022)Cao, Pruksachatkun, Chang, Gupta, Kumar, Dhamala, and Galstyan]{cao_2022}
Yang~Trista Cao, Yada Pruksachatkun, Kai-Wei Chang, Rahul Gupta, Varun Kumar, Jwala Dhamala, and Aram Galstyan.
\newblock On the intrinsic and extrinsic fairness evaluation metrics for contextualized language representations.
\newblock In Smaranda Muresan, Preslav Nakov, and Aline Villavicencio, editors, \emph{Proceedings of the 60th Annual Meeting of the Association for Computational Linguistics (Volume 2: Short Papers)}, pages 561--570, Dublin, Ireland, May 2022. Association for Computational Linguistics.
\newblock \doi{10.18653/v1/2022.acl-short.62}.
\newblock URL \url{https://aclanthology.org/2022.acl-short.62/}.

\bibitem[Sun et~al.(2022)Sun, He, Qiu, and Huang]{sun_2022}
Tianxiang Sun, Junliang He, Xipeng Qiu, and Xuanjing Huang.
\newblock {BERTS}core is unfair: On social bias in language model-based metrics for text generation.
\newblock In Yoav Goldberg, Zornitsa Kozareva, and Yue Zhang, editors, \emph{Proceedings of the 2022 Conference on Empirical Methods in Natural Language Processing}, pages 3726--3739, Abu Dhabi, United Arab Emirates, December 2022. Association for Computational Linguistics.
\newblock \doi{10.18653/v1/2022.emnlp-main.245}.
\newblock URL \url{https://aclanthology.org/2022.emnlp-main.245/}.

\bibitem[Samuel et~al.(2025)Samuel, DeGenaro, Guallar-Blasco, Sanders, Eisape, Spendlove, Reddy, Martin, Yates, Yang, Carpenter, Etter, Kayi, Wiesner, Murray, and Kriz]{samuel_2025}
Saron Samuel, Dan DeGenaro, Jimena Guallar-Blasco, Kate Sanders, Oluwaseun Eisape, Tanner Spendlove, Arun Reddy, Alexander Martin, Andrew Yates, Eugene Yang, Cameron Carpenter, David Etter, Efsun Kayi, Matthew Wiesner, Kenton Murray, and Reno Kriz.
\newblock Mmmorrf: Multimodal multilingual modularized reciprocal rank fusion, 2025.
\newblock URL \url{https://arxiv.org/abs/2503.20698}.

\bibitem[Arora et~al.(2020)Arora, Shterionov, Moriya, Kaushik, Dzendzik, and Jones]{arora_2020}
Piyush Arora, Dimitar Shterionov, Yasufumi Moriya, Abhishek Kaushik, Daria Dzendzik, and Gareth Jones.
\newblock An investigative study of multi-modal cross-lingual retrieval.
\newblock In Kathy McKeown, Douglas~W. Oard, Elizabeth, and Richard Schwartz, editors, \emph{Proceedings of the workshop on Cross-Language Search and Summarization of Text and Speech (CLSSTS2020)}, pages 58--67, Marseille, France, May 2020. European Language Resources Association.
\newblock ISBN 979-10-95546-55-9.
\newblock URL \url{https://aclanthology.org/2020.clssts-1.10/}.

\bibitem[Peng et~al.(2024{\natexlab{b}})Peng, Zhu, Liu, Bo, Shi, Hong, Zhang, and Tang]{peng_2024_grag}
Boci Peng, Yun Zhu, Yongchao Liu, Xiaohe Bo, Haizhou Shi, Chuntao Hong, Yan Zhang, and Siliang Tang.
\newblock Graph retrieval-augmented generation: A survey, 2024{\natexlab{b}}.
\newblock URL \url{https://arxiv.org/abs/2408.08921}.

\bibitem[Hu et~al.(2024)Hu, Lei, Zhang, Pan, Ling, and Zhao]{hu_2024_grag}
Yuntong Hu, Zhihan Lei, Zheng Zhang, Bo~Pan, Chen Ling, and Liang Zhao.
\newblock Grag: Graph retrieval-augmented generation, 2024.
\newblock URL \url{https://arxiv.org/abs/2405.16506}.

\bibitem[Procko and Ochoa(2024)]{Procko_2024_grag}
Tyler~Thomas Procko and Omar Ochoa.
\newblock Graph retrieval-augmented generation for large language models: A survey.
\newblock In \emph{2024 Conference on AI, Science, Engineering, and Technology (AIxSET)}, pages 166--169, 2024.
\newblock \doi{10.1109/AIxSET62544.2024.00030}.

\bibitem[Peng et~al.(2025)Peng, Hu, Chai, and Søgaard]{pengetal2025}
Qiwei Peng, Guimin Hu, Yekun Chai, and Anders Søgaard.
\newblock Debiasing multilingual llms in cross-lingual latent space, 2025.
\newblock URL \url{https://arxiv.org/abs/2508.17948}.

\bibitem[Abagyan et~al.(2025)Abagyan, Salamanca, Cruz-Salinas, Cao, Lin, Locatelli, Fadaee, Üstün, and Hooker]{abagyan2025tokenizerruleallemergent}
Diana Abagyan, Alejandro~R. Salamanca, Andres~Felipe Cruz-Salinas, Kris Cao, Hangyu Lin, Acyr Locatelli, Marzieh Fadaee, Ahmet Üstün, and Sara Hooker.
\newblock One tokenizer to rule them all: Emergent language plasticity via multilingual tokenizers, 2025.
\newblock URL \url{https://arxiv.org/abs/2506.10766}.

\bibitem[Rémy et~al.(2024)Rémy, Delobelle, Avetisyan, Khabibullina, de~Lhoneux, and Demeester]{remy2024transtokenization}
François Rémy, Pieter Delobelle, Hayastan Avetisyan, Alfiya Khabibullina, Miryam de~Lhoneux, and Thomas Demeester.
\newblock Trans-tokenization and cross-lingual vocabulary transfers: Language adaptation of llms for low-resource nlp.
\newblock \emph{arXiv preprint arXiv:2408.04303}, August 2024.
\newblock Accepted at COLM 2024.

\end{thebibliography}

%\begin{thebibliography}{1}
%\end{thebibliography}
\newpage

\end{document}